\documentclass[useAMS,usenatbib]{mn2e}
\usepackage{graphicx}
\usepackage{rotating}
\usepackage{lscape}
\usepackage{makeidx}
\usepackage{listings}
\usepackage{array}
\usepackage{aas_macros}
\usepackage{times}
\usepackage{longtable}
\usepackage{caption}
\usepackage{amssymb}

\title[AGN feedback dependance on galaxy mass and jet power]{Observational evidence that positive and negative AGN feedback depends on galaxy mass and jet power}
\author[Kalfountzou et al.]
{E. Kalfountzou$^{1,2}$\thanks{Email: ekalfountzou@sciops.esa.int},
J. A. Stevens$^{2}$, M. J. Jarvis$^{3,4}$, M. J. Hardcastle$^{2}$, D. Wilner$^{5}$, M. Elvis$^{5}$, 
\newauthor M. J. Page$^{6}$, M. Trichas$^{7}$,  D. J. B. Smith$^{2}$
\\
\footnotesize
$^{1}$XMM-Newton Science Operations Center, Science Operations Department of ESA, ESAC, 28691 Villanueva de la Ca$\tilde{n}$ada, Madrid, Spain \\
$^{2}$Centre for Astrophysics, Research School pf Physics, Astronomy and Mathematics, University of Hertfordshire, Hatfield, Herts, AL10 9AB, UK\\
$^{3}$Astrophysics, Department of Physics, Keble Road, Oxford, OX1 3RH, UK\\
$^{4}$Physics Department, University of the Western Cape, Private Bag X17, Bellville 7535, South Africa\\
$^{5}$Harvard-Smithsonian Center for Astrophysics, 60 Garden Street, Cambridge, MA 02138, USA \\
$^{6}$Mullard Space Science Laboratory, University College London, Holmbury St Mary, Dorking, Surrey RH5 6NT, UK \\
$^{7}$Airbus Defence \& Space, Gunnels Wood Road, Stevenage, Hertfordshire SG1 2AS, UK 
}

\begin{document}
\date{Received Month dd, yyyy; accepted Month dd, yyyy}
\pagerange{\pageref{firstpage}--\pageref{lastpage}} \pubyear{2012}

\maketitle
\label{firstpage}

\begin{abstract}

Several studies support the existence of a link between the AGN and star formation activity. Radio jets have been argued to be an ideal mechanism for direct interaction between the AGN and the host galaxy. A drawback of previous surveys of AGN is that they are fundamentally limited by the degeneracy between redshift and luminosity in flux-density limited samples. To overcome this limitation, we present far-infrared {\it Herschel} observations of 74 radio-loud quasars (RLQs), 72 radio-quiet quasars (RQQs) and 27 radio galaxies (RGs), selected at $0.9<z<1.1$ which span over two decades in optical luminosity. By decoupling luminosity from evolutionary effects, we investigate how the star formation rate (SFR) depends on AGN luminosity, radio-loudness and orientation. We find that: 1) the SFR shows a weak correlation with the bolometric luminosity for all AGN sub-samples, 2) the RLQs show a SFR excess of about a factor of 1.4 compared to the RQQs, matched in terms of black hole mass and bolometric luminosity, suggesting that either positive radio-jet feedback or radio AGN triggering are linked to star-formation triggering and 3) RGs have lower SFRs by a factor of 2.5 than the RLQ sub-sample with the same BH mass and bolometric luminosity. We suggest that there is some jet power threshold at which radio-jet feedback switches from enhancing star formation (by compressing gas) to suppressing it (by ejecting gas). This threshold depends on both galaxy mass and jet power. 

\end{abstract}


\section{INTRODUCTION}

In recent years the study of AGN has undergone a renaissance. This is due to the fact that AGN activity is now widely believed to be an important phase in the evolution of every massive galaxy in the Universe. There are a number of pieces of evidence that support a global evolutionary connection between the star formation and AGN activity, for example, 1) the differential redshift evolution of the AGN luminosity function, or ``AGN downsizing'' is also found for the star-forming galaxy population \citep[e.g.][]{Hasinger2005,Hopkins2006,Aird2010,Kalfountzou2014b}, 2) the redshift distribution of strongly star-forming galaxies follows that of powerful AGN \citep[e.g.][]{Willott2001,Chapman2005,Wardlow2011,Miyaji2015},~3) the star formation rate density as a function of redshift is broadly similar to the BH accretion rate density \citep[e.g.][]{Boyle1998,Merloni2004,Aird2010,Madau2014} and 4) a tight correlation is found between the BH and stellar mass of the host galaxy bulge \citep[e.g.][]{Magorrian1998, McConnell2013,Graham2013,Kormendy2013}. There are several examples of composite objects showing both AGN and star formation activity, in the literature \citep[e.g.][]{Page2001,Page2004, Alexander2005}, particularly at $z\approx 1$, close to the peak of the AGN luminosity density in the Universe \citep[e.g.][]{Barger2005, Hasinger2005}. However, the picture is still not clear, with investigations at different wavelengths producing many differences of opinion as to the amount of radiation that is absorbed and reprocessed by dust, how this is related to the host galaxy and whether the triggering mechanism behind the AGN activity is also responsible for massive star-formation activity \citep[e.g.][]{Harrison2012, Mullaney2012,Mullaney2012b,Rodighiero2015}. Moreover, it is also unclear how these processes depend on luminosity and radio-loudness and how they are observationally affected by orientation \citep[e.g.][]{Rosario2012,Page2012,Kalfountzou2012,Kalfountzou2014a,Chen2015}.

From a more theoretical perspective, semi-analytic and hydrodynamic models of galaxy formation suggest that the correlation between AGN and star-formation activity arises through AGN feedback processes between the galaxy and its accreting BH \citep[e.g.][]{DiMatteo2005,Hopkins2006,DiMatteo2008,HopkinsElvis2010}. However, it is still unclear what kind of AGN-driven feedback is the most important. The feedback process from a growing supermassive black hole (SMBH) can be split broadly into two types. Using the terminology of \cite{Croton2006}, these are ``quasar-mode" feedback, which comprises wide-angle, sub-relativistic outflows driven by radiation due to the efficient accretion of cold gas, and ``radio-mode" feedback, which are relativistic outflows that punch their way out of the host galaxy and into the surrounding inter-galactic medium (IGM), often but not exclusively due to radiatively inefficient accretion from a hot gas reservoir.

Quasar-mode feedback is considered to be driven by a wind created by the luminous accretion disk. In this case, the ignition of the nucleus in a star-forming galaxy heats up and removes the inter-stellar medium (ISM) gas from its host galaxy, thus reducing or even stopping star formation \citep[e.g.][]{Granato2001,Croton2006,Hopkins2010}. During this process, the flow of matter to the central SMBH can be reduced, lowering the accretion flow and eventually extinguishing the AGN. Once the gas cools down and starts to collapse into the nucleus again, a new AGN phase may begin and the cycle resumes. 

Radio-mode feedback is instead driven by relativistic jets. Direct observations show that jets can influence gas many tens of kpc from the centre of the parent host galaxy \citep[e.g.][]{Nesvadba2010,Nesvadba2011,Emonts2011}. Indeed, the brightest radio structures in radio-loud AGN are often observed on kpc scales and are produced by the coupling of the AGN outflow to its environment \citep[e.g.][]{Dicken2012}. Radio AGN energy output, in the form of heating, can prevent hot gas from cooling and falling into a galaxy to form stars \citep[e.g.][]{Croton2006}, especially in the more massive galaxies and at much smaller accretion rates than that of the quasar-mode feedback. The cooling of the hot gas onto the central region of the galaxy fuels intermittent AGN outbursts, which in turn heat the inflowing gas, perhaps stopping or slowing down the accretion inflow \citep[e.g.][]{Best2005}.

The role of radio jets in the evolution of galaxies, in particular with respect to star formation, has been widely discussed, with the observational consensus being mixed. Certainly, AGN jets have largely been assumed to effectively suppress or even quench star formation \citep[e.g.][]{Best2005,Croton2006,Karouzos2013,Hardcastle2013,Gurkan2015} because the jets warm up and ionize the gas they collide with, making collapse under self-gravity more difficult, or directly expel the molecular gas from the galaxy, effectively removing the ingredient for stars to form \citep[e.g.][]{Nesvadba2006,Nesvadba2011}. Interestingly, theoretical models \citep[e.g.][]{Silk2012}, recent simulations \citep[e.g.][]{Gaibler2012,Wagner2012} and observations \citep[e.g.][]{Kalfountzou2012,Kalfountzou2014a} reveal that jet activity can actually trigger star formation by generating some high density, low temperature cavities embedded in the cocoon around the jet \citep[e.g.][]{Antonuccio-Delogu2010,Silk2010,Silk2012}. The alignment effect seen in radio galaxies may also be a manifestation of this process \citep[e.g.][]{Eales1997,Inskip2005,Best2012}.

It is apparent that some form of feedback is needed to explain the observational results for black hole-galaxy co-evolution, but much still remains unclear. Many studies have attempted to determine the star-formation activity in quasar host galaxies using optical colours \citep[e.g.][]{Sanchez2004} or spectroscopy \citep[e.g.][]{Trichas2010,Kalfountzou2011,Trichas2012}. However, spectral diagnostics are not immune to AGN contamination, and optical diagnostics, in particular, are susceptible to the effects of reddening. The {\it Herschel Space Observatory} \citep{Pilbratt2010}, with its high FIR sensitivity and broad wavelength coverage, offers a powerful way of measuring the approximate SFR with minimal AGN contamination \citep[e.g.][]{Netzer2007,Hatziminaoglou2010,Mullaney2011,Bonfield2011,Hardcastle2013,Virdee2013}. However, a drawback of previous works is that they are fundamentally limited by the strong correlation between redshift and luminosity, i.e., only the most powerful sources are observed at high redshifts and, due to the much smaller volume probed, only the less luminous, more abundant populations are found at lower redshifts. While fundamental questions about the relation between radio-loud and radio-quiet AGN, and how they affect the host galaxy, are in principle soluble with multiwavelength surveys, with already available interesting results, most of them will remain intractable until we have a comprehensive AGN sample in which the influence of cosmological evolution and Malmquist bias have been decoupled from the effects of luminosity, radio-loudness and orientation. The sheer size of the Sloan Digital Sky Survey (SDSS) quasar sample \citep{Schneider2005} makes it possible to generate a homogeneous sample of quasars covering a large range in luminosity at a single epoch. The redshift range $0.9 < z < 1.1$ is ideal for this study because it allows us to probe over two decades in optical luminosity.

\begin{figure*}
\includegraphics[scale=0.45]{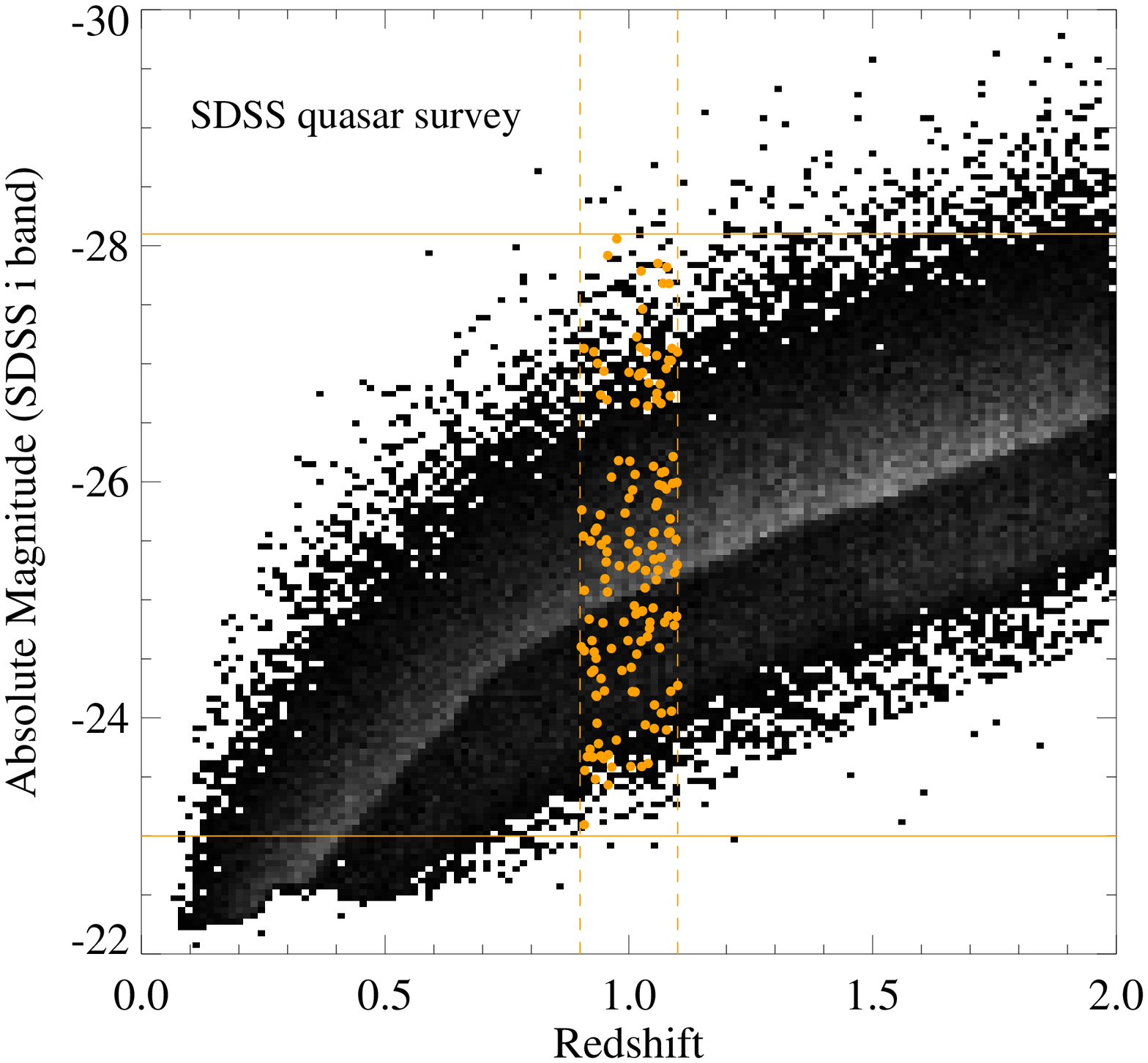}
\hspace{-1cm}
\includegraphics[scale=0.435]{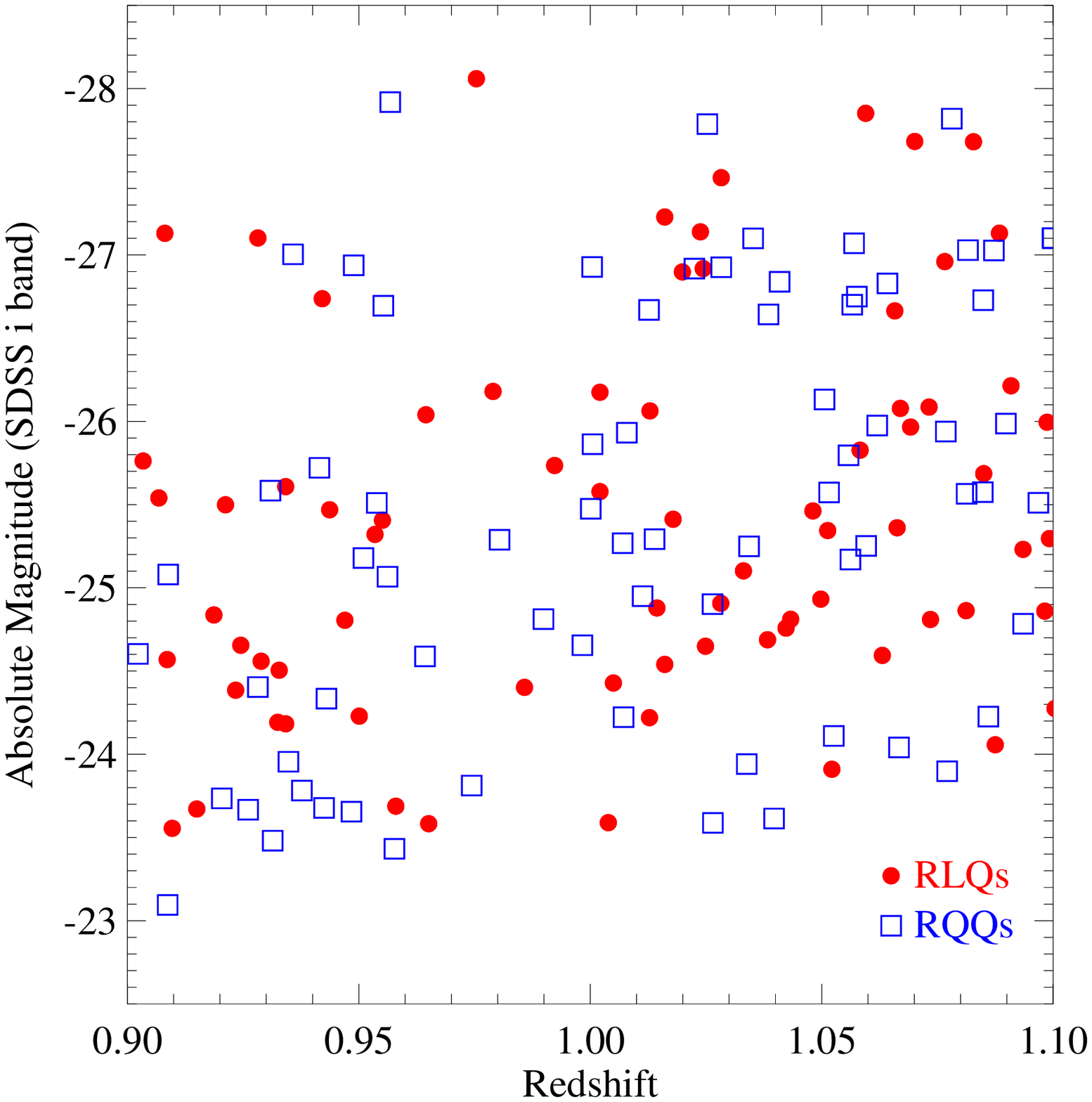}
\caption{Left: Optical (SDSS $i-$band) absolute magnitude density map of the SDSS quasar sample as a function of redshift. The orange dots are the RLQs and RQQs included in the $z\approx1$ sample. Selecting the sample at $z \approx 1$ ensures that we have the maximum coverage in luminosity while still probing enough volume to sample the bright end of the luminosity function, where most of the quasars at higher redshift lie. Right: Optical absolute magnitude (SDSS $i-$band) plotted against redshift for the quasars of our sample. RLQs are shown with red circles and RQQs with open blue squares.}
\label{fig:Mi_z}
\end{figure*}

In this paper, we present {\it Herschel} photometric observation using both {\it Photodetector Array Camera} (PACS) at 70 and 160 $\mu$m and the {\it Spectral and Photometric Imaging Receiver} (SPIRE) at 250, 350 and 500 $\mu$m for a $z\sim1$ benchmark sample of 173 AGN. We additionally present the SMA radio interferometer observations at 1300~$\mu$m of the RLQs sample in order to investigate the radio-jet synchrotron contamination of the FIR emission. The paper is structured as follows. In Section~\ref{z1_data} we describe the sample selection, the {\it Herschel} and SMA observations that we carried out, and the steps used for measuring the flux densities in the observed bands. BH and host galaxy properties and analysis are presented in Section~\ref{section:BH_HG_properties}. Sections~\ref{section:SFR_AGN}-\ref{section:RLQs_RGs} present our results on the star formation dependence on AGN luminosity, radio jets and orientation, respectively. In Section~\ref{section:Conclusions} we list and discuss our conclusions. Throughout the paper we use the cosmological parameters $H_{\rm 0} = 70~{\rm km~s^{-1}~Mpc^{-1}}$, $\Omega_{M} =0.3$ and $\Omega_{\Lambda} =0.7$, and we follow the conversion from FIR luminosities of \cite{Kennicutt1998} when deriving SFRs.

\section{DATA}  \label{z1_data}

The data presented in this paper consist of {\it Herschel}-PACS and SPIRE images of 173 AGN, along with millimetre images taken at 1300~$\mu$m with SMA for the RLQs. The sample is split into three sub-samples, all at the single cosmic epoch of $0.9<z<1.1$: 74 RLQs, 72 RQQs and 27 RGs. This redshift range is convenient because, as shown in Fig.~\ref{fig:Mi_z}, the SDSS survey allows us to probe over 5 magnitudes in quasar optical luminosity. This sample thus enables us to decouple luminosity generated effects from evolutionary ones, something which has plagued many other flux density limited studies in this area. 

This redshift is the minimum at which we have a large enough sample of high luminosity quasars ($M_{i}<-25.0$) which can be compared to the bright quasars found at higher redshifts. Observing both unobscured (type-1) AGN, in the form of quasars, and obscured (type-2) AGN, the RGs, allows us to test AGN unification schemes \citep[e.g.][]{Antonucci1993}. Details of the selection of the quasars are presented by \cite{Falder2010} while the RG selection is described by \cite{Fernandes2015}. The {\it Herschel} photometry is provided in Appendix A (Table~\ref{Table:Herschel_photometry}) while a summary of the main properties of the sample objects is given in Appendix A (Table~\ref{Table:full_properties}). In the next section we give a brief description of the sample criteria as they affect this paper.

\subsection{Sample selection}

The quasars were selected by their optical colours in the SDSS Quasar Survey \citep{Schneider2005}. The sheer size of the SDSS Quasar Survey allowed us to select a large enough initial sample to define matched samples of RLQs and RQQs. The initial quasar sample was then cross referenced with the NRAO VLA Sky Survey (NVSS; \citealp{Condon1998}), the VLA FIRST survey \citep{Becker1995} and the Westerbork Northern Sky Survey (WENSS; \citealp{Rengelink1997}) to pick out the RLQs and RQQs. Regarding the RLQs in the sample, the initial cross-match was done with the WENSS low-frequency survey (325 MHz). Therefore, the RLQs are selected based on optically thin extended emission, which means that the sample selection should be largely orientation independent. The RGs were selected from the low frequency, (178 or 151 MHz; orientation independent) radio samples of the 3CRR \citep{Laing1983}, 6CE \citep{Eales1985}, 7CRS \citep{Willott1998} and TOOT surveys \citep{Hill2003}. For the 6C objects the redshifts are taken from \cite{Best1996}, \cite{Rawlings2001} and \cite{Inskip2005}, and for the 6C* and TOOT objects from \cite{Jarvis2001} and \cite{Vardoulaki2010}, respectively. Combining these surveys, 27 RGs are found in the same $0.9 < z < 1.1$ redshift range as our quasars. The smaller RG sample arises from the limit of the known RG population at $z \approx 1$ at the time the samples were defined. 

\begin{figure}
\hspace{-0.8cm}
\includegraphics[scale=0.45]{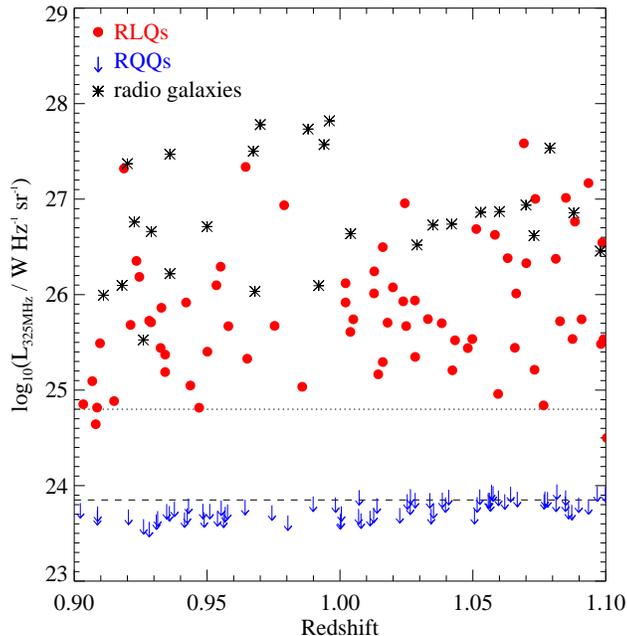}
\caption{325-MHz radio luminosity vs. redshift for our sample. RLQs are shown with red circles and RGs with black asterisks. For the RQQs, $5\sigma$ upper limits (extrapolated to rest-frame 325 MHz) from the FIRST survey are shown as blue upper limits. The dashed line shows the average $5\sigma$ limit of the WENSS survey, converted to a luminosity at $z \approx 1$ by assuming $\alpha = 0.7$. The dotted line shows the average $5\sigma$ limit of the FIRST survey, extrapolated to 325 MHz. The RQQs were selected to have a radio luminosity falling below this line. The assumed spectral indices for some conversions explains why some objects fall between the lines on this plot.}
\label{fig:Lradio_z}
\end{figure}

RLQs were chosen to have a low frequency WENSS (325 MHz) flux density of greater than 18 mJy, which is the $5\sigma$ limit of the survey (see Fig.~\ref{fig:Lradio_z}). This selection ensures that the vast majority of the RLQs included are characterized by steep radio spectra, avoiding flat radio spectrum quasars and blazars. Additionally, the low frequency radio flux selection allows us to compare the RLQs to the RGs without a severe orientation bias. \cite{Falder2010} presents a classification of the quasar population into radio-loud and radio-quiet based on the definition used by \cite{Ivezic2002}. With the exception of 4 objects all of our RLQs have $R_{i}>1$ where $R_{i}=\log_{10}(F_{\rm radio}/F_{i})$ and $F_{\rm radio}$ and $F_{i}$ are flux densities measured at 1.4 GHz and in the $i-$band respectively, so that the RLQ class we use here maps well onto traditional radio-loud/quiet definitions.

The RQQs were defined as being undetected by the FIRST survey at the $5\sigma$ level. FIRST was used for this definition because it provides a more sensitive flux density limit than WENSS. \cite{Falder2010} performed a stacking experiment to reveal the average value of the radio power at 1.4~GHz \citep[e.g.][]{White2007} for the RQQs in our sample. Using this technique they found an average flux density for the RQQs at 1.4 GHz of $0.10 \pm 0.02$ mJy (i.e. a 5-$\sigma$ detection). We extrapolate this estimate to a 325 MHz flux density of $0.30\pm0.06$ mJy assuming a spectral index of 0.7. At $z = 1$ this corresponds to a 325 MHz luminosity, $\log_{10}(L_{325{\rm MHz}}/{\rm W~Hz^{-1}~sr^{-1}}) = 23.82$.

74 RLQs and 72 RQQs matched in $i-$band magnitude and spanning 5 optical magnitudes were chosen for {\it Herschel} follow-up observations. The distribution of optical magnitudes as a function of redshift of the selected sources is shown in Fig.~\ref{fig:Mi_z}. Fig.~\ref{fig:Lradio_z} shows the radio luminosity distribution within the selected redshift range for RLQs and RGs. It is clear that, on average, the RGs are more radio-luminous than the RLQs, albeit with a significant overlap. The results of this selection are further discussed in Section~\ref{section:RLQs_RGs}. For the RQQs we have placed an upper limit on their radio luminosity (see Fig.~\ref{fig:Lradio_z}). In comparison to these limits, RLQs are at least one order of magnitude more radio-luminous than RQQs. The radio luminosity gap between the RLQs and RQQs (Fig.~\ref{fig:Lradio_z}) is due to our selection rather than a real radio power dichotomy, because of the different WENSS and FIRST survey depths from which the RLQs are selected.

\subsection{Herschel photometry} \label{section:Herschel_photometry}

The data for this work were obtained as part of the {\it Herschel} project `A benchmark study of active galactic nuclei' with 55.1 hours of observations allocated. SPIRE observations for 25 objects in our sample were obtained as part of other public {\it Herschel} projects (see Table~\ref{Table:Herschel_photometry}). The raw data for these objects were retrieved from the {\it Herschel} Science Archive (HSA), and the data reduction was performed as detailed below.

\subsubsection{PACS} \label{section:PACS}

PACS \citep{Poglitsch2010} photometric observations at 70 $\mu$m (5 arcsec angular resolution) and 160 $\mu$m (10 arcsec angular resolution) bands were carried out in the scan-map observational mode. A concatenated pair of small map scans of 4 arcmin length, each at two different orientations, was obtained for each source with a total integration time per source of 426-860 secs. The {\it Herschel Interactive Processing Environment} (HIPE, \citealp{Ott2010}, version 9.1.0) was used to perform the data reduction, following the standard procedures for deep field observations. The high-pass filtering method was applied to create the maps allowing us to minimize the point-source flux loss \citep{Popesso2012}. A preliminary map was created by combining the scan maps which were processed individually for each scan orientation. Using results from \cite{Popesso2012}, we choose a masking strategy based on circular patches at prior positions. This method avoids significant flux losses while any other kind of flux losses are independent of the PACS flux densities \citep{Popesso2012}. The final data reduction and mosaicing were then performed using the mask generated in the previous step.

Due to the fact that none of the sources show extended FIR emission and almost $\sim50$ per cent of the total sample is not detected at $>$3-$\sigma$ level we do not carry out aperture extraction of the FIR fluxes in order to consider all sources equally, even the ones with non-detections, rather than using their $3\sigma$ upper limits. Instead, we directly measure the FIR flux densities from the PSF-convolved images for both bands. We take the flux density to be the value in the image at the pixel closest to the optical position of our targets. We compared the direct flux density measurements to the aperture extraction for the FIR-detected sources and found an insignificant $<5$ per cent difference. The photometric uncertainties of each map were estimated from a set of 500 randomly selected positions \citep[e.g.][]{Lutz2011,Popesso2012}. The only requirement was that the measured pixels should have a total integration time at least 0.75 times the integration time of that of the source of interest in order to exclude the noisy map edges \citep[e.g.][]{Leipski2013}. The $1\sigma$ photometric uncertainty of the map is taken to be the $1\sigma$ value of the Gaussian fitted to the flux densities measured in these 500 random positions. Measured flux densities are provided in Appendix A (Table~\ref{Table:Herschel_photometry}).

\subsubsection{SPIRE} \label{section:SPIRE}

SPIRE \citep{Griffin2010} photometric observations at 250 (18.2 arcsec angular angular resolution), 350 (24.9 arcsec angular resolution) and 500 $\mu$m (36.3 arcsec angular resolution) were carried out in small scan-map observational mode. The total time per source was 487 secs. Similarly to the PACS data, we used the HIPE standard pipeline to reduce SPIRE data. The FIR flux densities in each band were directly measured from the PSF-convolved images at the pixel closest to the optical position of our targets.

As demonstrated from deep extragalactic observations \citep[e.g.][]{Nguyen2010}, SPIRE maps are dominated by confusion noise at the level of 6-7 mJy beam$^{-1}$. The method we have adopted in order to determine the photometric uncertainties in the SPIRE maps is fully described by \cite{Elbaz2011} and \cite{Pascale2011}. We have measured the noise level at the position of each source on the residual map produced by removing all individually-detected sources above the detection threshold, and then is convolved with the PSF \citep{Elbaz2011}. On this convolved residual we determined the dispersion of pixel values in a box, around each target, whose size is 8 times the PSF full width at half maximum (the PSF FWHM for the SPIRE passbands is: 18.2 arcsec, 24.9 arcsec, and 36.3 arcsec at 250, 350, and 500~$\mu$m, respectively) \citep[e.g][]{Elbaz2011,Leipski2013}. The box size was chosen as a compromise between appropriate sampling of local noise variations, surrounding the target, and avoiding inhomogeneities in the exposure time, such as noisy areas at the edges of the map. SPIRE flux densities and their associated errors are provided in Appendix A (Table~\ref{Table:Herschel_photometry}).

\subsection{SMA photometry}

\subsubsection{Synchrotron Contamination}

Radio-loud quasars are known to have strong non-thermal beamed core components which could possibly enhance the emission all the way through to the thermal-infrared and possibly the optical waveband \citep[e.g.][]{Blandford1974}. \cite{Archibald2001} proposed that high-frequency radio observations are needed to measure the contribution from non-thermal emission to the FIR waveband of radio sources. We expect that our RLQ sample should be dominated by steep-radio-spectrum sources as they are selected on optically thin lobe emission by using low frequency WENSS (325 MHz) observations. RGs are expected to have fainter flat-spectrum core components as a result of their larger angle to the direction of the observer. Given the lack of high-frequency radio observations, the best estimate assumes a spectral index based on the available low-frequency $(<1.4~{\rm GHz})$ radio data which could be conservative or a highly uncertain extrapolation to the SPIRE bands.

\subsubsection{The SMA sample}

For a RLQ to be considered as a candidate for synchrotron contamination at the SPIRE bands, we used the available 1.4 GHz radio observations or the additional data at higher frequencies from the literature as a reference point, and assuming the core spectral shape to be flat, we deem non-thermal contamination to be possible for those objects for which the highest available radio frequency flux falls close to (within the $3\sigma$ error) or above the 500 $\mu$m flux density. We emphasize that this is a very conservative estimate as other authors \citep[e.g.][]{Archibald2001, Shi2005, Cleary2007} fit a parabola, or multiple power-law fits, to the steep-spectrum components in order to take into account possible high-frequency steepening. High-frequency SMA observations at 1300~$\mu$m for the RLQ sample allow us to measure the possible contribution of the non-thermal components to the FIR emission accurately, and minimise the high uncertainties (1-2 orders of magnitudes) caused due to the use of different types of extrapolations (steep-spectrum or flat-spectrum components).

\begin{figure*}
\includegraphics[scale=0.56]{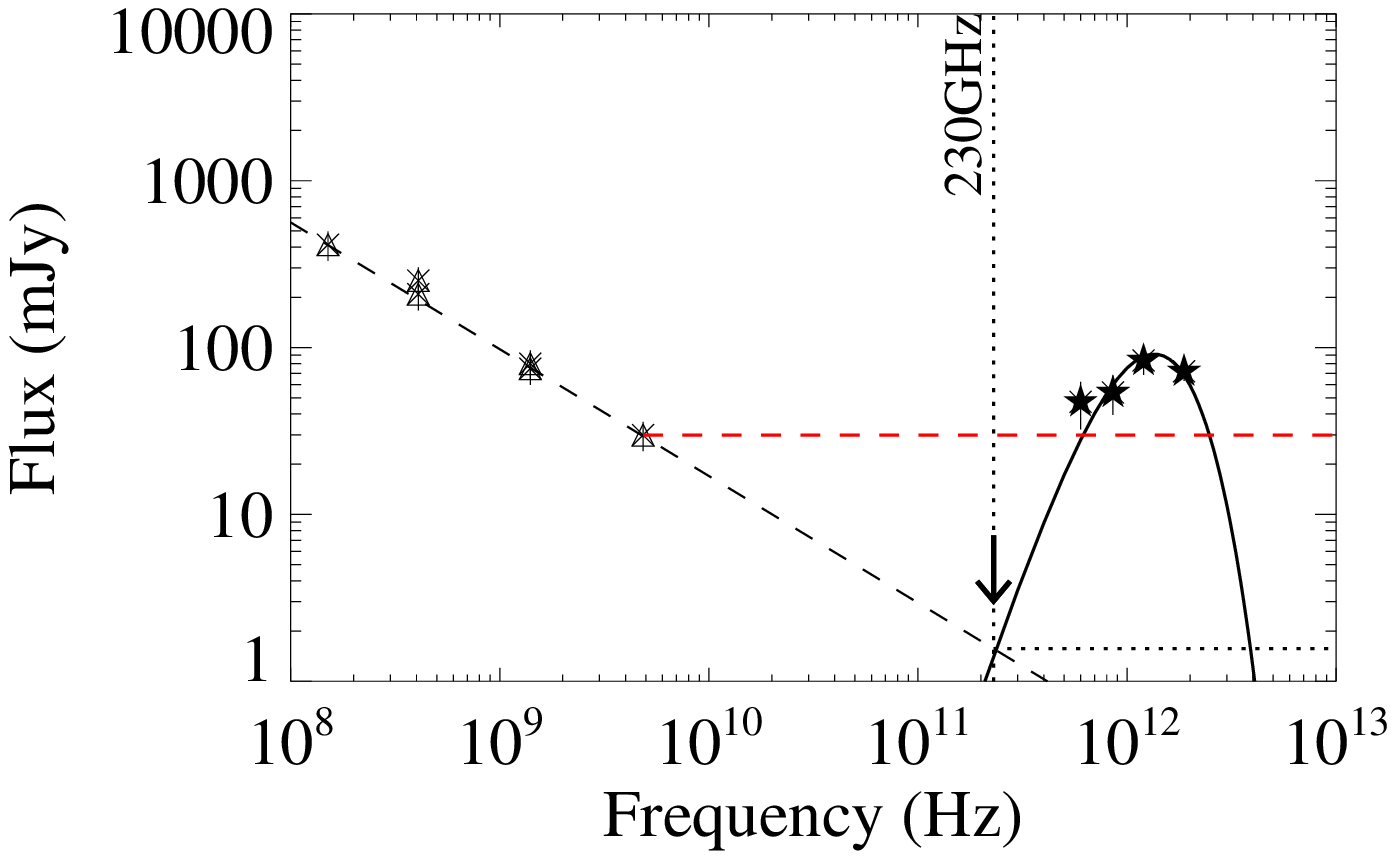}
%\hspace{-1cm}
\includegraphics[scale=0.56]{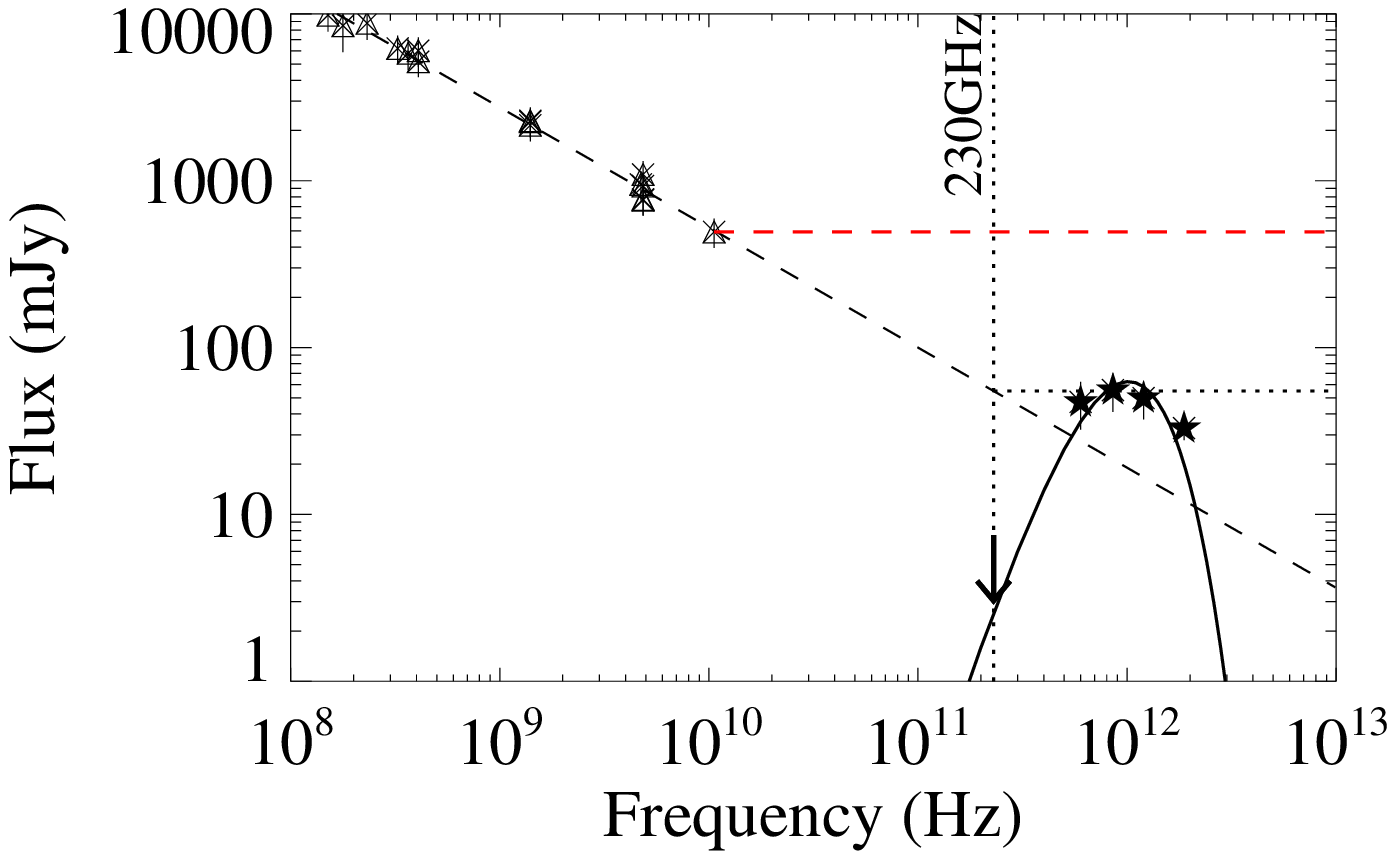}
\caption{Spectral energy distribution at radio and FIR wavelengths for a sample of the 44 RLQs. Filled black stars: the FIR data (\textit{Herschel}-SPIRE), Triangles: the radio data (FIRST, NVSS, WENSS and literature), Arrow: the maximum required flux at 1300~$\mu$m in order to not have a significant synchrotron contamination in the FIR bands. Error bars illustrate the $3\sigma$ errors. Black dashed line: Linear fit to radio data; red dashed line: flat radio spectrum; black solid line: grey-body fit; black dotted horizontal line: flat radio spectrum extrapolation at 1300~$\mu$m.}
\label{fig:SMA_SEDs_ex}
\end{figure*}

We initially used the existing radio data to assess the potential for synchrotron contamination. For each RLQ we have used an upper (flat-spectrum-dominated; red dashed line) and a lower limit (steep-spectrum-dominated; black dashed line; Fig.~\ref{fig:SMA_SEDs_ex}). We have found that 24 RLQs have potential contamination only when we assume a flat-spectrum core/jet component (Fig.~\ref{fig:SMA_SEDs_ex}; left), and 20 RLQs have potential contamination to their thermal FIR emission from either a steep-spectrum or a flat-spectrum component (Fig.~\ref{fig:SMA_SEDs_ex}; right). For each of these sources, using the 500 $\mu$m flux density as a reference and assuming the spectral shape to be flat, we have estimated the minimum flux density at 1300~$\mu$m in order to have a significant level of non-thermal contamination (Fig.~\ref{fig:SMA_SEDs_ex}; upper limit). For the vast majority of the sources this level is at $\sim7-10$~mJy.

\subsubsection{The SMA observations}

We used the SMA \citep{Ho2004} to observe the 44 RLQ candidates at wavelengths near 1300~$\mu$m (frequencies near 230~GHz) to assess the contribution from synchrotron emission to fluxes measured in the FIR bands. The SMA observations were performed in the 2014-2015 summer and winter semesters, typically in snapshots with 20 minutes on source bracketed by 2 minutes on nearby calibrators to determine complex gains. Many of the observations were executed in available short timeslots before or after other scheduled programmes and shared receiver tunings, correlator setups, as well as flux and passband calibrators. The total bandwidth available was 8 GHz, derived from two sidebands spanning $\pm(4 - 8)$ GHz from the local oscillator (LO) frequency. For each source, flux densities were measured by fitting a point source model to the visibilities using the task {\tt uvfit} in the Miriad software package. Each source was also imaged in order to confirm the visibility fit results. Table~\ref{Table:SMA_obs} lists the dates of observation, the characteristic atmospheric opacity during the observations, and the fitted flux densities. Variations in sensitivity are due to both weather conditions and the number of array antennas operating at the time of the observations. Overall, 15 sources were detected at the $>4\sigma$ level (a conservative threshold for these snapshot observations). The absolute flux scale has an estimated systematic uncertainty of $\sim20$ per cent. 

Using the SMA observations we have classified the 44 sources identified as having possible synchrotron contamination into two categories. In the first category, we have identified 14 sources with significant synchrotron contamination. All of these sources have been rejected from our sample and from further analysis. The vast majority of them (10) were detected at $>4\sigma$ with the SMA with some extreme cases reaching even $S_{\rm 1300\mu m}\approx200$~mJy. Some representative examples of the SEDs from this group are presented in Fig.~\ref{fig:SED_appendix}. In this category the SMA flux densities exceed the linear extrapolation from the lower-frequency radio data for 8 sources (three are upper limits), for three sources they follow the linear prediction, while for the last three sources they indicate the need of a steeper-spectrum radio component at the higher frequencies. However, even in the last two cases, the contamination to the FIR band is significant and therefore these sources have also been excluded from this work.

In the second category we have classified 30 sources without significant synchrotron contamination. For four cases there is a clear SMA detection at $>$4$\sigma$ while all the other observations indicate an upper limit. For this group of sources, the SMA data exceed the linear extrapolation in seven cases (all of them are upper limits) while in 18 cases they indicate the need of a steeper-spectrum radio component at the higher frequencies. Examples of the SEDs from this group are also presented in Fig.~\ref{fig:SED_appendix}.

Overall, we have rejected 17 RLQs from our sample, 14 based on their SMA observations, while three additional sources were classified as flat-spectrum RLQs or blazars based on literature radio observations and rejected ([HB89] 0906+015, SDSS J133749.63+550102.2, SDSS J161603.76+463225.2). As we describe in Section~\ref{sec:accretion_rate}, there are no particular trends for the sources excluded from our sample and they do not affect the sample matching between RLQs and RQQs. 

From our results, it is clear that high-frequency radio observations for similar studies are crucial as the linear extrapolation from lower frequencies works only for $\sim20$ per cent of the sources. Although most of the cases indicate that the steep-spectrum synchrotron component is likely to fall more quickly at higher frequencies, we find that in $\sim30$ per cent of the SMA observed sources a high-frequency core radio component is required to describe the radio spectrum. This would also agree with recent findings \citep[e.g.][]{Whittam2013,Whittam2015}. We note that almost half of these SMA observations are upper limits. Radio core variability might be responsible for some of these strong high-frequency components \citep[e.g][]{Barvainis2005}.

\section{The black hole and host galaxy properties} \label{section:BH_HG_properties}

In this section we describe how the key parameters for the analysis of this paper are derived, namely BH and stellar masses, Eddington ratios, bolometric luminosities and FIR luminosities. We further explore the importance of AGN contamination in the form of hot dust around the putative torus at FIR wavelengths comparing their FIR colours against normal galaxies. We finally study the correlation between the radio and FIR emission, examining at the same time whether some of the radio emission could be the result of star formation, rather than AGN activity.

\subsection{Stellar and black hole mass} \label{section:masses}

Early studies \citep[e.g.][]{Kormendy1995,Magorrian1998} suggest a correlation between galaxy bulge and its BH mass. The ratio of the so-called $M_{\rm BH}$-$M_{\rm bulge}$ relation \citep{Magorrian1998} was estimated to be approximately 0.6 per cent. In the same context, more recent studies using nearby galaxy samples \citep[e.g.][]{Haring2004} find that the median BH mass is $0.14\pm0.04$ per cent of the bulge mass.

For the quasars in this sample, the BH masses are computed using the virial estimator and the Mg{\sc ii} line at 2800 \AA~using SDSS spectroscopy, a technique described by \cite{McLure2002}, and based on work of \cite{McLure2004}. As the H$\beta$ line moves out of the optical window, we have to rely on the Mg{\sc ii} line for AGNs at $z >0.7$ \citep[e.g.][]{Wang2009}. BH masses for the quasars are given in Table~\ref{Table:full_properties}. We can use the BH mass of the quasars in the sample, along with the $M_{\rm BH}$-$M_{\rm bulge}$ relation to estimate the stellar mass of the galaxy. Despite the convenience of calibrating and using these virial estimators, one must keep in mind that the estimates of these lines are uncertain, potentially by as much as 0.4 dex \citep[e.g.][]{Shen2011}, due to the systematics involved in the calibration and usage \citep[e.g.][]{Jarvis2002,Jarvis2006,Marconi2008,Kelly2009}. We assume that there is no significant evolution of the $M_{\rm BH}$-$M_{\rm bulge}$ relation at $z\approx1$ from the local relation and thus use $M_{\rm BH}\sim0.0014M_{\rm bulge}$. Indeed, studies on $z\leq1$ RLAGN BH-bulge mass relation have found that the estimated ratio lies within the uncertainties of that found in the local Universe \citep[e.g.][]{McLure2006}. Although evolution in the $M_{\rm BH}$-$M_{\rm bulge}$ relation of about 0.2 dex at $z\approx1$ has been claimed in some papers \citep[e.g.][]{Merloni2010}, that would not significantly add to the uncertainties and would not affect the results of this work, as all of the AGN are selected in a very small redshift range.

\begin{figure*}
\includegraphics[scale=0.33]{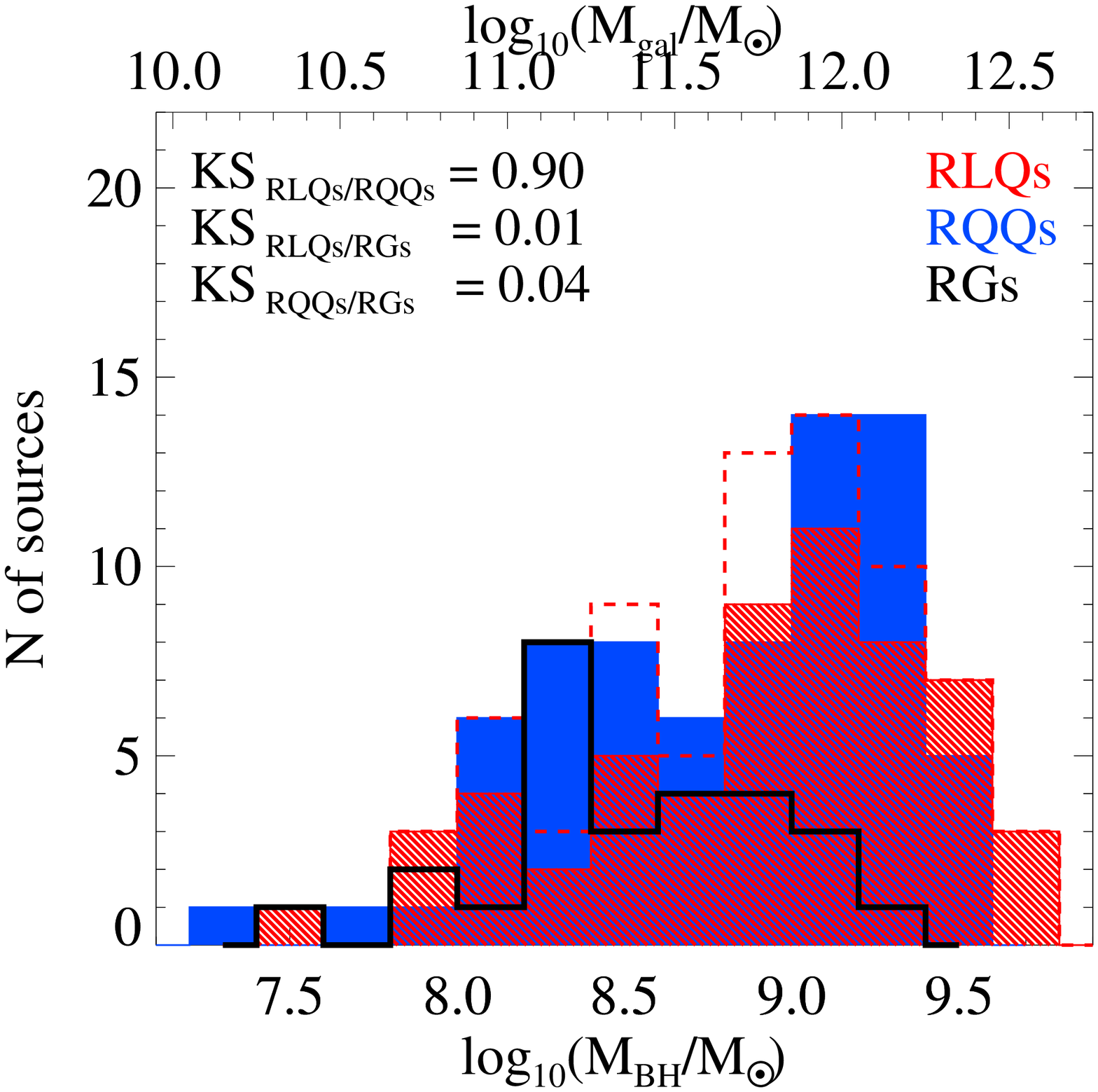}
\hspace{-1.45cm}
\includegraphics[scale=0.33]{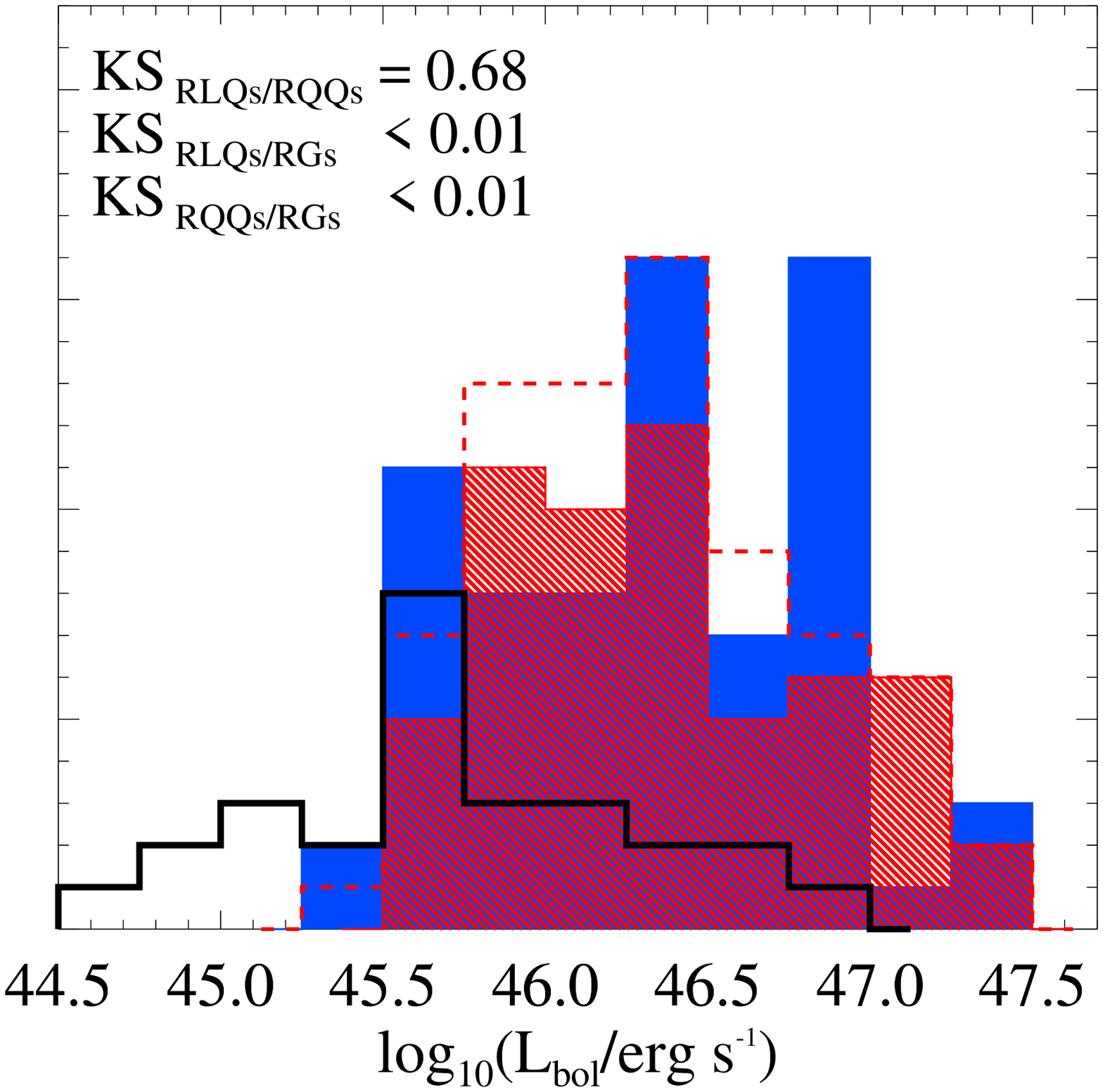}
\hspace{-1.45cm}
\includegraphics[scale=0.33]{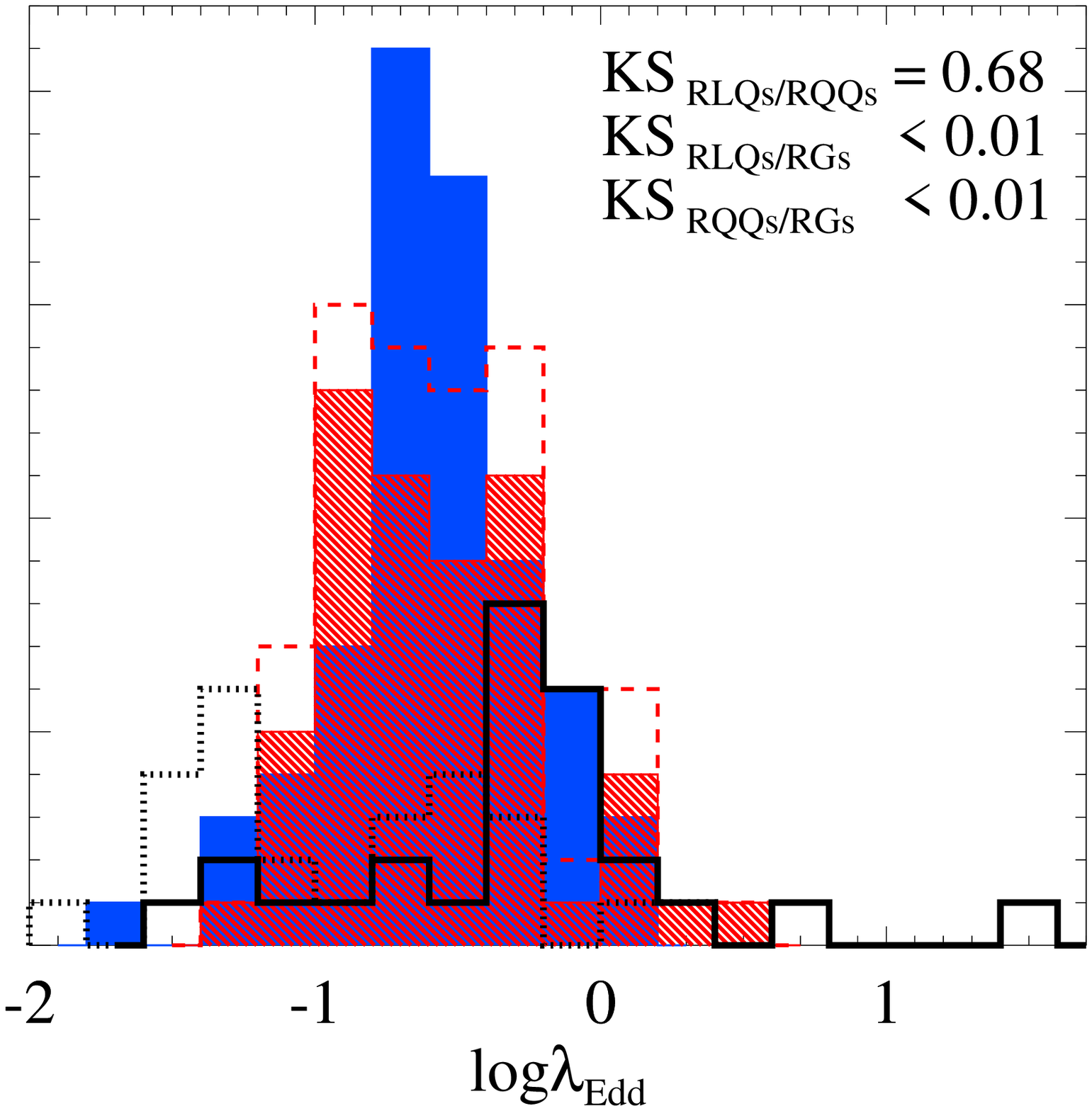}
\caption{Distributions of BH mass, $M_{\rm BH}$, bolometric luminosity, $L_{\rm bol}$ and Eddington ratio, $\lambda_{\rm Edd}$, for RLQs (red filled), RQQs (blue filled) and radio galaxies (black line). The total RLQ population, including the sources with significant synchrotron contamination is also presented with dashed red lines. In the last panel, the Eddington ratio distribution obtained with both methods of calculating for radio galaxies is presented (dotted black line considering only the accretion energy and solid black line including also the jet mechanical energy). Note the significant increase of the total accretion energy.}
\label{fig:distri_mass}
\end{figure*}

For the RGs in our sample, because the broad-line region is obscured, we do not have BH mass estimates as we did for the quasars. For this reason, we use the stellar mass of the galaxy, $M_{\rm gal}$, determined by the SED fitting of \cite{Fernandes2015} for the same radio galaxy sample as used in this work. \cite{Fernandes2015} used the same BH-bulge mass relation \citep{Haring2004} in order to calculate the BH mass of the radio galaxy sample. The implied $M_{\rm BH}$ are given in Table~\ref{Table:full_properties}. The radio galaxies in our sample have BH masses in the range $10^{7.5}-10^{9.4}~M_\odot$ (corresponding to $M_{\rm gal}=10^{10.3}-10^{12.0}~M_\odot$) while the quasars have $10^{7.2}-10^{9.7}~M_\odot$ (corresponding to $M_{\rm gal}=10^{10.1}-10^{12.4}~M_\odot$). These are consistent with the range of values found in the literature for similar objects \citep[e.g.][]{McLure2004,McLure2006,Seymour2007,Salviander2007}.

To test whether the BH and stellar mass distributions differ between the three populations we conducted a Kolmogorov-Smirnov (K-S) test for each pair. The test suggested that the BH masses for the RLQs and RQQs samples are not significantly different. The K-S test gives a result that corresponds to a probability, $p = 0.90$ under the null hypothesis (i.e. they are statistically indistinguishable). The mean BH masses are $\langle\log_{10}(M_{\rm BH}/M_\odot)\rangle=8.87\pm0.06$ for the RLQs and $\langle\log_{10}(M_{\rm BH}/M_\odot)\rangle=8.81\pm0.06$ for the RQQs so the means of the two samples are consistent and well within 1-$\sigma$ of each other. 

In contrast, the RG sample could not be selected to match the quasar sample in absolute optical magnitude. The RGs have nominally lower mean BH masses $\langle\log_{10}(M_{\rm BH}/M_\odot)\rangle=8.53\pm0.08$; a K-S test comparing to the quasar sample returns 0.01 probability under the null hypothesis, a marginally significant result. Selection effects might also contribute to the observed differences (e.g. RGs are selected from radio surveys without a pre-requisite to be optically bright). We further discuss these effects in Section~\ref{section:RLQs_RGs}. The distribution of BH masses is shown for all samples in Fig.~\ref{fig:distri_mass}.

\subsection{Accretion rate} \label{sec:accretion_rate}

In order to make an estimate of the AGN power for the radio galaxies an estimate of the bolometric radiative power of the AGN, $L_{\rm bol}$, is required. For the radio galaxy sample, we adopted the values of $L_{\rm bol}$ calculated by \cite{Fernandes2015} from the rest-frame 12$\mu$m luminosity, using a bolometric correction of 8.5 \citep[e.g.][]{Richards2006b}, $L_{\rm bol}=8.5\lambda L_{12\mu \rm m}$. The bolometric luminosity for the quasar sample has been computed from the 3000~\AA ~luminosity ($L_{3000}$) using the SDSS spectral fits and a bolometric correction of 5.15 from the composite SED in \cite{Richards2006b}, $L_{\rm bol}=5.15\lambda L_{3000}$. \cite{Fernandes2011} have computed the bolometric luminosity for the quasar sample based on the rest-frame 12$\mu$m luminosity, following the same method as the one applied for the radio galaxies. Their results suggest no systematics related to methodology or calibration when optical photometry is used for bolometric luminosity estimates.

The bolometric luminosity is proportional to the accretion rate of the BH, $\dot{M}$, and to the fraction of accreted mass that is radiated, i.e. the radiative efficiency, $\epsilon$, through the expression: 
\begin{equation}
L_{\rm bol}=\epsilon \dot{M} c^{2}.
\end{equation}
Assuming that $\epsilon$ takes the fiducial value of 0.1 \citep[e.g.][]{Marconi2004,Shankar2004,Martinez-Sansigre2009}, we determine the accretion rate of our sources using their estimated bolometric luminosity.

With both the BH mass and the accretion rate, we can estimate the Eddington ratio of the sources in our sample. The Eddington luminosity, $L_{\rm Edd}$, corresponds to a maximum accretion rate which a black hole can reach, without preventing further accretion onto it. This energy is a function of the black hole mass of the system and is given by 
$L_{\rm Edd}=1.3\times 10^{31}(\frac{M_{\rm BH}}{M_\odot})~{\rm W}$. The Eddington ratio, $\lambda$ is therefore simply,
\begin{equation}
\lambda \equiv \frac{L_{\rm bol}}{L_{\rm Edd}}.
\label{eq:lambda}
\end{equation}

Although for SMBHs accreting at a high fraction the Eddington ratio can be defined as in equation~\ref{eq:lambda}, for radio galaxies, especially those with SMBHs accreting at very low rates (e.g. low-excitation galaxies; LEGs), the contribution of the jet mechanical energy in the output of the accretion energy should be considered for the definition of the Eddington ratio. In this case, the total energy from the black hole accretion should equal the sum of the radiative luminosity and the jet mechanical luminosity \citep[e.g.][]{Hardcastle2007,Best2012}. Including the contribution of the jet power, $Q_{\rm jet}$, the Eddington ratio is given by
\begin{equation}
\lambda_{\rm rad+mec}=\frac{L_{\rm bol}+Q_{\rm jet}}{L_{\rm Edd}},
\label{eq:lambda_mec}
\end{equation}
where $\lambda_{\rm rad+mec}$ is the Eddington ratio accounting for both the radiative energy and the jet mechanical energy. We estimate the jet power using the relation 
\begin{equation}
Q_{\rm jet}\simeq3\times 10^{38}f^{3/2}(L_{\rm 151MHz}/10^{28})^{6/7}~{\rm W}
\label{eq:Q_jet}
\end{equation}
\citep{Willott1999}, where $1\leq f\leq20$ represents the combination of several uncertainty terms when estimating $Q_{\rm jet}$ from $L_{\rm 151MHz}$. Following \cite{Fernandes2015}, we chose $f=10$ as this is the expectation value of a flat prior in natural space. We note that the $Q_{\rm jet}$ contributes significantly to the total power only in the radio galaxies of our sample, which is derived from the $L_{12\mu \rm m}$, and not in the RLQs ($<10$ per cent). The use of any derived radio-luminosity -- jet-power relation should be treated with caution, especially for the derivation of the kinematic luminosity function, as they may depend sensitively on selection effects \citep[e.g.][]{Shabala2013}.

The distribution of bolometric luminosity and Eddington ratio are shown for all samples in Fig.~\ref{fig:distri_mass}. The solid black line is for $\lambda = (L_{\rm bol} +Q_{\rm jet})/L_{\rm Edd}$ and the dotted line for $\lambda = L_{\rm bol} /L_{\rm Edd}$. The Eddington ratio for radio galaxies is significantly higher in the first case, where $\lambda = (L_{\rm bol} +Q_{\rm jet})/L_{\rm Edd}$, and this trend is dominated by high-excitation galaxies (HEGs; see \citealp{Fernandes2015}; Fig. 7). The red shaded histograms in Fig.~\ref{fig:distri_mass} represent the RLQ sample after excluding the synchrotron contaminated sources. The total RLQ population is also overplotted (red dashed lines) to stress that no selection biases are introduced in our sample after removing synchrotron contaminated RLQs. No particular trends are observed in any of the distributions between the RLQs and the RQQs as a result of the original matching in absolute optical magnitude and colours.

\subsection{FIR emission in RLQs, RQQs and RGs} \label{Herschel fluxes}

For each of the quasars in our sample we derive the FIR flux densities in the two PACS and the three SPIRE bands directly from the PSF-convolved images measuring the value at the image pixel closest to the optical position of our targets. The errors are estimated as described in Sections~\ref{section:PACS} and \ref{section:SPIRE}. We find that about 33 per cent (43/149) of the QSOs and 8 per cent (2/27) of the RGs in our sample have robust PACS and SPIRE detections. These detection rates are translated to ULIRG-like star formation luminosities suggesting SFRs of hundreds of solar masses per year.

\begin{figure*}
\includegraphics[scale=0.8]{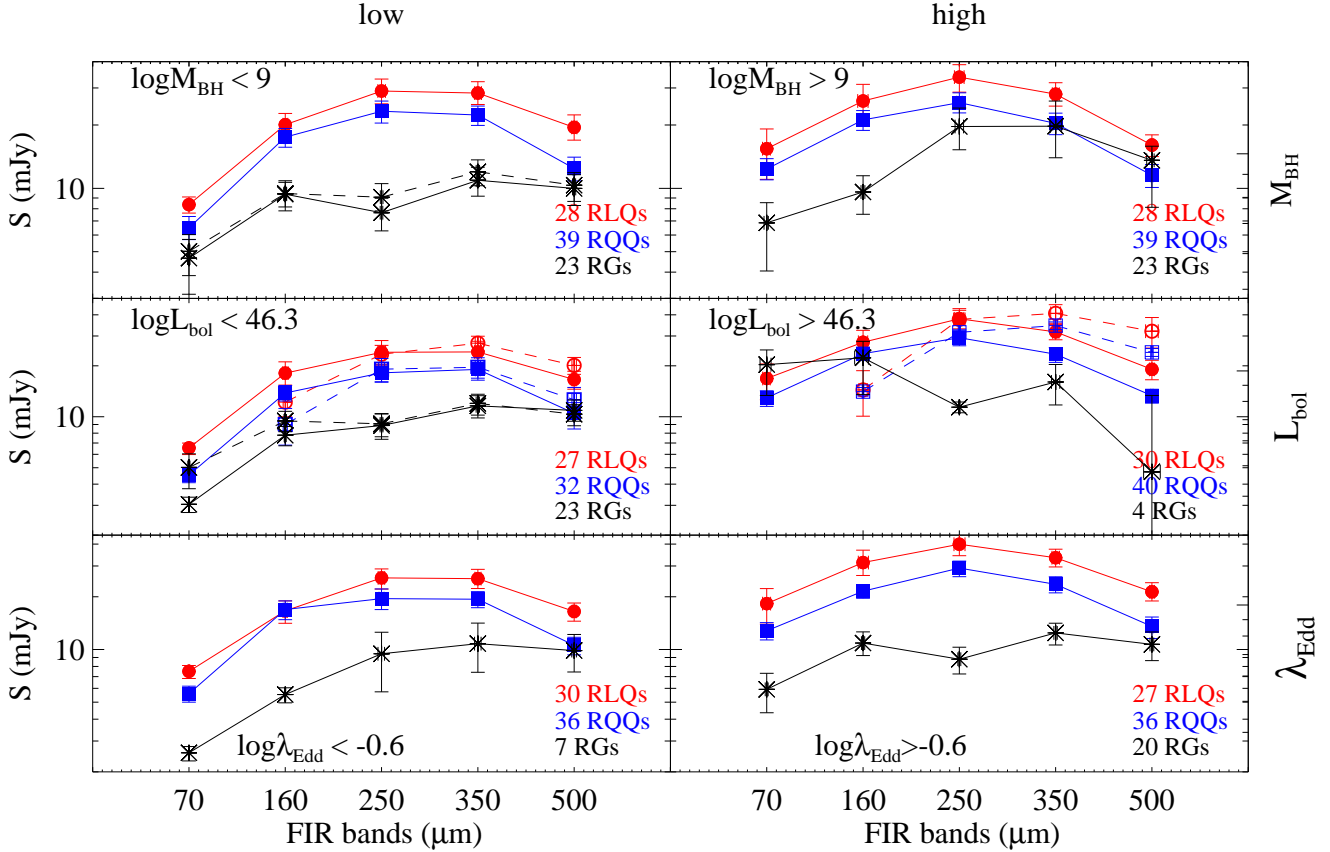}

\caption{The PACS and SPIRE mean flux densities for RLQS (red filled circles), RQQs (blue filled squares) and RGs (black stars) as a function of BH mass (top panel), bolometric luminosity (middle panel) and Eddington ratio (bottom panel). For low BH mass and bolometric luminosity bins we also present the mean flux density of the total RG population (dashed black line) in order to indicate the contribution of the only four sources found at high bins. We compare our measurements to Kalfountzou et al. (2014b) RLQs (red open circles) and RQQs (blue open squares) with similar bolometric luminosities but different redshift in the middle panels. Table 1 of Kalfountzou et al. (2014b) provides the mean flux densities over their total RLQ sample. Here, for comparison reasons, we present their mean flux densities after removing the RLQs with potential synchrotron contamination.} 
\label{fig:mean_flux}
\end{figure*}

\begin{table*}
\begin{center}
\caption{The RLQs, RQQs and RGs FIR average stacked fluxes in the 70, 160, 250, 350 and 500 $\mu$m bandpasses as described in Sectrion 3.3. The AGN populations have been separated into bolometric luminosity bins. The number of objects within each stack is also given.}
\centering
\begin{tabular} {c c c c c c c c }

\hline 

Class	&	$\log(L_{\rm bol}/{\rm erg~s^{-1}})$		&		$N$ per	bin	&	\multicolumn{5}{c}{Mean flux density (mJy)}		\\
		&	 	&		 	& 70 $\mu$m	&	160 $\mu$m	&		250 $\mu$m	&	350 $\mu$m	&	500 $\mu$m		\\

\hline

\hline
RLQs	&	$<46.3$ 	&	27	&	$6.55^{+0.44}_{-0.57}$	&	$18.10^{+3.00}_{-2.72}$	&	$24.04^{+4.15}_{-3.86}$	&	$24.15^{+3.76}_{-3.53}$	&  $16.59^{+1.97}_{-2.04}$	\\
	&	$\geq46.3$ 	&	30	&	$16.83^{+3.68}_{-4.22}$	&	$27.55^{+4.85}_{-4.85}$	&	$37.89^{+4.65}_{-4.82}$	&	$31.72^{+3.41}_{-3.24}$	&  $19.05^{+2.50}_{-2.51}$\\
	    
\hline
RQQs	&	$<46.3$ 	&	32	&	$4.53^{+0.31}_{-0.45}$	&	$13.83^{+1.48}_{-1.43}$	&	$18.17^{+2.06}_{-2.17}$	&  	$19.04^{+2.44}_{-2.60}$	&	$10.48^{+1.64}_{-2.01}$	\\
	&	$\geq46.3$ 	&	40	&	$12.93^{+1.31}_{-1.42}$	&	$23.58^{+2.20}_{-2.11}$	&	$29.28^{+2.96}_{-3.02}$	&	$23.42^{+2.17}_{-2.47}$	&	$13.33^{+1.21}_{-1.30}$\\
\hline

RGs	&	$<46.3$		&	23	&	$3.04^{+0.31}_{-0.32}$	&	$7.78^{+0.98}_{-1.04}$	&	$8.88^{+1.50}_{-1.52}$	&	$11.60^{+1.72}_{-1.76}$	&	$10.90^{+1.67}_{-1.49}$	\\
	&	$\geq46.3$	&	4	&	$20.32^{+4.50}_{-6.96}$	&	$22.30^{+5.58}_{-8.75}$	&	$11.39^{+0.83}_{-0.83}$	&	$16.05^{+4.31}_{-4.31}$	&	$4.73^{+8.63}_{-8.63}$	\\
\hline

\end{tabular}
\label{Table:mean_fluxes}
\end{center}
\end{table*}

We have separated the RLQ and RQQ samples in bolometric luminosity, BH mass and Eddington ratio bins to examine whether the fluxes vary. Within each bin we stack PACS and SPIRE residual maps at the optical position of the AGN from which we derive the mean flux densiy in all FIR bands. We then average the stacked fluxes with the fluxes of the detected ($>3\sigma$ significance level) sources, weighting by the number of sources \citep[e.g.][]{Elbaz2011,Santini2012,Rosario2013}. The estimates for each band and bin are shown in Fig.~\ref{fig:mean_flux}. Errors have been estimated by applying the bootstrap technique using randomly selected galaxies from within each bin. The advantage of bootstrapping is that no assumption is made on the shape of the flux distribution. Radio galaxies have significantly lower mean flux densities compared to RLQs and RQQs with a K-S test probability of $p<0.05$. The only exception is the 500 $\mu$m band which might indicate some contribution from synchrotron contamination, or confusion bias, or a combination of them in the case of RGs. This contamination may extend to even lower wavelength bands (e.g. 350 $\mu$m and 250 $\mu$m). Regarding the total quasar sample, the mean flux density appears to increase at high $M_{\rm BH}$, $L_{\rm Bol}$ and $\lambda_{\rm Edd}$. Comparing the RLQs to the RQQs we see that at low $M_{\rm BH}$, $L_{\rm Bol}$ and $\lambda_{\rm Edd}$ RLQs have higher flux densities in all SPIRE bands and all bins, these differences seem to become more significant at the high $M_{\rm BH}$, $L_{\rm Bol}$ and $\lambda_{\rm Edd}$. As no obvious differences are found for the RLQs and RQQs between bolometric luminosity, BH mass and Eddington ratio bins, we give in Table~\ref{Table:mean_fluxes} the mean flux estimations from each band and population only for high and low bolometric luminosity bins. We used a black-body modified by frequency-dependent emissivity component (see Section 3.5) to convert the the mean FIR fluxes from our stacked images to mean integated 8-1000$\mu$m far-infrared luminosities for different bolometric luminosity bins. The results are presented in Teble B1.

The stacking method assumes implicitly that the sources in the map are not clustered. It has been shown that this might not be the case for wide PSF \citep[e.g.][]{Bethermin2010b,Penner2011} with various stacking methods taking this into account also for {\it Herschel} beams \citep[e.g.][]{Magnelli2014}. However, with the {\it Spitzer} and {\it Herschel} beams, it has been shown that the effects of clustering on the stacking are not important (less than 15 per cent; \citealp{Bavouzet2008,FernandezConde2008, FernandezConde2010,CaoOrjales2012}). In addition, Cao Orjales et al. (in prep.), found on average a small overdensity of {\it Herschel} detected star-forming galaxies for the same sample of sources. Once these sources are accounted for, consist of $\sim 0.4$ star-forming galaxies in every AGN field. This overdensity appears to be relatively uniform for both RLQs and RQQs and extends out to the Mpc-scale.

How do these results fit with our previous work? For purposes of comparison we have overplotted in Fig.~\ref{fig:mean_flux} the mean flux densities obtained by \cite{Kalfountzou2014a}, hereafter K14b, (dashed lines) for low and high optical luminosity RLQs (red circles) and RQQs (blue squares). We note two main differences between our current results and those of K14b. Although in the low bolometric luminosity sample of K14b the mean redshift is $z\approx0.9$, so that we do not expect the evolution effects to significantly change the mean properties, almost all the quasars with high bolometric luminosities have $z>1.0$ up to $z\approx3$. Therefore, we have converted the mean fluxes of the K14b to the $z\approx1$ rest-frame. A ratio method was applied in order to derive the k-corrections between the FIR flux densities of a Mrk 231 greybody template ($T=44.75$~K, $\beta=1.55$, constraining 32 data points so to exclude a contribution from AGN-heated dust emission) placed at the redshift of the QSO and the FIR flux density of the QSO. The new flux densities were stacked as described above. Additionally, due to the much larger sample of RQQs in K14b ($>10$ times larger than this work) the uncertainties of this sub-class are expected to be higher in the present paper. Fig.~\ref{fig:mean_flux} suggests that at low bolometric luminosities, our results are in excellent agreement, at least for the SPIRE bands. The disagreement between our PACS flux densities and those used by K14b is not unexpected since H-ATLAS PACS observations are about 5 times less sensitive than our observations \citep{Ibar2010}. Despite the similar trends, the differences between the RLQ and RQQ populations were more significant in K14b due to the smaller uncertainties for the RQQs. On the other hand, for the high bolometric luminosity bin, both H-ATLAS/SDSS RLQs and RQQs show significantly higher flux densities than the sample in this work, especially at 350 $\mu$m, with a characteristic shift of the the mean peak to the 350 $\mu$m band, indicating colder dust temperatures. These differences provide evidence for the evolution of the FIR emission between $z\approx2.0$ and $z\approx1.0$ high bolometric luminosity quasars. That would be expected if QSOs' host galaxies are evolving with cosmic time in the same way as the general galaxy population \citep[e.g.][]{Madau2014}.

\begin{figure}
\includegraphics[scale=0.42]{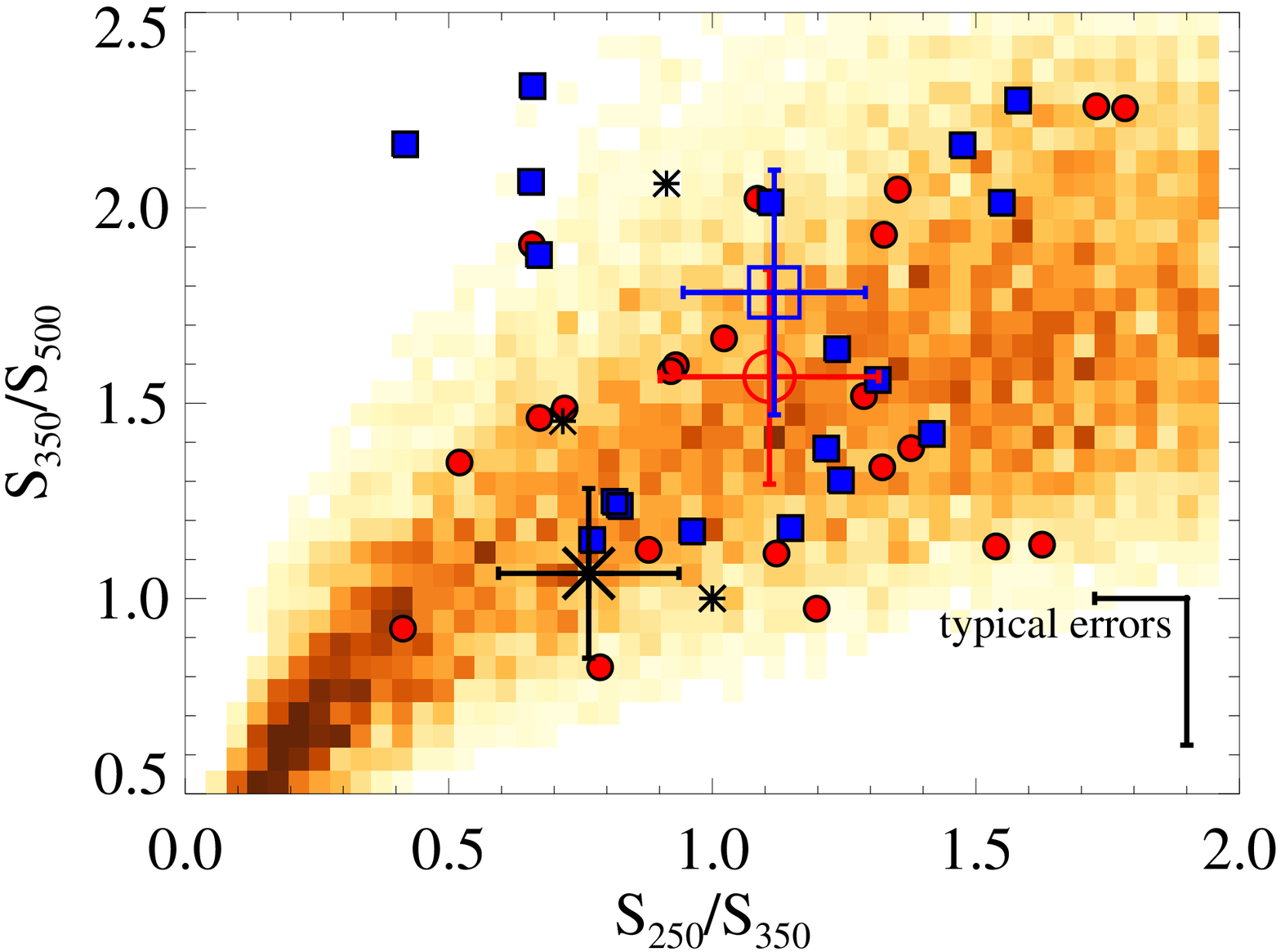}
\includegraphics[scale=0.42]{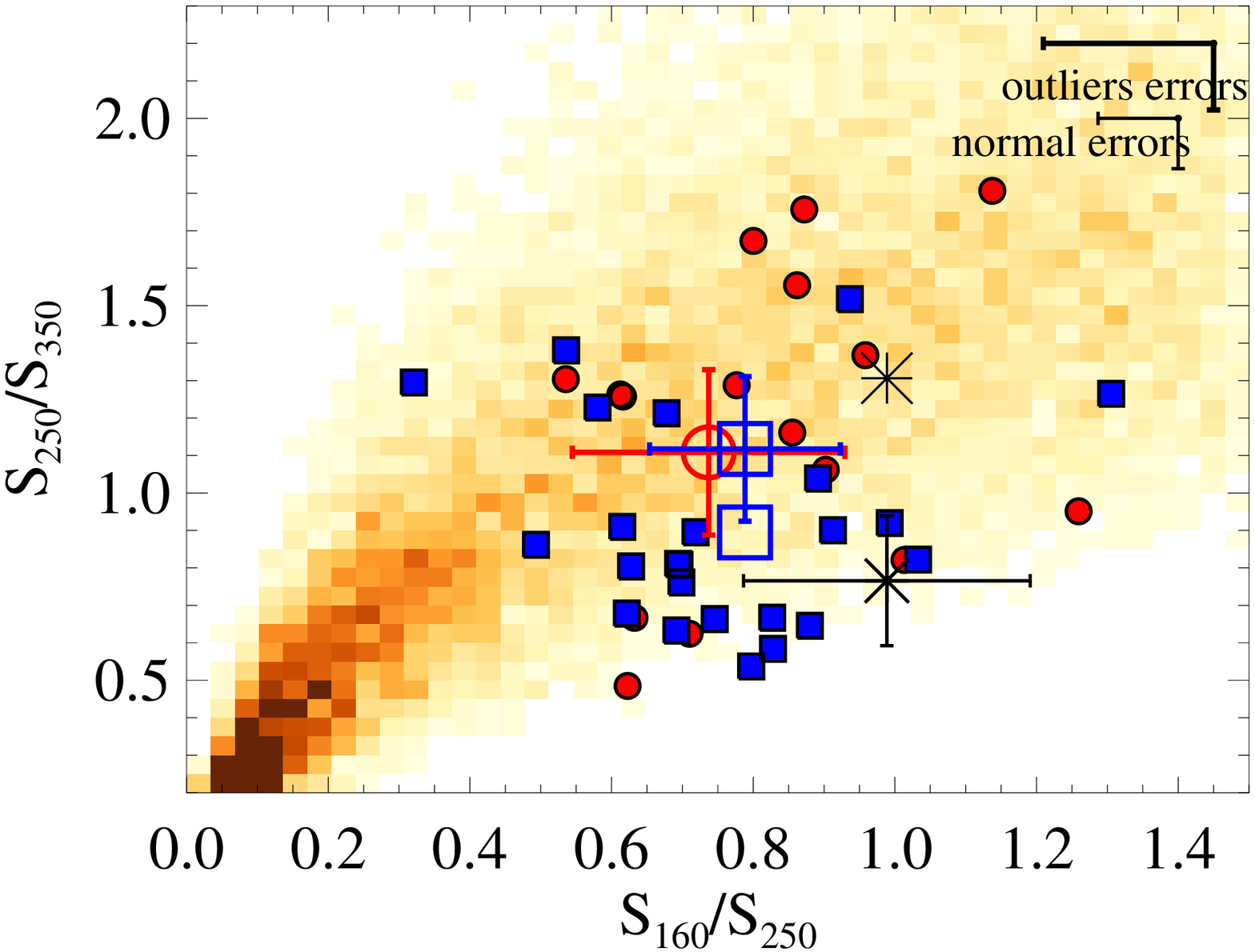}
\includegraphics[scale=0.42]{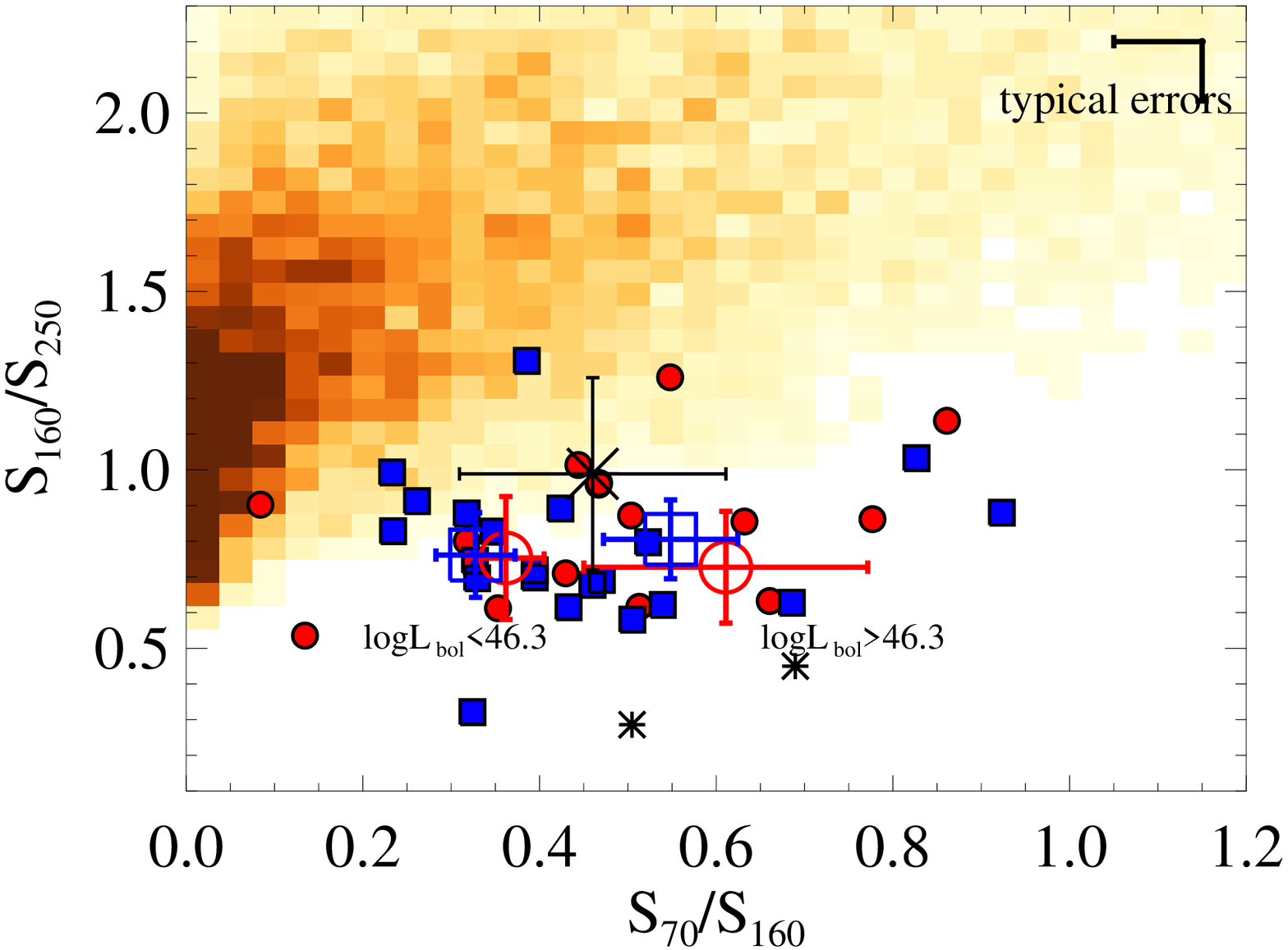}
\caption{SPIRE and PACS colour-colour diagrams of the AGN in our sample. Small symbols indicate the FIR detected AGN (RLQs = red circle, RQQs = blue squares, RGs = black stars). Detected sources have been selected by imposing a $3\sigma$ cut in each band, excepting 70 and 500 $\mu$m where we use a $2\sigma$ cut due to the low detection rates. Their typical $1\sigma$ errors are also presented in each panel. Stacked measurements for all AGN classes are shown as large symbols with their $1\sigma$ errors. The background density map indicates the colour-colour spaces of our $10^{6}$ randomly generated model SEDs. The darker colours of the density map correspond to denser regions. At the bottom panel, the mean values correspond to the low and high bolometric bins as indicated.}
\label{fig:FIR_colors}
\end{figure}

\subsection{The FIR colours of RLQs, RQQs and RGs} 

We now investigate the FIR colours of our sample of AGN. A straightforward approach towards exploring the effect of AGN light on FIR emission is to compare the FIR colours of AGNs against a control sample consisting of galaxies not hosting AGNs. AGN radiation field can heat the dust resulting in systematically warmer temperatures and causing the SED to flatten out at long IR wavelengths which, in turn, leads to bluer FIR colour in galaxies with a significant AGN contamination in the FIR.

In Fig.~\ref{fig:FIR_colors}, we compare the FIR colours of the detected AGN sub-sample and the stacked values of the total sub-samples (large symbols) to the FIR colours of $10^{6}$ randomly generated  modified black-body, single dust temperature $T_{\rm d}$ spectra models, with a frequency dependence of $\epsilon_{\nu}\propto \nu^{\beta}$. In generating these models, we follow the method of \cite{Amblard2010} considering uniformly distributed dust temperatures from 10 to 60 K, emissivity parameter $0<\beta<2$ and redshift range similar to our sample ($0.9<z<1.1$). In order to consider for flux uncertainties in the colour-colour diagram, we have broadened the SED tracks by adding an extra Gaussian standard deviation of 10 per cent to the model fluxes. Thus, the choice of emissivity parameter would make just a minor difference.

As shown in Fig.~\ref{fig:FIR_colors} (top), we find that in the SPIRE-only colour diagram the colours of the sources are well within the limits defined by the randomly generated model. This is the case for all AGN sub-classes of our sample and also for the individually SPIRE-detected AGN and the mean values. We find no significant dependence of SPIRE colours on any of the AGN associated parameters (e.g. BH mass, bolometric luminosity, Eddington ratio) for each of the AGN sub-classes, so we only present the mean colour-colour values for the total RLQ, RQQs and RG populations. This result, along with the similarity between the AGN SPIRE colours and the model, indicates that SPIRE bands are not significantly affected by emission from the torus (or hot dust surrounding the AGN). Although both quasars' and radio galaxies' mean colours lie inside the model tracks, the mean colour of the RGs is shifted from that of the bulk of the model galaxies and the quasars, indicating that it is possible that RGs are associated with redder colours, and therefore cooler dust, or be affected by synchrotron contamination.

Similarly, Fig.~\ref{fig:FIR_colors} (middle) shows that the 160-$\mu$m band does not suffer from torus emission contamination, as the quasars' and RGs' $S_{160}/S_{250}$ colours are similar to those of the models. We find that a few individually detected sources lie outside the model set of tracks. However, these outlier sources might be caused by the fractionally larger flux errors of the PACS band, or some of them (mainly RQQs; see blue outliers top panel) are associated with very strong 350-$\mu$m emission suggesting colder dust temperatures than the mean QSO population. By and large, most AGNs can safely be assumed to be dominated by cool dust emission in the SPIRE and 160 $\mu$m FIR bands. As in the top panel, the mean $S_{250}/S_{350}$ for the RGs indicates colder dust temperatures. Again, we find no significant dependence of $S_{160}/S_{250}$ on any of the AGN associated parameters.

In contrast, when we examine the PACS 70~$\mu$m colour, we find that most of the individually FIR detected AGN and the stacked mean colours lie outside th limits defined by our model, suggesting that the PACS 70-$\mu$m band may be significantly contaminated by AGN emission. In the Fig.~\ref{fig:FIR_colors} (bottom) the $S_{160}/S_{70}$ - $S_{250}/S_{160}$ colour-colour diagram for SPIRE 250~$\mu$m and PACS bands of our sample are shown. Although the fractionally larger PACS flux errors could explain some of these outliers, it is possible that some of these sources require a second, warmer dust component \citep[e.g.][]{Dunne2001}. For the low and high bolometric luminosity sub-samples there is a clear separation, despite the large error bars, in the $S_{160}/S_{70}$ colours. This difference seems to arise from the AGN contamination at 70~$\mu$m ($\sim35~\mu$m at the rest-frame). Indeed, at the redshifts of our sample, the PACS 70~$\mu$m contains the longer wavelengths of the torus emission \citep[e.g.][]{Mullaney2010,Xue2010}. If this is the case, then the strong correlation between the 70-$\mu$m emission and AGN emission found for powerful AGN \citep[e.g.][]{Dicken2009} could be explained by the heavy torus contamination. For this reason, the 70-$\mu$m emission is not used for the FIR luminosity calculation (see Section~\ref{section:4_SED_fitting}). On the other hand, the $S_{250}/S_{160}$ ratio seems to be unaffected by the AGN emission, indicating that 160-$\mu$m emission is largely generated by cold dust, heated by star formation.

\subsection{SED fitting} \label{section:4_SED_fitting}

As discussed in the previous section, we expect that the rest-frame FIR emission (160-500~$\mu$m) is mainly generated by cold dust heated by star formation in the AGN host galaxy. Therefore, we interpret the FIR emission as being powered by star formation \citep[e.g.][]{Rowan1995,Schweitzer2006,Netzer2007}, and we represent it with a black-body modified by frequency-dependent emissivity component \citep{Hildebrand1983}, given by
\begin{equation}
S_{\nu}\propto B_{\nu}(T)\nu^{\beta},
\end{equation}
where $B_{\nu}$ is the Planck function, $T$ is the effective dust temperature and $\beta$ is the dust emissivity index. Since $T$ and $\beta$ are degenerate for sparsely sampled SEDs, we reduced the numbers of free parameters by fixing the dust emissivity. Using a range $1.4<\beta<2.2$ (see \citealp[e.g.][]{Dye2010,Hardcastle2010,Smith2013}) we find that the best-fitting model returns lower $\chi^{2}$ values for a fixed $\beta=1.6$ dust emissivity for all AGN populations in the sample. The selection of $\beta=1.6$ is consistent with the work of \cite{Dye2010}. The remaining two free parameters are the cold dust temperature, which we have varied over the range $10<T({\rm K})<60$ and the flux normalization of the modified black-body component.

\begin{figure}
\includegraphics[scale=0.415]{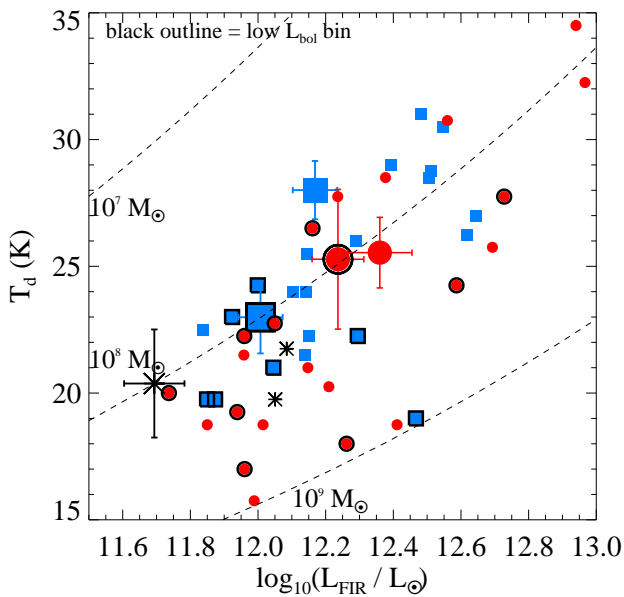}
\caption{FIR luminosity (${\rm L_{FIR}}$) versus dust temperature ($T_{\rm d}$) for individually FIR detected AGN and weighted mean values when the RLQs and RQQs are divided into bolometric luminosity bins. For RGs, we present the weighted mean values for the total population as all the sources but 4 belong to the low bolometric luminosity bin. Black outline indicates the sources and the weighted mean values for the low bolometric luminosity bin. Colours and symbols are similar to Fig.~\ref{fig:FIR_colors}. The black lines correspond to the dust mass ($M_{\rm d}$) estimates based on the ${\rm L_{FIR}}$ - $T_{\rm d}$ relation ${\rm L_{FIR}}\propto M_{\rm d}T_{\rm d}^{4+\beta}$, assuming $\beta=1.6$, for dust masses of $10^{7}$, $10^{8}$ and $10^{9}~{\rm M_{\odot}}$.}
\label{fig:all_together}
\end{figure}

For each source we estimated the integrated FIR luminosity ($8 - 1000~\mu$m) using a modified black-body fitting with the best fit temperature. The dust temperature was obtained from the best fit model derived from minimization of the $\chi^{2}$ values. The uncertainty in the measurement was obtained by mapping the $\Delta \chi^{2}$ error ellipse, allowing the individual photometric measurements to vary within their $1\sigma$ ranges of uncertainty. In addition to the integrated FIR luminosity we calculate the mass of the FIR emitting dust component using
\begin{equation}
M_{\rm dust}=\frac{1}{1+z}\frac{S_{250}D_{L}^{2}}{\kappa B(\nu,T)},
\end{equation}
where $S_{250}$ is the 250 $\mu$m observed flux, $D_{L}$ is the luminosity distance, $\kappa$ is the dust mass absorption coefficient, which \cite{Dunne2011} take to be $0.89~{\rm m^{2}~kg^{-1}}$ and $B(\nu, T)$ is the Planck function.

In the case of RGs and RLQs, we also extend the modified black-body model to the radio bands with either a power-law slope $S_{\nu}\propto \nu ^{-\alpha}$, with $\alpha$ estimated from 325 MHz and 1.4 GHz radio observations or, a broken power-law for the RLQs with available SMA observations at 1300~$\mu$m. In the second case, the broken point is fixed at the 1.4 GHz. Examples of the SED fits are presented in Fig.~\ref{fig:SED_appendix}.

As the majority of the sources are undetected at the $3\sigma$ limit in all {\it Herschel} bands, in addition to probing the properties of the individually FIR-detected objects, we carry out two different stacking approaches for the estimation of the FIR luminosities. In the first approach, we follow the method of \cite{Hardcastle2010,Hardcastle2013} regarding the consideration of the undetected sources ($<3\sigma$) in our sample. We determine the luminosity of each source from the {\it Herschel} flux densities (excluding 70 $\mu$m), even if negative, on the grounds that this is the maximum-likelihood estimator of the true luminosity, withouth making any assumption for their distribution in contrast to \cite{Hardcastle2010}. We then take the weighted mean of the parameter we are interested in within each bin. For the mean calculation, the luminosity is weighted using the errors calculated from $\Delta \chi^{2}= 2.3$ and the errors on the stacked parameters are determined using the bootstrap method. We use the same bins across the AGN sub-classes in order to facilitate comparisons. In the second approach, we consider the FIR upper limits of each source as tentative detections, and estimate upper limits for the ${\rm L_{FIR}}$ using the procedure adopted for the objects detected in {\it Herschel} bands. The motivation for the second approach is the comparison of our results with recent works that follow similar statistical analysis \citep[e.g.][]{Drouart2014,Podigachoski2015}. The mean far-infrared luminosities for both stacking and statistical methods are given in Appendix B. We found that our main results are consistent with the results we obtain when using the direct stacking analysis (Section 3.3). As also found in \cite{Kalfountzou2014a}, we found small but insignificant differences between the two methods, so for convenience we present here only the results of the weighted first approach. For the estimation of the mean FIR luminosity we again use two approaches. The first one is a weighted mean, each FIR luminosity is using the errors calculated from the $\Delta \chi^{2}= 2.3$ of the fitting. The second one is a simple median. Both are in a good agreement as we present in the following plots.

Fig.~\ref{fig:all_together} shows the FIR luminosity and dust temperature ($T_{\rm d}$) plane divided into dust mass ($M_{\rm d}$) regions based on the $L_{\rm FIR}\propto M_{\rm d}T_{\rm d}^{4+\beta}$, assuming $\beta=1.6$, for the FIR-detected AGN of our sample (similar cuts to Fig.~\ref{fig:FIR_colors} top) and the weighted mean values for the total sample and for each sub-class. The sources have been additionally divided into bolometric luminosity bins as specified in Table~\ref{Table:mean_fluxes}. Both types of quasars show high FIR luminosity with most of the detected sources and the weighted mean values having $L_{\rm FIR}>10^{12}~{\rm L_{\odot}}$, characterizing them as ultra-luminous infrared galaxies (ULIRGs). The weighted mean FIR luminosity of the RGs is significantly lower, even compared to the low bolometric luminosity quasars. Similar differences are also found for low BH mass and Eddington ratio bins. Comparing the FIR luminosity of the RLQs and RQQs it is notable that RLQs have higher weighted mean FIR luminosity than RQQs in both bolometric luminosity bins at $>1\sigma$ level with a significance of $p=0.014$. Similar trends are also found for BH masses and Eddington ratio.

As already indicated from the colour-colour plots, RGs show lower dust temperatures than both RLQs and RQQs (by about 5K) at a significance level of $p=0.036$ and $p<0.001$, respectively, under a K-S test. For all AGN subclasses and bins, the weighted mean values follow the $10^{8}~{\rm M_{\odot}}$ dust mass curve, with the exception of high bolometric luminosity RQQs which have slightly lower weighted mean dust mass (and higher dust temperature). Most of the FIR-detected RLQs lie between the $10^{8-9}~{\rm M_{\odot}}$ dust mass curves. This mass range is comparable to that obtained for submillimetre galaxies \citep[e.g.][]{Santini2010} at similar redshifts to our sample.

\subsection{FIR-radio correlation} \label{section:FIR-radio correlation}

In this section we determine whether some of the radio emission could be the result of star formation, rather than AGN activity, by comparing the observed radio flux with that predicted from the FIR/radio correlation. As the high detection rates and the weighted mean FIR flux densities in the RLQ sample indicate, almost 50 per cent of the population is expected to have high star formation activity. High star formation activity, at the level of ${\rm L_{FIR}}>10^{11}L_{\odot}$, could result in radio emission up to $10^{24}~{\rm W~Hz^{-1}}$ at 1.4 GHz, which is the detection level of our RLQs. We additionally investigate whether radio excess (i.e. radio emission associated with radio jets) correlates with star formation as one would expect assuming a jet-induced star formation (positive feedback) model.

We calculate the ratio between the IR and radio emission ($q$) using the definition given by \cite{Helou1985}
\begin{equation}
q=\log[f_{\rm FIR}/(3.75\times 10^{12}~{\rm Hz})]-\log[S_{\nu}({\rm 1.4~GHz})]
\end{equation}
where $f_{\rm FIR}$ is in units of ${\rm W~m^{-2}}$, determined from the {\it Herschel} photometry and $S_{\nu}(1.4~{\rm GHz})$ is rest-frame 1.4 GHz radio flux density in units of ${\rm W~m^{-2}~Hz^{-1}}$. We extrapolate the above relation to 325 MHz using the power-law slope $S_{\nu}\propto \nu^{-\alpha}$, with $\alpha=0.7$, typical for star-forming galaxies \citep[e.g.][]{Ibar2009,Ibar2010,Condon2013}.

\begin{figure}
\includegraphics[scale=0.415]{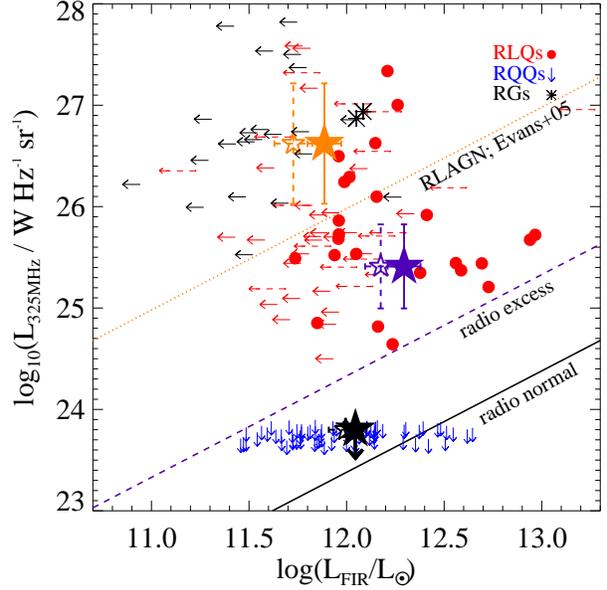}
\caption{FIR luminosity ($L_{\rm FIR}$) vs. radio 350 MHz luminosity ($L_{\rm 325MHz}$); the orange dotted line corresponds to $q=-0.38$ (average for a sample of radio-loud AGN from \citealp{Evans2005}); the black solid line corresponds to $q=2.2$ (average for `radio-normal' sources) and the purple dashed line corresponds to $q=1.68$, our selection limit for radio-excess sources. The large filled stars are the weighted mean values of all sources in each region while small open stars with dashed error bars are the weighted median values. No significant differences are found between mean and median estimations. Colours are associated to the lines. For individually FIR detected sources, colours and symbols are similar to Fig.~\ref{fig:FIR_colors}. FIR-undetected sources are presented as upper limits.  We also include FIR luminosity upper limits (dashed red upper limits) for the 17 RLQs that have been excluded from the sample due to the significant synchrotron contamination at the FIR bands.}
\label{fig:radio_FIR}
\end{figure}

In Fig.~\ref{fig:radio_FIR} we show the FIR and the radio 325-MHz luminosities for all of the RQQs (blue upper limits), RLQs (red circles and upper limits for FIR-undetected sources) and RGs (black point stars and upper limits for FIR-undetected sources) in our sample. The diagonal lines represent the mean $q=2.2$ value typically obtained for star-forming/starburst galaxies \citep[e.g.][]{Helou1985} and also typical radio-quiet AGN \citep[e.g.][]{Padovani2011,Sargent2010} and the mean $q=-0.38$ for a sample of radio-loud AGN from \cite{Evans2005}. For the separation between `radio-normal' and radio-excess sources we have picked the mean $q=1.2$ value that perfectly separates the RLQs and RQQs in our sample. We note that this is a conservative value compared to previous works (e.g. $q=1.68$; \citealp{Delmoro2013}, $q_{\rm max}=1.5$; \citealp{Hardcastle2010}, $q_{\rm max}=1.1$; \citealp{Jarvis2010}) indicating that above this limit we are predominantly detecting genuine radio-loud AGN.

The average upper limit $q$ for the RQQs lies near to the `radio-normal' diagonal line, taking into account that all the RQQs radio luminosities shown are the $5\sigma$ limits. All but one of the RGs in our sample is found above the RLAGN $q=-0.38$ diagonal line (orange dotted line) while about 70 percent of the RLQs lie in a region between those occupied by RGs and RQQs. This is consistent with the selection of the RG and RLQ samples. It is clear that the level of radio emission from star formation is insignificant for both RLQs and RGs. All radio-sources are found above the $q=-1.68$ diagonal line suggesting that the radio emission associated with star formation may contribute by a maximum of 10 per cent for the least radio-luminous RLQs.

For each region in Fig.~\ref{fig:radio_FIR}, we have estimated the weighted mean FIR luminosity, represented by the large filled stars; orange for the sources in the RLAGN region, purple for the sources in the radio-excess region and black for the radio normal region. As expected based on Fig.~\ref{fig:all_together} the objects in the radio-excess region, which consists only of RLQs, show a higher FIR luminosity. In contrast, the weighted mean FIR luminosity at the RLAGN region is lower than that in both the `radio-normal' and radio-excess regions. Although the RLQs in this region are associated with higher FIR luminosities compared to the RGs, and have about a 50 per cent detection rate, their individually measured FIR luminosities do not exceed the weighted mean FIR luminosity of the radio excess region. Weighted mean values of the total radio population, including both RGs and RLQs, show an anti-correlation between FIR and radio luminosity. Sources with higher radio luminosity show weaker star formation. We can investigate the apparent anti-correlation further considering the individual sources, although the numerous upper limits might affect the the establishment of such a correlation. In order to consider also the sources with FIR upper limits, we use Kendall's Tau statistical test. For this, the IRAF statistics package, which implements the Astronomical Survival Analysis programs (see \citealp{Feigelson1985,Lavalley1992}), was used. This test examines the null hypothesis that no correlation is present between the two variables being tested. For the total radio sample the generalized Kendall's correlation coefficient is $\tau=-0.13$ with a null hypothesis probability of $p=0.12$, implying no significant correlation. The same trend is observed even if we use, instead, a more outlier-resistant averaging such as the median (open stars). 

These results are not affected by the exclusion of the 17 RLQs with strong synchrotron contamination, as it is clear that they follow similar trends with the general RLQs population (see dashed red upper limits in Fig.~\ref{fig:radio_FIR}). In addition the mean radio luminosity of the rejected RLQs is $10^{26.10\pm0.47}~{\rm W~Hz^{-1}~sr^{-1}}$, very similar to the included RLQs population ($10^{25.84\pm0.41}~{\rm W~Hz^{-1}~sr^{-1}}$). Thus, we do not expect that the rejected sample would affect differently the two regions separated by the \cite{Evans2005} line. Although the upper limit FIR luminosity estimation for the excluded RLQs would be insufficient to draw firm conclusion, especially on account of the strong blazar variability, we could follow a different approach in order to insure that we do not introduce any selection biases rejecting these sources. Assuming that the dust temperature distribution of the excluded sample should be similar to the included RLQs, we use the $250\mu$m flux density, which should be the least contaminated from the synchrotron emission, as a proxy of the FIR luminosity. We found that both the included and the excluded RLQs have very similar median $250\mu$m flux densities, $20.06\pm4.14$~mJy and $18.46\pm9.12$~mJy, respectively. If we also consider the excluded RLQs sample for the estimation of the median $250\mu$m flux densities at the `RGs+RLQs' and `RLQs' regions the radio-luminous sources above the \cite{Evans2005} line show significantly lower median $250~\mu$m flux densities, $10.55\pm1.90$~mJy and $19.53\pm6.51$~mJy, respectively.

To check for the robustness of the differences in the mean FIR luminosity values between the different classes, we carried out the non-parametric Mann-Whitney U test (hereafter referred to as the M-W U test). The M-W U test allows the comparison of two groups without the underlying distribution of the data being necessarily normal. The FIR luminosities of the two groups are significantly different at a $>98.5$ per cent confidence level. In order to account for the upper limits in our sample, we also use statistical methods that are often generalizations of these classical non-parametric test. We use the astronomical survival analysis package (ASURV; \citealp{Feigelson1985}). Using three different tests, the Gehan’s Generalized Wilcoxon test; the log-rank test; and the Peto and Peto Generalized Wilcoxon test, the difference in the RLs and RLQs distributions of the FIR luminosity is confirmed at $>98.9$, $>99.3$, and $>98.9$ per cent confidence level, respectively.

The fact that high radio luminosity RGs and RLQs (see orange star; Fig.~\ref{fig:radio_FIR}), are associated with lower FIR luminosity compared to lower radio luminosity RLQs (purple star; Fig.~\ref{fig:radio_FIR}) may indicate two possible physical scenarios. In the first scenario, we can assume that there is a radio-jet power limit above which radio jets suppress the star formation in the host galaxy. That would be consistent with the negative radio-jet feedback scenario \citep[e.g.][]{Croton2006}. In contrast, lower power radio-jets might enhance the star formation (positive feedback) and that would explain the FIR excess between RQQs and RLQs with intermediate radio luminosity, the ones found in the radio-excess region. However, we should expect that these processes are controlled by the gas availability (i.e. galaxy mass). Indeed, RLQs with similar radio luminosities to the RGs have higher FIR luminosities and higher black hole masses. Therefore, we expect that they are hosted by galaxies with larger masses, assuming that the Maggorian relation holds. Although this interpretation could explain the observed differences, the effects of the radio jets cannot be so straightforwardly understood unless we control for galaxy mass. We discuss this scenario further in Section~\ref{section:RLQs_RGs}. Another important parameter is the environment of these sources, which can lead to a second possible scenario. Taking into consideration that the RGs have been selected from radio surveys whereas the RLQs are optically selected, we might have picked the two populations in either different evolutionary stages or different environments (see the discussion in Section~\ref{section:RLQs_RGs}). This fact could drive the apparent lower FIR luminosity when we consider both RGs and RLQs.

\section{The star formation dependence on AGN activity} \label{section:SFR_AGN}

\begin{figure}
\includegraphics[scale=0.415]{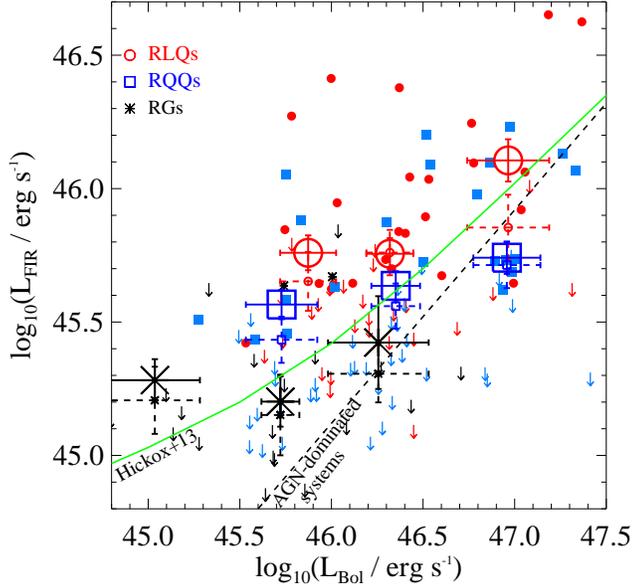}
\caption{FIR luminosity ($L_{\rm FIR}$) vs. bolometric luminosity ($L_{\rm bol}$) for each AGN class in 3 $L_{\rm bol}$ bins. For RLQs and RQQs each bin contains about 18 and 23 sources, respectively, while for RGs about 9 sources. The large symbols (red circles, blue squares and black stars) are the weighted mean values of each bin for RLQs, RQQs and RGs, respectively. The small open symbols with dashed error bars are the weighed median values. No significant differences are found when the median values are used, apart from the last RLQs bin. For individual sources, colours and symbols are similar to Fig.~\ref{fig:FIR_colors}. FIR undetected sources are presented as upper limits. The dashed black line is the correlation line shown by AGN-dominated systems in Netzer et al. (2009). The solid green line show the predictions of the Hickox et al. (2014) fiducial model on BH variability and star formation-AGN connection.}
\label{fig:LFIRvsLAGN}
\end{figure}

Using measurements of FIR luminosity, we will now study the relationship between FIR emission and SMBH accretion. In Fig.~\ref{fig:LFIRvsLAGN}, the FIR luminosity is plotted against bolometric luminosity $L_{\rm bol}$ with symbols representing both FIR luminosities for individually FIR-detected sources (small symbols) and weighted mean (large open symbols) and median (small open symbols with dashed error bars) values. The $L_{\rm bol}$ is the mean value for the objects in each bin with their associated $1\sigma$ error bars. Different colours are used to represent the different AGN classes. A crucial point of our results is that about 30 per cent of our QSOs are FIR-detected, indicating high FIR luminosities at the level of $L_{\rm FIR} \gtrsim 10^{12} L_{\odot}$. The high FIR emission suggests that starburst activity in 30 per cent of our QSOs has not been quenched yet. These results argue for a scenario in which powerful quasars, on average, have not yet suppressed the star formation in the host galaxy (see \citealp[e.g.][]{Harrison2012,Rosario2013,Stanley2015} but also \citealp[e.g.][]{Page2012}).

To search for possible trends between bolometric and FIR luminosity we performed a correlation analysis on each of the sub-samples. In order to take account of the sources with FIR upper limits, we use the Kendall's Tau statistical test as described in Section~\ref{section:FIR-radio correlation}. This test examines the null hypothesis that no correlation is present between the two variables being tested. The correlation analysis returns $\tau=0.34$ ($p=0.02$), $\tau=0.28$ ($p=0.02$) and $\tau=0.15$ ($p=0.32$) for RLQs, RQQs and RGs, respectively, suggesting a moderately significant correlation over more than 2 orders of magnitude in $L_{\rm bol}$ for both RLQs and RQQs. For RGs no significant correlation is observed over $\lesssim 2$ orders of magnitude.

A correlation between AGN luminosity and host galaxy star formation rate has been reported by several studies of high redshift AGNs and QSOs \citep[e.g][]{Lutz2008,Shao2010,Rosario2012,Rovilos2012}. \cite{Netzer2007} found for luminous PG QSOs that this relationship  has a slope of $\alpha\approx0.8$ (see black dotted line in Fig.~\ref{fig:LFIRvsLAGN}). Consistent slopes have also been suggested by other authors for mm-bright QSOs at $z\sim2$ \citep[e.g.][]{Lutz2008} and X-ray AGN \citep[e.g.][]{Rosario2012} at least for high AGN luminosities. We note that \cite{Rosario2012} suggested a flatter or even zero slope at low AGN luminosities ($L_{\rm AGN} <10^{44}~{\rm erg~s^{-1}}$). As these works have selected their AGN samples without any use of radio information, we expect that they are dominated by radio-quiet AGN. Radio-loud AGN are expected to make up to 10 per cent of uniformly selected AGN samples, so they should not significantly affect the estimation of these works.

The RQQs of our sample are very similar to that of \cite{Netzer2007}, with a good overlap on AGN luminosity up to $L_{\rm AGN} \sim 10^{46.5}~{\rm erg~s^{-1}}$, while our sample extends to about an order of magnitude higher in AGN luminosity. The correlation between the FIR and AGN luminosity based on the \cite{Netzer2007} QSO sample is presented in Fig.~\ref{fig:LFIRvsLAGN}. One important difference is the redshift range of the two samples, with the QSOs of \cite{Netzer2007} having $z<0.3$. Notwithstanding this difference, the selection of our sample in a narrow redshift range decouples the evolution effect and makes it perfect for comparison to either lower or higher redshift samples. One can immediately notice from Fig.~\ref{fig:LFIRvsLAGN} that the $L_{\rm AGN}$-$L_{\rm FIR}$ correlation is much weaker and flatter than the one proposed by \cite{Netzer2007}. The correlation slope for the RQQs of this work is found to be $\alpha\approx0.26\pm0.06$. Specifically, the lower $L_{\rm AGN}$ sources in our sample show a weighted mean $L_{\rm FIR}$ of one order of magnitude higher than that implied by the correlation of \cite{Netzer2007} while at higher $L_{\rm AGN}$ they are in better agreement. Such an increase of the FIR luminosity at a fixed AGN luminosity bin with redshift has been suggested by other authors (\citealp{Rosario2012}; about 0.7 dex from $z\sim0.3$ to $z\sim1$ AGN) and it would explain the FIR luminosity difference between our sample and that of \cite{Netzer2007} in fixed $L_{\rm AGN}$ bins. On the other hand, QSO selection at lower redshifts (e.g. $z<0.3$), where the star formation density in the universe is very low, might be affected by Malmquist bias. A similar trend for shallower slope ($\alpha=0.58\pm0.18$) at similar $L_{\rm AGN}$ and redshift but for X-ray AGN was suggested by \cite{Rosario2012} although the quality of the fit is rather poor. Even in this case, our data suggest a much shallower slope ($\alpha\approx0.26\pm0.06$) for the RQQs. Note that even if we include the RLQs the estimated slope can reach a maximum of $\sim0.32$.

Selection effects that arise from flux limited surveys could influence the observed AGN - FIR luminosity slope \citep[e.g.][]{Shen2008,Schulze2010,Steinhardt2010} as the black hole mass is correlated with the stellar mass and this, in turn, with the SFR. The lower luminosity QSOs in our sample contain more systems with lower black hole masses, and thus lower stellar masses which might explain the lower FIR luminosity.

Luminous, high-redshift quasars typically yield lifetimes for luminous accretion of $\sim 10^{6}-10^{7}$ years \citep[e.g.][]{Hopkins2005,Goncalves2008,Shankar2010}. On the other hand, galactic-scale star formation has a dynamical time of around $10^{8}$ yr. In addition, as FIR emission arises mostly from dust that can be heated by both young and old stars, it can average over timescales of tens to hundreds of Myr, especially in galaxies with star formation at a relatively steady rate over their lifetime. Thus, the weak observed correlation between star formation and BH accretion might be attributed to the timescale difference between the AGN accretion efficiency and star formation variability \citep[e.g.][]{Mullaney2012,Chen2013,Hickox2014}.

From a theoretical point of view, \cite{Hickox2014} suggested a simple model in which accretion and star formation are perfectly connected, but this connection is `hidden' by short-timescale AGN variability over a large dynamic range (see Fig.~\ref{fig:LFIRvsLAGN} green solid line). Despite the fact that the model goes through our data points for the individual QSOs, the mean measurements are systematically off-set. Although the model of \cite{Hickox2014} describes well the lack of a strong correlation between $L_{\rm FIR}$ and $L_{\rm AGN}$ for moderate-luminosity AGN and the shift to higher $L_{\rm FIR}$ with redshift as suggested by observational results, it suggests a strong correlation between $L_{\rm FIR}$ and $L_{\rm AGN}$ at high luminosities, in contrast to our results. However, the apparent disagreement could arise from limitations in the simplistic AGN variability model (for a discussion see \citealp{Hickox2014}) or from the fact that our sample contains exclusively powerful QSOs with high accretion rates. An alternative model, suggested by \cite{Aird2013}, assumes that the probability of a galaxy hosting an AGN is determined by a universal specific accretion rate distribution that is independent of host stellar mass or star formation properties. This model would be consistent with the observed weak AGN/star-formation correlation of this work even in the most luminous QSOs.

\begin{figure}
\includegraphics[scale=0.415]{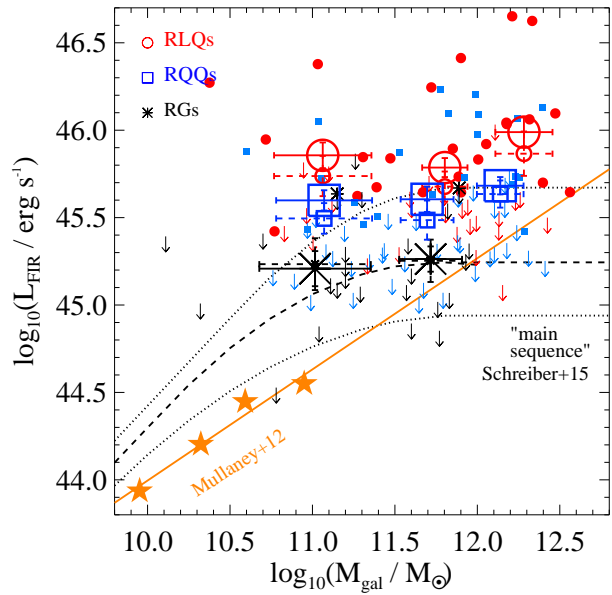}
\caption{FIR luminosity ($L_{\rm FIR}$) vs. galaxy mass ($M_{\rm gal}$) for each AGN class in 3 $L_{\rm bol}$ bins. Colours and symbols are similar to Fig.~\ref{fig:LFIRvsLAGN}. The large symbols (red circles, blue squares and black stars) are the weighted mean values of each bin for RLQs, RQQs and RGs, respectively. The small open symbols with dashed error bars are the weighed median values. No significant differences are found when the median values are used. Similar results are obtained even when the median values are used for each bin. The solid orange line and orange stars correspond to the average FIR luminosity and stellar mass for Mullaney et al. (2012) $z\sim1$ sample of star-forming galaxies. The black dashed line with the associated scatter (dotted black lines) corresponds to the expected ${\rm L_{FIR}} - M_{\rm gal}$ relation for $z\sim1$ as defined by Schreiber et al. (2015).}
\label{fig:FIR_Mgal}
\end{figure}

In Fig.~\ref{fig:FIR_Mgal} we present the average FIR luminosity of each AGN population, as a function of $M_{\rm gal}$. We have to note that the galaxy masses are not actual stellar mass measurements for our QSOs but they have been estimated based on the black hole measurements assuming a Magorrian relation. This fact may introduce high uncertainties. As expected under the assumption of a hidden QSO – star-formation correlation due to the different timescales of the two phases, we find no correlation for any AGN sub-class between FIR luminosity and stellar mass, in contrast to the result of \cite{Mullaney2012}. The most luminous AGN, like the ones in our sample, are generally missed from small field surveys. However, at redshift $0.9<z<1.1$ they make up to 10 per cent \citep[e.g.][]{Aird2010} of the total AGN population ($L_{\rm X}>10^{42}~{\rm erg~s^{-1}}$). Despite their large FIR excess, an order of magnitude in FIR luminosity, their rarity means that they might not significantly change the results found by \cite{Mullaney2012}.

Comparing the average SFRs of this sample to the observed relationship between SFR and stellar mass of normal star forming galaxies, which is known as the ``main sequence'' \citep[e.g.][]{Elbaz2007,Elbaz2011, Schreiber2015,Johnston2015} we can examine whether QSOs have SFRs that are consistent with being selected from the overall star forming galaxy population. To make this comparison we use the \cite{Schreiber2015} definition of the `main sequence' at $z\sim 1$ (see Eq. 9 of \citealp{Schreiber2015}). They found evidence for a flattening of the main sequence at high masses ($\log_{10}(M_{\rm gal}/M_{\odot}) > 10.5$), similar to the one observed for the sources in our sample. Note that they use stellar masses up to $\log_{10}(M_{\rm gal}/M_{\odot}) \approx 11.5$ to extract their model. Although the weighted mean FIR luminosity of the RGs (large point stars) is consistent with that of star forming galaxies of the same redshift and mass, the weighted mean FIR luminosity for QSOs is systematically higher than the higher end of the FIR luminosity region covered by `main sequence' galaxies. Similar results have recently been reported for luminous, optically selected quasars. This supports the statement that luminous AGNs are more likely to be associated with major mergers \citep[e.g.][]{Ma2015,Dong2016}.

\cite{Santini2012} have also reported that, on average, X-ray AGN hosts show somewhat enhanced star-formation activity with respect to a control sample of inactive galaxies, although they found them to be consistent with star forming galaxies. While different interpretations are possible, our findings are consistent with a scenario whereby periods of enhanced AGN activity and star-forming bursts are induced by major mergers \citep[e.g.][]{DeBreuck2005, Elbaz2011, Sargent2013}.

X-ray and FIR observations have been widely used for the investigation and comparison of the star formation activity in distant AGN and star-forming galaxies \citep[e.g.][]{Shao2010,Harrison2012,Mullaney2012,Rovilos2012,Santini2012,Rosario2013}. They suggest that the sSFRs of AGN are consistent with those of star-forming galaxies, with possible exceptions the luminous AGN \citep[e.g.][]{Rovilos2012,Treister2012}. Most of these X-ray works investigate the mean SFRs of AGNs that are less luminous than those studied here, and should also include a high fraction of narrow-emission line AGN. \cite{Rosario2013} suggested a baseline model for X-ray broad-emission line QSOs based on which moderate luminosity QSOs are hosted by galaxies that lie on the star-forming mass sequence. 

In the case of the RQQs in this work, which are similar only to the most luminous sources of \cite{Rosario2013}, we have found a mean positive offset of $\sim0.4$~dex in $\log L_{\rm FIR}$, which corresponds to the upper limit of the region covered by ``main sequence'' galaxies. As we see in Fig.~\ref{fig:FIR_Mgal}, the FIR-detected quasars at the $3\sigma$ level (about 30\%) are mainly responsible for the SF enhancements compared to the star-forming galaxies, while the upper limits of the FIR-undetected quasars lie well inside the ``main sequence''. A possible explanation for the differences between our observations and \cite{Rosario2013} baseline model could arise from the fact that a significant fraction of our RQQs is preferentially in ``special'' populations such as starburst or major mergers that are associated with higher star formation efficiency \citep[e.g][]{Daddi2010,Genzel2010}.

Indeed as some studies have suggested, the fraction of quasars hosted by mergers and/or interacting system is about 30 per cent from unreddended quasars \citep[e.g.][]{Dunlop2003, Floyd2004}, while for red quasars the merger fraction increases to $\sim85$ per cent. Assuming that the merger fraction might rises with bolometric luminosity \citep[e.g.][]{Hopkins2006,Urrutia2008, Somerville2008,Treister2012} that would explain why the mean SFR for our sample, which is more luminous than \cite{Rosario2013}, is higher than the one for the inactive galaxies. 

These studies imply that the host galaxies of moderate luminous AGN and the most luminous AGN might evolve along different paths. The low and moderate luminous AGNs are fueled by secular processes \citep[e.g.][]{Hopkins2006,Jogee2006,Younger2008}, while high luminosity AGNs evolve through major mergers and might have a direct link between the black hole growth and bulge growth.

\section{The dependence of star formation on radio jets}	\label{section:RLQs_RQQs}

\begin{figure}
\includegraphics[scale=0.415]{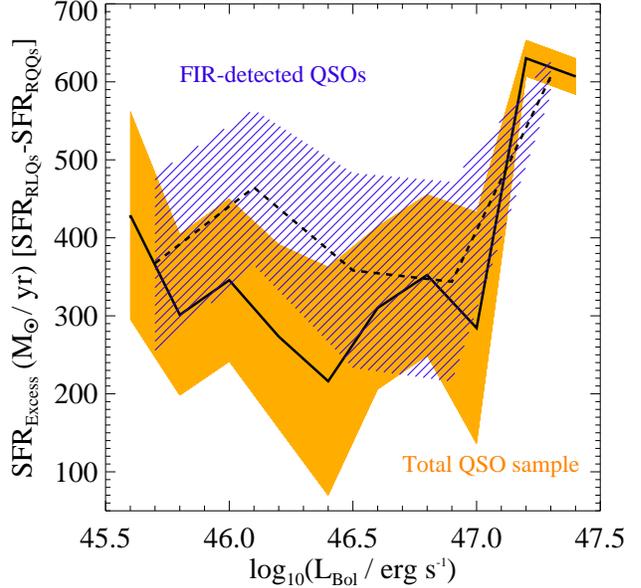}
\caption{SFR excess between RLQs and RQQs as a function of bolometric luminosity for the total QSO sample (orange) and the FIR detected QSO sample (blue). The solid and dashed lines represent the weighted mean SFR excess and the coloured areas the $1\sigma$ error for the total QSO sample and the FIR detected QSO sample, respectively.}
\label{fig:SFRexcess}
\end{figure}

In this section we will discuss the effect of the presence of radio jets in a QSO on star formation activity. As is already clear from the previous section (see Fig.~\ref{fig:LFIRvsLAGN}), RLQs are associated with higher FIR luminosity than RQQs. This excess is almost constant and independent of AGN properties. Fig.~\ref{fig:SFRexcess} shows the SFR excess, defined as the SFR difference between RLQs and RQQs, for the individual sources in each $L_{\rm bol}$ bin, taking into account the total population (orange filled area) or only the FIR detected QSOs (purple shaded area). Apart from the highest bolometric luminosity bin, where only a few sources are found, the SFR excess is almost constant with $\Delta{\rm SFR}\approx 315 ~{\rm M_{\odot}/yr}$ for the total sample and $\Delta{\rm SFR}\approx 380 ~{\rm M_{\odot}/yr}$ for the FIR-detected QSOs. This excess corresponds to about a factor of two. A similar increase in SFR due to the onset of radio jets has been suggested also by simulations of massive, gas-rich, high-redshift galaxies \citep{Silk2010,Gaibler2012}.

SFR-enhancing phases in RLQs can be caused due to the formation of bow shocks generated by the jet which compresses the interstellar medium (ISM). Jets create cocoons of turbulent gas surrounding the jet leading to a much more efficient clumping of molecular hydrogen and thus accelerated star formation \citep[e.g.][]{vanBreugel2004,Gaibler2012,Wagner2012,Ishibashi2012}. Simulations have shown that, although powerful jets' interaction with the ISM might be volume limited, the resulting pressure can impact the galactic disk also at larger radii and eventually all of the galaxy (see \citealp{Gaibler2012}). Thermal or kinetic AGN feedback is often thought to heat and expel most residual gas from the galaxy \citep[e.g.][]{Springel2005, Croton2006,Bower2008, Dubois2012}, reducing the SFR. On the contrary, our results suggest an entirely opposite effect, indicating the formation of an additional population of stars, compared to the RQQs. The need for additional enhancement of star formation has been recently suggested by \cite{Khochfar2011} for high-redshift galaxies ($z > 5$) who introduced stochastic boosts in star formation in order to reproduce the observations. Such enhancement could indeed be triggered by the radio jets in gas-rich galaxies; however, there are very few radio galaxies at $z > 5$ \citep[e.g.][]{Jarvis2000,Jarvis2001c,Wall2005,Rigby2011,Rigby2015}. Therefore, it is important to understand at which epochs and under which conditions radio jets can efficiently boost the host galaxy star formation. Near future synergies between optical spectroscopy (WHT Enhanced Area Velocity Explorer, WEAVE; \citealp{Dalton2012}) and radio continuum (e.g. Low Frequency Array, LOFAR; \citealp{vanHaarlem2013}) surveys will provide much greater sample sizes for radio AGN allowing more stringent constraints on the evolution of the radio population, out to greater redshifts (e.g. WEAVE-LOFAR; \citealp{Smith2015}).

In our previous work \citep{Kalfountzou2014a} we compared the SFR between RLQs and RQQs over a wide redshift range, up to $z\sim3$ with a couple of QSOs at even higher redshifts and we found an excess of $\leq100 ~{\rm M_{\odot}/yr}$ for RLQs with low bolometric luminosity and no difference at high bolometric luminosities. This excess corresponds to more than a factor of 2, but to much lower SFRs than the ones found here. We note that the vast majority of low bolometric QSOs in the K14b sample have $z<1.0$ while the high bolometric luminosity QSOs are found at much higher redshifts. The differences between these two studies give some evidence regarding the evolution of the jet-induced star-formation efficiency. As, in this work, we do not find any effect of bolometric luminosity on SFR excess, we assume that the results of K14b are associated with redshift evolution. In this case it would be possible that radio jets' positive feedback efficiency evolves with redshift, peaking at $z\approx1.0$ where we find the maximum SFR excess. However, both RLQs and RQQs may have more star formation at higher redshifts due to the same process as in normal galaxies \citep[e.g.][]{Madau2014}. Therefore, the enhancement of SFR due to radio jets to the normal SFR might be smaller and harder to detect at higher redshifts. Galaxies in which the conditions for positive feedback by radio-jets may be optimal at $z\approx1.0$ might be associated with the radio-AGN evolution that shows a monotonic increase in space density with redshift out to $\sim1.0$ (with a radio luminosity dependence; \citealp{Rigby2011}), in line with the increasing space density of cosmic SFR \citep[e.g.][]{Best2014,Madau2014}. Indeed, a consistent picture emerges whereby the availability of a cold gas supply regulates both the radiative-mode AGN and star formation activity \citep[e.g.][]{Hardcastle2007,Heckman2014}.

While this work is consistent with positive feedback, we should be aware of selection effects and the conditions under which radio jets would enhance the star-formation. For example, our sample consists of very massive QSOs with high SFR even in the case of RQQs. The high SFRs would suggest that these QSOs might have gone through recent, major, gas-rich merger events indicating high gas supplies. Especially for RLQs, minor merger events might be more common as they are often associated with high density environments \citep[e.g.][]{Venemans2007,Falder2010,Kuiper2011}. Under these assumptions, radio jet feedback might depend on gas availability associated with the environment and cold gas supplies.

\section{Star formation in RG and RLQ} \label{section:RLQs_RGs}

It has been suspected from submillimetre studies that the hosts of powerful radio-loud AGN undergo brief episodes of intense star formation which increase with redshift \citep[e.g.][]{Archibald2001}. Using {\it Herschel} data, \cite{Seymour2011} found a mean SFR range of 80 to 600 ${\rm M_{\odot}~yr^{-1}}$ for $1.2<z<3.0$ radio-selected AGN. In the same context, \cite{Drouart2014} estimated SFRs of a few hundred to a few thousand solar masses per year for $1<z<5$ radio galaxies. Recently, \cite{Podigachoski2015}, comparing the SFR of 3C radio-loud AGN and radio quasars at $z>1$, found similar SFRs for the two classes and at the same levels with the previous works. The idea that the hosts of high-$z$ radio-loud AGN can form stars at high rates is consistent with the jet-induced star formation model. 

In this work, while we find that RLQs are associated with vigorous star formation activity, the RGs of this sample have significantly lower SFRs of about a factor of 2.5 for the same BH masses with only two FIR detected sources. \cite{Priddey2003} found quite similar differences (about a factor $\sim2$) using submillimetre observations of $1.5<z<3$ RQQs and RGs drawn from SCUBA surveys \citep{Archibald2001}. On the other hand, \cite{Isaak2002} suggested that these differences are far less marked at $z>4$.

The FIR-radio luminosity plane is presented in Fig.~\ref{fig:radio_FIR}. The RGs in our sample are associated with higher radio luminosities than RLQs (see Fig.~\ref{fig:Lradio_z}; almost all RGs have $\log_{10}(L_{325{\rm MHz}}/{\rm W~Hz^{-1}~sr^{-1}}) > 26.0$). Assuming that both RGs and RLQs emanate from the same parent population, we find that the FIR luminosities of the most radio luminous sources in Fig.~\ref{fig:radio_FIR} (see RLAGN region) are significantly lower than the radio sources with lower radio luminosities. However, no significant evidence is found regarding an anti-correlation between FIR and radio luminosity.

A possible interpretation of this result would be that star formation enhancement efficiency depends on the radio power, with powerful radio jets associated with negative feedback reducing the star formation in the host galaxy. In fact, radio jet pressure can be sufficiently large to expel significant quantities of gas from the galaxies \citep{Nesvadba2006}, thereby quenching the star formation \citep{Croton2006}. However, this interpretation should also depend on galaxy mass. Indeed, the fraction of radio-loud AGN is a strong function of stellar mass and redshift \citep[e.g.][]{Jiang2007,Donoso2009} suggesting that radio jet feedback predominantly occur in massive halos. Thus, we might expect that its influence will have the clearest signature in massive galaxies. Observational studies on this issue return controversial results \citep[e.g.][]{Nesvadba2010,Papadopoulos2010} with positive feedback being directly observed in a few local \citep[e.g.][]{Croft2006}, intermediate \citep[e.g.][]{Inskip2008} and high redshift sources \citep[e.g.][]{Dey1997,Bicknell2000}. The observed differences could be explained by the fact that galaxy masses vary strongly with redshift but also amongst radio galaxies at similar redshifts \citep[e.g.][]{Kauffmann2008}. In Fig.~\ref{fig:radio_FIR} we compare the SFR between RLQs and RGs with similar radio power (see RLAGN region) and it is clear that RLQs have higher FIR luminosity than RGs. That can be explained as a consequence of the RLQs in the RLAGN region apparently having higher galaxy masses than the RGs (e.g. see Fig.~\ref{fig:distri_mass} where RLQs are associated with higher black hole masses). Assuming the ${\rm L_{FIR}} - M_{\rm gal}$ relation for $z\sim1$ as defined by \cite{Schreiber2015} for the RLQs ($\langle M_{\rm gal}\rangle=10^{12.1}~M_{\odot}$) and the RGs ($\langle M_{\rm gal}\rangle=10^{11.6}~M_{\odot}$) in the RLAGN region we would expect a similar level of FIR luminosity. However, the large uncertainties of the ${\rm L_{FIR}} - M_{\rm gal}$ relation at $ M_{\rm gal}>10^{11.5}~M_{\odot}$, about $\pm0.3$ in $\log{\rm L_{FIR}}$, could explain the offset in FIR luminosity we observe in the RLAGN region. This conclusion arises from the assumption that the Magorrian relation holds both ways around.

The RG selection from radio surveys favours objects with the highest values of radio luminosity (i.e. jet power) explaining why the RGs in our sample are more radio luminous than the RLQs. On the other hand, the RLQ selection from both optical and radio surveys favours objects with both high jet power and bolometric luminosity (e.g. accretion rate). However, in all likelihood the quasars are probably biased towards bigger black holes due to the optical selection, as we are selecting on BH properties rather than host galaxy properties.

In order to explain the observed differences regarding the SFR in the two populations taking into account both the galaxy mass and jet power, we suggest a `toy model' in which there is some jet power threshold at which radio-jet feedback switches from enhancing star formation (by compressing gas) to suppressing it (by ejecting gas). Then that threshold will be dependent on both galaxy mass and jet power. In this model, the SFR enhancement (i.e. the level of SFR excess compared to a control sample of radio-quiet AGN with the same bolometric luminosity and galaxy mass) starts from zero for AGN without radio jets, has a mass-dependent peak as jet power increases, and then decreases gradually for higher jet power. 

The value of the model is that it can explain the differences between the SFRs estimates in different radio-power sources selected in different ways in recent studies \citep[e.g.][]{Seymour2011,Dicken2012,Magliocchetti2014,Podigachoski2015,Magliocchetti2016}. It also confirms and extends the high star-forming activity observed in the hosts of radio-active AGN selected by different methods \citep[e.g.][]{Hatziminaoglou2010,Santini2012,Rosario2013,Drouart2014}. For instance, \cite{Podigachoski2015} use a sample of radio-loud objects that is similar in many ways to the dataset in this work, though they target more radio luminous systems. In term of source selection, both \cite{Podigachoski2015} RGs and RLQs samples and our RGs are selected from bright radio surveys (e.g. the Revised Third Cambridge Catalogue of radio sources, hereafter 3CR; \citealp{Spinrad1985}). Comparing their radio luminosities, the 3CR sample have a $L_{325\rm{MHz}}\geq10^{26}~{\rm W~Hz^{-1}~sr^{-1}}$ limit which is similar to the one of the radio galaxies in this work (see Fig.~\ref{fig:Lradio_z}). Thus, our SFR estimations for the RG population are in perfect agreement ($L_{\rm FIR}\sim10^{11.7}~L_\odot$) to the ones found by \cite{Podigachoski2015}, especially for their FIR-undetected sample if we consider that almost all of our RGs are FIR-undetected. In addition, our RG systems are also found to have SFRs that are quite similar to inactive galaxies selected from the deeper {\it Herschel} surveys \citep[e.g.][]{Rosario2013}. However, this is not the case of the RLQs in this work. As we show in Fig.~\ref{fig:Lradio_z}, the RLQs' radio luminosity goes down to $L_{325\rm{MHz}}\sim10^{24.5}~{\rm W~Hz^{-1}~sr^{-1}}$ with the majority of the sources having $10^{25}-10^{26}~{\rm W~Hz^{-1}~sr^{-1}}$. As we suggest in our model, at the highest radio powers negative feedback could lead to an overall suppression in SFR. In this case, \cite{Podigachoski2015} results are in agreement with our model as both RGs and RLQs with similar radio luminosities share very similar SFRs, just like the RGs in this work. The fact that our RLQs could be characterized as moderate radio systems, at least compared to the RGs, can possible explain the reported SFRs differences. Apart for the radio-jet positive feedback which could have increased the SFRs in these systems, compared to the RGs and the inactive galaxies, the galaxy mass could also control somehow these results. Specifically, \cite{Podigachoski2015} assume a stellar mass range of $1.5\times10^{11}-6\times10^{11}~M_\odot$. This range is similar to the RGs of our sample (see Fig.~\ref{fig:distri_mass}) but our QSOs extend to higher stellar masses with a mean of $7.5\times10^{11}~M_\odot$.	

For the same high radio luminosity regime and $z<0.7$, \cite{Dicken2012} did not find a close link between starbursts and powerful radio-loud AGN using {\it Spitzer}/Infrared spectroscopy. On the other hand, \cite{Magliocchetti2016} recently found an intense star-forming activity in the majority of less luminous radio-selected AGN ($L_{\rm 1.4~GHz}<10^{25}~{\rm W~HZ^{-1}~sr^{-1}}$). A comparison to this work might be hard not only because the different source selection and redshift distribution but also because \cite{Magliocchetti2016} results arise from FIR-detected radio-selected AGN. However, our suggested model seems to be applicable even in this case. \cite{Magliocchetti2016} found that the IR luminosity distribution of their sources peaks at around $L_{\rm IR}=10^{12.5}~L_\odot$, slightly higher than our mean value for sources with similar stellar mass found in the radio-excess region in Fig.~\ref{fig:radio_FIR}.

In order to investigate how our observations fit in this `toy model', we use the RQQs as a control sample. We have separated the RQQs into 4 bolometric luminosity bins, with about the same number of sources ($\sim18$), and for each bin we estimated the weighted mean specific star-formation rate (sSFR) where the stellar masses are calculated as described in Section~\ref{section:masses}. From low to high bolometric luminosity bins we found $1.472\pm0.554$, $0.314\pm0.103$, $0.388\pm0.089$ and $0.180\pm0.039~{\rm Gyr^{-1}}$. Then, the sSFR of each RLQ and RG in our sample was normalized by the weighted mean sSFR from the RQQ control sample, depending on the bolometric luminosity of each source, in order to estimate the sSFR enchantment fraction associated with the radio jets. We prefer the use of sSFR instead of the SFR in order to account for the galaxy mass dependence in the `toy model'. We have excluded six RGs from this analysis with bolometric luminosities lower than the lower RQQ bolometric luminosity bin ($L_{\rm Bol}<10^{45.3}~{\rm erg~s^{-1}}$, see Fig.~\ref{fig:distri_mass} middle panel). We note that five of these RGs have been classified as LERGs by \cite{Fernandes2015}.

\begin{figure}
\centering
\includegraphics[scale=0.44]{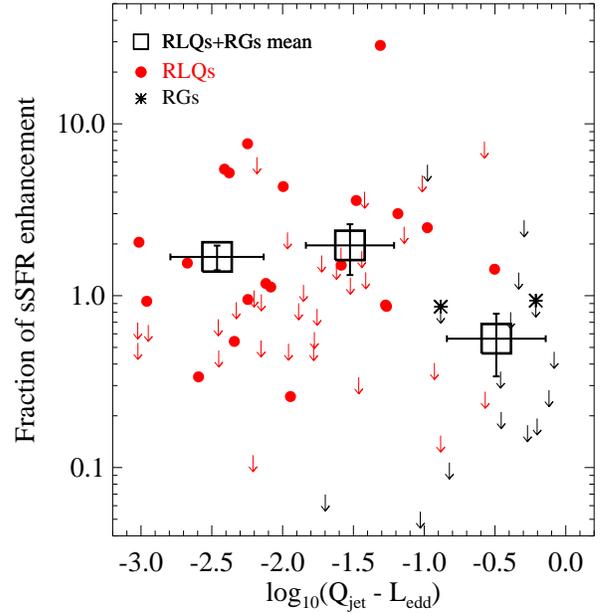}
\caption{Fraction of sSFR enhancement for RLQs and RGs, normalized to the control sample of RQQs AGN with the same bolometric luminosity, as a function of $Q_{\rm jet}/L_{\rm Edd}$ 
(jet power over Eddington luminosity). The large black squares are the weighted mean values for 
the three $Q_{\rm jet}/L_{\rm Edd}$ bins taking into account both RLQs and RGs in each bin. 
For individual sources, colours and symbols are similar to Fig.~\ref{fig:FIR_colors}. 
FIR undetected sources are presented as upper limits.}
\label{fig:model}
\end{figure}

In Fig.~\ref{fig:model}, we present the fraction of sSFR enhancement due to the radio jets as a function of jet power. As described above, we expect a mass-dependent peak, therefore we normalize the jet power to the Eddington luminosity (i.e. black hole or galaxy mass) in order to control for this dependence. Higher-mass galaxies will be able to hold on to their gas better for a given jet power, so there will be some mass-dependent threshold in jet power beyond which jets tend to have an increasingly suppressing effect on star formation. It seems that our observations follow the suggested `toy model' with sources at the low and intermediate jet power found at the peak of the star-formation enhancement, while at the highest jet power the radio sources have passed the jet power threshold at which radio-jet feedback switches from enhancing star formation to suppressing it. Indeed, the estimated mean sSFR suppresses fraction is $<1$ suggesting that powerful jets for a given galaxy mass suppress the star formation in the host galaxy compared to a radio-quiet source. Larger RG samples, covering a wide range of galaxy masses and radio luminosities, would provide us with additional observational constraints for our model.

Although the suggested model seems to explain the observations, we have to keep in mind that the star formation in the host galaxies of these RLQ and RG systems might be controlled by many additional parameters, like the environment and merger activity, which we expect to be quite common especially for the quasars in our sample \citep[e.g.][]{Santini2010,Kartaltepe2012,RamosAlmeida2012}. For instance, \cite{Stevens2003} presented submillimetre imaging of seven high-redshift RGs, several of which present spatially extended massive star-formation activity ($\sim30-150$ kpc), co-spatial in same cases with similarly extended UV emission \citep[e.g.][]{Hatch2008}. This suggests that the brightest submillimetre companions trace to the high-redshift RGs may trace a large-scale structure which would contain the densest cross-sections of gas. In this case, the very brightest radio sources in our sample, dominated by RGs due to the selection method, might be physically associated with over-dense regions. Therefore, the high jet power sources of our sample might have formed their stars at earlier epochs and we now observe them at a passive evolutionary stage.

\section{Conclusions} \label{section:Conclusions}

We have presented {\it Herschel} photometry of RLQs, RQQs and RGs selected at a single epoch, $z\approx1$. Combining the {\it Herschel} observations with SMA observations we performed a full radio-FIR SED analysis to investigate the non-thermal contamination to the FIR bands. SDSS data for the QSOs and mid-IR data for the RGs in our sample were used to estimate the AGN luminosity of each source. The FIR observations were used to estimate the SFR for the individually FIR detected sources and the stacked SFR for a variety of AGN and radio properties. We summarize the results below:

\begin{enumerate}
\item About 33 per cent (43/149) of the QSOs and 8 per cent (2/27) of the RGs have robust PACS and SPIRE detections. These detection rates are translated to ULIRG-like star formation luminosities suggesting SFRs of hundreds of solar masses per year.

\item SMA 1300 $\mu$m observations lead us to reject 17 RLQs in which the 500 $\mu$m flux may suffer significant synchrotron contamination. 

\item We find that about 40 per cent (22/57) of RLQs have robust FIR detections and 30 per cent (21/72) of RQQs. The SFRs of the FIR detected QSOs are higher than a simulated mass-matched galaxy sample supporting the scenario of a merger induced star formation activity. Additionally, the high SFRs and detection rates suggest that there is no clear evidence that the star formation has been quenched in the hosts of these powerful QSOs compared to the non-AGN galaxies. Although radio-jets can enhance the SFR in the RLQs compared to the RQQs, they are not the likely cause of the star formation as RQQ systems as still found with significantly high star formation activity.

\item The FIR luminosity does not show a strong correlation with the AGN luminosity or the stellar mass for any of the three sub-samples in contrast to what is expected for AGN-dominated systems. The lack of dependence on AGN luminosity might suggest that neither the QSO continuum is the cause of star formation activity in any of the AGN systems we studied in this work. A multi-wavelength SED for the measurement of the bolometric luminosity would improve the uncertainties arise from the $L_{12\mu{\rm m}}$ and $L_{3000}$ use for the $L_{\rm bol}$ calculation and their associated bolometric corrections, in order to confirm our results.

\item The RLQs are found to have a SFR excess of about $300~{\rm M_{\odot}~yr^{-1}}$ (a factor of 2.5) over RQQs of the same bolometric luminosity, similar to the one suggested from simulations in gas-rich radio-loud AGN \citep{Gaibler2012}.

\item Merger induced star formation activity is a possible mechanism leading to the SFRs obtained for  RQQs while radio-jet triggered star formation seems to be the likely cause for the SFR excess in RLQs compare to the AGN luminosity matched RQQ sample. It is expected that RGs' low detection rates are associated with the radio selection of the sample, suggesting the existence of a jet power threshold below which the radio jets enhance the star formation and above which they suppress the star formation in the host galaxy by ejecting gas.

\end{enumerate}

\section*{Acknowledgments}
The Submillimeter Array is a joint project between the Smithsonian Astrophysical Observatory and the Academia Sinica Institute of Astronomy and Astrophysics and is funded by the Smithsonian Institution and the Academia Sinica. MJH acknowledges support from the UK's Science and Technology Facilities Council [grant number ST/M001008/1].

\bibliography{QSOs_z1_arXiv.bib}

\begin{thebibliography}{218}
\expandafter\ifx\csname natexlab\endcsname\relax\def\natexlab#1{#1}\fi

\bibitem[{{Aird} {et~al}\mbox{.}(2013){Aird}, {Coil}, {Moustakas},
  {Diamond-Stanic}, {Blanton}, {Cool}, {Eisenstein}, {Wong}, \&
  {Zhu}}]{Aird2013}
{Aird} J. {et~al.}, 2013, \apj, 775, 41

\bibitem[{{Aird} {et~al}\mbox{.}(2010){Aird}, {Nandra}, {Laird}, {Georgakakis},
  {Ashby}, {Barmby}, {Coil}, {Huang}, {Koekemoer}, {Steidel}, \&
  {Willmer}}]{Aird2010}
{Aird} J. {et~al.}, 2010, \mnras, 401, 2531

\bibitem[{{Alexander} {et~al}\mbox{.}(2005){Alexander}, {Bauer}, {Chapman},
  {Smail}, {Blain}, {Brandt}, \& {Ivison}}]{Alexander2005}
{Alexander} D.~M., {Bauer} F.~E., {Chapman} S.~C., {Smail} I., {Blain} A.~W.,
  {Brandt} W.~N., {Ivison} R.~J., 2005, \apj, 632, 736

\bibitem[{{Amblard} {et~al}\mbox{.}(2010){Amblard}, {Cooray}, {Serra}, {Temi},
  {Barton}, {Negrello}, {Auld}, {Baes}, {Baldry}, {Bamford}, {Blain}, {Bock},
  {Bonfield}, {Burgarella}, {Buttiglione}, {Cameron}, {Cava}, {Clements},
  {Croom}, {Dariush}, {de Zotti}, {Driver}, {Dunlop}, {Dunne}, {Dye}, {Eales},
  {Frayer}, {Fritz}, {Gardner}, {Gonzalez-Nuevo}, {Herranz}, {Hill}, {Hopkins},
  {Hughes}, {Ibar}, {Ivison}, {Jarvis}, {Jones}, {Kelvin}, {Lagache}, {Leeuw},
  {Liske}, {Lopez-Caniego}, {Loveday}, {Maddox}, {Micha{\l}owski}, {Norberg},
  {Parkinson}, {Peacock}, {Pearson}, {Pascale}, {Pohlen}, {Popescu},
  {Prescott}, {Robotham}, {Rigby}, {Rodighiero}, {Samui}, {Sansom}, {Scott},
  {Serjeant}, {Sharp}, {Sibthorpe}, {Smith}, {Thompson}, {Tuffs}, {Valtchanov},
  {van Kampen}, {van der Werf}, {Verma}, {Vieira}, \& {Vlahakis}}]{Amblard2010}
{Amblard} A. {et~al.}, 2010, \aap, 518, L9

\bibitem[{{Antonucci}(1993)}]{Antonucci1993}
{Antonucci} R., 1993, \araa, 31, 473

\bibitem[{{Antonuccio-Delogu} \& {Silk}(2010)}]{Antonuccio-Delogu2010}
{Antonuccio-Delogu} V., {Silk} J., 2010, \mnras, 405, 1303

\bibitem[{{Archibald} {et~al}\mbox{.}(2001){Archibald}, {Dunlop}, {Hughes},
  {Rawlings}, {Eales}, \& {Ivison}}]{Archibald2001}
{Archibald} E.~N., {Dunlop} J.~S., {Hughes} D.~H., {Rawlings} S., {Eales}
  S.~A., {Ivison} R.~J., 2001, \mnras, 323, 417

\bibitem[{{Barger} {et~al}\mbox{.}(2005){Barger}, {Cowie}, {Mushotzky}, {Yang},
  {Wang}, {Steffen}, \& {Capak}}]{Barger2005}
{Barger} A.~J., {Cowie} L.~L., {Mushotzky} R.~F., {Yang} Y., {Wang} W.-H.,
  {Steffen} A.~T., {Capak} P., 2005, \aj, 129, 578

\bibitem[{{Barvainis} {et~al}\mbox{.}(2005){Barvainis}, {Leh{\'a}r},
  {Birkinshaw}, {Falcke}, \& {Blundell}}]{Barvainis2005}
{Barvainis} R., {Leh{\'a}r} J., {Birkinshaw} M., {Falcke} H., {Blundell} K.~M.,
  2005, \apj, 618, 108

\bibitem[{{Bavouzet} {et~al}\mbox{.}(2008){Bavouzet}, {Dole}, {Le Floc'h},
  {Caputi}, {Lagache}, \& {Kochanek}}]{Bavouzet2008}
{Bavouzet} N., {Dole} H., {Le Floc'h} E., {Caputi} K.~I., {Lagache} G.,
  {Kochanek} C.~S., 2008, \aap, 479, 83

\bibitem[{{Becker} {et~al}\mbox{.}(1995){Becker}, {White}, \&
  {Helfand}}]{Becker1995}
{Becker} R.~H., {White} R.~L., {Helfand} D.~J., 1995, \apj, 450, 559

\bibitem[{{Best} \& {Heckman}(2012)}]{Best2012}
{Best} P.~N., {Heckman} T.~M., 2012, \mnras, 421, 1569

\bibitem[{{Best} {et~al}\mbox{.}(2005){Best}, {Kauffmann}, {Heckman}, \&
  {Ivezi{\'c}}}]{Best2005}
{Best} P.~N., {Kauffmann} G., {Heckman} T.~M., {Ivezi{\'c}} {\v Z}., 2005,
  \mnras, 362, 9

\bibitem[{{Best} {et~al}\mbox{.}(2014){Best}, {Ker}, {Simpson}, {Rigby}, \&
  {Sabater}}]{Best2014}
{Best} P.~N., {Ker} L.~M., {Simpson} C., {Rigby} E.~E., {Sabater} J., 2014,
  \mnras, 445, 955

\bibitem[{{Best} {et~al}\mbox{.}(1996){Best}, {Longair}, \&
  {Rottgering}}]{Best1996}
{Best} P.~N., {Longair} M.~S., {Rottgering} H.~J.~A., 1996, \mnras, 280, L9

\bibitem[{{B{\'e}thermin} {et~al}\mbox{.}(2010){B{\'e}thermin}, {Dole},
  {Cousin}, \& {Bavouzet}}]{Bethermin2010b}
{B{\'e}thermin} M., {Dole} H., {Cousin} M., {Bavouzet} N., 2010, \aap, 516, A43

\bibitem[{{Bicknell} {et~al}\mbox{.}(2000){Bicknell}, {Sutherland}, {van
  Breugel}, {Dopita}, {Dey}, \& {Miley}}]{Bicknell2000}
{Bicknell} G.~V., {Sutherland} R.~S., {van Breugel} W.~J.~M., {Dopita} M.~A.,
  {Dey} A., {Miley} G.~K., 2000, \apj, 540, 678

\bibitem[{{Blandford} \& {Rees}(1974)}]{Blandford1974}
{Blandford} R.~D., {Rees} M.~J., 1974, \mnras, 169, 395

\bibitem[{{Bonfield} {et~al}\mbox{.}(2011){Bonfield}, {Jarvis}, {Hardcastle},
  {Cooray}, {Hatziminaoglou}, {Ivison}, {Page}, {Stevens}, {de Zotti}, {Auld},
  {Baes}, {Buttiglione}, {Cava}, {Dariush}, {Dunlop}, {Dunne}, {Dye}, {Eales},
  {Fritz}, {Hopwood}, {Ibar}, {Maddox}, {Micha{\l}owski}, {Pascale}, {Pohlen},
  {Rigby}, {Rodighiero}, {Serjeant}, {Smith}, {Temi}, \& {van der
  Werf}}]{Bonfield2011}
{Bonfield} D.~G. {et~al.}, 2011, \mnras, 416, 13

\bibitem[{{Bower} {et~al}\mbox{.}(2008){Bower}, {McCarthy}, \&
  {Benson}}]{Bower2008}
{Bower} R.~G., {McCarthy} I.~G., {Benson} A.~J., 2008, \mnras, 390, 1399

\bibitem[{{Boyle} \& {Terlevich}(1998)}]{Boyle1998}
{Boyle} B.~J., {Terlevich} R.~J., 1998, \mnras, 293, L49

\bibitem[{{Cao Orjales} {et~al}\mbox{.}(2012){Cao Orjales}, {Stevens},
  {Jarvis}, {Smith}, {Hardcastle}, {Auld}, {Baes}, {Cava}, {Clements},
  {Cooray}, {Coppin}, {Dariush}, {De Zotti}, {Dunne}, {Dye}, {Eales},
  {Hopwood}, {Hoyos}, {Ibar}, {Ivison}, {Maddox}, {Page}, \&
  {Valiante}}]{CaoOrjales2012}
{Cao Orjales} J.~M. {et~al.}, 2012, \mnras, 427, 1209

\bibitem[{{Chapman} {et~al}\mbox{.}(2005){Chapman}, {Blain}, {Smail}, \&
  {Ivison}}]{Chapman2005}
{Chapman} S.~C., {Blain} A.~W., {Smail} I., {Ivison} R.~J., 2005, \apj, 622,
  772

\bibitem[{{Chen} {et~al}\mbox{.}(2013){Chen}, {Hickox}, {Alberts}, {Brodwin},
  {Jones}, {Murray}, {Alexander}, {Assef}, {Brown}, {Dey}, {Forman}, {Gorjian},
  {Goulding}, {Le Floc'h}, {Jannuzi}, {Mullaney}, \& {Pope}}]{Chen2013}
{Chen} C.-T.~J. {et~al.}, 2013, \apj, 773, 3

\bibitem[{{Chen} {et~al}\mbox{.}(2015){Chen}, {Hickox}, {Alberts}, {Harrison},
  {Alexander}, {Assef}, {Brodwin}, {Brown}, {Del Moro}, {Forman}, {Gorjian},
  {Goulding}, {Hainline}, {Jones}, {Kochanek}, {Murray}, {Pope}, {Rovilos}, \&
  {Stern}}]{Chen2015}
{Chen} C.-T.~J. {et~al.}, 2015, \apj, 802, 50

\bibitem[{{Cleary} {et~al}\mbox{.}(2007){Cleary}, {Lawrence}, {Marshall},
  {Hao}, \& {Meier}}]{Cleary2007}
{Cleary} K., {Lawrence} C.~R., {Marshall} J.~A., {Hao} L., {Meier} D., 2007,
  \apj, 660, 117

\bibitem[{{Condon} {et~al}\mbox{.}(1998){Condon}, {Cotton}, {Greisen}, {Yin},
  {Perley}, {Taylor}, \& {Broderick}}]{Condon1998}
{Condon} J.~J., {Cotton} W.~D., {Greisen} E.~W., {Yin} Q.~F., {Perley} R.~A.,
  {Taylor} G.~B., {Broderick} J.~J., 1998, \aj, 115, 1693

\bibitem[{{Condon} {et~al}\mbox{.}(2013){Condon}, {Kellermann}, {Kimball},
  {Ivezi{\'c}}, \& {Perley}}]{Condon2013}
{Condon} J.~J., {Kellermann} K.~I., {Kimball} A.~E., {Ivezi{\'c}} {\v Z}.,
  {Perley} R.~A., 2013, \apj, 768, 37

\bibitem[{{Croft} {et~al}\mbox{.}(2006){Croft}, {van Breugel}, {de Vries},
  {Dopita}, {Martin}, {Morganti}, {Neff}, {Oosterloo}, {Schiminovich},
  {Stanford}, \& {van Gorkom}}]{Croft2006}
{Croft} S. {et~al.}, 2006, \apj, 647, 1040

\bibitem[{{Croton} {et~al}\mbox{.}(2006){Croton}, {Springel}, {White}, {De
  Lucia}, {Frenk}, {Gao}, {Jenkins}, {Kauffmann}, {Navarro}, \&
  {Yoshida}}]{Croton2006}
{Croton} D.~J. {et~al.}, 2006, \mnras, 365, 11

\bibitem[{{Daddi} {et~al}\mbox{.}(2010){Daddi}, {Bournaud}, {Walter},
  {Dannerbauer}, {Carilli}, {Dickinson}, {Elbaz}, {Morrison}, {Riechers},
  {Onodera}, {Salmi}, {Krips}, \& {Stern}}]{Daddi2010}
{Daddi} E. {et~al.}, 2010, \apj, 713, 686

\bibitem[{{Dalton} {et~al}\mbox{.}(2012){Dalton}, {Trager}, {Abrams}, {Carter},
  {Bonifacio}, {Aguerri}, {MacIntosh}, {Evans}, {Lewis}, {Navarro}, {Agocs},
  {Dee}, {Rousset}, {Tosh}, {Middleton}, {Pragt}, {Terrett}, {Brock}, {Benn},
  {Verheijen}, {Cano Infantes}, {Bevil}, {Steele}, {Mottram}, {Bates},
  {Gribbin}, {Rey}, {Rodriguez}, {Delgado}, {Guinouard}, {Walton}, {Irwin},
  {Jagourel}, {Stuik}, {Gerlofsma}, {Roelfsma}, {Skillen}, {Ridings},
  {Balcells}, {Daban}, {Gouvret}, {Venema}, \& {Girard}}]{Dalton2012}
{Dalton} G. {et~al.}, 2012, in Society of Photo-Optical Instrumentation
  Engineers (SPIE) Conference Series, Vol. 8446, Society of Photo-Optical
  Instrumentation Engineers (SPIE) Conference Series, p. 84460P

\bibitem[{{De Breuck} {et~al}\mbox{.}(2005){De Breuck}, {Downes}, {Neri}, {van
  Breugel}, {Reuland}, {Omont}, \& {Ivison}}]{DeBreuck2005}
{De Breuck} C., {Downes} D., {Neri} R., {van Breugel} W., {Reuland} M., {Omont}
  A., {Ivison} R., 2005, \aap, 430, L1

\bibitem[{{Del Moro} {et~al}\mbox{.}(2013){Del Moro}, {Alexander}, {Mullaney},
  {Daddi}, {Pannella}, {Bauer}, {Pope}, {Dickinson}, {Elbaz}, {Barthel},
  {Garrett}, {Brandt}, {Charmandaris}, {Chary}, {Dasyra}, {Gilli}, {Hickox},
  {Hwang}, {Ivison}, {Juneau}, {Le Floc'h}, {Luo}, {Morrison}, {Rovilos},
  {Sargent}, \& {Xue}}]{Delmoro2013}
{Del Moro} A. {et~al.}, 2013, \aap, 549, A59

\bibitem[{{Dey} {et~al}\mbox{.}(1997){Dey}, {van Breugel}, {Vacca}, \&
  {Antonucci}}]{Dey1997}
{Dey} A., {van Breugel} W., {Vacca} W.~D., {Antonucci} R., 1997, \apj, 490, 698

\bibitem[{{Di Matteo} {et~al}\mbox{.}(2008){Di Matteo}, {Colberg}, {Springel},
  {Hernquist}, \& {Sijacki}}]{DiMatteo2008}
{Di Matteo} T., {Colberg} J., {Springel} V., {Hernquist} L., {Sijacki} D.,
  2008, \apj, 676, 33

\bibitem[{{Di Matteo} {et~al}\mbox{.}(2005){Di Matteo}, {Springel}, \&
  {Hernquist}}]{DiMatteo2005}
{Di Matteo} T., {Springel} V., {Hernquist} L., 2005, \nat, 433, 604

\bibitem[{{Dicken} {et~al}\mbox{.}(2009){Dicken}, {Tadhunter}, {Axon},
  {Morganti}, {Inskip}, {Holt}, {Gonz{\'a}lez Delgado}, \&
  {Groves}}]{Dicken2009}
{Dicken} D., {Tadhunter} C., {Axon} D., {Morganti} R., {Inskip} K.~J., {Holt}
  J., {Gonz{\'a}lez Delgado} R., {Groves} B., 2009, \apj, 694, 268

\bibitem[{{Dicken} {et~al}\mbox{.}(2012){Dicken}, {Tadhunter}, {Axon},
  {Morganti}, {Robinson}, {Kouwenhoven}, {Spoon}, {Kharb}, {Inskip}, {Holt},
  {Ramos Almeida}, \& {Nesvadba}}]{Dicken2012}
{Dicken} D. {et~al.}, 2012, \apj, 745, 172

\bibitem[{{Dong} \& {Wu}(2016)}]{Dong2016}
{Dong} X.~Y., {Wu} X.-B., 2016, \apj, 824, 70

\bibitem[{{Donoso} {et~al}\mbox{.}(2009){Donoso}, {Best}, \&
  {Kauffmann}}]{Donoso2009}
{Donoso} E., {Best} P.~N., {Kauffmann} G., 2009, \mnras, 392, 617

\bibitem[{{Drouart} {et~al}\mbox{.}(2014){Drouart}, {De Breuck}, {Vernet},
  {Seymour}, {Lehnert}, {Barthel}, {Bauer}, {Ibar}, {Galametz}, {Haas},
  {Hatch}, {Mullaney}, {Nesvadba}, {Rocca-Volmerange}, {R{\"o}ttgering},
  {Stern}, \& {Wylezalek}}]{Drouart2014}
{Drouart} G. {et~al.}, 2014, \aap, 566, A53

\bibitem[{{Dubois} {et~al}\mbox{.}(2012){Dubois}, {Devriendt}, {Slyz}, \&
  {Teyssier}}]{Dubois2012}
{Dubois} Y., {Devriendt} J., {Slyz} A., {Teyssier} R., 2012, \mnras, 420, 2662

\bibitem[{{Dunlop} {et~al}\mbox{.}(2003){Dunlop}, {McLure}, {Kukula}, {Baum},
  {O'Dea}, \& {Hughes}}]{Dunlop2003}
{Dunlop} J.~S., {McLure} R.~J., {Kukula} M.~J., {Baum} S.~A., {O'Dea} C.~P.,
  {Hughes} D.~H., 2003, \mnras, 340, 1095

\bibitem[{{Dunne} \& {Eales}(2001)}]{Dunne2001}
{Dunne} L., {Eales} S.~A., 2001, \mnras, 327, 697

\bibitem[{{Dunne} {et~al}\mbox{.}(2011){Dunne}, {Gomez}, {da Cunha}, {Charlot},
  {Dye}, {Eales}, {Maddox}, {Rowlands}, {Smith}, {Auld}, {Baes}, {Bonfield},
  {Bourne}, {Buttiglione}, {Cava}, {Clements}, {Coppin}, {Cooray}, {Dariush},
  {de Zotti}, {Driver}, {Fritz}, {Geach}, {Hopwood}, {Ibar}, {Ivison},
  {Jarvis}, {Kelvin}, {Pascale}, {Pohlen}, {Popescu}, {Rigby}, {Robotham},
  {Rodighiero}, {Sansom}, {Serjeant}, {Temi}, {Thompson}, {Tuffs}, {van der
  Werf}, \& {Vlahakis}}]{Dunne2011}
{Dunne} L. {et~al.}, 2011, \mnras, 417, 1510

\bibitem[{{Dye} {et~al}\mbox{.}(2010){Dye}, {Dunne}, {Eales}, {Smith},
  {Amblard}, {Auld}, {Baes}, {Baldry}, {Bamford}, {Blain}, {Bonfield},
  {Bremer}, {Burgarella}, {Buttiglione}, {Cameron}, {Cava}, {Clements},
  {Cooray}, {Croom}, {Dariush}, {de Zotti}, {Driver}, {Dunlop}, {Frayer},
  {Fritz}, {Gardner}, {Gomez}, {Gonzalez-Nuevo}, {Herranz}, {Hill}, {Hopkins},
  {Ibar}, {Ivison}, {Jarvis}, {Jones}, {Kelvin}, {Lagache}, {Leeuw}, {Liske},
  {Lopez-Caniego}, {Loveday}, {Maddox}, {Micha{\l}owski}, {Negrello},
  {Norberg}, {Page}, {Parkinson}, {Pascale}, {Peacock}, {Pohlen}, {Popescu},
  {Prescott}, {Rigopoulou}, {Robotham}, {Rigby}, {Rodighiero}, {Samui},
  {Scott}, {Serjeant}, {Sharp}, {Sibthorpe}, {Temi}, {Thompson}, {Tuffs},
  {Valtchanov}, {van der Werf}, {van Kampen}, \& {Verma}}]{Dye2010}
{Dye} S. {et~al.}, 2010, \aap, 518, L10

\bibitem[{{Eales} {et~al}\mbox{.}(1997){Eales}, {Rawlings}, {Law-Green},
  {Cotter}, \& {Lacy}}]{Eales1997}
{Eales} S., {Rawlings} S., {Law-Green} D., {Cotter} G., {Lacy} M., 1997,
  \mnras, 291, 593

\bibitem[{{Eales}(1985)}]{Eales1985}
{Eales} S.~A., 1985, \mnras, 217, 149

\bibitem[{{Elbaz} {et~al}\mbox{.}(2007){Elbaz}, {Daddi}, {Le Borgne},
  {Dickinson}, {Alexander}, {Chary}, {Starck}, {Brandt}, {Kitzbichler},
  {MacDonald}, {Nonino}, {Popesso}, {Stern}, \& {Vanzella}}]{Elbaz2007}
{Elbaz} D. {et~al.}, 2007, \aap, 468, 33

\bibitem[{{Elbaz} {et~al}\mbox{.}(2011){Elbaz}, {Dickinson}, {Hwang},
  {D{\'{\i}}az-Santos}, {Magdis}, {Magnelli}, {Le Borgne}, {Galliano},
  {Pannella}, {Chanial}, {Armus}, {Charmandaris}, {Daddi}, {Aussel}, {Popesso},
  {Kartaltepe}, {Altieri}, {Valtchanov}, {Coia}, {Dannerbauer}, {Dasyra},
  {Leiton}, {Mazzarella}, {Alexander}, {Buat}, {Burgarella}, {Chary}, {Gilli},
  {Ivison}, {Juneau}, {Le Floc'h}, {Lutz}, {Morrison}, {Mullaney}, {Murphy},
  {Pope}, {Scott}, {Brodwin}, {Calzetti}, {Cesarsky}, {Charlot}, {Dole},
  {Eisenhardt}, {Ferguson}, {F{\"o}rster Schreiber}, {Frayer}, {Giavalisco},
  {Huynh}, {Koekemoer}, {Papovich}, {Reddy}, {Surace}, {Teplitz}, {Yun}, \&
  {Wilson}}]{Elbaz2011}
{Elbaz} D. {et~al.}, 2011, \aap, 533, A119

\bibitem[{{Emonts} {et~al}\mbox{.}(2011){Emonts}, {Feain}, {Mao}, {Norris},
  {Miley}, {Ekers}, {Villar-Mart{\'{\i}}n}, {R{\"o}ttgering}, {Sadler}, {Rees},
  {Morganti}, {Saikia}, {Oosterloo}, {Stevens}, \& {Tadhunter}}]{Emonts2011}
{Emonts} B.~H.~C. {et~al.}, 2011, \apjl, 734, L25

\bibitem[{{Evans} {et~al}\mbox{.}(2005){Evans}, {Mazzarella}, {Surace},
  {Frayer}, {Iwasawa}, \& {Sanders}}]{Evans2005}
{Evans} A.~S., {Mazzarella} J.~M., {Surace} J.~A., {Frayer} D.~T., {Iwasawa}
  K., {Sanders} D.~B., 2005, \apjs, 159, 197

\bibitem[{{Falder} {et~al}\mbox{.}(2010){Falder}, {Stevens}, {Jarvis},
  {Hardcastle}, {Lacy}, {McLure}, {Hatziminaoglou}, {Page}, \&
  {Richards}}]{Falder2010}
{Falder} J.~T. {et~al.}, 2010, \mnras, 405, 347

\bibitem[{{Feigelson} \& {Nelson}(1985)}]{Feigelson1985}
{Feigelson} E.~D., {Nelson} P.~I., 1985, \apj, 293, 192

\bibitem[{{Fernandes} {et~al}\mbox{.}(2015){Fernandes}, {Jarvis},
  {Mart{\'{\i}}nez-Sansigre}, {Rawlings}, {Afonso}, {Hardcastle}, {Lacy},
  {Stevens}, \& {Vardoulaki}}]{Fernandes2015}
{Fernandes} C.~A.~C. {et~al.}, 2015, \mnras, 447, 1184

\bibitem[{{Fernandes} {et~al}\mbox{.}(2011){Fernandes}, {Jarvis}, {Rawlings},
  {Mart{\'{\i}}nez-Sansigre}, {Hatziminaoglou}, {Lacy}, {Page}, {Stevens}, \&
  {Vardoulaki}}]{Fernandes2011}
{Fernandes} C.~A.~C. {et~al.}, 2011, \mnras, 411, 1909

\bibitem[{{Fernandez-Conde} {et~al}\mbox{.}(2008){Fernandez-Conde}, {Lagache},
  {Puget}, \& {Dole}}]{FernandezConde2008}
{Fernandez-Conde} N., {Lagache} G., {Puget} J.-L., {Dole} H., 2008, \aap, 481,
  885

\bibitem[{{Fernandez-Conde} {et~al}\mbox{.}(2010){Fernandez-Conde}, {Lagache},
  {Puget}, \& {Dole}}]{FernandezConde2010}
{Fernandez-Conde} N., {Lagache} G., {Puget} J.-L., {Dole} H., 2010, \aap, 515,
  A48

\bibitem[{{Floyd} {et~al}\mbox{.}(2004){Floyd}, {Kukula}, {Dunlop}, {McLure},
  {Miller}, {Percival}, {Baum}, \& {O'Dea}}]{Floyd2004}
{Floyd} D.~J.~E., {Kukula} M.~J., {Dunlop} J.~S., {McLure} R.~J., {Miller} L.,
  {Percival} W.~J., {Baum} S.~A., {O'Dea} C.~P., 2004, \mnras, 355, 196

\bibitem[{{Gaibler} {et~al}\mbox{.}(2012){Gaibler}, {Khochfar}, {Krause}, \&
  {Silk}}]{Gaibler2012}
{Gaibler} V., {Khochfar} S., {Krause} M., {Silk} J., 2012, \mnras, 425, 438

\bibitem[{{Genzel} {et~al}\mbox{.}(2010){Genzel}, {Tacconi}, {Gracia-Carpio},
  {Sternberg}, {Cooper}, {Shapiro}, {Bolatto}, {Bouch{\'e}}, {Bournaud},
  {Burkert}, {Combes}, {Comerford}, {Cox}, {Davis}, {Schreiber},
  {Garcia-Burillo}, {Lutz}, {Naab}, {Neri}, {Omont}, {Shapley}, \&
  {Weiner}}]{Genzel2010}
{Genzel} R. {et~al.}, 2010, \mnras, 407, 2091

\bibitem[{{Gon{\c c}alves} {et~al}\mbox{.}(2008){Gon{\c c}alves}, {Steidel}, \&
  {Pettini}}]{Goncalves2008}
{Gon{\c c}alves} T.~S., {Steidel} C.~C., {Pettini} M., 2008, \apj, 676, 816

\bibitem[{{Graham} \& {Scott}(2013)}]{Graham2013}
{Graham} A.~W., {Scott} N., 2013, \apj, 764, 151

\bibitem[{{Granato} {et~al}\mbox{.}(2001){Granato}, {Silva}, {Monaco},
  {Panuzzo}, {Salucci}, {De Zotti}, \& {Danese}}]{Granato2001}
{Granato} G.~L., {Silva} L., {Monaco} P., {Panuzzo} P., {Salucci} P., {De
  Zotti} G., {Danese} L., 2001, \mnras, 324, 757

\bibitem[{{Griffin} {et~al}\mbox{.}(2010){Griffin}, {Abergel}, {Abreu}, {Ade},
  {Andr{\'e}}, {Augueres}, {Babbedge}, {Bae}, {Baillie}, {Baluteau}, {Barlow},
  {Bendo}, {Benielli}, {Bock}, {Bonhomme}, {Brisbin}, {Brockley-Blatt},
  {Caldwell}, {Cara}, {Castro-Rodriguez}, {Cerulli}, {Chanial}, {Chen},
  {Clark}, {Clements}, {Clerc}, {Coker}, {Communal}, {Conversi}, {Cox},
  {Crumb}, {Cunningham}, {Daly}, {Davis}, {de Antoni}, {Delderfield}, {Devin},
  {di Giorgio}, {Didschuns}, {Dohlen}, {Donati}, {Dowell}, {Dowell}, {Duband},
  {Dumaye}, {Emery}, {Ferlet}, {Ferrand}, {Fontignie}, {Fox}, {Franceschini},
  {Frerking}, {Fulton}, {Garcia}, {Gastaud}, {Gear}, {Glenn}, {Goizel},
  {Griffin}, {Grundy}, {Guest}, {Guillemet}, {Hargrave}, {Harwit}, {Hastings},
  {Hatziminaoglou}, {Herman}, {Hinde}, {Hristov}, {Huang}, {Imhof}, {Isaak},
  {Israelsson}, {Ivison}, {Jennings}, {Kiernan}, {King}, {Lange}, {Latter},
  {Laurent}, {Laurent}, {Leeks}, {Lellouch}, {Levenson}, {Li}, {Li},
  {Lilienthal}, {Lim}, {Liu}, {Lu}, {Madden}, {Mainetti}, {Marliani}, {McKay},
  {Mercier}, {Molinari}, {Morris}, {Moseley}, {Mulder}, {Mur}, {Naylor},
  {Nguyen}, {O'Halloran}, {Oliver}, {Olofsson}, {Olofsson}, {Orfei}, {Page},
  {Pain}, {Panuzzo}, {Papageorgiou}, {Parks}, {Parr-Burman}, {Pearce},
  {Pearson}, {P{\'e}rez-Fournon}, {Pinsard}, {Pisano}, {Podosek}, {Pohlen},
  {Polehampton}, {Pouliquen}, {Rigopoulou}, {Rizzo}, {Roseboom}, {Roussel},
  {Rowan-Robinson}, {Rownd}, {Saraceno}, {Sauvage}, {Savage}, {Savini},
  {Sawyer}, {Scharmberg}, {Schmitt}, {Schneider}, {Schulz}, {Schwartz},
  {Shafer}, {Shupe}, {Sibthorpe}, {Sidher}, {Smith}, {Smith}, {Smith},
  {Spencer}, {Stobie}, {Sudiwala}, {Sukhatme}, {Surace}, {Stevens}, {Swinyard},
  {Trichas}, {Tourette}, {Triou}, {Tseng}, {Tucker}, {Turner}, {Vaccari},
  {Valtchanov}, {Vigroux}, {Virique}, {Voellmer}, {Walker}, {Ward}, {Waskett},
  {Weilert}, {Wesson}, {White}, {Whitehouse}, {Wilson}, {Winter}, {Woodcraft},
  {Wright}, {Xu}, {Zavagno}, {Zemcov}, {Zhang}, \& {Zonca}}]{Griffin2010}
{Griffin} M.~J. {et~al.}, 2010, \aap, 518, L3

\bibitem[{{Gurkan} {et~al}\mbox{.}(2015){Gurkan}, {Hardcastle}, {Jarvis},
  {Smith}, {Bourne}, {Dunne}, {Maddox}, {Ivison}, \& {Fritz}}]{Gurkan2015}
{Gurkan} G. {et~al.}, 2015, ArXiv e-prints

\bibitem[{{Hardcastle} {et~al}\mbox{.}(2013){Hardcastle}, {Ching}, {Virdee},
  {Jarvis}, {Croom}, {Sadler}, {Mauch}, {Smith}, {Stevens}, {Baes}, {Baldry},
  {Brough}, {Cooray}, {Dariush}, {De Zotti}, {Driver}, {Dunne}, {Dye}, {Eales},
  {Hopwood}, {Liske}, {Maddox}, {Micha{\l}owski}, {Rigby}, {Robotham},
  {Steele}, {Thomas}, \& {Valiante}}]{Hardcastle2013}
{Hardcastle} M.~J. {et~al.}, 2013, \mnras, 429, 2407

\bibitem[{{Hardcastle} {et~al}\mbox{.}(2007){Hardcastle}, {Evans}, \&
  {Croston}}]{Hardcastle2007}
{Hardcastle} M.~J., {Evans} D.~A., {Croston} J.~H., 2007, \mnras, 376, 1849

\bibitem[{{Hardcastle} {et~al}\mbox{.}(2010){Hardcastle}, {Virdee}, {Jarvis},
  {Bonfield}, {Dunne}, {Rawlings}, {Stevens}, {Christopher}, {Heywood},
  {Mauch}, {Rigopoulou}, {Verma}, {Baldry}, {Bamford}, {Buttiglione}, {Cava},
  {Clements}, {Cooray}, {Croom}, {Dariush}, {de Zotti}, {Eales}, {Fritz},
  {Hill}, {Hughes}, {Hopwood}, {Ibar}, {Ivison}, {Jones}, {Loveday}, {Maddox},
  {Micha{\l}owski}, {Negrello}, {Norberg}, {Pohlen}, {Prescott}, {Rigby},
  {Robotham}, {Rodighiero}, {Scott}, {Sharp}, {Smith}, {Temi}, \& {van
  Kampen}}]{Hardcastle2010}
{Hardcastle} M.~J. {et~al.}, 2010, \mnras, 409, 122

\bibitem[{{H{\"a}ring} \& {Rix}(2004)}]{Haring2004}
{H{\"a}ring} N., {Rix} H.-W., 2004, \apjl, 604, L89

\bibitem[{{Harrison} {et~al}\mbox{.}(2012){Harrison}, {Alexander}, {Mullaney},
  {Altieri}, {Coia}, {Charmandaris}, {Daddi}, {Dannerbauer}, {Dasyra}, {Del
  Moro}, {Dickinson}, {Hickox}, {Ivison}, {Kartaltepe}, {Le Floc'h}, {Leiton},
  {Magnelli}, {Popesso}, {Rovilos}, {Rosario}, \& {Swinbank}}]{Harrison2012}
{Harrison} C.~M. {et~al.}, 2012, \apjl, 760, L15

\bibitem[{{Hasinger} {et~al}\mbox{.}(2005){Hasinger}, {Miyaji}, \&
  {Schmidt}}]{Hasinger2005}
{Hasinger} G., {Miyaji} T., {Schmidt} M., 2005, \aap, 441, 417

\bibitem[{{Hatch} {et~al}\mbox{.}(2008){Hatch}, {Overzier}, {R{\"o}ttgering},
  {Kurk}, \& {Miley}}]{Hatch2008}
{Hatch} N.~A., {Overzier} R.~A., {R{\"o}ttgering} H.~J.~A., {Kurk} J.~D.,
  {Miley} G.~K., 2008, \mnras, 383, 931

\bibitem[{{Hatziminaoglou} {et~al}\mbox{.}(2010){Hatziminaoglou}, {Omont},
  {Stevens}, {Amblard}, {Arumugam}, {Auld}, {Aussel}, {Babbedge}, {Blain},
  {Bock}, {Boselli}, {Buat}, {Burgarella}, {Castro-Rodr{\'{\i}}guez}, {Cava},
  {Chanial}, {Clements}, {Conley}, {Conversi}, {Cooray}, {Dowell}, {Dwek},
  {Dye}, {Eales}, {Elbaz}, {Farrah}, {Fox}, {Franceschini}, {Gear}, {Glenn},
  {Gonz{\'a}lez Solares}, {Griffin}, {Halpern}, {Ibar}, {Isaak}, {Ivison},
  {Lagache}, {Levenson}, {Lu}, {Madden}, {Maffei}, {Mainetti}, {Marchetti},
  {Mortier}, {Nguyen}, {O'Halloran}, {Oliver}, {Page}, {Panuzzo},
  {Papageorgiou}, {Pearson}, {P{\'e}rez-Fournon}, {Pohlen}, {Rawlings},
  {Rigopoulou}, {Rizzo}, {Roseboom}, {Rowan-Robinson}, {Sanchez Portal},
  {Schulz}, {Scott}, {Seymour}, {Shupe}, {Smith}, {Symeonidis}, {Trichas},
  {Tugwell}, {Vaccari}, {Valtchanov}, {Vigroux}, {Wang}, {Ward}, {Wright},
  {Xu}, \& {Zemcov}}]{Hatziminaoglou2010}
{Hatziminaoglou} E. {et~al.}, 2010, \aap, 518, L33

\bibitem[{{Heckman} \& {Best}(2014)}]{Heckman2014}
{Heckman} T.~M., {Best} P.~N., 2014, \araa, 52, 589

\bibitem[{{Helou} {et~al}\mbox{.}(1985){Helou}, {Soifer}, \&
  {Rowan-Robinson}}]{Helou1985}
{Helou} G., {Soifer} B.~T., {Rowan-Robinson} M., 1985, \apjl, 298, L7

\bibitem[{{Hickox} {et~al}\mbox{.}(2014){Hickox}, {Mullaney}, {Alexander},
  {Chen}, {Civano}, {Goulding}, \& {Hainline}}]{Hickox2014}
{Hickox} R.~C., {Mullaney} J.~R., {Alexander} D.~M., {Chen} C.-T.~J., {Civano}
  F.~M., {Goulding} A.~D., {Hainline} K.~N., 2014, \apj, 782, 9

\bibitem[{{Hildebrand}(1983)}]{Hildebrand1983}
{Hildebrand} R.~H., 1983, \qjras, 24, 267

\bibitem[{{Hill} \& {Rawlings}(2003)}]{Hill2003}
{Hill} G.~J., {Rawlings} S., 2003, nature, 47, 373

\bibitem[{{Ho} {et~al}\mbox{.}(2004){Ho}, {Moran}, \& {Lo}}]{Ho2004}
{Ho} P.~T.~P., {Moran} J.~M., {Lo} K.~Y., 2004, \apjl, 616, L1

\bibitem[{{Hopkins} \& {Elvis}(2010)}]{HopkinsElvis2010}
{Hopkins} P.~F., {Elvis} M., 2010, \mnras, 401, 7

\bibitem[{{Hopkins} \& {Hernquist}(2006)}]{Hopkins2006}
{Hopkins} P.~F., {Hernquist} L., 2006, \apjs, 166, 1

\bibitem[{{Hopkins} {et~al}\mbox{.}(2005){Hopkins}, {Hernquist}, {Cox}, {Di
  Matteo}, {Robertson}, \& {Springel}}]{Hopkins2005}
{Hopkins} P.~F., {Hernquist} L., {Cox} T.~J., {Di Matteo} T., {Robertson} B.,
  {Springel} V., 2005, \apj, 630, 716

\bibitem[{{Hopkins} {et~al}\mbox{.}(2010){Hopkins}, {Younger}, {Hayward},
  {Narayanan}, \& {Hernquist}}]{Hopkins2010}
{Hopkins} P.~F., {Younger} J.~D., {Hayward} C.~C., {Narayanan} D., {Hernquist}
  L., 2010, \mnras, 402, 1693

\bibitem[{{Ibar} {et~al}\mbox{.}(2009){Ibar}, {Ivison}, {Biggs}, {Lal}, {Best},
  \& {Green}}]{Ibar2009}
{Ibar} E., {Ivison} R.~J., {Biggs} A.~D., {Lal} D.~V., {Best} P.~N., {Green}
  D.~A., 2009, \mnras, 397, 281

\bibitem[{{Ibar} {et~al}\mbox{.}(2010){Ibar}, {Ivison}, {Cava}, {Rodighiero},
  {Buttiglione}, {Temi}, {Frayer}, {Fritz}, {Leeuw}, {Baes}, {Rigby}, {Verma},
  {Serjeant}, {M{\"u}ller}, {Auld}, {Dariush}, {Dunne}, {Eales}, {Maddox},
  {Panuzzo}, {Pascale}, {Pohlen}, {Smith}, {de Zotti}, {Vaccari}, {Hopwood},
  {Cooray}, {Burgarella}, \& {Jarvis}}]{Ibar2010}
{Ibar} E. {et~al.}, 2010, \mnras, 409, 38

\bibitem[{{Inskip} {et~al}\mbox{.}(2005){Inskip}, {Best}, {Longair}, \&
  {R{\"o}ttgering}}]{Inskip2005}
{Inskip} K.~J., {Best} P.~N., {Longair} M.~S., {R{\"o}ttgering} H.~J.~A., 2005,
  \mnras, 359, 1393

\bibitem[{{Inskip} {et~al}\mbox{.}(2008){Inskip}, {Villar-Mart{\'{\i}}n},
  {Tadhunter}, {Morganti}, {Holt}, \& {Dicken}}]{Inskip2008}
{Inskip} K.~J., {Villar-Mart{\'{\i}}n} M., {Tadhunter} C.~N., {Morganti} R.,
  {Holt} J., {Dicken} D., 2008, \mnras, 386, 1797

\bibitem[{{Isaak} {et~al}\mbox{.}(2002){Isaak}, {Priddey}, {McMahon}, {Omont},
  {Peroux}, {Sharp}, \& {Withington}}]{Isaak2002}
{Isaak} K.~G., {Priddey} R.~S., {McMahon} R.~G., {Omont} A., {Peroux} C.,
  {Sharp} R.~G., {Withington} S., 2002, \mnras, 329, 149

\bibitem[{{Ishibashi} \& {Fabian}(2012)}]{Ishibashi2012}
{Ishibashi} W., {Fabian} A.~C., 2012, \mnras, 427, 2998

\bibitem[{{Ivezi{\'c}} {et~al}\mbox{.}(2002){Ivezi{\'c}}, {Menou}, {Knapp},
  {Strauss}, {Lupton}, {Vanden Berk}, {Richards}, {Tremonti}, {Weinstein},
  {Anderson}, {Bahcall}, {Becker}, {Bernardi}, {Blanton}, {Eisenstein}, {Fan},
  {Finkbeiner}, {Finlator}, {Frieman}, {Gunn}, {Hall}, {Kim}, {Kinkhabwala},
  {Narayanan}, {Rockosi}, {Schlegel}, {Schneider}, {Strateva}, {SubbaRao},
  {Thakar}, {Voges}, {White}, {Yanny}, {Brinkmann}, {Doi}, {Fukugita},
  {Hennessy}, {Munn}, {Nichol}, \& {York}}]{Ivezic2002}
{Ivezi{\'c}} {\v Z}. {et~al.}, 2002, \aj, 124, 2364

\bibitem[{{Jarvis} \& {McLure}(2002)}]{Jarvis2002}
{Jarvis} M.~J., {McLure} R.~J., 2002, \mnras, 336, L38

\bibitem[{{Jarvis} \& {McLure}(2006)}]{Jarvis2006}
{Jarvis} M.~J., {McLure} R.~J., 2006, \mnras, 369, 182

\bibitem[{{Jarvis} \& {Rawlings}(2000)}]{Jarvis2000}
{Jarvis} M.~J., {Rawlings} S., 2000, \mnras, 319, 121

\bibitem[{{Jarvis} {et~al}\mbox{.}(2001{\natexlab{a}}){Jarvis}, {Rawlings},
  {Eales}, {Blundell}, {Bunker}, {Croft}, {McLure}, \& {Willott}}]{Jarvis2001c}
{Jarvis} M.~J., {Rawlings} S., {Eales} S., {Blundell} K.~M., {Bunker} A.~J.,
  {Croft} S., {McLure} R.~J., {Willott} C.~J., 2001{\natexlab{a}}, \mnras, 326,
  1585

\bibitem[{{Jarvis} {et~al}\mbox{.}(2001{\natexlab{b}}){Jarvis}, {Rawlings},
  {Willott}, {Blundell}, {Eales}, \& {Lacy}}]{Jarvis2001}
{Jarvis} M.~J., {Rawlings} S., {Willott} C.~J., {Blundell} K.~M., {Eales} S.,
  {Lacy} M., 2001{\natexlab{b}}, \mnras, 327, 907

\bibitem[{{Jarvis} {et~al}\mbox{.}(2010){Jarvis}, {Smith}, {Bonfield},
  {Hardcastle}, {Falder}, {Stevens}, {Ivison}, {Auld}, {Baes}, {Baldry},
  {Bamford}, {Bourne}, {Buttiglione}, {Cava}, {Cooray}, {Dariush}, {de Zotti},
  {Dunlop}, {Dunne}, {Dye}, {Eales}, {Fritz}, {Hill}, {Hopwood}, {Hughes},
  {Ibar}, {Jones}, {Kelvin}, {Lawrence}, {Leeuw}, {Loveday}, {Maddox},
  {Micha{\l}owski}, {Negrello}, {Norberg}, {Pohlen}, {Prescott}, {Rigby},
  {Robotham}, {Rodighiero}, {Scott}, {Sharp}, {Temi}, {Thompson}, {van der
  Werf}, {van Kampen}, {Vlahakis}, \& {White}}]{Jarvis2010}
{Jarvis} M.~J. {et~al.}, 2010, \mnras, 409, 92

\bibitem[{{Jiang} {et~al}\mbox{.}(2007){Jiang}, {Fan}, {Ivezi{\'c}},
  {Richards}, {Schneider}, {Strauss}, \& {Kelly}}]{Jiang2007}
{Jiang} L., {Fan} X., {Ivezi{\'c}} {\v Z}., {Richards} G.~T., {Schneider}
  D.~P., {Strauss} M.~A., {Kelly} B.~C., 2007, \apj, 656, 680

\bibitem[{{Jogee}(2006)}]{Jogee2006}
{Jogee} S., 2006, in Lecture Notes in Physics, Berlin Springer Verlag, Vol.
  693, Physics of Active Galactic Nuclei at all Scales, {Alloin} D., ed., p.
  143

\bibitem[{{Johnston} {et~al}\mbox{.}(2015){Johnston}, {Vaccari}, {Jarvis},
  {Smith}, {Giovannoli}, {H{\"a}u{\ss}ler}, \& {Prescott}}]{Johnston2015}
{Johnston} R., {Vaccari} M., {Jarvis} M., {Smith} M., {Giovannoli} E.,
  {H{\"a}u{\ss}ler} B., {Prescott} M., 2015, \mnras, 453, 2540

\bibitem[{{Kalfountzou} {et~al}\mbox{.}(2014{\natexlab{a}}){Kalfountzou},
  {Civano}, {Elvis}, {Trichas}, \& {Green}}]{Kalfountzou2014b}
{Kalfountzou} E., {Civano} F., {Elvis} M., {Trichas} M., {Green} P.,
  2014{\natexlab{a}}, \mnras, 445, 1430

\bibitem[{{Kalfountzou} {et~al}\mbox{.}(2012){Kalfountzou}, {Jarvis},
  {Bonfield}, \& {Hardcastle}}]{Kalfountzou2012}
{Kalfountzou} E., {Jarvis} M.~J., {Bonfield} D.~G., {Hardcastle} M.~J., 2012,
  \mnras, 427, 2401

\bibitem[{{Kalfountzou} {et~al}\mbox{.}(2014{\natexlab{b}}){Kalfountzou},
  {Stevens}, {Jarvis}, {Hardcastle}, {Smith}, {Bourne}, {Dunne}, {Ibar},
  {Eales}, {Ivison}, {Maddox}, {Smith}, {Valiante}, \& {de
  Zotti}}]{Kalfountzou2014a}
{Kalfountzou} E. {et~al.}, 2014{\natexlab{b}}, \mnras, 442, 1181

\bibitem[{{Kalfountzou} {et~al}\mbox{.}(2011){Kalfountzou}, {Trichas},
  {Rowan-Robinson}, {Clements}, {Babbedge}, \& {Seiradakis}}]{Kalfountzou2011}
{Kalfountzou} E., {Trichas} M., {Rowan-Robinson} M., {Clements} D., {Babbedge}
  T., {Seiradakis} J.~H., 2011, \mnras, 413, 249

\bibitem[{{Karouzos} {et~al}\mbox{.}(2013){Karouzos}, {Im}, {Trichas}, {Ruiz},
  {Goto}, {Malkan}, {Jeon}, {Kim}, {Lee}, {Kim}, {Oi}, {Matsuhara}, {Takagi},
  {Murata}, {Wada}, {Wada}, {Shim}, {Hanami}, {Serjeant}, {White}, {Pearson},
  \& {Ohyama}}]{Karouzos2013}
{Karouzos} M. {et~al.}, 2013, ArXiv e-prints

\bibitem[{{Kartaltepe} {et~al}\mbox{.}(2012){Kartaltepe}, {Dickinson},
  {Alexander}, {Bell}, {Dahlen}, {Elbaz}, {Faber}, {Lotz}, {McIntosh},
  {Wiklind}, {Altieri}, {Aussel}, {Bethermin}, {Bournaud}, {Charmandaris},
  {Conselice}, {Cooray}, {Dannerbauer}, {Dav{\'e}}, {Dunlop}, {Dekel},
  {Ferguson}, {Grogin}, {Hwang}, {Ivison}, {Kocevski}, {Koekemoer}, {Koo},
  {Lai}, {Leiton}, {Lucas}, {Lutz}, {Magdis}, {Magnelli}, {Morrison}, {Mozena},
  {Mullaney}, {Newman}, {Pope}, {Popesso}, {van der Wel}, {Weiner}, \&
  {Wuyts}}]{Kartaltepe2012}
{Kartaltepe} J.~S. {et~al.}, 2012, \apj, 757, 23

\bibitem[{{Kauffmann} {et~al}\mbox{.}(2008){Kauffmann}, {Heckman}, \&
  {Best}}]{Kauffmann2008}
{Kauffmann} G., {Heckman} T.~M., {Best} P.~N., 2008, \mnras, 384, 953

\bibitem[{{Kelly} {et~al}\mbox{.}(2009){Kelly}, {Vestergaard}, \&
  {Fan}}]{Kelly2009}
{Kelly} B.~C., {Vestergaard} M., {Fan} X., 2009, \apj, 692, 1388

\bibitem[{{Kennicutt}(1998)}]{Kennicutt1998}
{Kennicutt}, Jr. R.~C., 1998, \apj, 498, 541

\bibitem[{{Khochfar} \& {Silk}(2011)}]{Khochfar2011}
{Khochfar} S., {Silk} J., 2011, \mnras, 410, L42

\bibitem[{{Kormendy} \& {Ho}(2013)}]{Kormendy2013}
{Kormendy} J., {Ho} L.~C., 2013, \araa, 51, 511

\bibitem[{{Kormendy} \& {Richstone}(1995)}]{Kormendy1995}
{Kormendy} J., {Richstone} D., 1995, \araa, 33, 581

\bibitem[{{Kuiper} {et~al}\mbox{.}(2011){Kuiper}, {Hatch}, {Miley}, {Nesvadba},
  {R{\"o}ttgering}, {Kurk}, {Lehnert}, {Overzier}, {Pentericci}, {Schaye}, \&
  {Venemans}}]{Kuiper2011}
{Kuiper} E. {et~al.}, 2011, \mnras, 415, 2245

\bibitem[{{Laing} {et~al}\mbox{.}(1983){Laing}, {Riley}, \&
  {Longair}}]{Laing1983}
{Laing} R.~A., {Riley} J.~M., {Longair} M.~S., 1983, \mnras, 204, 151

\bibitem[{{Lavalley} {et~al}\mbox{.}(1992){Lavalley}, {Isobe}, \&
  {Feigelson}}]{Lavalley1992}
{Lavalley} M., {Isobe} T., {Feigelson} E., 1992, in Astronomical Society of the
  Pacific Conference Series, Vol.~25, Astronomical Data Analysis Software and
  Systems I, {Worrall} D.~M., {Biemesderfer} C., {Barnes} J., eds., p. 245

\bibitem[{{Leipski} {et~al}\mbox{.}(2013){Leipski}, {Meisenheimer}, {Walter},
  {Besel}, {Dannerbauer}, {Fan}, {Haas}, {Klaas}, {Krause}, \&
  {Rix}}]{Leipski2013}
{Leipski} C. {et~al.}, 2013, \apj, 772, 103

\bibitem[{{Lutz} {et~al}\mbox{.}(2011){Lutz}, {Poglitsch}, {Altieri},
  {Andreani}, {Aussel}, {Berta}, {Bongiovanni}, {Brisbin}, {Cava}, {Cepa},
  {Cimatti}, {Daddi}, {Dominguez-Sanchez}, {Elbaz}, {F{\"o}rster Schreiber},
  {Genzel}, {Grazian}, {Gruppioni}, {Harwit}, {Le Floc'h}, {Magdis},
  {Magnelli}, {Maiolino}, {Nordon}, {P{\'e}rez Garc{\'{\i}}a}, {Popesso},
  {Pozzi}, {Riguccini}, {Rodighiero}, {Saintonge}, {Sanchez Portal}, {Santini},
  {Shao}, {Sturm}, {Tacconi}, {Valtchanov}, {Wetzstein}, \&
  {Wieprecht}}]{Lutz2011}
{Lutz} D. {et~al.}, 2011, \aap, 532, A90

\bibitem[{{Lutz} {et~al}\mbox{.}(2008){Lutz}, {Sturm}, {Tacconi}, {Valiante},
  {Schweitzer}, {Netzer}, {Maiolino}, {Andreani}, {Shemmer}, \&
  {Veilleux}}]{Lutz2008}
{Lutz} D. {et~al.}, 2008, \apj, 684, 853

\bibitem[{{Ma} \& {Yan}(2015)}]{Ma2015}
{Ma} Z., {Yan} H., 2015, \apj, 811, 58

\bibitem[{{Madau} \& {Dickinson}(2014)}]{Madau2014}
{Madau} P., {Dickinson} M., 2014, \araa, 52, 415

\bibitem[{{Magliocchetti} {et~al}\mbox{.}(2014){Magliocchetti}, {Lutz},
  {Rosario}, {Berta}, {Le Floc'h}, {Magnelli}, {Pozzi}, {Riguccini}, \&
  {Santini}}]{Magliocchetti2014}
{Magliocchetti} M. {et~al.}, 2014, \mnras, 442, 682

\bibitem[{{Magliocchetti} {et~al}\mbox{.}(2016){Magliocchetti}, {Lutz},
  {Santini}, {Salvato}, {Popesso}, {Berta}, \& {Pozzi}}]{Magliocchetti2016}
{Magliocchetti} M., {Lutz} D., {Santini} P., {Salvato} M., {Popesso} P.,
  {Berta} S., {Pozzi} F., 2016, \mnras, 456, 431

\bibitem[{{Magnelli} {et~al}\mbox{.}(2014){Magnelli}, {Lutz}, {Saintonge},
  {Berta}, {Santini}, {Symeonidis}, {Altieri}, {Andreani}, {Aussel},
  {B{\'e}thermin}, {Bock}, {Bongiovanni}, {Cepa}, {Cimatti}, {Conley}, {Daddi},
  {Elbaz}, {F{\"o}rster Schreiber}, {Genzel}, {Ivison}, {Le Floc'h}, {Magdis},
  {Maiolino}, {Nordon}, {Oliver}, {Page}, {P{\'e}rez Garc{\'{\i}}a},
  {Poglitsch}, {Popesso}, {Pozzi}, {Riguccini}, {Rodighiero}, {Rosario},
  {Roseboom}, {Sanchez-Portal}, {Scott}, {Sturm}, {Tacconi}, {Valtchanov},
  {Wang}, \& {Wuyts}}]{Magnelli2014}
{Magnelli} B. {et~al.}, 2014, \aap, 561, A86

\bibitem[{{Magorrian} {et~al}\mbox{.}(1998){Magorrian}, {Tremaine},
  {Richstone}, {Bender}, {Bower}, {Dressler}, {Faber}, {Gebhardt}, {Green},
  {Grillmair}, {Kormendy}, \& {Lauer}}]{Magorrian1998}
{Magorrian} J. {et~al.}, 1998, \aj, 115, 2285

\bibitem[{{Marconi} {et~al}\mbox{.}(2008){Marconi}, {Axon}, {Maiolino},
  {Nagao}, {Pastorini}, {Pietrini}, {Robinson}, \& {Torricelli}}]{Marconi2008}
{Marconi} A., {Axon} D.~J., {Maiolino} R., {Nagao} T., {Pastorini} G.,
  {Pietrini} P., {Robinson} A., {Torricelli} G., 2008, \apj, 678, 693

\bibitem[{{Marconi} {et~al}\mbox{.}(2004){Marconi}, {Risaliti}, {Gilli},
  {Hunt}, {Maiolino}, \& {Salvati}}]{Marconi2004}
{Marconi} A., {Risaliti} G., {Gilli} R., {Hunt} L.~K., {Maiolino} R., {Salvati}
  M., 2004, \mnras, 351, 169

\bibitem[{{Mart{\'{\i}}nez-Sansigre} \& {Taylor}(2009)}]{Martinez-Sansigre2009}
{Mart{\'{\i}}nez-Sansigre} A., {Taylor} A.~M., 2009, \apj, 692, 964

\bibitem[{{McConnell} \& {Ma}(2013)}]{McConnell2013}
{McConnell} N.~J., {Ma} C.-P., 2013, \apj, 764, 184

\bibitem[{{McLure} \& {Dunlop}(2004)}]{McLure2004}
{McLure} R.~J., {Dunlop} J.~S., 2004, \mnras, 352, 1390

\bibitem[{{McLure} \& {Jarvis}(2002)}]{McLure2002}
{McLure} R.~J., {Jarvis} M.~J., 2002, \mnras, 337, 109

\bibitem[{{McLure} {et~al}\mbox{.}(2006){McLure}, {Jarvis}, {Targett},
  {Dunlop}, \& {Best}}]{McLure2006}
{McLure} R.~J., {Jarvis} M.~J., {Targett} T.~A., {Dunlop} J.~S., {Best} P.~N.,
  2006, Astronomische Nachrichten, 327, 213

\bibitem[{{Merloni} {et~al}\mbox{.}(2010){Merloni}, {Bongiorno}, {Bolzonella},
  {Brusa}, {Civano}, {Comastri}, {Elvis}, {Fiore}, {Gilli}, {Hao}, {Jahnke},
  {Koekemoer}, {Lusso}, {Mainieri}, {Mignoli}, {Miyaji}, {Renzini}, {Salvato},
  {Silverman}, {Trump}, {Vignali}, {Zamorani}, {Capak}, {Lilly}, {Sanders},
  {Taniguchi}, {Bardelli}, {Carollo}, {Caputi}, {Contini}, {Coppa}, {Cucciati},
  {de la Torre}, {de Ravel}, {Franzetti}, {Garilli}, {Hasinger}, {Impey},
  {Iovino}, {Iwasawa}, {Kampczyk}, {Kneib}, {Knobel}, {Kova{\v c}},
  {Lamareille}, {Le Borgne}, {Le Brun}, {Le F{\`e}vre}, {Maier}, {Pello},
  {Peng}, {Perez Montero}, {Ricciardelli}, {Scodeggio}, {Tanaka}, {Tasca},
  {Tresse}, {Vergani}, \& {Zucca}}]{Merloni2010}
{Merloni} A. {et~al.}, 2010, \apj, 708, 137

\bibitem[{{Merloni} {et~al}\mbox{.}(2004){Merloni}, {Rudnick}, \& {Di
  Matteo}}]{Merloni2004}
{Merloni} A., {Rudnick} G., {Di Matteo} T., 2004, \mnras, 354, L37

\bibitem[{{Miyaji} {et~al}\mbox{.}(2015){Miyaji}, {Hasinger}, {Salvato},
  {Brusa}, {Cappelluti}, {Civano}, {Puccetti}, {Elvis}, {Brunner},
  {Fotopoulou}, {Ueda}, {Griffiths}, {Koekemoer}, {Akiyama}, {Comastri},
  {Gilli}, {Lanzuisi}, {Merloni}, \& {Vignali}}]{Miyaji2015}
{Miyaji} T. {et~al.}, 2015, \apj, 804, 104

\bibitem[{{Mullaney} {et~al}\mbox{.}(2011){Mullaney}, {Alexander}, {Goulding},
  \& {Hickox}}]{Mullaney2011}
{Mullaney} J.~R., {Alexander} D.~M., {Goulding} A.~D., {Hickox} R.~C., 2011,
  \mnras, 414, 1082

\bibitem[{{Mullaney} {et~al}\mbox{.}(2010){Mullaney}, {Alexander}, {Huynh},
  {Goulding}, \& {Frayer}}]{Mullaney2010}
{Mullaney} J.~R., {Alexander} D.~M., {Huynh} M., {Goulding} A.~D., {Frayer} D.,
  2010, \mnras, 401, 995

\bibitem[{{Mullaney} {et~al}\mbox{.}(2012{\natexlab{a}}){Mullaney}, {Daddi},
  {B{\'e}thermin}, {Elbaz}, {Juneau}, {Pannella}, {Sargent}, {Alexander}, \&
  {Hickox}}]{Mullaney2012b}
{Mullaney} J.~R. {et~al.}, 2012{\natexlab{a}}, \apjl, 753, L30

\bibitem[{{Mullaney} {et~al}\mbox{.}(2012{\natexlab{b}}){Mullaney}, {Pannella},
  {Daddi}, {Alexander}, {Elbaz}, {Hickox}, {Bournaud}, {Altieri}, {Aussel},
  {Coia}, {Dannerbauer}, {Dasyra}, {Dickinson}, {Hwang}, {Kartaltepe},
  {Leiton}, {Magdis}, {Magnelli}, {Popesso}, {Valtchanov}, {Bauer}, {Brandt},
  {Del Moro}, {Hanish}, {Ivison}, {Juneau}, {Luo}, {Lutz}, {Sargent}, {Scott},
  \& {Xue}}]{Mullaney2012}
{Mullaney} J.~R. {et~al.}, 2012{\natexlab{b}}, \mnras, 419, 95

\bibitem[{{Nesvadba} {et~al}\mbox{.}(2010){Nesvadba}, {Boulanger},
  {Salom{\'e}}, {Guillard}, {Lehnert}, {Ogle}, {Appleton}, {Falgarone}, \&
  {Pineau Des Forets}}]{Nesvadba2010}
{Nesvadba} N.~P.~H. {et~al.}, 2010, \aap, 521, A65

\bibitem[{{Nesvadba} {et~al}\mbox{.}(2011){Nesvadba}, {De Breuck}, {Lehnert},
  {Best}, {Binette}, \& {Proga}}]{Nesvadba2011}
{Nesvadba} N.~P.~H., {De Breuck} C., {Lehnert} M.~D., {Best} P.~N., {Binette}
  L., {Proga} D., 2011, \aap, 525, A43

\bibitem[{{Nesvadba} {et~al}\mbox{.}(2006){Nesvadba}, {Lehnert}, {Eisenhauer},
  {Gilbert}, {Tecza}, \& {Abuter}}]{Nesvadba2006}
{Nesvadba} N.~P.~H., {Lehnert} M.~D., {Eisenhauer} F., {Gilbert} A., {Tecza}
  M., {Abuter} R., 2006, \apj, 650, 693

\bibitem[{{Netzer} {et~al}\mbox{.}(2007){Netzer}, {Lutz}, {Schweitzer},
  {Contursi}, {Sturm}, {Tacconi}, {Veilleux}, {Kim}, {Rupke}, {Baker},
  {Dasyra}, {Mazzarella}, \& {Lord}}]{Netzer2007}
{Netzer} H. {et~al.}, 2007, \apj, 666, 806

\bibitem[{{Nguyen} {et~al}\mbox{.}(2010){Nguyen}, {Schulz}, {Levenson},
  {Amblard}, {Arumugam}, {Aussel}, {Babbedge}, {Blain}, {Bock}, {Boselli},
  {Buat}, {Castro-Rodriguez}, {Cava}, {Chanial}, {Chapin}, {Clements},
  {Conley}, {Conversi}, {Cooray}, {Dowell}, {Dwek}, {Eales}, {Elbaz}, {Fox},
  {Franceschini}, {Gear}, {Glenn}, {Griffin}, {Halpern}, {Hatziminaoglou},
  {Ibar}, {Isaak}, {Ivison}, {Lagache}, {Lu}, {Madden}, {Maffei}, {Mainetti},
  {Marchetti}, {Marsden}, {Marshall}, {O'Halloran}, {Oliver}, {Omont}, {Page},
  {Panuzzo}, {Papageorgiou}, {Pearson}, {Perez Fournon}, {Pohlen}, {Rangwala},
  {Rigopoulou}, {Rizzo}, {Roseboom}, {Rowan-Robinson}, {Scott}, {Seymour},
  {Shupe}, {Smith}, {Stevens}, {Symeonidis}, {Trichas}, {Tugwell}, {Vaccari},
  {Valtchanov}, {Vigroux}, {Wang}, {Ward}, {Wiebe}, {Wright}, {Xu}, \&
  {Zemcov}}]{Nguyen2010}
{Nguyen} H.~T. {et~al.}, 2010, \aap, 518, L5

\bibitem[{{Ott}(2010)}]{Ott2010}
{Ott} S., 2010, in Astronomical Society of the Pacific Conference Series, Vol.
  434, Astronomical Data Analysis Software and Systems XIX, {Mizumoto} Y.,
  {Morita} K.-I., {Ohishi} M., eds., p. 139

\bibitem[{{Padovani} {et~al}\mbox{.}(2011){Padovani}, {Miller}, {Kellermann},
  {Mainieri}, {Rosati}, \& {Tozzi}}]{Padovani2011}
{Padovani} P., {Miller} N., {Kellermann} K.~I., {Mainieri} V., {Rosati} P.,
  {Tozzi} P., 2011, \apj, 740, 20

\bibitem[{{Page} {et~al}\mbox{.}(2004){Page}, {Stevens}, {Ivison}, \&
  {Carrera}}]{Page2004}
{Page} M.~J., {Stevens} J.~A., {Ivison} R.~J., {Carrera} F.~J., 2004, \apjl,
  611, L85

\bibitem[{{Page} {et~al}\mbox{.}(2001){Page}, {Stevens}, {Mittaz}, \&
  {Carrera}}]{Page2001}
{Page} M.~J., {Stevens} J.~A., {Mittaz} J.~P.~D., {Carrera} F.~J., 2001,
  Science, 294, 2516

\bibitem[{{Page} {et~al}\mbox{.}(2012){Page}, {Symeonidis}, {Vieira},
  {Altieri}, {Amblard}, {Arumugam}, {Aussel}, {Babbedge}, {Blain}, {Bock},
  {Boselli}, {Buat}, {Castro-Rodr{\'{\i}}guez}, {Cava}, {Chanial}, {Clements},
  {Conley}, {Conversi}, {Cooray}, {Dowell}, {Dubois}, {Dunlop}, {Dwek}, {Dye},
  {Eales}, {Elbaz}, {Farrah}, {Fox}, {Franceschini}, {Gear}, {Glenn},
  {Griffin}, {Halpern}, {Hatziminaoglou}, {Ibar}, {Isaak}, {Ivison}, {Lagache},
  {Levenson}, {Lu}, {Madden}, {Maffei}, {Mainetti}, {Marchetti}, {Nguyen},
  {O'Halloran}, {Oliver}, {Omont}, {Panuzzo}, {Papageorgiou}, {Pearson},
  {P{\'e}rez-Fournon}, {Pohlen}, {Rawlings}, {Rigopoulou}, {Riguccini},
  {Rizzo}, {Rodighiero}, {Roseboom}, {Rowan-Robinson}, {Portal}, {Schulz},
  {Scott}, {Seymour}, {Shupe}, {Smith}, {Stevens}, {Trichas}, {Tugwell},
  {Vaccari}, {Valtchanov}, {Viero}, {Vigroux}, {Wang}, {Ward}, {Wright}, {Xu},
  \& {Zemcov}}]{Page2012}
{Page} M.~J. {et~al.}, 2012, \nat, 485, 213

\bibitem[{{Papadopoulos} {et~al}\mbox{.}(2010){Papadopoulos}, {van der Werf},
  {Isaak}, \& {Xilouris}}]{Papadopoulos2010}
{Papadopoulos} P.~P., {van der Werf} P., {Isaak} K., {Xilouris} E.~M., 2010,
  \apj, 715, 775

\bibitem[{{Pascale} {et~al}\mbox{.}(2011){Pascale}, {Auld}, {Dariush}, {Dunne},
  {Eales}, {Maddox}, {Panuzzo}, {Pohlen}, {Smith}, {Buttiglione}, {Cava},
  {Clements}, {Cooray}, {Dye}, {de Zotti}, {Fritz}, {Hopwood}, {Ibar},
  {Ivison}, {Jarvis}, {Leeuw}, {L{\'o}pez-Caniego}, {Rigby}, {Rodighiero},
  {Scott}, {Smith}, {Temi}, {Vaccari}, \& {Valtchanov}}]{Pascale2011}
{Pascale} E. {et~al.}, 2011, \mnras, 415, 911

\bibitem[{{Penner} {et~al}\mbox{.}(2011){Penner}, {Pope}, {Chapin}, {Greve},
  {Bertoldi}, {Brodwin}, {Chary}, {Conselice}, {Coppin}, {Giavalisco},
  {Hughes}, {Ivison}, {Perera}, {Scott}, {Scott}, \& {Wilson}}]{Penner2011}
{Penner} K. {et~al.}, 2011, \mnras, 410, 2749

\bibitem[{{Pilbratt} {et~al}\mbox{.}(2010){Pilbratt}, {Riedinger}, {Passvogel},
  {Crone}, {Doyle}, {Gageur}, {Heras}, {Jewell}, {Metcalfe}, {Ott}, \&
  {Schmidt}}]{Pilbratt2010}
{Pilbratt} G.~L. {et~al.}, 2010, \aap, 518, L1

\bibitem[{{Podigachoski} {et~al}\mbox{.}(2015){Podigachoski}, {Barthel},
  {Haas}, {Leipski}, {Wilkes}, {Kuraszkiewicz}, {Westhues}, {Willner}, {Ashby},
  {Chini}, {Clements}, {Fazio}, {Labiano}, {Lawrence}, {Meisenheimer},
  {Peletier}, {Siebenmorgen}, \& {Verdoes Kleijn}}]{Podigachoski2015}
{Podigachoski} P. {et~al.}, 2015, \aap, 575, A80

\bibitem[{{Poglitsch} {et~al}\mbox{.}(2010){Poglitsch}, {Waelkens}, {Geis},
  {Feuchtgruber}, {Vandenbussche}, {Rodriguez}, {Krause}, {Renotte}, {van
  Hoof}, {Saraceno}, {Cepa}, {Kerschbaum}, {Agn{\`e}se}, {Ali}, {Altieri},
  {Andreani}, {Augueres}, {Balog}, {Barl}, {Bauer}, {Belbachir}, {Benedettini},
  {Billot}, {Boulade}, {Bischof}, {Blommaert}, {Callut}, {Cara}, {Cerulli},
  {Cesarsky}, {Contursi}, {Creten}, {De Meester}, {Doublier}, {Doumayrou},
  {Duband}, {Exter}, {Genzel}, {Gillis}, {Gr{\"o}zinger}, {Henning},
  {Herreros}, {Huygen}, {Inguscio}, {Jakob}, {Jamar}, {Jean}, {de Jong},
  {Katterloher}, {Kiss}, {Klaas}, {Lemke}, {Lutz}, {Madden}, {Marquet},
  {Martignac}, {Mazy}, {Merken}, {Montfort}, {Morbidelli}, {M{\"u}ller},
  {Nielbock}, {Okumura}, {Orfei}, {Ottensamer}, {Pezzuto}, {Popesso},
  {Putzeys}, {Regibo}, {Reveret}, {Royer}, {Sauvage}, {Schreiber}, {Stegmaier},
  {Schmitt}, {Schubert}, {Sturm}, {Thiel}, {Tofani}, {Vavrek}, {Wetzstein},
  {Wieprecht}, \& {Wiezorrek}}]{Poglitsch2010}
{Poglitsch} A. {et~al.}, 2010, \aap, 518, L2

\bibitem[{{Popesso} {et~al}\mbox{.}(2012){Popesso}, {Magnelli}, {Buttiglione},
  {Lutz}, {Poglitsch}, {Berta}, {Nordon}, {Altieri}, {Aussel}, {Billot},
  {Gastaud}, {Ali}, {Balog}, {Cava}, {Feuchtgruber}, {Gonzalez Garcia}, {Geis},
  {Kiss}, {Klaas}, {Linz}, {Liu}, {Moor}, {Morin}, {Muller}, {Nielbock},
  {Okumura}, {Osterhage}, {Ottensamer}, {Paladini}, {Pezzuto}, {Dublier
  Pritchard}, {Regibo}, {Rodighiero}, {Royer}, {Sauvage}, {Sturm}, {Wetzstein},
  {Wieprecht}, \& {Wiezorrek}}]{Popesso2012}
{Popesso} P. {et~al.}, 2012, ArXiv e-prints

\bibitem[{{Priddey} {et~al}\mbox{.}(2003){Priddey}, {Isaak}, {McMahon}, \&
  {Omont}}]{Priddey2003}
{Priddey} R.~S., {Isaak} K.~G., {McMahon} R.~G., {Omont} A., 2003, \mnras, 339,
  1183

\bibitem[{{Ramos Almeida} {et~al}\mbox{.}(2012){Ramos Almeida}, {Bessiere},
  {Tadhunter}, {P{\'e}rez-Gonz{\'a}lez}, {Barro}, {Inskip}, {Morganti}, {Holt},
  \& {Dicken}}]{RamosAlmeida2012}
{Ramos Almeida} C. {et~al.}, 2012, \mnras, 419, 687

\bibitem[{{Rawlings} {et~al}\mbox{.}(2001){Rawlings}, {Eales}, \&
  {Lacy}}]{Rawlings2001}
{Rawlings} S., {Eales} S., {Lacy} M., 2001, \mnras, 322, 523

\bibitem[{{Rengelink} {et~al}\mbox{.}(1997){Rengelink}, {Tang}, {de Bruyn},
  {Miley}, {Bremer}, {Roettgering}, \& {Bremer}}]{Rengelink1997}
{Rengelink} R.~B., {Tang} Y., {de Bruyn} A.~G., {Miley} G.~K., {Bremer} M.~N.,
  {Roettgering} H.~J.~A., {Bremer} M.~A.~R., 1997, \aaps, 124, 259

\bibitem[{{Richards} {et~al}\mbox{.}(2006){Richards}, {Strauss}, {Fan}, {Hall},
  {Jester}, {Schneider}, {Vanden Berk}, {Stoughton}, {Anderson}, {Brunner},
  {Gray}, {Gunn}, {Ivezi{\'c}}, {Kirkland}, {Knapp}, {Loveday}, {Meiksin},
  {Pope}, {Szalay}, {Thakar}, {Yanny}, {York}, {Barentine}, {Brewington},
  {Brinkmann}, {Fukugita}, {Harvanek}, {Kent}, {Kleinman}, {Krzesi{\'n}ski},
  {Long}, {Lupton}, {Nash}, {Neilsen}, {Nitta}, {Schlegel}, \&
  {Snedden}}]{Richards2006b}
{Richards} G.~T. {et~al.}, 2006, \aj, 131, 2766

\bibitem[{{Rigby} {et~al}\mbox{.}(2015){Rigby}, {Argyle}, {Best}, {Rosario}, \&
  {R{\"o}ttgering}}]{Rigby2015}
{Rigby} E.~E., {Argyle} J., {Best} P.~N., {Rosario} D., {R{\"o}ttgering}
  H.~J.~A., 2015, \aap, 581, A96

\bibitem[{{Rigby} {et~al}\mbox{.}(2011){Rigby}, {Best}, {Brookes}, {Peacock},
  {Dunlop}, {R{\"o}ttgering}, {Wall}, \& {Ker}}]{Rigby2011}
{Rigby} E.~E., {Best} P.~N., {Brookes} M.~H., {Peacock} J.~A., {Dunlop} J.~S.,
  {R{\"o}ttgering} H.~J.~A., {Wall} J.~V., {Ker} L., 2011, \mnras, 416, 1900

\bibitem[{{Rodighiero} {et~al}\mbox{.}(2015){Rodighiero}, {Brusa}, {Daddi},
  {Negrello}, {Mullaney}, {Delvecchio}, {Lutz}, {Renzini}, {Franceschini},
  {Baronchelli}, {Pozzi}, {Gruppioni}, {Strazzullo}, {Cimatti}, \&
  {Silverman}}]{Rodighiero2015}
{Rodighiero} G. {et~al.}, 2015, \apjl, 800, L10

\bibitem[{{Rosario} {et~al}\mbox{.}(2012){Rosario}, {Santini}, {Lutz}, {Shao},
  {Maiolino}, {Alexander}, {Altieri}, {Andreani}, {Aussel}, {Bauer}, {Berta},
  {Bongiovanni}, {Brandt}, {Brusa}, {Cepa}, {Cimatti}, {Cox}, {Daddi}, {Elbaz},
  {Fontana}, {F{\"o}rster Schreiber}, {Genzel}, {Grazian}, {Le Floch},
  {Magnelli}, {Mainieri}, {Netzer}, {Nordon}, {P{\'e}rez Garcia}, {Poglitsch},
  {Popesso}, {Pozzi}, {Riguccini}, {Rodighiero}, {Salvato}, {Sanchez-Portal},
  {Sturm}, {Tacconi}, {Valtchanov}, \& {Wuyts}}]{Rosario2012}
{Rosario} D.~J. {et~al.}, 2012, \aap, 545, A45

\bibitem[{{Rosario} {et~al}\mbox{.}(2013){Rosario}, {Trakhtenbrot}, {Lutz},
  {Netzer}, {Trump}, {Silverman}, {Schramm}, {Lusso}, {Berta}, {Bongiorno},
  {Brusa}, {F{\"o}rster-Schreiber}, {Genzel}, {Lilly}, {Magnelli}, {Mainieri},
  {Maiolino}, {Merloni}, {Mignoli}, {Nordon}, {Popesso}, {Salvato}, {Santini},
  {Tacconi}, \& {Zamorani}}]{Rosario2013}
{Rosario} D.~J. {et~al.}, 2013, \aap, 560, A72

\bibitem[{{Rovilos} {et~al}\mbox{.}(2012){Rovilos}, {Comastri}, {Gilli},
  {Georgantopoulos}, {Ranalli}, {Vignali}, {Lusso}, {Cappelluti}, {Zamorani},
  {Elbaz}, {Dickinson}, {Hwang}, {Charmandaris}, {Ivison}, {Merloni}, {Daddi},
  {Carrera}, {Brandt}, {Mullaney}, {Scott}, {Alexander}, {Del Moro},
  {Morrison}, {Murphy}, {Altieri}, {Aussel}, {Dannerbauer}, {Kartaltepe},
  {Leiton}, {Magdis}, {Magnelli}, {Popesso}, \& {Valtchanov}}]{Rovilos2012}
{Rovilos} E. {et~al.}, 2012, \aap, 546, A58

\bibitem[{{Rowan-Robinson}(1995)}]{Rowan1995}
{Rowan-Robinson} M., 1995, \mnras, 272, 737

\bibitem[{{Salviander} {et~al}\mbox{.}(2007){Salviander}, {Shields},
  {Gebhardt}, \& {Bonning}}]{Salviander2007}
{Salviander} S., {Shields} G.~A., {Gebhardt} K., {Bonning} E.~W., 2007, \apj,
  662, 131

\bibitem[{{S{\'a}nchez} {et~al}\mbox{.}(2004){S{\'a}nchez}, {Jahnke},
  {Wisotzki}, {McIntosh}, {Bell}, {Barden}, {Beckwith}, {Borch}, {Caldwell},
  {H{\"a}ussler}, {Jogee}, {Meisenheimer}, {Peng}, {Rix}, {Somerville}, \&
  {Wolf}}]{Sanchez2004}
{S{\'a}nchez} S.~F. {et~al.}, 2004, \apj, 614, 586

\bibitem[{{Santini} {et~al}\mbox{.}(2010){Santini}, {Maiolino}, {Magnelli},
  {Silva}, {Grazian}, {Altieri}, {Andreani}, {Aussel}, {Berta}, {Bongiovanni},
  {Brisbin}, {Calura}, {Cava}, {Cepa}, {Cimatti}, {Daddi}, {Dannerbauer},
  {Dominguez-Sanchez}, {Elbaz}, {Fontana}, {F{\"o}rster Schreiber}, {Genzel},
  {Granato}, {Gruppioni}, {Lutz}, {Magdis}, {Magliocchetti}, {Matteucci},
  {Nordon}, {P{\'e}rez Garcia}, {Poglitsch}, {Popesso}, {Pozzi}, {Riguccini},
  {Rodighiero}, {Saintonge}, {Sanchez-Portal}, {Shao}, {Sturm}, {Tacconi}, \&
  {Valtchanov}}]{Santini2010}
{Santini} P. {et~al.}, 2010, \aap, 518, L154

\bibitem[{{Santini} {et~al}\mbox{.}(2012){Santini}, {Rosario}, {Shao}, {Lutz},
  {Maiolino}, {Alexander}, {Altieri}, {Andreani}, {Aussel}, {Bauer}, {Berta},
  {Bongiovanni}, {Brandt}, {Brusa}, {Cepa}, {Cimatti}, {Daddi}, {Elbaz},
  {Fontana}, {F{\"o}rster Schreiber}, {Genzel}, {Grazian}, {Le Floc'h},
  {Magnelli}, {Mainieri}, {Nordon}, {P{\'e}rez Garcia}, {Poglitsch}, {Popesso},
  {Pozzi}, {Riguccini}, {Rodighiero}, {Salvato}, {Sanchez-Portal}, {Sturm},
  {Tacconi}, {Valtchanov}, \& {Wuyts}}]{Santini2012}
{Santini} P. {et~al.}, 2012, \aap, 540, A109

\bibitem[{{Sargent} {et~al}\mbox{.}(2013){Sargent}, {Patel}, {Meixner},
  {Otsuka}, {Riebel}, \& {Srinivasan}}]{Sargent2013}
{Sargent} B.~A., {Patel} N.~A., {Meixner} M., {Otsuka} M., {Riebel} D.,
  {Srinivasan} S., 2013, \apj, 765, 20

\bibitem[{{Sargent} {et~al}\mbox{.}(2010){Sargent}, {Schinnerer}, {Murphy},
  {Aussel}, {Le Floc'h}, {Frayer}, {Mart{\'{\i}}nez-Sansigre}, {Oesch},
  {Salvato}, {Smol{\v c}i{\'c}}, {Zamorani}, {Brusa}, {Cappelluti}, {Carilli},
  {Carollo}, {Ilbert}, {Kartaltepe}, {Koekemoer}, {Lilly}, {Sanders}, \&
  {Scoville}}]{Sargent2010}
{Sargent} M.~T. {et~al.}, 2010, \apjs, 186, 341

\bibitem[{{Schneider} {et~al}\mbox{.}(2005){Schneider}, {Hall}, {Richards},
  {Vanden Berk}, {Anderson}, {Fan}, {Jester}, {Stoughton}, {Strauss},
  {SubbaRao}, {Brandt}, {Gunn}, {Yanny}, {Bahcall}, {Barentine}, {Blanton},
  {Boroski}, {Brewington}, {Brinkmann}, {Brunner}, {Csabai}, {Doi},
  {Eisenstein}, {Frieman}, {Fukugita}, {Gray}, {Harvanek}, {Heckman},
  {Ivezi{\'c}}, {Kent}, {Kleinman}, {Knapp}, {Kron}, {Krzesinski}, {Long},
  {Loveday}, {Lupton}, {Margon}, {Munn}, {Neilsen}, {Newberg}, {Newman},
  {Nichol}, {Nitta}, {Pier}, {Rockosi}, {Saxe}, {Schlegel}, {Snedden},
  {Szalay}, {Thakar}, {Uomoto}, {Voges}, \& {York}}]{Schneider2005}
{Schneider} D.~P. {et~al.}, 2005, \aj, 130, 367

\bibitem[{{Schreiber} {et~al}\mbox{.}(2015){Schreiber}, {Pannella}, {Elbaz},
  {B{\'e}thermin}, {Inami}, {Dickinson}, {Magnelli}, {Wang}, {Aussel}, {Daddi},
  {Juneau}, {Shu}, {Sargent}, {Buat}, {Faber}, {Ferguson}, {Giavalisco},
  {Koekemoer}, {Magdis}, {Morrison}, {Papovich}, {Santini}, \&
  {Scott}}]{Schreiber2015}
{Schreiber} C. {et~al.}, 2015, \aap, 575, A74

\bibitem[{{Schulze} \& {Wisotzki}(2010)}]{Schulze2010}
{Schulze} A., {Wisotzki} L., 2010, \aap, 516, A87

\bibitem[{{Schweitzer} {et~al}\mbox{.}(2006){Schweitzer}, {Lutz}, {Sturm},
  {Contursi}, {Tacconi}, {Lehnert}, {Dasyra}, {Genzel}, {Veilleux}, {Rupke},
  {Kim}, {Baker}, {Netzer}, {Sternberg}, {Mazzarella}, \&
  {Lord}}]{Schweitzer2006}
{Schweitzer} M. {et~al.}, 2006, \apj, 649, 79

\bibitem[{{Seymour} {et~al}\mbox{.}(2007){Seymour}, {Stern}, {De Breuck},
  {Vernet}, {Rettura}, {Dickinson}, {Dey}, {Eisenhardt}, {Fosbury}, {Lacy},
  {McCarthy}, {Miley}, {Rocca-Volmerange}, {R{\"o}ttgering}, {Stanford},
  {Teplitz}, {van Breugel}, \& {Zirm}}]{Seymour2007}
{Seymour} N. {et~al.}, 2007, \apjs, 171, 353

\bibitem[{{Seymour} {et~al}\mbox{.}(2011){Seymour}, {Symeonidis}, {Page},
  {Amblard}, {Arumugam}, {Aussel}, {Blain}, {Bock}, {Boselli}, {Buat},
  {Castro-Rodr{\'{\i}}guez}, {Cava}, {Chanial}, {Clements}, {Conley},
  {Conversi}, {Cooray}, {Dowell}, {Dwek}, {Eales}, {Elbaz}, {Franceschini},
  {Glenn}, {Solares}, {Griffin}, {Hatziminaoglou}, {Ibar}, {Isaak}, {Ivison},
  {Lagache}, {Levenson}, {Lu}, {Madden}, {Maffei}, {Mainetti}, {Marchetti},
  {Nguyen}, {O'Halloran}, {Oliver}, {Omont}, {Panuzzo}, {Papageorgiou},
  {Pearson}, {P{\'e}rez-Fournon}, {Pohlen}, {Rawlings}, {Rizzo}, {Roseboom},
  {Rowan-Robinson}, {Schulz}, {Scott}, {Shupe}, {Smith}, {Stevens}, {Trichas},
  {Tugwell}, {Vaccari}, {Valtchanov}, {Vigroux}, {Wang}, {Wright}, {Xu}, \&
  {Zemcov}}]{Seymour2011}
{Seymour} N. {et~al.}, 2011, \mnras, 413, 1777

\bibitem[{{Shabala} \& {Godfrey}(2013)}]{Shabala2013}
{Shabala} S.~S., {Godfrey} L.~E.~H., 2013, \apj, 769, 129

\bibitem[{{Shankar}(2010)}]{Shankar2010}
{Shankar} F., 2010, in IAU Symposium, Vol. 267, IAU Symposium, pp. 248--253

\bibitem[{{Shankar} {et~al}\mbox{.}(2004){Shankar}, {Salucci}, {Granato}, {De
  Zotti}, \& {Danese}}]{Shankar2004}
{Shankar} F., {Salucci} P., {Granato} G.~L., {De Zotti} G., {Danese} L., 2004,
  \mnras, 354, 1020

\bibitem[{{Shao} {et~al}\mbox{.}(2010){Shao}, {Lutz}, {Nordon}, {Maiolino},
  {Alexander}, {Altieri}, {Andreani}, {Aussel}, {Bauer}, {Berta},
  {Bongiovanni}, {Brandt}, {Brusa}, {Cava}, {Cepa}, {Cimatti}, {Daddi},
  {Dominguez-Sanchez}, {Elbaz}, {F{\"o}rster Schreiber}, {Geis}, {Genzel},
  {Grazian}, {Gruppioni}, {Magdis}, {Magnelli}, {Mainieri}, {P{\'e}rez
  Garc{\'{\i}}a}, {Poglitsch}, {Popesso}, {Pozzi}, {Riguccini}, {Rodighiero},
  {Rovilos}, {Saintonge}, {Salvato}, {Sanchez Portal}, {Santini}, {Sturm},
  {Tacconi}, {Valtchanov}, {Wetzstein}, \& {Wieprecht}}]{Shao2010}
{Shao} L. {et~al.}, 2010, \aap, 518, L26

\bibitem[{{Shen} {et~al}\mbox{.}(2008){Shen}, {Greene}, {Strauss}, {Richards},
  \& {Schneider}}]{Shen2008}
{Shen} Y., {Greene} J.~E., {Strauss} M.~A., {Richards} G.~T., {Schneider}
  D.~P., 2008, \apj, 680, 169

\bibitem[{{Shen} {et~al}\mbox{.}(2011){Shen}, {Richards}, {Strauss}, {Hall},
  {Schneider}, {Snedden}, {Bizyaev}, {Brewington}, {Malanushenko},
  {Malanushenko}, {Oravetz}, {Pan}, \& {Simmons}}]{Shen2011}
{Shen} Y. {et~al.}, 2011, \apjs, 194, 45

\bibitem[{{Shi} {et~al}\mbox{.}(2005){Shi}, {Rieke}, {Hines}, {Neugebauer},
  {Blaylock}, {Rigby}, {Egami}, {Gordon}, \& {Alonso-Herrero}}]{Shi2005}
{Shi} Y. {et~al.}, 2005, \apj, 629, 88

\bibitem[{{Silk} {et~al}\mbox{.}(2012){Silk}, {Antonuccio-Delogu}, {Dubois},
  {Gaibler}, {Haas}, {Khochfar}, \& {Krause}}]{Silk2012}
{Silk} J., {Antonuccio-Delogu} V., {Dubois} Y., {Gaibler} V., {Haas} M.~R.,
  {Khochfar} S., {Krause} M., 2012, \aap, 545, L11

\bibitem[{{Silk} \& {Nusser}(2010)}]{Silk2010}
{Silk} J., {Nusser} A., 2010, \apj, 725, 556

\bibitem[{{Smith}(2015)}]{Smith2015}
{Smith} D.~J.~B., 2015, ArXiv e-prints

\bibitem[{{Smith} {et~al}\mbox{.}(2013){Smith}, {Hardcastle}, {Jarvis},
  {Maddox}, {Dunne}, {Bonfield}, {Eales}, {Serjeant}, {Thompson}, {Baes},
  {Clements}, {Cooray}, {De Zotti}, {Gonz{\`a}lez-Nuevo}, {van der Werf},
  {Virdee}, {Bourne}, {Dariush}, {Hopwood}, {Ibar}, \& {Valiante}}]{Smith2013}
{Smith} D.~J.~B. {et~al.}, 2013, \mnras, 436, 2435

\bibitem[{{Somerville} {et~al}\mbox{.}(2008){Somerville}, {Hopkins}, {Cox},
  {Robertson}, \& {Hernquist}}]{Somerville2008}
{Somerville} R.~S., {Hopkins} P.~F., {Cox} T.~J., {Robertson} B.~E.,
  {Hernquist} L., 2008, \mnras, 391, 481

\bibitem[{{Spinrad} {et~al}\mbox{.}(1985){Spinrad}, {Marr}, {Aguilar}, \&
  {Djorgovski}}]{Spinrad1985}
{Spinrad} H., {Marr} J., {Aguilar} L., {Djorgovski} S., 1985, \pasp, 97, 932

\bibitem[{{Springel} {et~al}\mbox{.}(2005){Springel}, {White}, {Jenkins},
  {Frenk}, {Yoshida}, {Gao}, {Navarro}, {Thacker}, {Croton}, {Helly},
  {Peacock}, {Cole}, {Thomas}, {Couchman}, {Evrard}, {Colberg}, \&
  {Pearce}}]{Springel2005}
{Springel} V. {et~al.}, 2005, \nat, 435, 629

\bibitem[{{Stanley} {et~al}\mbox{.}(2015){Stanley}, {Harrison}, {Alexander},
  {Swinbank}, {Aird}, {Del Moro}, {Hickox}, \& {Mullaney}}]{Stanley2015}
{Stanley} F., {Harrison} C.~M., {Alexander} D.~M., {Swinbank} A.~M., {Aird}
  J.~A., {Del Moro} A., {Hickox} R.~C., {Mullaney} J.~R., 2015, \mnras, 453,
  591

\bibitem[{{Steinhardt} \& {Elvis}(2010)}]{Steinhardt2010}
{Steinhardt} C.~L., {Elvis} M., 2010, \mnras, 402, 2637

\bibitem[{{Stevens} {et~al}\mbox{.}(2003){Stevens}, {Ivison}, {Dunlop},
  {Smail}, {Percival}, {Hughes}, {R{\"o}ttgering}, {van Breugel}, \&
  {Reuland}}]{Stevens2003}
{Stevens} J.~A. {et~al.}, 2003, \nat, 425, 264

\bibitem[{{Treister} {et~al}\mbox{.}(2012){Treister}, {Schawinski}, {Urry}, \&
  {Simmons}}]{Treister2012}
{Treister} E., {Schawinski} K., {Urry} C.~M., {Simmons} B.~D., 2012, \apjl,
  758, L39

\bibitem[{{Trichas} {et~al}\mbox{.}(2012){Trichas}, {Green}, {Silverman},
  {Aldcroft}, {Barkhouse}, {Cameron}, {Constantin}, {Ellison}, {Foltz},
  {Haggard}, {Jannuzi}, {Kim}, {Marshall}, {Mossman}, {P{\'e}rez},
  {Romero-Colmenero}, {Ruiz}, {Smith}, {Smith}, {Torres}, {Wik}, {Wilkes}, \&
  {Wolfgang}}]{Trichas2012}
{Trichas} M. {et~al.}, 2012, \apjs, 200, 17

\bibitem[{{Trichas} {et~al}\mbox{.}(2010){Trichas}, {Rowan-Robinson},
  {Georgakakis}, {Valtchanov}, {Nandra}, {Farrah}, {Morrison}, {Clements}, \&
  {Waddington}}]{Trichas2010}
{Trichas} M. {et~al.}, 2010, \mnras, 405, 2243

\bibitem[{{Urrutia} {et~al}\mbox{.}(2008){Urrutia}, {Lacy}, \&
  {Becker}}]{Urrutia2008}
{Urrutia} T., {Lacy} M., {Becker} R.~H., 2008, \apj, 674, 80

\bibitem[{{van Breugel} {et~al}\mbox{.}(2004){van Breugel}, {Fragile},
  {Anninos}, \& {Murray}}]{vanBreugel2004}
{van Breugel} W., {Fragile} C., {Anninos} P., {Murray} S., 2004, in IAU
  Symposium, Vol. 217, Recycling Intergalactic and Interstellar Matter, {Duc}
  P.-A., {Braine} J., {Brinks} E., eds., p. 472

\bibitem[{{van Haarlem} {et~al}\mbox{.}(2013){van Haarlem}, {Wise}, {Gunst},
  {Heald}, {McKean}, {Hessels}, {de Bruyn}, {Nijboer}, {Swinbank}, {Fallows},
  {Brentjens}, {Nelles}, {Beck}, {Falcke}, {Fender}, {H{\"o}randel},
  {Koopmans}, {Mann}, {Miley}, {R{\"o}ttgering}, {Stappers}, {Wijers},
  {Zaroubi}, {van den Akker}, {Alexov}, {Anderson}, {Anderson}, {van Ardenne},
  {Arts}, {Asgekar}, {Avruch}, {Batejat}, {B{\"a}hren}, {Bell}, {Bell}, {van
  Bemmel}, {Bennema}, {Bentum}, {Bernardi}, {Best}, {B{\^i}rzan}, {Bonafede},
  {Boonstra}, {Braun}, {Bregman}, {Breitling}, {van de Brink}, {Broderick},
  {Broekema}, {Brouw}, {Br{\"u}ggen}, {Butcher}, {van Cappellen}, {Ciardi},
  {Coenen}, {Conway}, {Coolen}, {Corstanje}, {Damstra}, {Davies}, {Deller},
  {Dettmar}, {van Diepen}, {Dijkstra}, {Donker}, {Doorduin}, {Dromer}, {Drost},
  {van Duin}, {Eisl{\"o}ffel}, {van Enst}, {Ferrari}, {Frieswijk}, {Gankema},
  {Garrett}, {de Gasperin}, {Gerbers}, {de Geus}, {Grie{\ss}meier}, {Grit},
  {Gruppen}, {Hamaker}, {Hassall}, {Hoeft}, {Holties}, {Horneffer}, {van der
  Horst}, {van Houwelingen}, {Huijgen}, {Iacobelli}, {Intema}, {Jackson},
  {Jelic}, {de Jong}, {Juette}, {Kant}, {Karastergiou}, {Koers}, {Kollen},
  {Kondratiev}, {Kooistra}, {Koopman}, {Koster}, {Kuniyoshi}, {Kramer},
  {Kuper}, {Lambropoulos}, {Law}, {van Leeuwen}, {Lemaitre}, {Loose}, {Maat},
  {Macario}, {Markoff}, {Masters}, {McFadden}, {McKay-Bukowski}, {Meijering},
  {Meulman}, {Mevius}, {Middelberg}, {Millenaar}, {Miller-Jones}, {Mohan},
  {Mol}, {Morawietz}, {Morganti}, {Mulcahy}, {Mulder}, {Munk}, {Nieuwenhuis},
  {van Nieuwpoort}, {Noordam}, {Norden}, {Noutsos}, {Offringa}, {Olofsson},
  {Omar}, {Orr{\'u}}, {Overeem}, {Paas}, {Pandey-Pommier}, {Pandey}, {Pizzo},
  {Polatidis}, {Rafferty}, {Rawlings}, {Reich}, {de Reijer}, {Reitsma},
  {Renting}, {Riemers}, {Rol}, {Romein}, {Roosjen}, {Ruiter}, {Scaife}, {van
  der Schaaf}, {Scheers}, {Schellart}, {Schoenmakers}, {Schoonderbeek},
  {Serylak}, {Shulevski}, {Sluman}, {Smirnov}, {Sobey}, {Spreeuw}, {Steinmetz},
  {Sterks}, {Stiepel}, {Stuurwold}, {Tagger}, {Tang}, {Tasse}, {Thomas},
  {Thoudam}, {Toribio}, {van der Tol}, {Usov}, {van Veelen}, {van der Veen},
  {ter Veen}, {Verbiest}, {Vermeulen}, {Vermaas}, {Vocks}, {Vogt}, {de Vos},
  {van der Wal}, {van Weeren}, {Weggemans}, {Weltevrede}, {White}, {Wijnholds},
  {Wilhelmsson}, {Wucknitz}, {Yatawatta}, {Zarka}, {Zensus}, \& {van
  Zwieten}}]{vanHaarlem2013}
{van Haarlem} M.~P. {et~al.}, 2013, \aap, 556, A2

\bibitem[{{Vardoulaki} {et~al}\mbox{.}(2010){Vardoulaki}, {Rawlings}, {Hill},
  {Mauch}, {Inskip}, {Riley}, {Brand}, {Croft}, \& {Willott}}]{Vardoulaki2010}
{Vardoulaki} E. {et~al.}, 2010, \mnras, 401, 1709

\bibitem[{{Venemans} {et~al}\mbox{.}(2007){Venemans}, {R{\"o}ttgering},
  {Miley}, {van Breugel}, {de Breuck}, {Kurk}, {Pentericci}, {Stanford},
  {Overzier}, {Croft}, \& {Ford}}]{Venemans2007}
{Venemans} B.~P. {et~al.}, 2007, \aap, 461, 823

\bibitem[{{Virdee} {et~al}\mbox{.}(2013){Virdee}, {Hardcastle}, {Rawlings},
  {Rigopoulou}, {Mauch}, {Jarvis}, {Verma}, {Smith}, {Heywood}, {White},
  {Baes}, {Cooray}, {Zotti}, {Eales}, {Micha{\l}owski}, {Bourne}, {Dariush},
  {Dunne}, {Hopwood}, {Ibar}, {Maddox}, {Smith}, \& {Valiante}}]{Virdee2013}
{Virdee} J.~S. {et~al.}, 2013, \mnras, 432, 609

\bibitem[{{Wagner} {et~al}\mbox{.}(2012){Wagner}, {Bicknell}, \&
  {Umemura}}]{Wagner2012}
{Wagner} A.~Y., {Bicknell} G.~V., {Umemura} M., 2012, \apj, 757, 136

\bibitem[{{Wall} {et~al}\mbox{.}(2005){Wall}, {Jackson}, {Shaver}, {Hook}, \&
  {Kellermann}}]{Wall2005}
{Wall} J.~V., {Jackson} C.~A., {Shaver} P.~A., {Hook} I.~M., {Kellermann}
  K.~I., 2005, \aap, 434, 133

\bibitem[{{Wang} {et~al}\mbox{.}(2009){Wang}, {Dong}, {Wang}, {Ho}, {Yuan},
  {Wang}, {Zhang}, {Zhang}, \& {Zhou}}]{Wang2009}
{Wang} J.-G. {et~al.}, 2009, \apj, 707, 1334

\bibitem[{{Wardlow} {et~al}\mbox{.}(2011){Wardlow}, {Smail}, {Coppin},
  {Alexander}, {Brandt}, {Danielson}, {Luo}, {Swinbank}, {Walter}, {Wei{\ss}},
  {Xue}, {Zibetti}, {Bertoldi}, {Biggs}, {Chapman}, {Dannerbauer}, {Dunlop},
  {Gawiser}, {Ivison}, {Knudsen}, {Kov{\'a}cs}, {Lacey}, {Menten}, {Padilla},
  {Rix}, \& {van der Werf}}]{Wardlow2011}
{Wardlow} J.~L. {et~al.}, 2011, \mnras, 415, 1479

\bibitem[{{White} {et~al}\mbox{.}(2007){White}, {Helfand}, {Becker}, {Glikman},
  \& {de Vries}}]{White2007}
{White} R.~L., {Helfand} D.~J., {Becker} R.~H., {Glikman} E., {de Vries} W.,
  2007, \apj, 654, 99

\bibitem[{{Whittam} {et~al}\mbox{.}(2013){Whittam}, {Riley}, {Green}, {Jarvis},
  {Prandoni}, {Guglielmino}, {Morganti}, {R{\"o}ttgering}, \&
  {Garrett}}]{Whittam2013}
{Whittam} I.~H. {et~al.}, 2013, \mnras, 429, 2080

\bibitem[{{Whittam} {et~al}\mbox{.}(2015){Whittam}, {Riley}, {Green}, {Jarvis},
  \& {Vaccari}}]{Whittam2015}
{Whittam} I.~H., {Riley} J.~M., {Green} D.~A., {Jarvis} M.~J., {Vaccari} M.,
  2015, \mnras, 453, 4244

\bibitem[{{Willott} {et~al}\mbox{.}(1998){Willott}, {Rawlings}, {Blundell}, \&
  {Lacy}}]{Willott1998}
{Willott} C.~J., {Rawlings} S., {Blundell} K.~M., {Lacy} M., 1998, \mnras, 300,
  625

\bibitem[{{Willott} {et~al}\mbox{.}(1999){Willott}, {Rawlings}, {Blundell}, \&
  {Lacy}}]{Willott1999}
{Willott} C.~J., {Rawlings} S., {Blundell} K.~M., {Lacy} M., 1999, \mnras, 309,
  1017

\bibitem[{{Willott} {et~al}\mbox{.}(2001){Willott}, {Rawlings}, {Blundell},
  {Lacy}, \& {Eales}}]{Willott2001}
{Willott} C.~J., {Rawlings} S., {Blundell} K.~M., {Lacy} M., {Eales} S.~A.,
  2001, \mnras, 322, 536

\bibitem[{{Xue} {et~al}\mbox{.}(2010){Xue}, {Brandt}, {Luo}, {Rafferty},
  {Alexander}, {Bauer}, {Lehmer}, {Schneider}, \& {Silverman}}]{Xue2010}
{Xue} Y.~Q. {et~al.}, 2010, \apj, 720, 368

\bibitem[{{Younger} {et~al}\mbox{.}(2008){Younger}, {Fazio}, {Wilner}, {Ashby},
  {Blundell}, {Gurwell}, {Huang}, {Iono}, {Peck}, {Petitpas}, {Scott},
  {Wilson}, \& {Yun}}]{Younger2008}
{Younger} J.~D. {et~al.}, 2008, \apj, 688, 59

\end{thebibliography}

\appendix

\section{Catalogues of the AGN sample and SEDs of the RLQs} 

In this appendix the best-fit radio to FIR SED plots for a representative sample of RLQs are reported (Fig.~\ref{fig:SED_appendix}). Table~\ref{Table:Herschel_photometry} lists the {\it Herschel} photometry of the objects studied in this work with their $1\sigma$ photometric uncertainties (cols. 7, 8, 10, 11, 12) obtained as described in Sec.~\ref{section:Herschel_photometry}.  Their PACS and SPIRE observation IDs are given in cols. 6 and 9 indicating also the cases that the observations obtained from public data.

We also present the properties of the total AGN sample in Table~\ref{Table:full_properties}. The latter lists the following information for each source: right ascension and declination (J2000.0) in degrees (cols. 1 and 2), type classification (col. 3), redshift (col. 4), logarithmic bolometric luminosity measured as described in Sec.~\ref{sec:accretion_rate} with the associated $1\sigma$ uncertainty (col. 5), logarithmic 325 MHz radio luminosity including the upper limit estimates for RQQs (col. 6), synchrotron contamination at the FIR bands (Y: for sources with contamination that have been rejected from the sample; col. 7), logarithmic black hole mass measured as described in Sec.~\ref{section:BH_HG_properties} with the associated $1\sigma$ uncertainty (col. 8), logarithmic ${\rm L_{FIR}}$ luminosity measured as described in Sec.~\ref{section:4_SED_fitting} with the associated $1\sigma$ uncertainty (col. 9).

In Table~\ref{Table:SMA_obs} we present the SMA observations of the 44 RLQs candidates for synchrotron contamination. The following information are given for each RLQs: right ascension and declination (J2000.0) in degrees (cols. 1 and 2), atmosheric opacity (col. 3), date of observations (year, month, day; col. 4), observed frequency (col. 5), number of antennas used (col. 6), 1300~$\mu$m flux with the $1\sigma$ error (col. 7). For undetected sources we provide a $4\sigma$ upper limit. Note that for seven sources there are multiple observations due to the poor weather conditions.

%\begin{table*}
\begin{sidewaystable*}
 \vspace{-18.0cm}
\begin{center}
\caption{{\bf Table A1.} {\it Herschel} photometry of the objects studied in this work. Photometric uncertainties are $1\sigma$ values.
}
\centering
\begin{tabular}{|l r r r r r r r r  r r r|}
\hline
  \multicolumn{1}{|c}{name} &
  \multicolumn{1}{c}{RA} &
  \multicolumn{1}{c}{DEC} &
  \multicolumn{1}{c}{type} &
  \multicolumn{1}{c}{$z$} &
  \multicolumn{1}{c}{PACS\_ID} &
  \multicolumn{1}{c}{$S_{70\mu{\rm m}}$} &
  \multicolumn{1}{c}{$S_{160\mu{\rm m}}$} &
  \multicolumn{1}{c}{SPIRE\_ID} &
  \multicolumn{1}{c}{$S_{250\mu{\rm m}}$} &
  \multicolumn{1}{c}{$S_{350\mu{\rm m}}$} &
  \multicolumn{1}{c|}{$S_{500\mu{\rm m}}$} \\

  \multicolumn{1}{|c}{ } &
  \multicolumn{1}{c}{(deg)} &
  \multicolumn{1}{c}{(deg)} &
  \multicolumn{1}{c}{ } &
  \multicolumn{1}{c}{ } &
  \multicolumn{1}{c}{ } &
  \multicolumn{1}{c}{(mJy)} &
  \multicolumn{1}{c}{(mJy)} &
  \multicolumn{1}{c}{ } &
  \multicolumn{1}{c}{(mJy)} &
  \multicolumn{1}{c}{(mJy)} &
  \multicolumn{1}{c|}{(mJy)} \\
\hline
  3C356 	& 261.0790 	& 50.9612 & RG & 1.079 &   ...	    & ...   		& ...	 	& 1342206197	& $12.21\pm5.97$ 	& $20.31\pm5.01$ 	& $-4.02\pm7.73$	\\
  5C7.23 	& 123.4110 	& 29.2749 & RG & 1.098 & 1342230090 & $0.40\pm1.90$ 	& $0.57\pm4.31$ 	& 1342244174 	& $4.45\pm5.80$ 	& $3.73\pm5.22$ 	& $7.36\pm8.26$  \\
  5C7.17 	& 123.2474	& 26.8626 & RG & 0.936 & 1342232054 & $-0.02\pm1.79$ 	& $0.58\pm4.37$ 	& 1342230764 	& $4.03\pm6.21$ 	& $0.30\pm5.22$ 	& $6.92\pm7.00$		\\
  3C268.1 	& 180.0995	& 73.0124 & RG & 0.970 & 1342245706 & $2.88\pm1.95$ 	& $1.30\pm3.57$ 	& 1342229628 	& $2.78\pm5.97$ 	& $16.03\pm5.09$ 	& $8.49\pm7.34$ \\
  3C184 	& 114.8511 	& 70.3864 & RG & 0.994 & 1342243742 & $6.49\pm1.99$ 	& $17.97\pm3.64$ 	& 1342229126 	& $19.05\pm6.21$ 	& $7.95\pm5.92$ 	& $14.73\pm7.73$ \\
  3C175.1 	& 108.5196 	& 14.6063 & RG & 0.920 & 1342242694 & $2.77\pm2.18$ 	& $9.15\pm4.93$ 	& 1342230780 	& $-2.23\pm9.00$ 	& $3.87\pm7.19$ 	& $6.05\pm10.02$ \\
\hline
\end{tabular}
\label{Table:Herschel_photometry}
%\end{table*}
\end{center}
%\end{sidewaystable}
\end{sidewaystable*}%table*}

%\begin{table*}
\begin{sidewaystable*}
 \vspace{18.0cm}
\begin{center}
\caption*{A1. continued}
\centering
\begin{tabular}{|l r r r r r r r r  r r r|}
\hline
  \multicolumn{1}{|c}{name} &
  \multicolumn{1}{c}{RA} &
  \multicolumn{1}{c}{DEC} &
  \multicolumn{1}{c}{type} &
  \multicolumn{1}{c}{z} &
  \multicolumn{1}{c}{PACS\_ID} &
  \multicolumn{1}{c}{$S_{70\mu{\rm m}}$} &
  \multicolumn{1}{c}{$S_{160\mu{\rm m}}$} &
  \multicolumn{1}{c}{SPIRE\_ID} &
  \multicolumn{1}{c}{$S_{250\mu{\rm m}}$} &
  \multicolumn{1}{c}{$S_{350\mu{\rm m}}$} &
  \multicolumn{1}{c|}{$S_{500\mu{\rm m}}$} \\

  \multicolumn{1}{|c}{ } &
  \multicolumn{1}{c}{(deg)} &
  \multicolumn{1}{c}{(deg)} &
  \multicolumn{1}{c}{ } &
  \multicolumn{1}{c}{ } &
  \multicolumn{1}{c}{ } &
  \multicolumn{1}{c}{(mJy)} &
  \multicolumn{1}{c}{(mJy)} &
  \multicolumn{1}{c}{ } &
  \multicolumn{1}{c}{(mJy)} &
  \multicolumn{1}{c}{(mJy)} &
  \multicolumn{1}{c|}{(mJy)} \\
\hline
  SDSS093303.5+460440.2 	& 143.2645 & 46.0777 	& RQQ & 1.090 & 1342245676 & $6.26\pm2.33$ 	& $15.88\pm4.86$ 	& 1342230741 		& $22.10\pm6.16$ 	& $19.78\pm5.43$ 	& $14.35\pm7.75$ \\
  SDSS104114.2+590219.4 	& 160.3091 & 59.0387 	& RQQ & 1.094 & 1342245196 & $4.17\pm2.72$ 	& $5.99\pm5.44$ 	& L-SWIRE$^\alpha$ 	& $10.06\pm5.95$ 	& $4.77\pm5.75$ 	& $2.85\pm8.24$ \\
  SDSS103829.7+585204.1 	& 159.6239 & 58.8678 	& RQQ & 0.935 & 1342245194 & $6.99\pm2.85$ 	& $17.26\pm5.83$ 	& L-SWIRE$^\alpha$ 	& $32.41\pm6.17$ 	& $22.88\pm5.21$ 	& $16.10\pm6.88$\\
  SDSS090153.4+065915.6 	& 135.4726 & 6.9876 	& RQQ & 1.082 & 1342245188 & $7.61\pm2.49$ 	& $15.07\pm4.77$ 	& 1342230785 		& $25.98\pm6.50$ 	& $31.90\pm5.55$ 	& $25.57\pm7.58$ \\
  SDSS084723.7+011010.4 	& 131.8485 & 1.1695 	& RQQ & 1.081 & 1342244276 & $3.75\pm2.30$ 	& $13.86\pm5.74$ 	& H-ATLAS$^b$\let\thefootnote\relax\footnotetext{$^b$ Source overlaps with {\it Herschel}-ATLAS fields.} 	& $17.24\pm7.69$ 	& $18.07\pm5.11$ 	& $5.80\pm7.17$ \\
  SDSS074729.2+434607.5 	& 116.8715 & 43.7688 	& RQQ & 1.086 & 1342243754 & $3.46\pm2.56$ 	& $8.21\pm4.76$ 	& 1342229647 		& $-1.70\pm6.36$ 	& $10.25\pm5.11$ 	& $6.73\pm6.26$  \\
\hline
\end{tabular}
%\end{table*}
\end{center}
%\end{sidewaystable}
\end{sidewaystable*}%table*}

%\begin{table*}
\begin{sidewaystable*}
 \vspace{-18.0cm}
\begin{center}
\caption*{A1. continued}
\centering
\begin{tabular}{|l r r r r r r r r  r r r|}
\hline
  \multicolumn{1}{|c}{name} &
  \multicolumn{1}{c}{RA} &
  \multicolumn{1}{c}{DEC} &
  \multicolumn{1}{c}{type} &
  \multicolumn{1}{c}{z} &
  \multicolumn{1}{c}{PACS\_ID} &
  \multicolumn{1}{c}{$S_{70\mu{\rm m}}$} &
  \multicolumn{1}{c}{$S_{160\mu{\rm m}}$} &
  \multicolumn{1}{c}{SPIRE\_ID} &
  \multicolumn{1}{c}{$S_{250\mu{\rm m}}$} &
  \multicolumn{1}{c}{$S_{350\mu{\rm m}}$} &
  \multicolumn{1}{c|}{$S_{500\mu{\rm m}}$} \\

  \multicolumn{1}{|c}{ } &
  \multicolumn{1}{c}{(deg)} &
  \multicolumn{1}{c}{(deg)} &
  \multicolumn{1}{c}{ } &
  \multicolumn{1}{c}{ } &
  \multicolumn{1}{c}{ } &
  \multicolumn{1}{c}{(mJy)} &
  \multicolumn{1}{c}{(mJy)} &
  \multicolumn{1}{c}{ } &
  \multicolumn{1}{c}{(mJy)} &
  \multicolumn{1}{c}{(mJy)} &
  \multicolumn{1}{c|}{(mJy)} \\
\hline
  SDSS092829.9+504836.6 & 142.1241 & 50.8101 & RQQ & 1.034 & 1342232401 & $2.42\pm2.73$ 	& $8.72\pm5.31$  & 1342230874 & $-5.27\pm6.54$ & $5.71\pm5.22$  & $0.72\pm7.48$  \\
  SDSS133733.3+590622.6 & 204.3887 & 59.1063 & RQQ & 1.087 & 1342232192 & $16.15\pm2.23$ 	& $14.01\pm6.31$ & 1342229512 & $25.71\pm5.81$ & $22.39\pm5.00$ & $18.97\pm6.72$  \\
  SDSS133713.1+610749.0 & 204.3042 & 61.1303 & RQQ & 0.926 & 1342232190 & $3.07\pm2.33$ 	& $6.44\pm6.01$  & 1342229510 & $7.20\pm6.54$  & $7.40\pm5.14$  & $6.86\pm8.04$  \\
  SDSS122832.9+603735.1 & 187.1372 & 60.6264 & RQQ & 1.040 & 1342232184 & $2.48\pm2.51$ 	& $6.13\pm5.84$  & 1342229504 & $6.62\pm5.34$  & $9.46\pm4.87$  & $5.73\pm6.88$  \\
  SDSS093023.3+403111.0 & 142.5970 & 40.5197 & RQQ & 1.097 & 1342232034 & $6.05\pm2.49$ 	& $19.05\pm5.20$ & 1342230747 & $21.71\pm6.58$ & $11.73\pm5.27$ & $9.06\pm6.59$  \\
  SDSS073802.4+383116.3 & 114.5098 & 38.5213 & RQQ & 1.023 & 1342231556 & $35.02\pm2.46$ 	& $64.91\pm4.67$ & 1342229479 & $104.45\pm6.32$ & $70.82\pm5.15$ & $32.79\pm7.39$  \\
\hline
\end{tabular}
%%%%%%%%%%%%%%%%%%%%%%%%\label{Table:SMA_obs}
%\end{table*}
\end{center}
%\end{sidewaystable}
\end{sidewaystable*}%table*}

%\begin{table*}
\begin{sidewaystable*}
 \vspace{18.0cm}
\begin{center}
\caption*{A1. continued}
\centering
\begin{tabular}{|l r r r r r r r r  r r r|}
\hline
  \multicolumn{1}{|c}{name} &
  \multicolumn{1}{c}{RA} &
  \multicolumn{1}{c}{DEC} &
  \multicolumn{1}{c}{type} &
  \multicolumn{1}{c}{z} &
  \multicolumn{1}{c}{PACS\_ID} &
  \multicolumn{1}{c}{$S_{70\mu{\rm m}}$} &
  \multicolumn{1}{c}{$S_{160\mu{\rm m}}$} &
  \multicolumn{1}{c}{SPIRE\_ID} &
  \multicolumn{1}{c}{$S_{250\mu{\rm m}}$} &
  \multicolumn{1}{c}{$S_{350\mu{\rm m}}$} &
  \multicolumn{1}{c|}{$S_{500\mu{\rm m}}$} \\

  \multicolumn{1}{|c}{ } &
  \multicolumn{1}{c}{(deg)} &
  \multicolumn{1}{c}{(deg)} &
  \multicolumn{1}{c}{ } &
  \multicolumn{1}{c}{ } &
  \multicolumn{1}{c}{ } &
  \multicolumn{1}{c}{(mJy)} &
  \multicolumn{1}{c}{(mJy)} &
  \multicolumn{1}{c}{ } &
  \multicolumn{1}{c}{(mJy)} &
  \multicolumn{1}{c}{(mJy)} &
  \multicolumn{1}{c|}{(mJy)} \\
\hline
  SDSS160516.1+313620.8 & 241.3169 & 31.6058 & RLQ & 1.028 & 1342235404 & $3.85\pm2.41$ 	& $11.00\pm5.12$ 	& 1342229552 & $1.44\pm5.91$ 	& $-5.59\pm5.26$ 	& $-8.40\pm6.80$ \\
  SDSS152949.8+394509.6 & 232.4573 & 39.7527 & RLQ & 1.081 & 1342235396 & $3.11\pm2.76$ 	& $13.69\pm5.32$ 	& 1342229534 & $12.02\pm6.03$ 	& $4.28\pm4.51$ 	& $2.03\pm6.88$ \\
  SDSS155729.9+330446.9 & 239.3747 & 33.0797 & RLQ & 0.954 & 1342235114 & $3.83\pm2.64$ 	& $2.79\pm5.05$ 	& 1342229549 & $24.43\pm6.43$ 	& $63.98\pm6.05$ 	& $69.34\pm7.35$ \\
  SDSS163403.0+390000.6 & 248.5125 & 39.0002 & RLQ & 1.085 & 1342234446 & $6.02\pm2.47$ 	& $18.73\pm5.22$ 	& 1342229542 & $31.54\pm5.99$ 	& $20.44\pm4.69$ 	& $18.76\pm7.21$ \\
  SDSS150031.8+483646.8 & 225.1323 & 48.6131 & RLQ & 1.028 & 1342234357 & $23.01\pm2.88$ 	& $36.41\pm5.72$ 	& 1342229521 & $42.58\pm6.68$ 	& $36.67\pm6.10$ 	& $14.00\pm6.86$ \\
  SDSS161806.3+422532.1 & 244.5264 & 42.4256 & RLQ & 0.934 & 1342234347 & $7.11\pm2.75$ 	& $5.10\pm5.09$ 	& 1342229545 & $9.33\pm6.12$ 	& $-0.62\pm5.07$ 	& $5.11\pm6.75$ \\
\hline
\end{tabular}
%%%%%%%%%%%%%%%%%%%%%%%%\label{Table:SMA_obs}
%\end{table*}
\end{center}
%\end{sidewaystable}
\end{sidewaystable*}%table*}

%\begin{table*}
\begin{sidewaystable*}
 \vspace{-18.0cm}
\begin{center}
\caption*{A1. continued}
\centering
\begin{tabular}{|l r r r r r r r r  r r r|}
\hline
  \multicolumn{1}{|c}{name} &
  \multicolumn{1}{c}{RA} &
  \multicolumn{1}{c}{DEC} &
  \multicolumn{1}{c}{type} &
  \multicolumn{1}{c}{z} &
  \multicolumn{1}{c}{PACS\_ID} &
  \multicolumn{1}{c}{$S_{70\mu{\rm m}}$} &
  \multicolumn{1}{c}{$S_{160\mu{\rm m}}$} &
  \multicolumn{1}{c}{SPIRE\_ID} &
  \multicolumn{1}{c}{$S_{250\mu{\rm m}}$} &
  \multicolumn{1}{c}{$S_{350\mu{\rm m}}$} &
  \multicolumn{1}{c|}{$S_{500\mu{\rm m}}$} \\

  \multicolumn{1}{|c}{ } &
  \multicolumn{1}{c}{(deg)} &
  \multicolumn{1}{c}{(deg)} &
  \multicolumn{1}{c}{ } &
  \multicolumn{1}{c}{ } &
  \multicolumn{1}{c}{ } &
  \multicolumn{1}{c}{(mJy)} &
  \multicolumn{1}{c}{(mJy)} &
  \multicolumn{1}{c}{ } &
  \multicolumn{1}{c}{(mJy)} &
  \multicolumn{1}{c}{(mJy)} &
  \multicolumn{1}{c|}{(mJy)} \\
\hline
  SDSS074815.4+220059.5 & 117.0643 & 22.0166 & RLQ & 1.060 & 1342230049 & $14.03\pm2.60$ 	& $11.90\pm4.83$ 	& 1342229467 & $11.75\pm6.84$ 	& $4.09\pm6.02$ 	& $4.53\pm8.19$ \\
  SDSS075928.3+301028.3 & 119.8679 & 30.1746 & RLQ & 1.002 & 1342230043 & $10.40\pm2.40$ 	& $20.28\pm4.75$ 	& 1342229474 & $32.91\pm6.65$ 	& $26.17\pm5.81$ 	& $6.51\pm7.72$ \\
  SDSS165920.0+374332.7 & 254.8333 & 37.7258 & RLQ & 1.025 & 1342228447 & $5.61\pm2.56$ 	& $5.82\pm4.67$ 	& 1342229598 & $7.13\pm5.82$ 	& $6.67\pm4.86$ 	& $2.29\pm6.07$ \\
  SDSS165943.1+375422.7 & 254.9295 & 37.9063 & RLQ & 1.038 & 1342225445 & $2.47\pm2.45$ 	& $11.15\pm4.88$ 	& 1342229597 & $-0.53\pm5.86$ 	& $-3.25\pm5.41$ 	& $1.91\pm7.53$ \\
  SDSS164054.2+314329.9 & 250.2257 & 31.7250 & RLQ & 0.958 & 1342225439 & $3.36\pm2.64$ 	& $11.81\pm6.03$ 	& 1342229592 & $18.87\pm6.53$ 	& $23.46\pm6.28$ 	& $21.91\pm7.49$ \\
  SDSS170949.2+303259.2 & 257.4552 & 30.5498 & RLQ & 1.043 & 1342225272 & $4.02\pm2.66$ 	& $9.12\pm6.51$ 	& 1342229589 & $18.23\pm6.02$ 	& $23.01\pm5.48$ 	& $20.46\pm7.13$ \\
\hline
\end{tabular}
%%%%%%%%%%%%%%%%%%%%%%%%\label{Table:SMA_obs}
%\end{table*}
\end{center}
%\end{sidewaystable}
\end{sidewaystable*}%table*}

\begin{table*}
%\normalsize
\caption{Properties of the $z \approx 1$ AGN sample.}
%\centering
\begin{minipage}{20.3cm}
\begin{tabular}{|l r l l l l r r|}

\hline

  \multicolumn{1}{|l}{name} &
  \multicolumn{1}{c}{type} &
  \multicolumn{1}{c}{$z$} &
  \multicolumn{1}{c}{$L_{\rm Bol}$\footnote{The logarithmic bolometric luminosity in units of ${\rm erg ~s^{-1}}$ with the associated $1\sigma$ errors.}} &
  \multicolumn{1}{c}{$L_{\rm rad}$\footnote{The logarithmic 325 MHz radio luminosity in units of ${\rm W~Hz^{-1}~sr^{-1}}$ with the associated $1\sigma$ errors.}} &
  \multicolumn{1}{c}{cont.\footnote{Synchrotron contamination at the FIR bands}} &
  \multicolumn{1}{c}{$M_{\rm BH}$\footnote{The logarithmic black hole mass in units of $M_{\odot}$ with the associated $1\sigma$ errors.}} &
  \multicolumn{1}{c|}{$L_{\rm FIR}$\footnote{The logarithmic FIR luminosity in units of ${\rm erg ~s^{-1}}$ with the associated $1\sigma$ errors.}} \\
\hline

  TOOT1140 & RG & 0.911 & 44.779$\pm$0.294 & 25.9934 & N & 8.916$\pm$0.255 & 44.853$\pm$0.885 \\
  TOOT1066 & RG & 0.926 & 45.136$\pm$0.130 & 25.5255 & N & 8.436$\pm$0.340 & 45.105$\pm$0.199 \\
  TOOT1267 & RG & 0.968 & 45.743$\pm$0.049 & 26.0336 & N & 9.076$\pm$0.255 & 45.288$\pm$0.037\\
  3C22 & RG & 0.936 & 46.813$\pm$0.010 & 27.4695 & ? & 9.366$\pm$0.425 & -45.259$\pm$0.745\\
  6C0128+394 & RG & 0.929 & 44.795$\pm$0.328 & 26.6611 & ? & 8.246$\pm$0.849 & 45.153$\pm$0.000\\
  6C0133+486 & RG & 1.029 & 44.690$\pm$0.702 & 26.5199 & N & 7.846$\pm$0.849 & 45.409$\pm$0.617\\
  5C6.24 & RG & 1.073 & 45.682$\pm$0.072 & 26.6181 & N & 8.906$\pm$0.425 & 45.018$\pm$0.000\\
\hline
\end{tabular}
\label{Table:full_properties}
\end{minipage}
\end{table*}

\begin{table*}
%\normalsize
\caption*{Properties of the $z \approx 1$ AGN sample.}
%\centering
\begin{minipage}{20.3cm}
\begin{tabular}{|l r l l l l r r|}

\hline
  \multicolumn{1}{|l}{name} &
  \multicolumn{1}{c}{type} &
  \multicolumn{1}{c}{$z$} &
  \multicolumn{1}{c}{$L_{\rm Bol}$} &
  \multicolumn{1}{c}{$L_{\rm rad}$} &
  \multicolumn{1}{c}{cont.} &
  \multicolumn{1}{c}{$M_{\rm BH}$} &
  \multicolumn{1}{c|}{$L_{\rm FIR}$} \\
\hline

  SDSS145506.1+562935.6 & RQQ & 1.038 & 46.730$\pm$0.002 & 23.836 & N & 9.220$\pm$0.025 & 45.729$\pm$0.131\\
  SDSS164617.2+364509.4 & RQQ & 0.958 & 45.585$\pm$0.015 & 23.749 & N & 8.114$\pm$0.317 & 45.434$\pm$0.105\\
  SDSS102006.0+033308.5 & RQQ & 0.936 & 46.938$\pm$0.002 & 23.724 & N & 9.431$\pm$0.029 & 45.621$\pm$0.521\\
  SDSS123059.7+101624.8 & RQQ & 1.056 & 46.126$\pm$0.005 & 23.854 & N & 9.325$\pm$0.281 & 45.354$\pm$0.193\\
  SDSS112317.5+051804.0 & RQQ & 1.000 & 46.988$\pm$0.001 & 23.795 & N & 9.250$\pm$0.054 & 45.586$\pm$0.228\\
  SDSS084723.7+011010.4 & RQQ & 1.081 & 46.398$\pm$0.007 & 23.879 & N & 9.312$\pm$0.083 & 45.590$\pm$0.163\\
  SDSS115027.3+665848.0 & RQQ & 1.035 & 46.961$\pm$0.001 & 23.832 & N & 9.395$\pm$0.019 & 45.425$\pm$0.158\\
\hline
\end{tabular}
\end{minipage}
\end{table*}

\begin{table*}
%\normalsize
\caption*{Properties of the $z \approx 1$ AGN sample.}
%\centering
\begin{minipage}{20.3cm}
\begin{tabular}{|l r l l l l r r|}
\hline
  \multicolumn{1}{|l}{name} &
  \multicolumn{1}{c}{type} &
  \multicolumn{1}{c}{$z$} &
  \multicolumn{1}{c}{$L_{\rm Bol}$} &
  \multicolumn{1}{c}{$L_{\rm rad}$} &
  \multicolumn{1}{c}{cont.} &
  \multicolumn{1}{c}{$M_{\rm BH}$} &
  \multicolumn{1}{c|}{$L_{\rm FIR}$} \\
\hline
  SDSS152556.2+591659.5 & RLQ & 0.955 & 46.329$\pm$0.003 & 26.292 & N & 9.544$\pm$0.047 & 45.600$\pm$0.024\\
  SDSS094740.0+515456.8 & RLQ & 1.063 & 45.996$\pm$0.006 & 26.382 & N & 9.138$\pm$0.393 & 45.215$\pm$0.195\\
  SDSS160913.2+535429.6 & RLQ & 0.992 & 46.449$\pm$0.008 & 25.698 & N & 9.299$\pm$0.119 & 45.014$\pm$0.180\\
  SDSS144837.5+501448.9 & RLQ & 1.074 & 46.031$\pm$0.008 & 27.002 & N & 7.863$\pm$0.134 & 45.847$\pm$0.028\\
  SDSS104542.2+525112.6 & RLQ & 1.058 & 46.403$\pm$0.002 & 26.626 & N & 9.151$\pm$0.219 & 45.732$\pm$0.035\\
  SDSS081520.6+273617.0 & RLQ & 0.908 & 47.036$\pm$0.002 & 24.643 & N & 9.200$\pm$0.045 & 45.821$\pm$0.034\\
  SDSS090142.4+425631.0 & RLQ & 1.014 & 46.171$\pm$0.003 & 25.166 & N & 9.088$\pm$0.033 & 45.502$\pm$0.039\\
\hline
\end{tabular}
\end{minipage}
\end{table*}

\begin{table*}
\begin{center}
\caption{SMA observations of the 44 RLQs candidates for synchrotron contamination}
\centering
\begin{tabular} {|l c c c c c |}

\hline 

name	&	($\tau$)	&	Dates	&	Observed frequency	&	Number of antennas	&	$S_{1300\mu{\rm m}}$ 	\\
	&	Atmosheric opacity		&		&	(GHz)			&		&	(mJy)				\\

\hline

\hline

SDSS075928.3+301028.3		&	0.05		&	150130	&	225.436	&	3	&	$<4\times3.8$		\\
				&	0.12		&	150219	&	225.435	&	6	&	$6.94\pm2.90$		\\
				&	0.2		&	150220	&	219.137	&	6	&	$6.01\pm1.80$		\\
SDSS080915.9+321041.6		&	0.05		&	150130	&	225.436	&	3	&	$5.13\pm3.4$		\\
				&	0.12		&	150219	&	225.435	&	6	&	$7.13\pm3.00$		\\
				&	0.2		&	150220	&	219.137	&	6	&	$3.79\pm1.80$	\\
\hline

\end{tabular}
\label{Table:SMA_obs}
%\end{table*}
\end{center}
%\end{sidewaystable}
\end{table*}

\section{The mean far-infrared luminosity calculations} 

In this Appendix we present our results for the different methods that we performed to estimate any systematic errors for the calculation of the mean far-infrared luminosity values. In Section 3.3. we describe the method we used to calculate the mean far-infrared luminosity from the direct stacking on {\it Herschel} maps. In Section 3.5 we describe the different statistical methods we have used to calculate the mean far-infrared luminosity of each bolometric luminosity bin. In Table~\ref{Table:mean_L_stat} we present our calculations for each of the methods and for different statistical mean estimations. We find that the results for each method are broadly consistent and do not affect our conclusions allowing us to compare our results with previous studies.

\begin{table*}
%\begin{center}
\caption{Calculation of the mean far-infrared luminosity values for the different methods presented in Sections 3.3 and 3.5 using the same bolomotric luminosity bins as in Figure 9.}
\centering
 \smallskip
 \begin{minipage}{15.5cm}
\centering
\begin{tabular} {|c c c c c c c |}

\hline 

Class	&	$\log(L_{\rm bol}/{\rm erg~s^{-1}})$ 	&	\multicolumn{2}{c}{$<L_{\rm FIR}>~({L_\odot})$ Method A\footnote{We determine the luminosity of each source from the {\it Herschel} flux densities (excluding $70~\mu$m), even if negative, on the grounds that this is the maximum-likelihood estimator of the true luminosity \citep[e.g.][]{Hardcastle2010,Hardcastle2013}.}}	&	\multicolumn{2}{c}{$<L_{\rm FIR}>~({L_\odot})$ Method B\footnote{We took the FIR upper limits for each source as tentative detections, and estimated upper limits for the ${\rm L_{FIR}}$ using the procedure adopted for the objects detected in {\it Herschel} bands.}} & $<L_{\rm FIR}>~({L_\odot})$ Method C\footnote{We determine the mean far-infrared luminosity of each bin using the direct stacks of the {\it Herschel} maps (Section 3.3).}	\\

	&	range	&		Weighted mean	& Median	& 	Weighted mean	& Median	&  Direct Stacking\\
\hline

\hline
RLQs	&	45.50 - 46.10 	& $45.76\pm0.07$	&  $45.65\pm0.11$	&  $45.80\pm0.10$	&  $45.67\pm0.15$	&	$45.61\pm0.16$	\\
	&	46.10 - 46.55 	& $45.76\pm0.05$	&  $45.75\pm0.09$	&  $45.82\pm0.09$	&  $45.80\pm0.13$	&	$45.74\pm0.12$\\
	&	46.55 - 47.40 	& $46.11\pm0.08$	&  $46.01\pm0.09$	&  $45.17\pm0.13$	&  $46.06\pm0.14$	&	$46.04\pm0.13$\\
	    
\hline
RQQs	&	45.60 - 46.10 	& $45.57\pm0.06$	&  $45.43\pm0.09$	&  $45.61\pm0.10$	&  $45.37\pm0.14$	&	$45.49\pm0.15$ \\
	&	46.10 - 46.55 	& $45.64\pm0.05$	&  $45.56\pm0.08$	&  $45.72\pm0.12$	&  $45.64\pm0.15$	&	$45.58\pm0.12$  \\
	&	46.55 - 47.40	& $45.74\pm0.07$	&  $45.71\pm0.09$	&  $45.81\pm0.10$	&  $45.79\pm0.14$	&	$45.78\pm0.14$\\
\hline

RGs	&	44.69 - 45.50	&  $45.28\pm0.09$	&  $45.21\pm0.13$	&  $<45.41$		&  $<45.32$		& 	$45.25\pm0.18$	\\
	&	45.50 - 45.90	&  $45.20\pm0.10$	&  $45.15\pm0.15$	&  $45.10\pm0.14$	&  $45.24\pm0.21$	&	$45.14\pm0.23$ \\
	&	45.90 - 46.70	&  $45.31\pm0.11$	&  $45.42\pm0.17$	&  $<45.43$		&	$<45.55$		&	$45.36\pm0.49$\\
\hline

\end{tabular}
 \vspace{-0.75\skip\footins}
\label{Table:mean_L_stat}
%\end{table*}
%%%\end{center}
%\end{sidewaystable}
 \end{minipage}
\end{table*}

\clearpage

\begin{figure*}
\begin{tabular}{c c l l l}
	&
\includegraphics[trim=25 50 27 1, clip, scale=0.3]{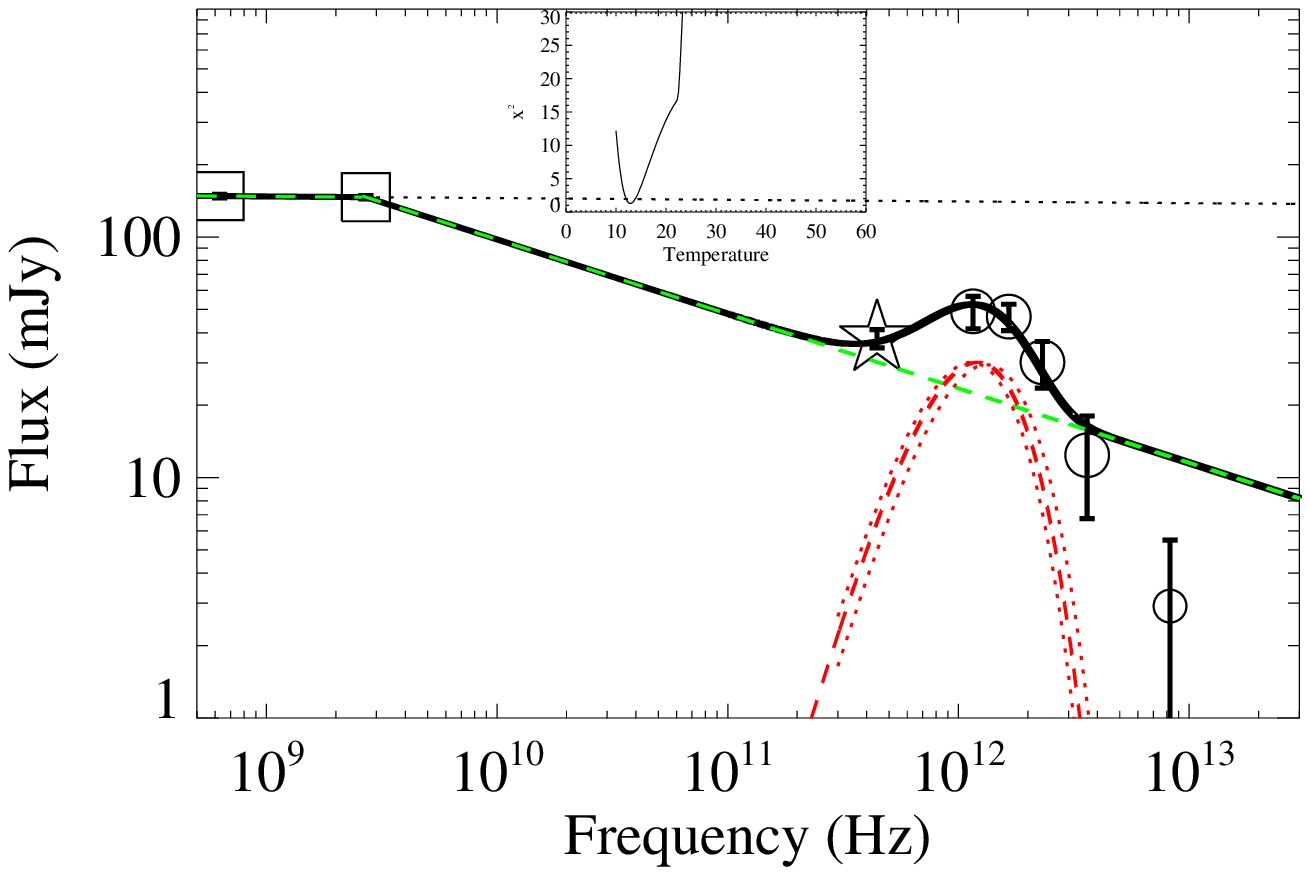} & 
\includegraphics[trim=46 50 27 1, clip,scale=0.3]{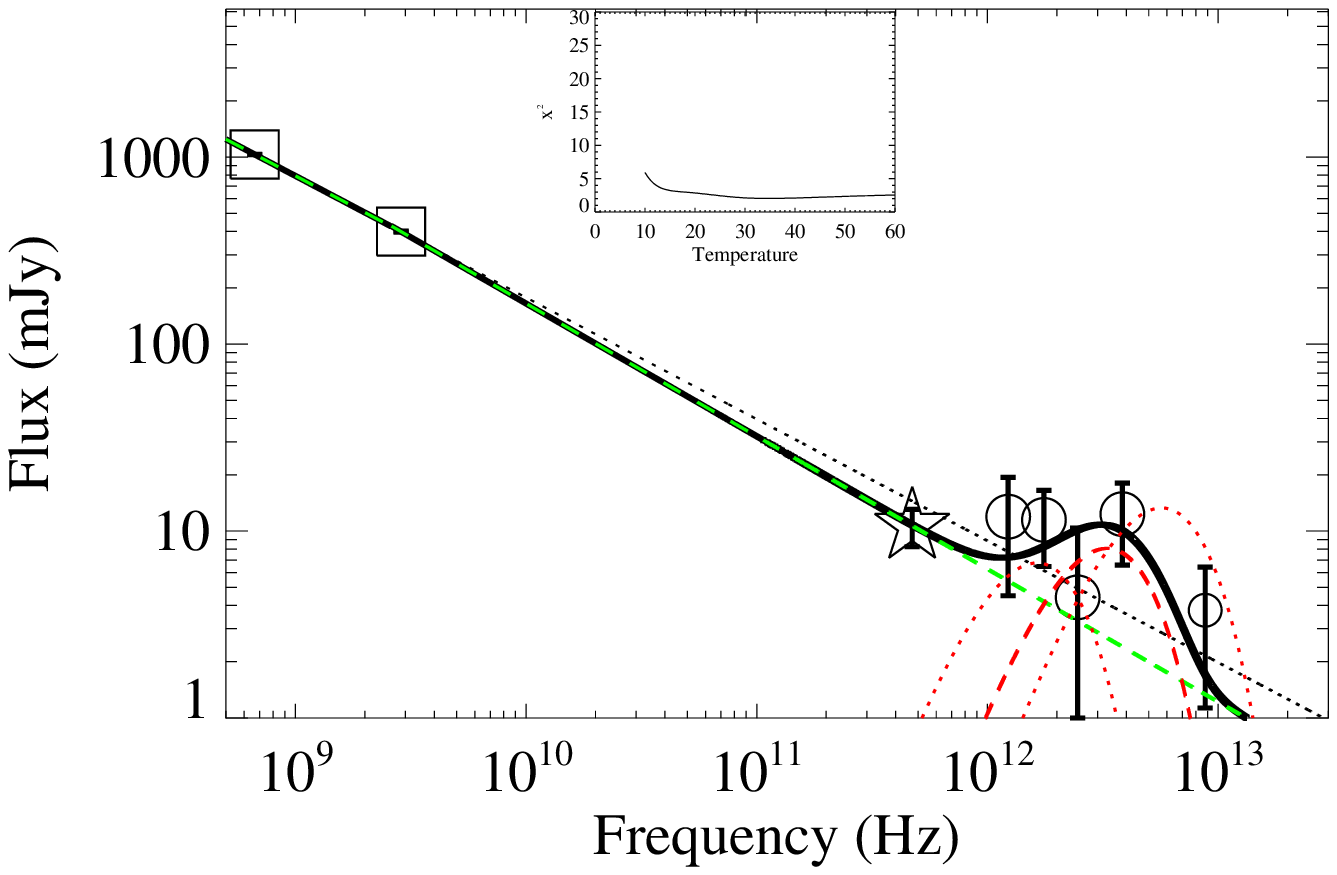} & 
\includegraphics[trim=46 60 27 1, clip,scale=0.3]{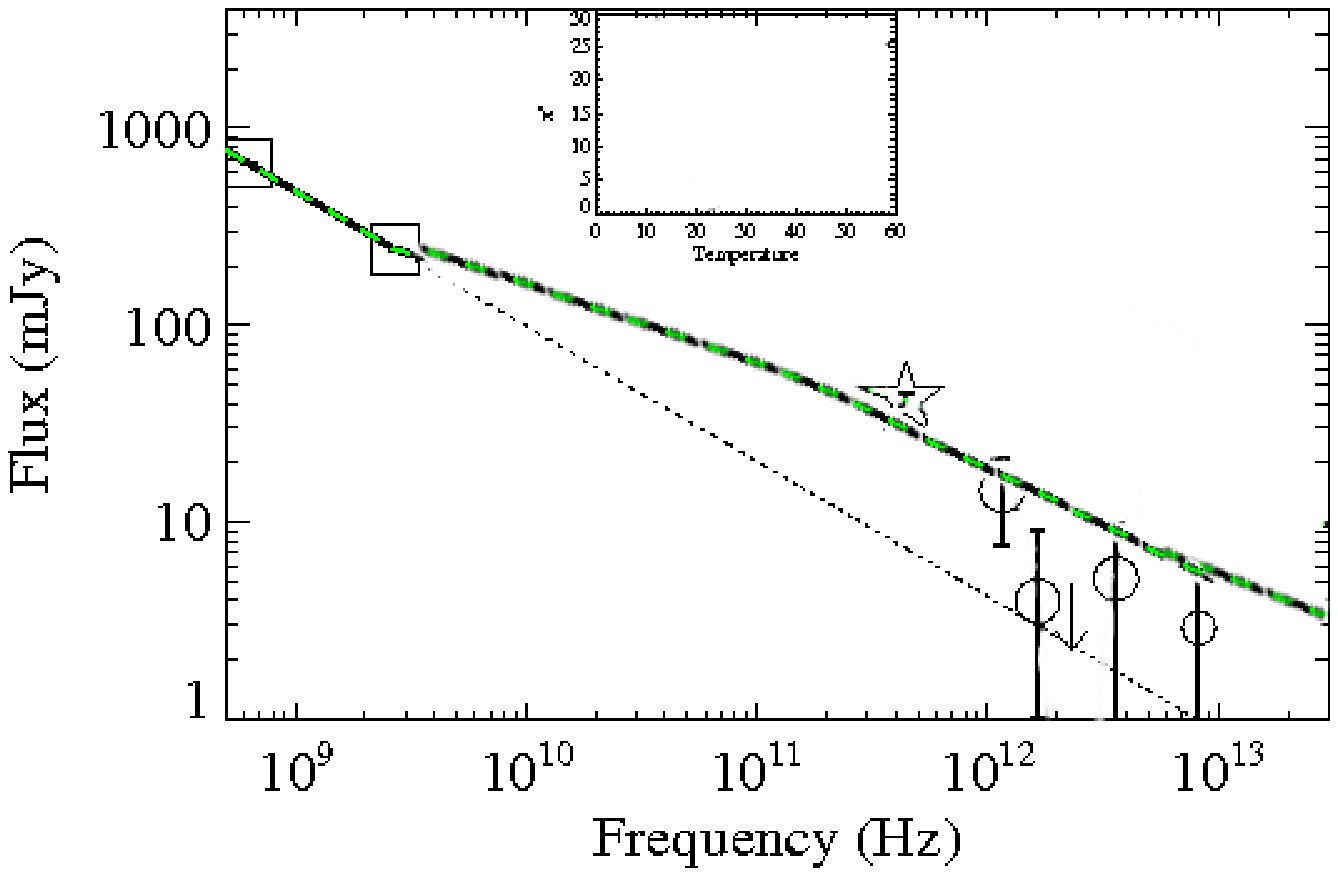} &
\includegraphics[trim=47 50 26 1, clip, scale=0.3]{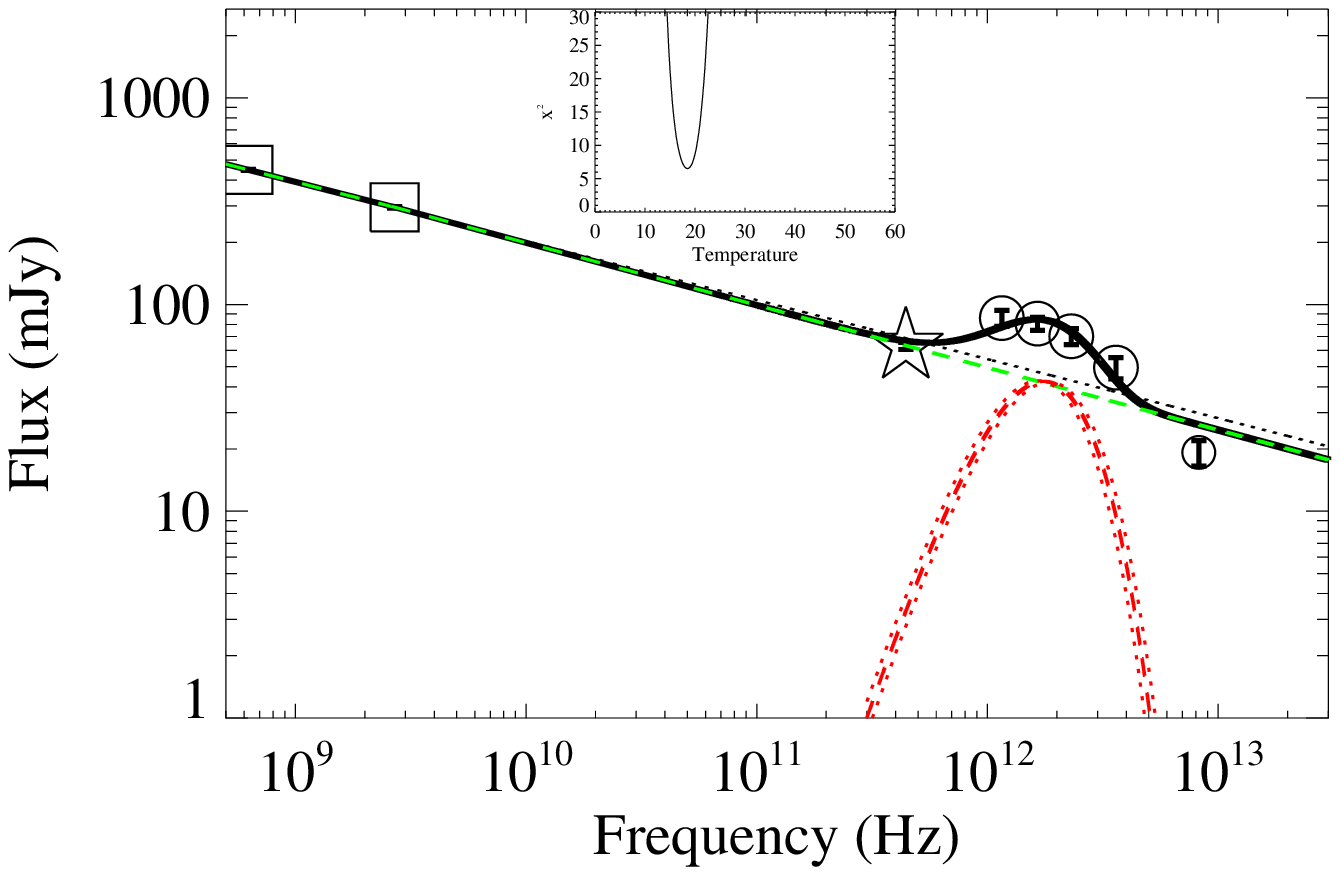} \\

\rotatebox{90}{\bf{Rejected}}		&
\includegraphics[trim=25 50 27 25, clip,scale=0.3]{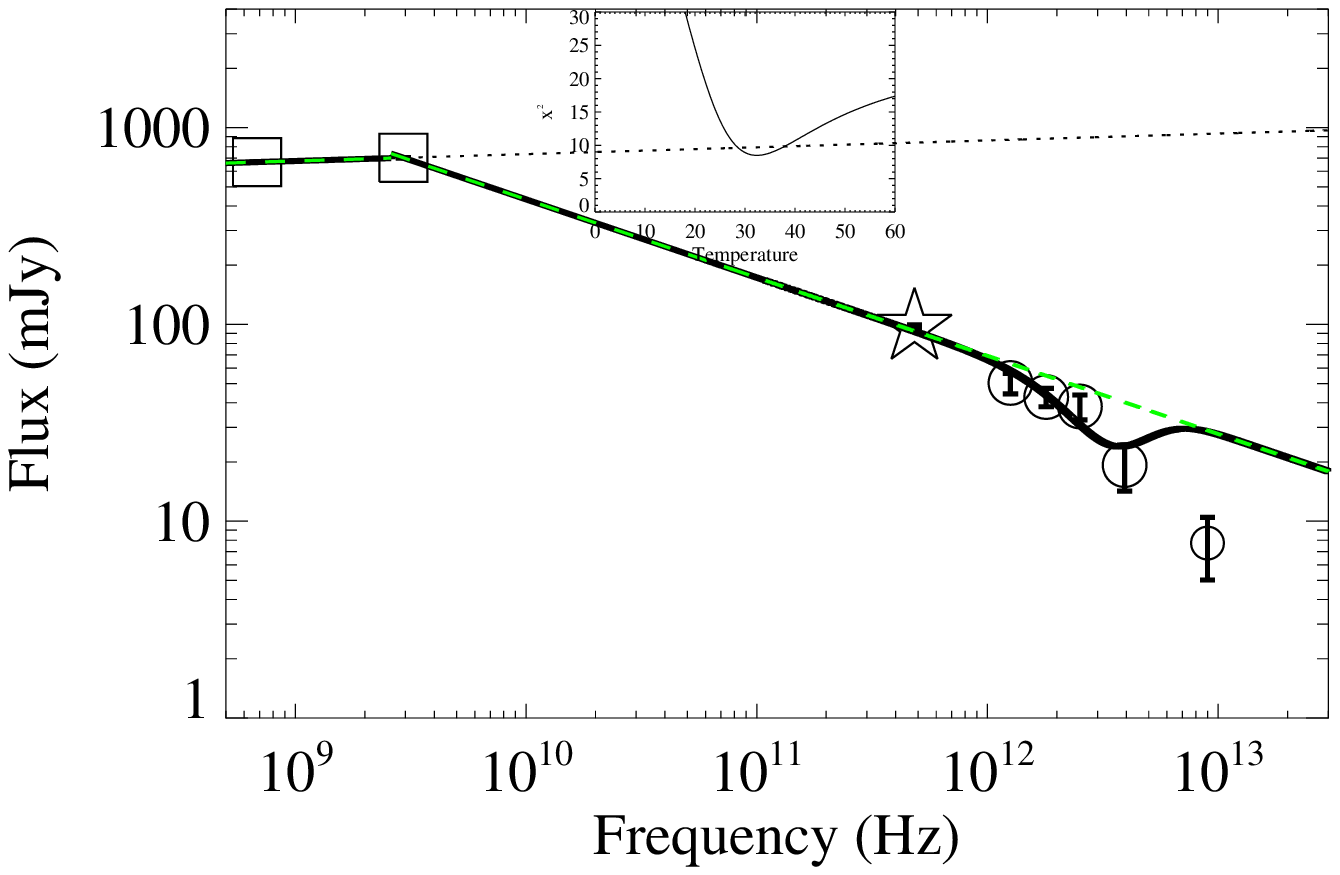} & 
\includegraphics[trim=46 50 27 25, clip,scale=0.3]{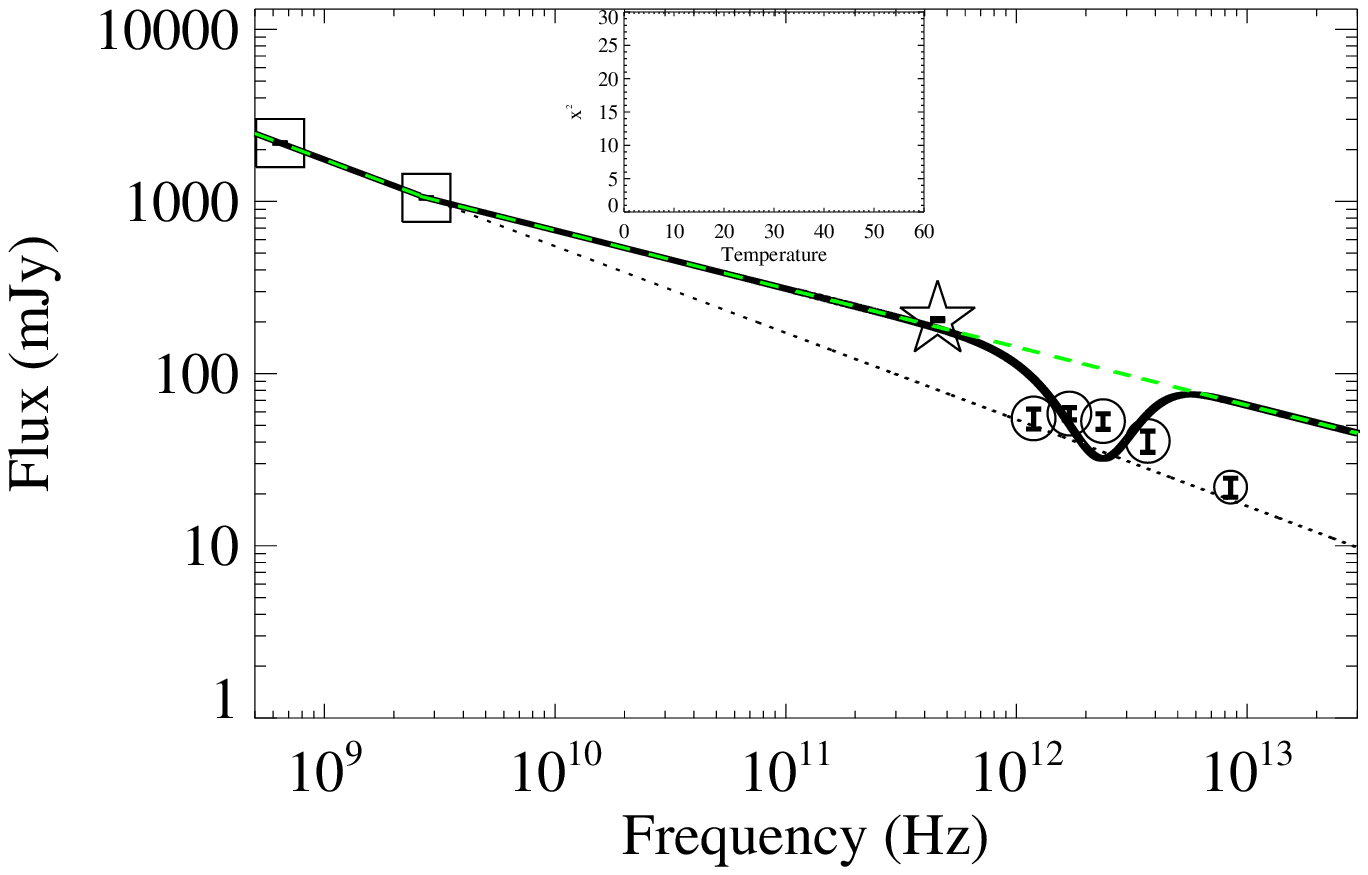} &
\includegraphics[trim=46 50 27 25, clip, scale=0.3]{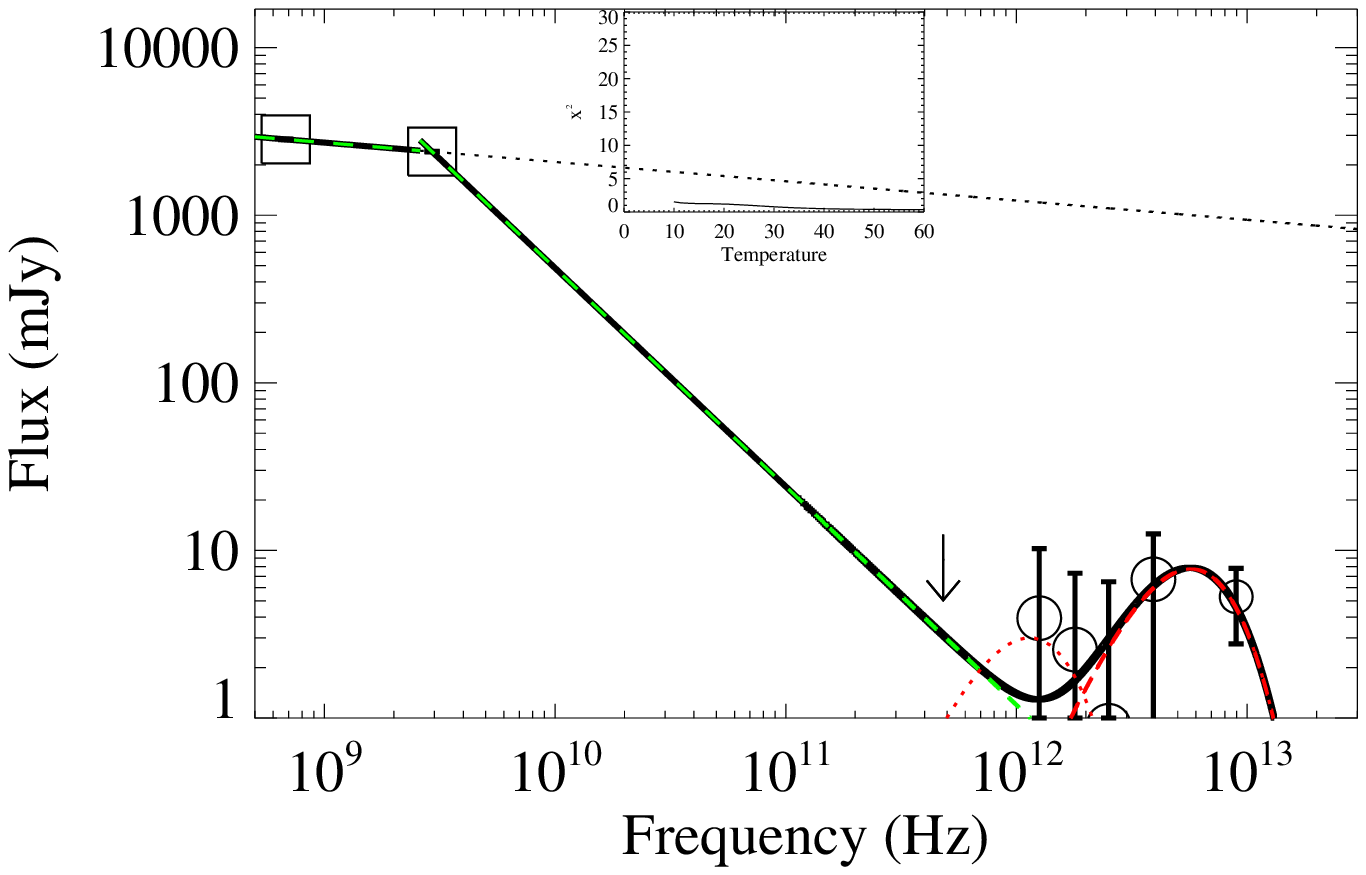} & 
\includegraphics[trim=47 50 26 25, clip, scale=0.3]{SED219.686_62.1985.ps}  \\

	&
\includegraphics[trim=25 50 27 25, clip,scale=0.3]{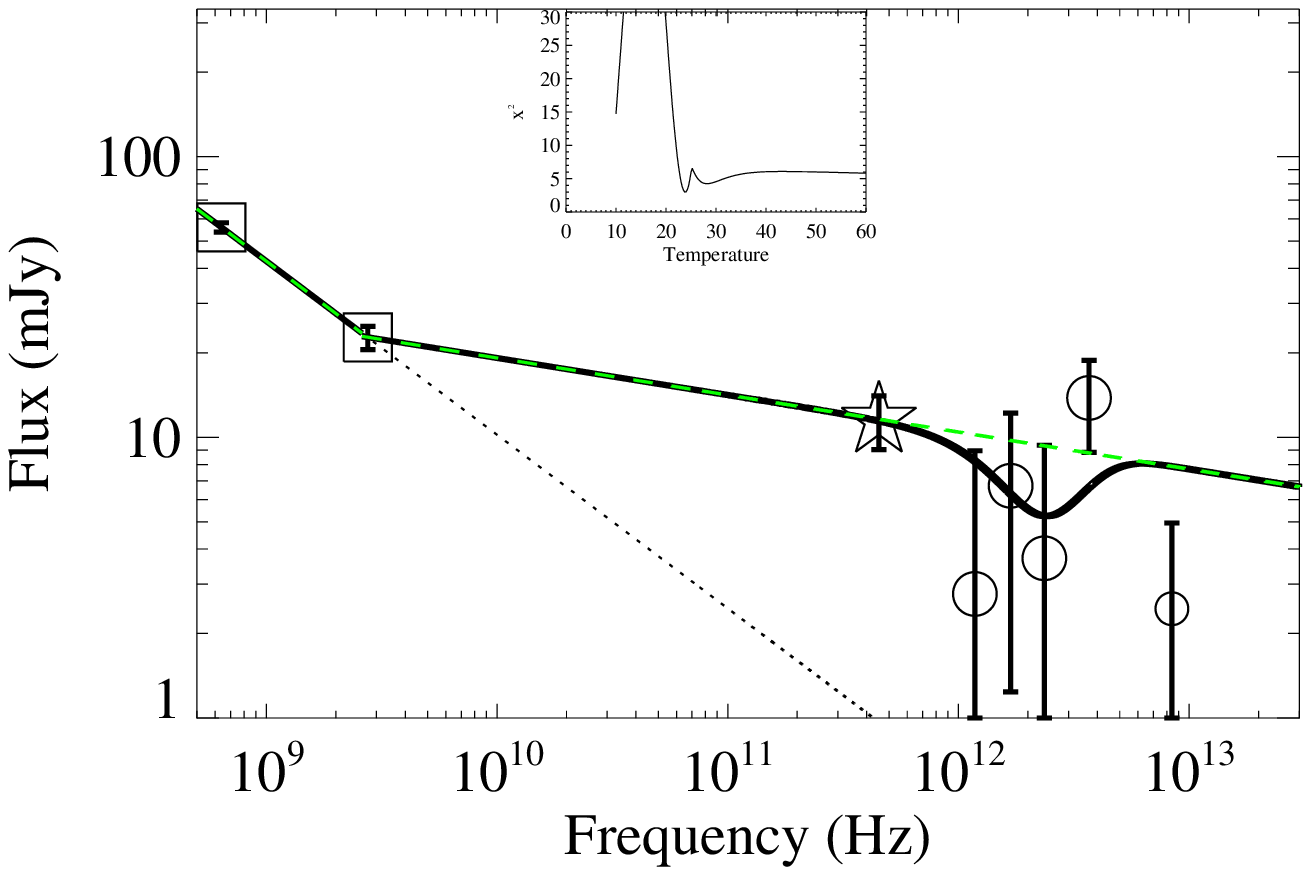} & 
\includegraphics[trim=46 50 27 25, clip,scale=0.3]{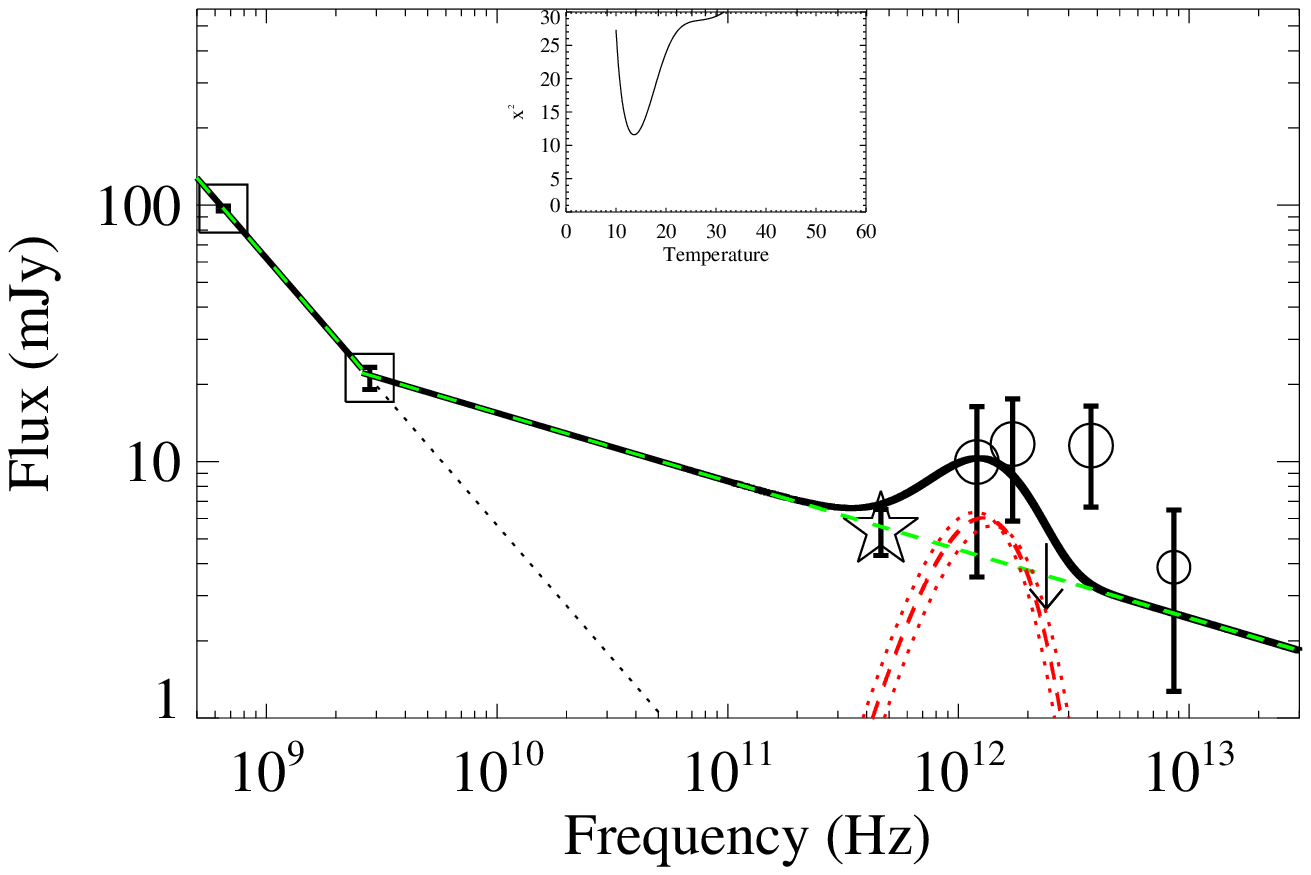} &
\includegraphics[trim=46 50 27 25, clip, scale=0.3]{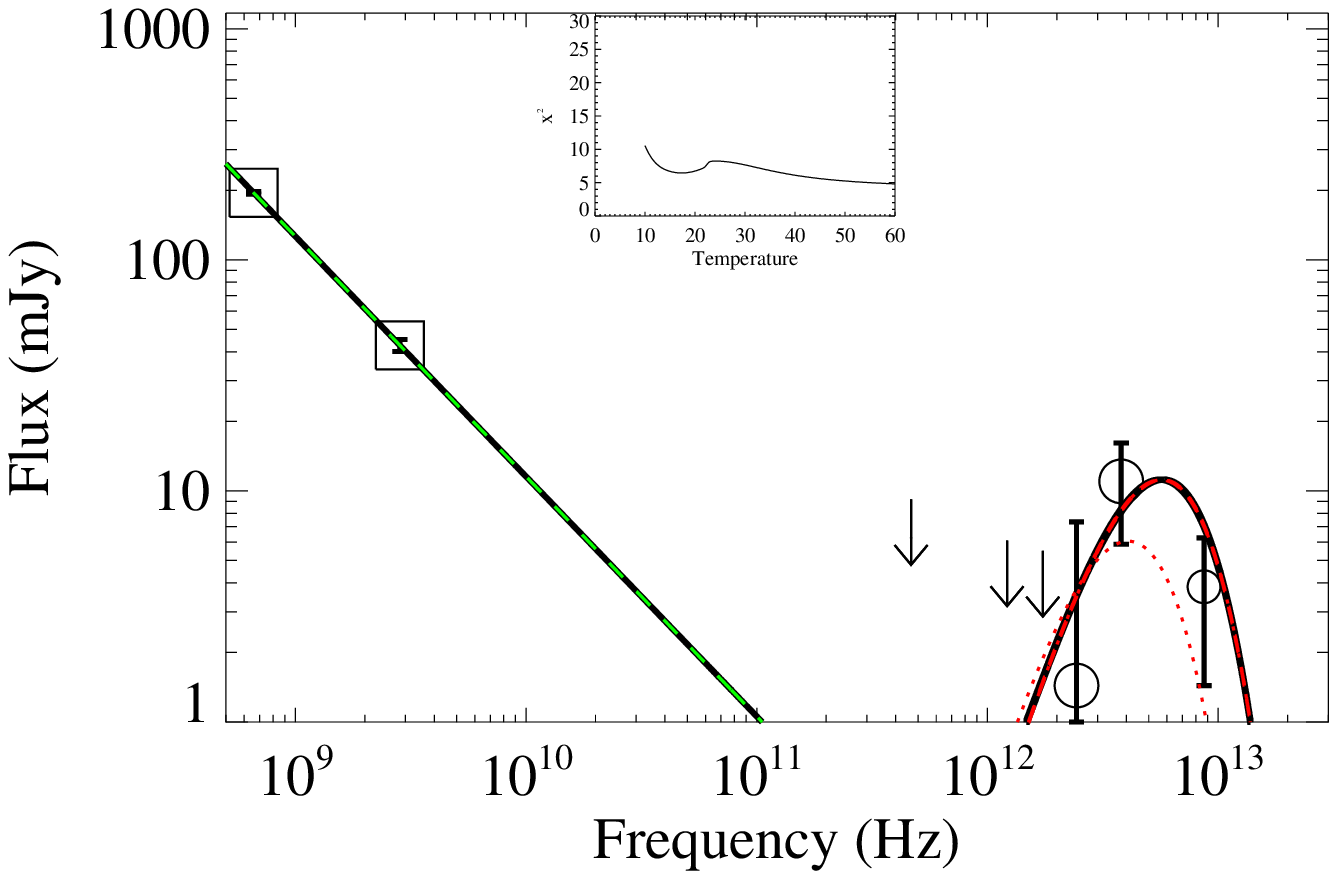} & 
\includegraphics[trim=47 50 26 25, clip,scale=0.3]{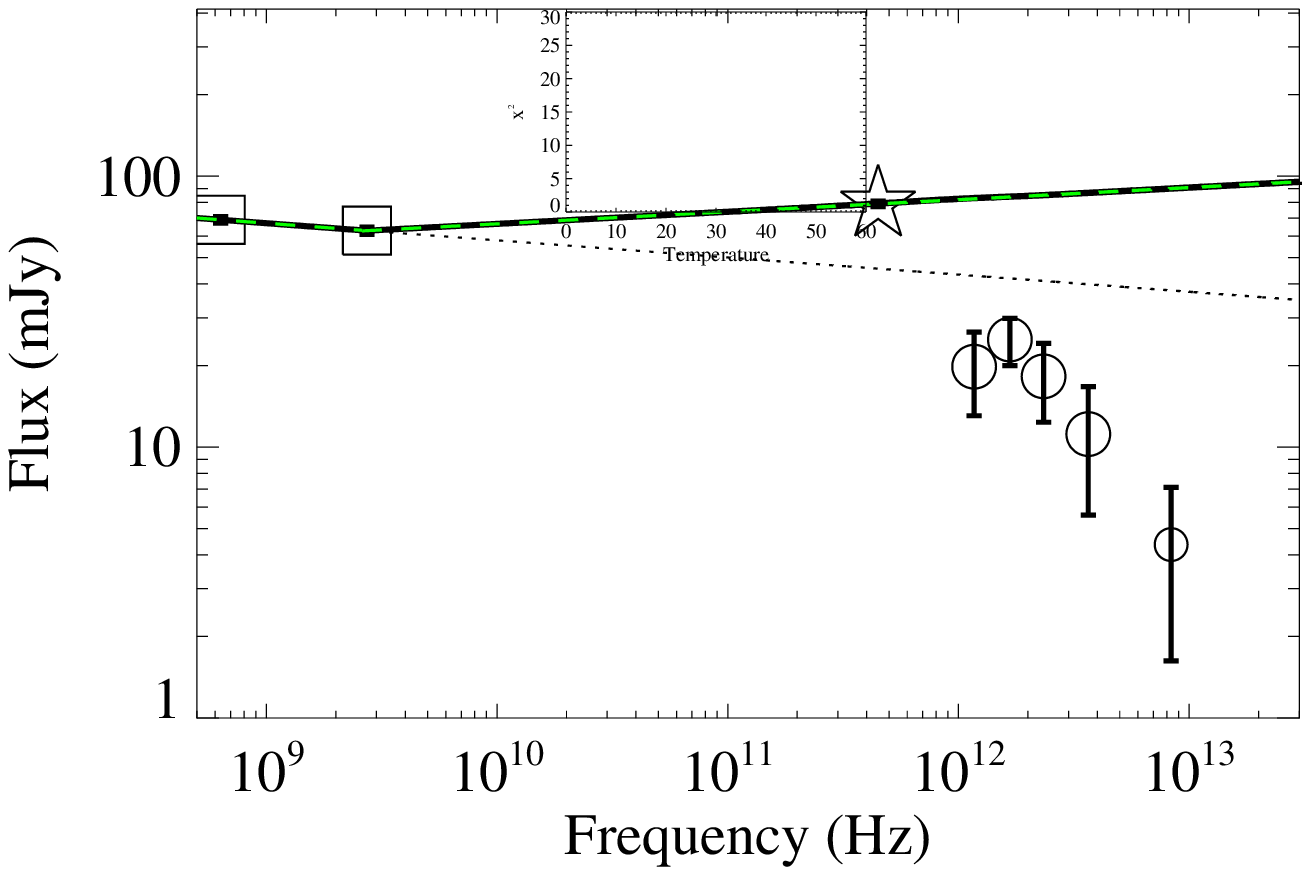} \\

	&
\includegraphics[trim=25 1 25 25, clip,scale=0.3]{SED244.016_46.5404.ps}& 
\includegraphics[trim=46 1 27 25, clip,scale=0.3]{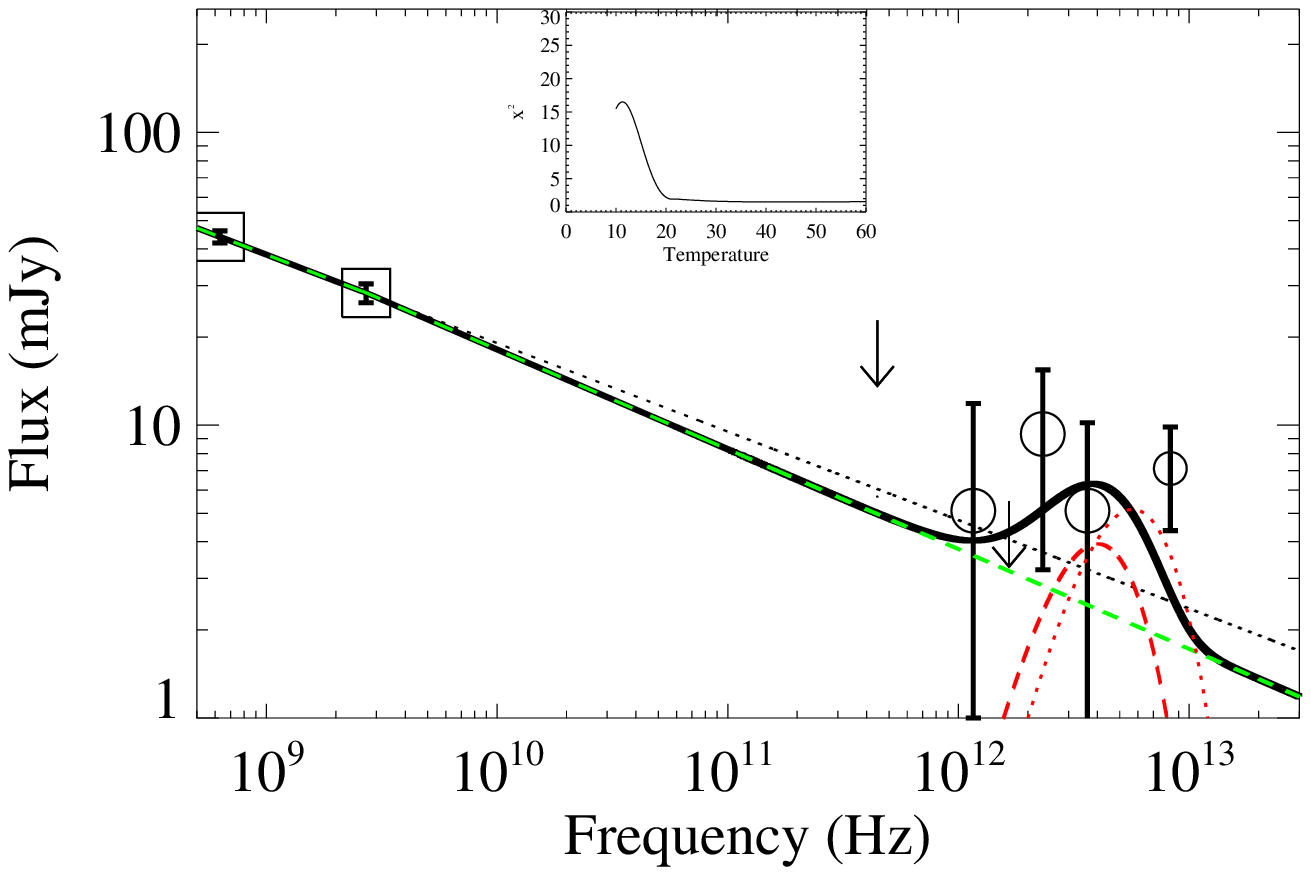} &
\includegraphics[trim=46 1 27 25, clip, scale=0.3]{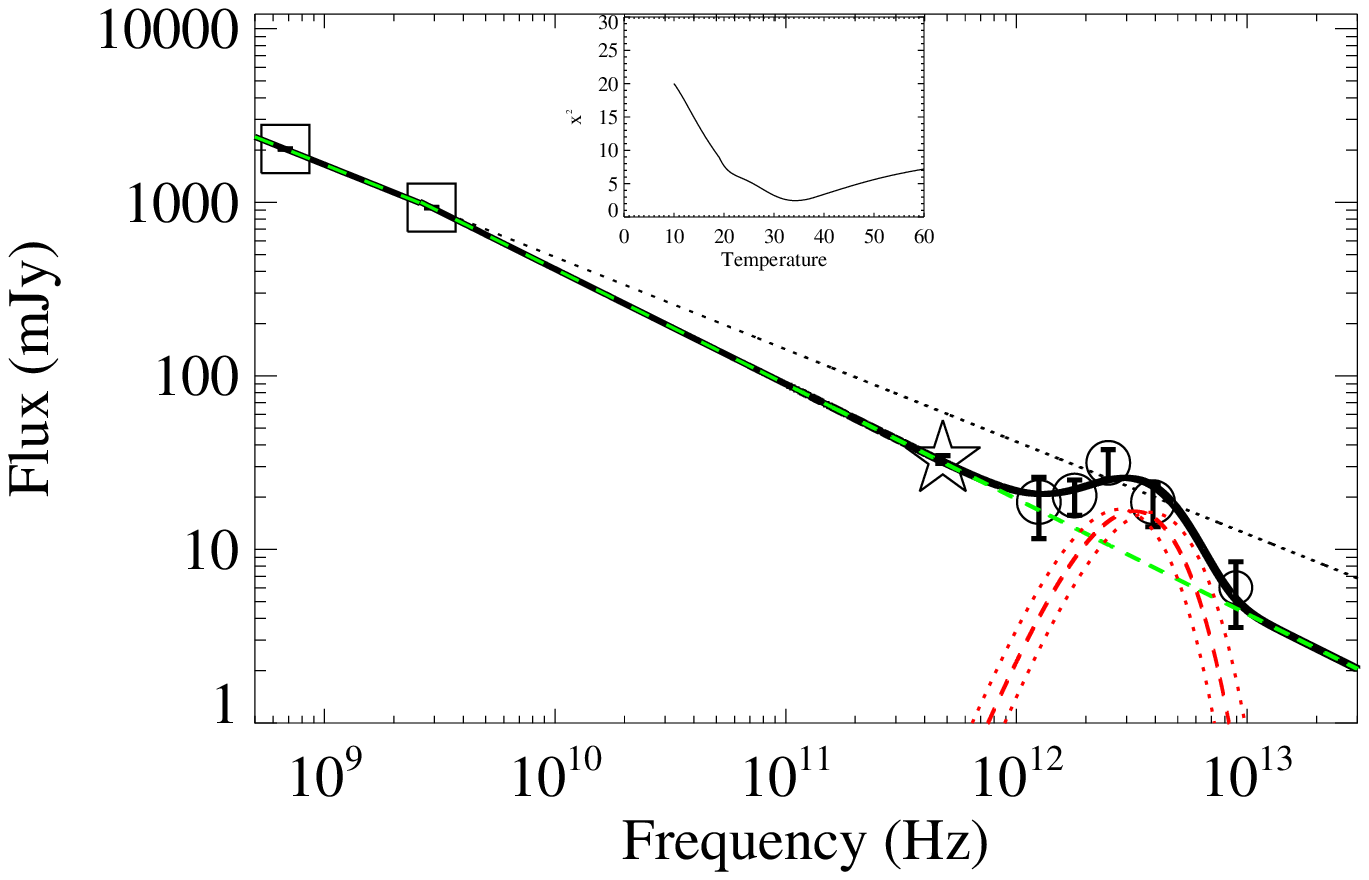} &
\\
%%%%%%%%%%%%%%%%%%%%%%%%%%%%%%%%%%%%%%%5

	&
\includegraphics[trim=25 50 27 1, clip, scale=0.3]{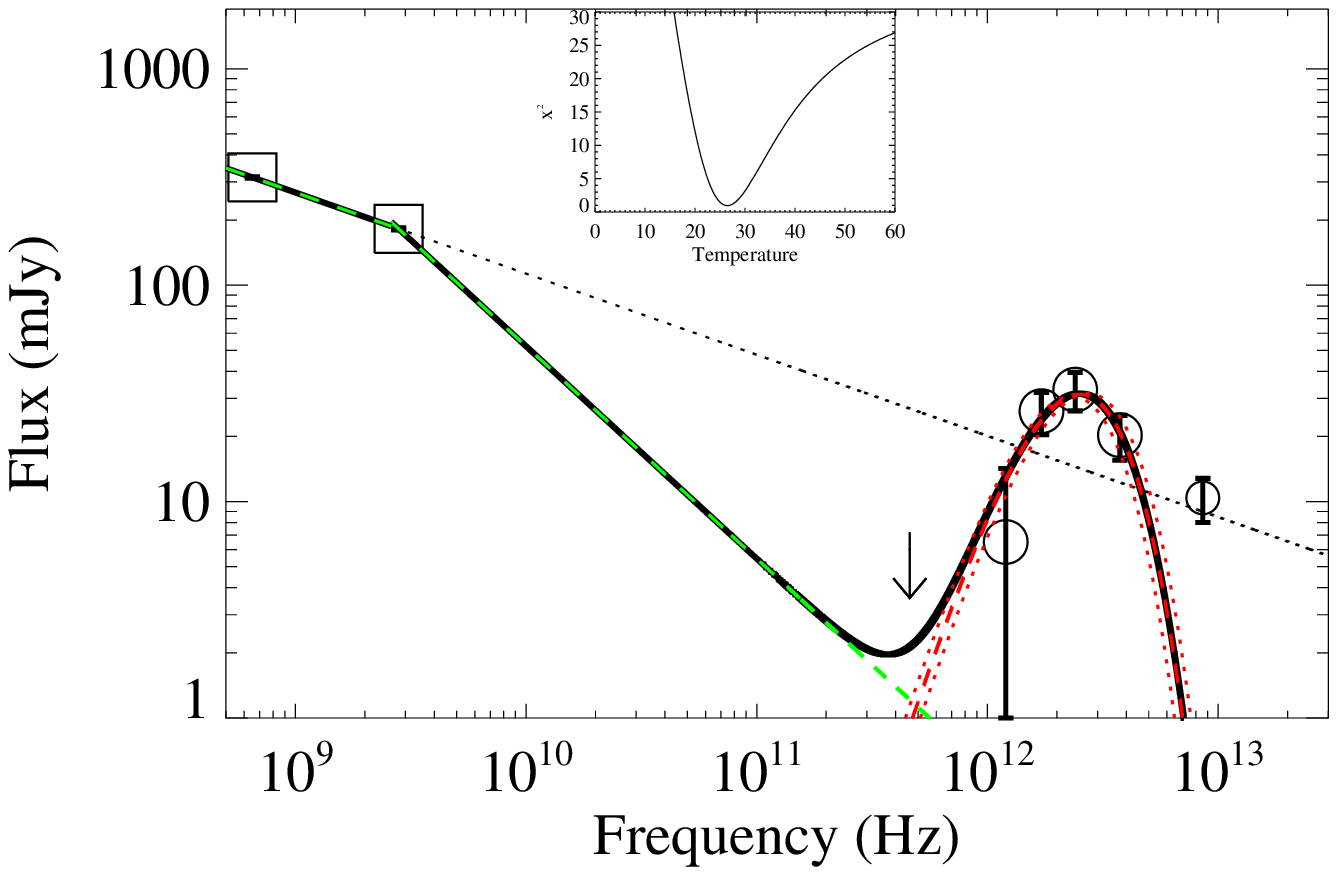} & 
\includegraphics[trim=46 50 27 1, clip,scale=0.3]{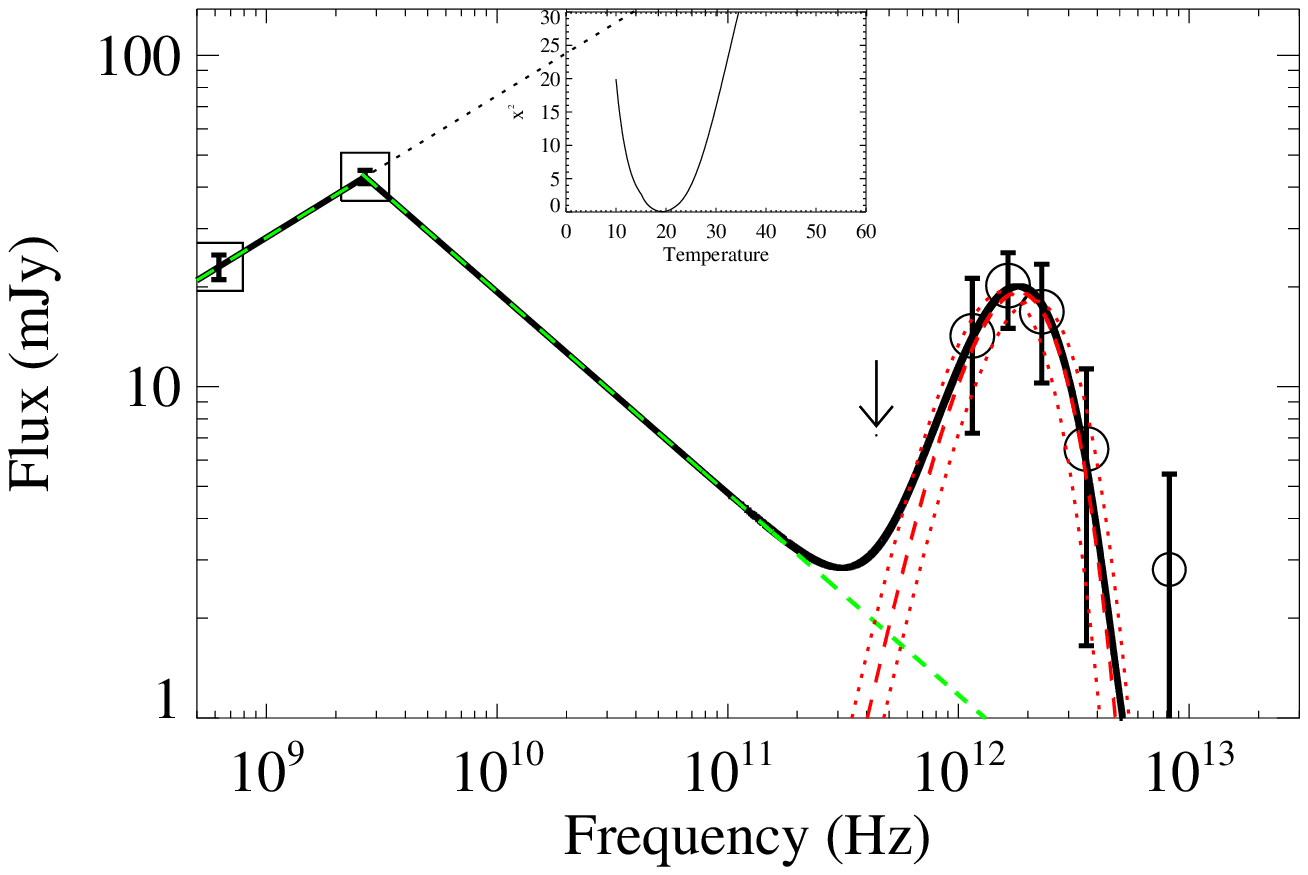} & 
\includegraphics[trim=46 50 27 1, clip,scale=0.3]{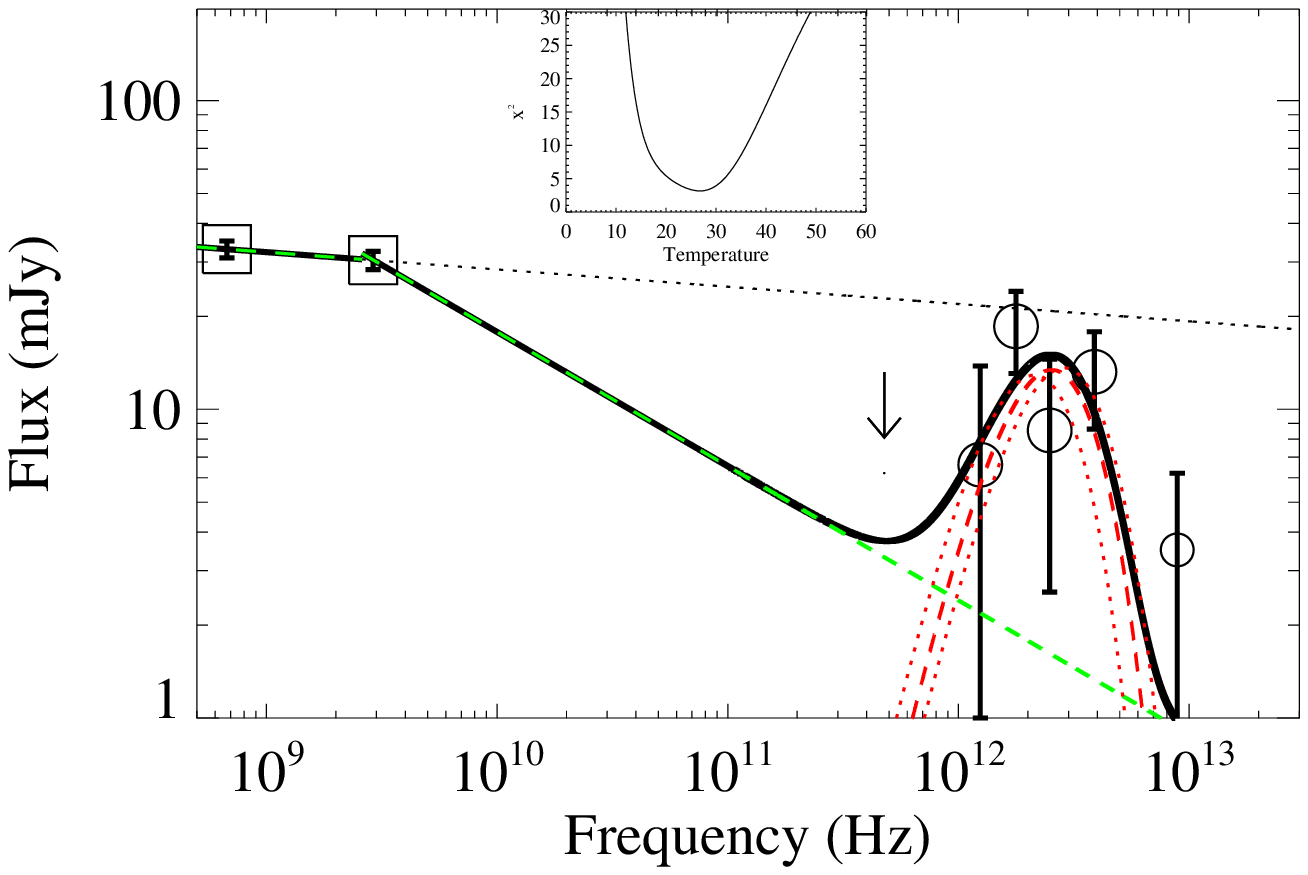} &
\includegraphics[trim=47 50 27 1, clip, scale=0.3]{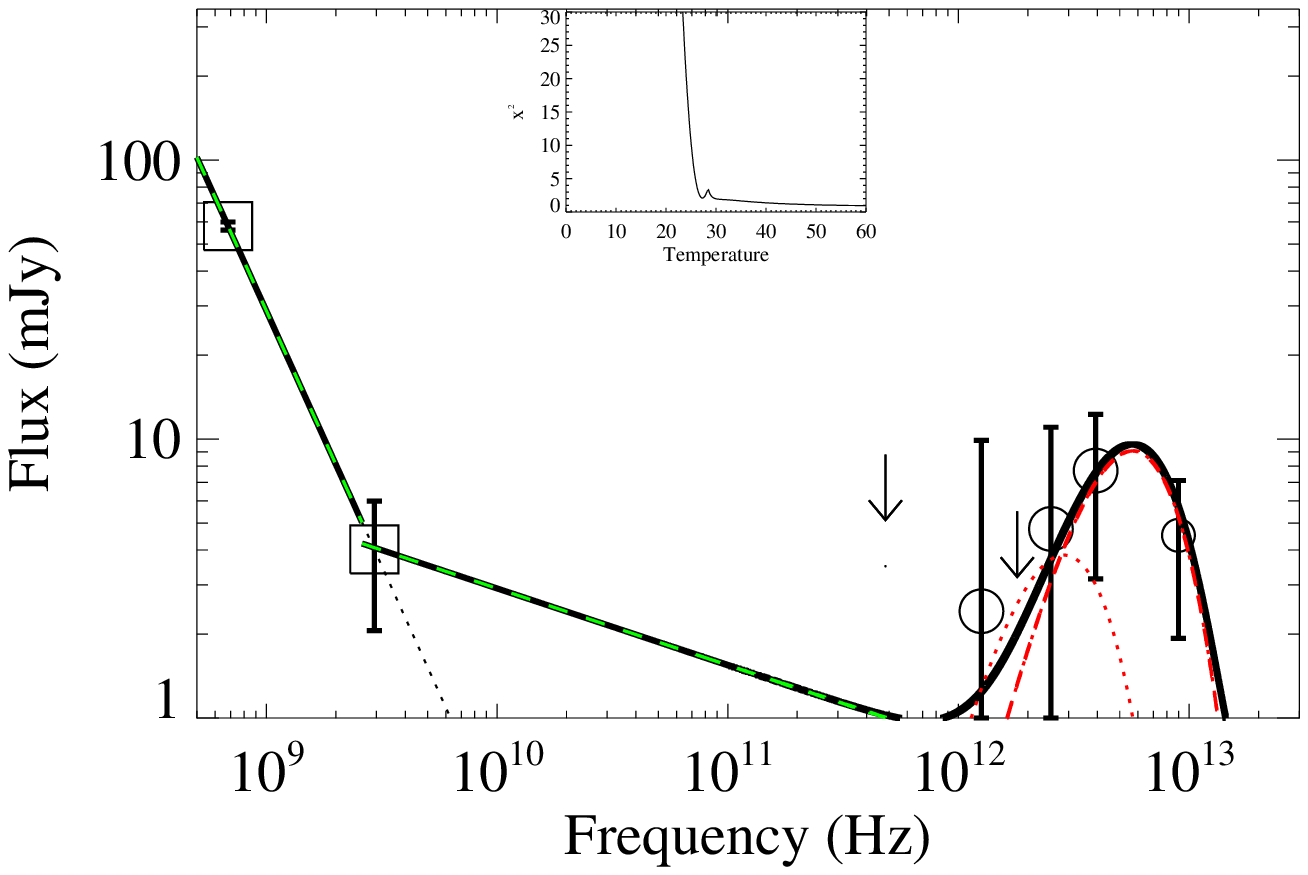} \\

	&
\includegraphics[trim=25 50 27 25, clip, scale=0.3]{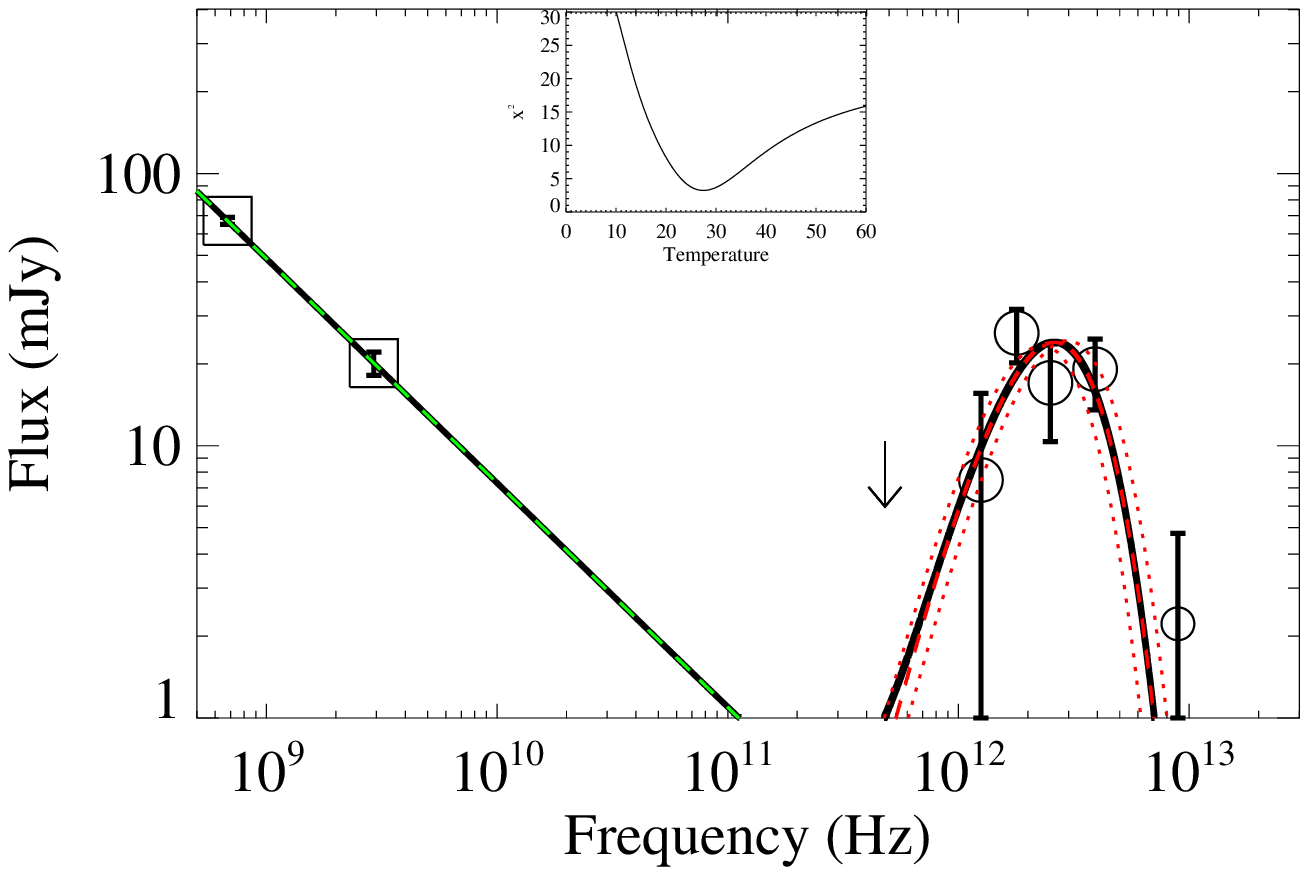} & 
\includegraphics[trim=46 50 27 25, clip,scale=0.3]{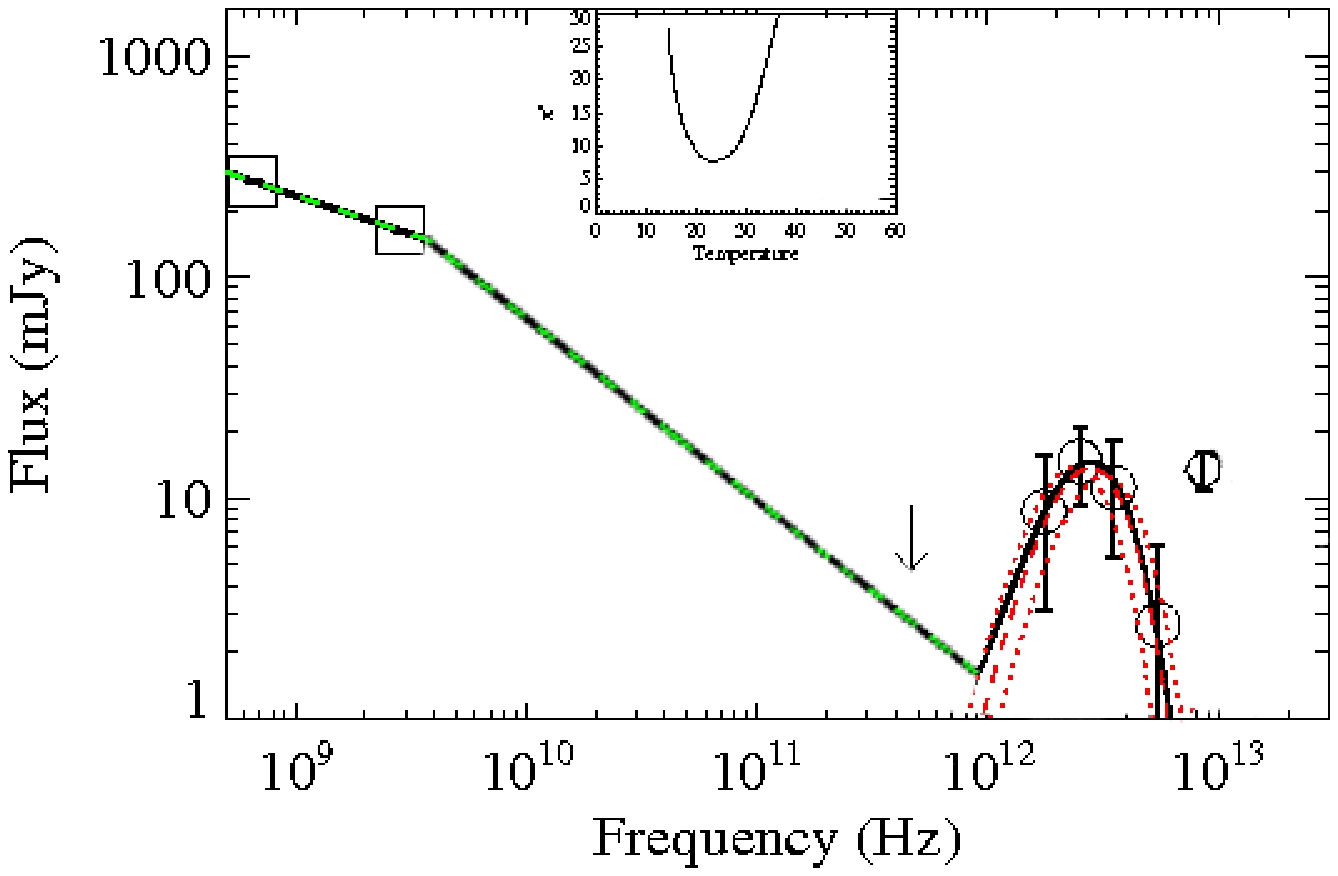} & 
\includegraphics[trim=46 50 27 25, clip,scale=0.3]{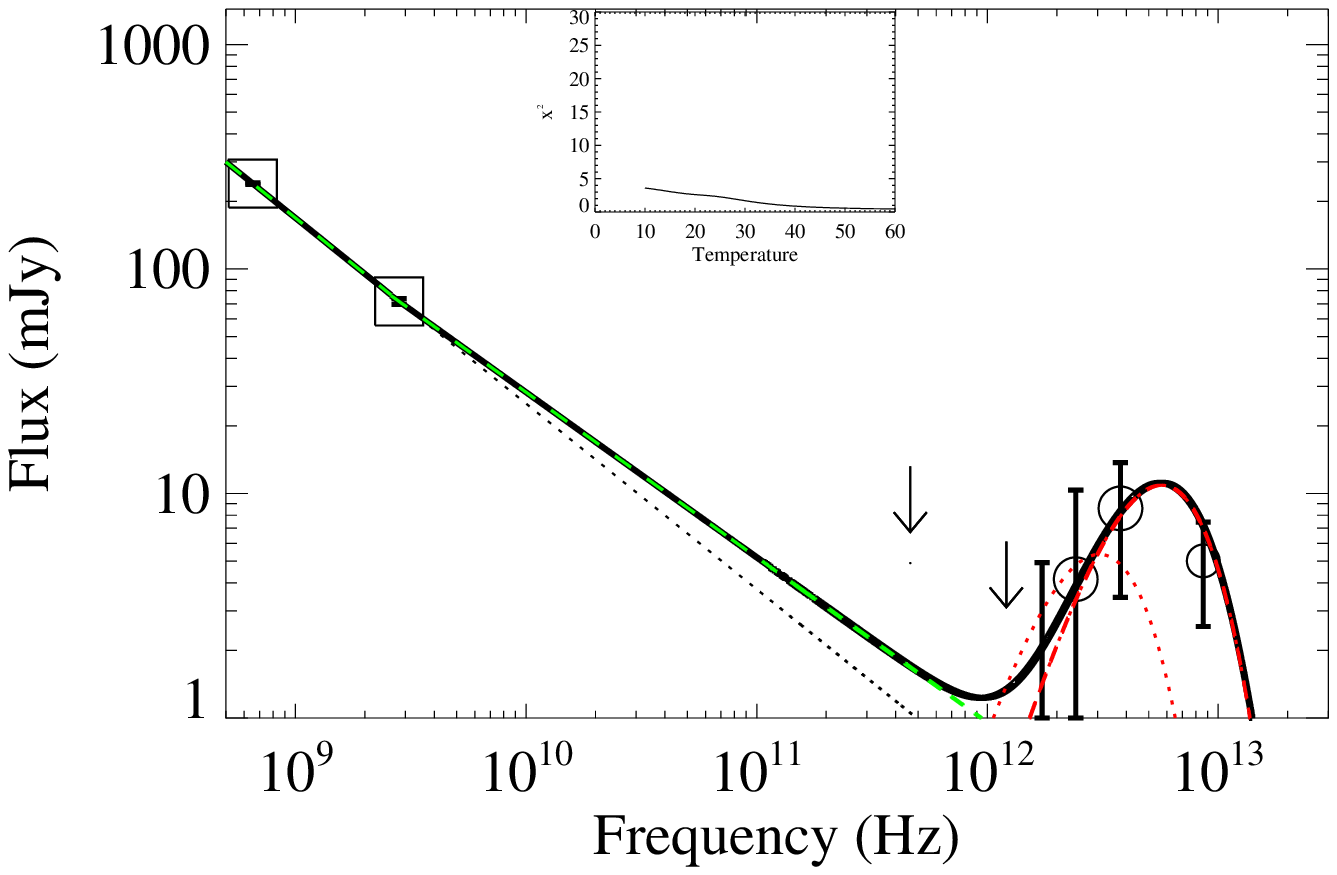} &
\includegraphics[trim=47 50 27 25, clip, scale=0.3]{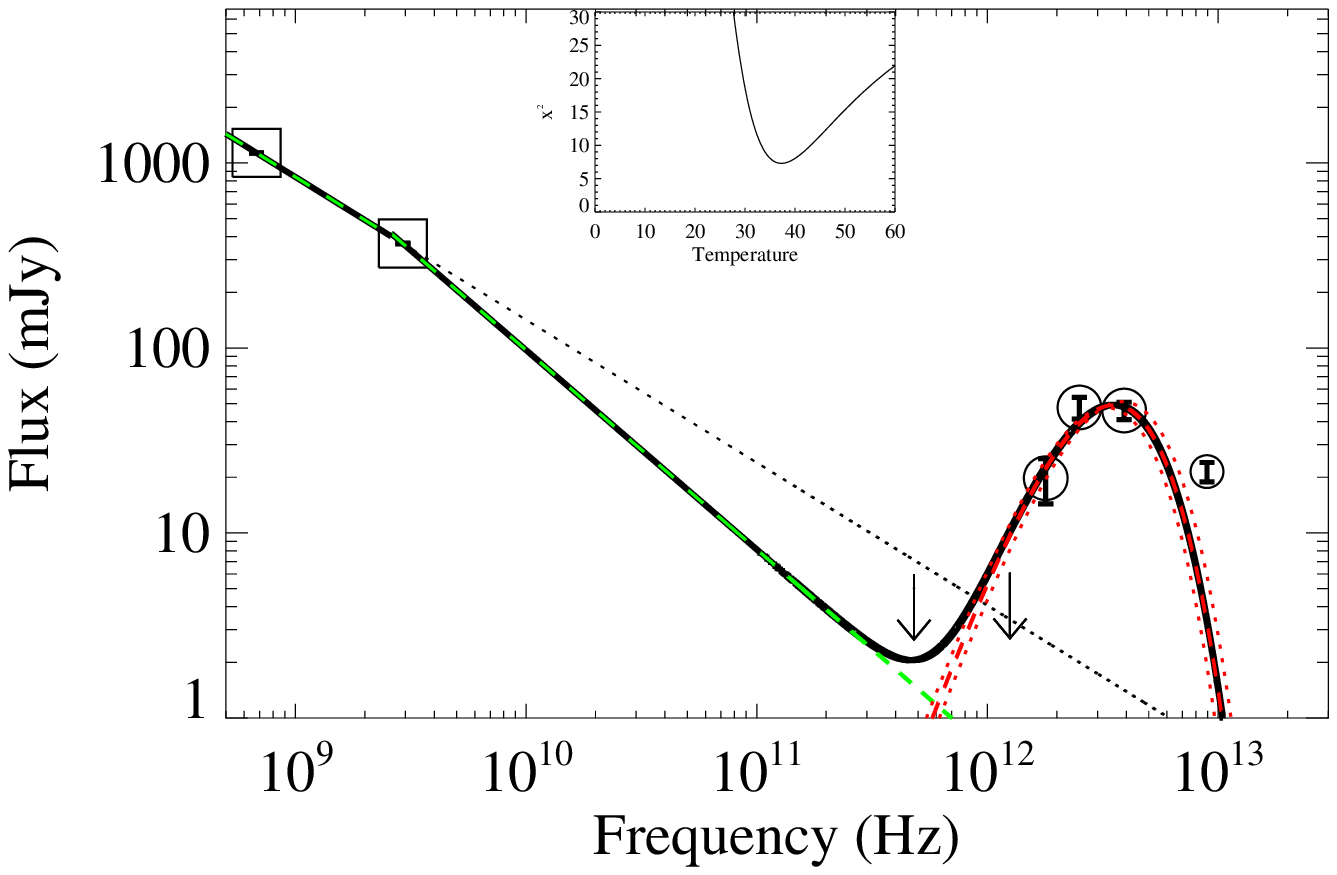} \\

\rotatebox{90}{\bf{Without contam.}}	&
\includegraphics[trim=25 50 27 25, clip, scale=0.3]{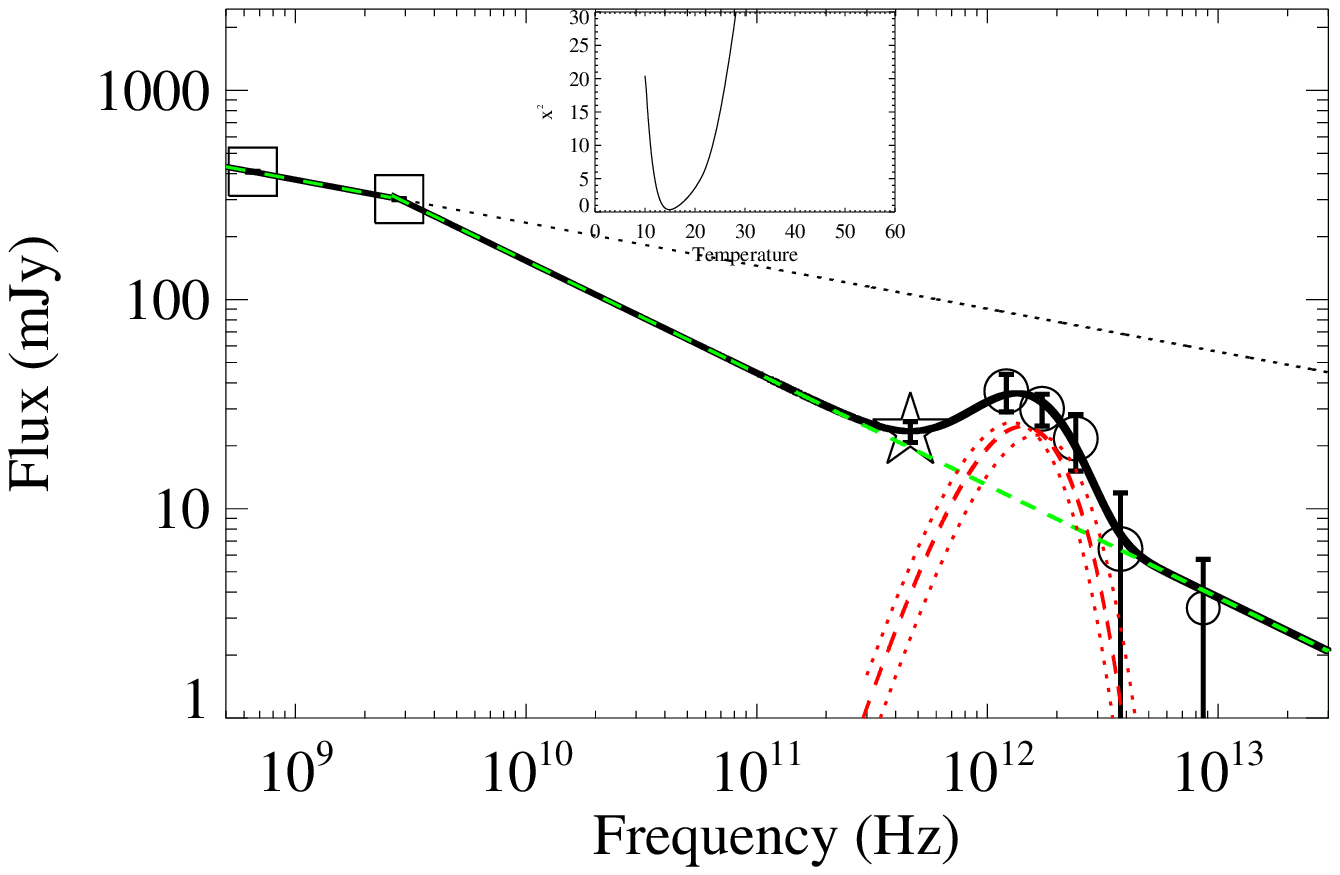} & 
\includegraphics[trim=46 50 27 25, clip,scale=0.3]{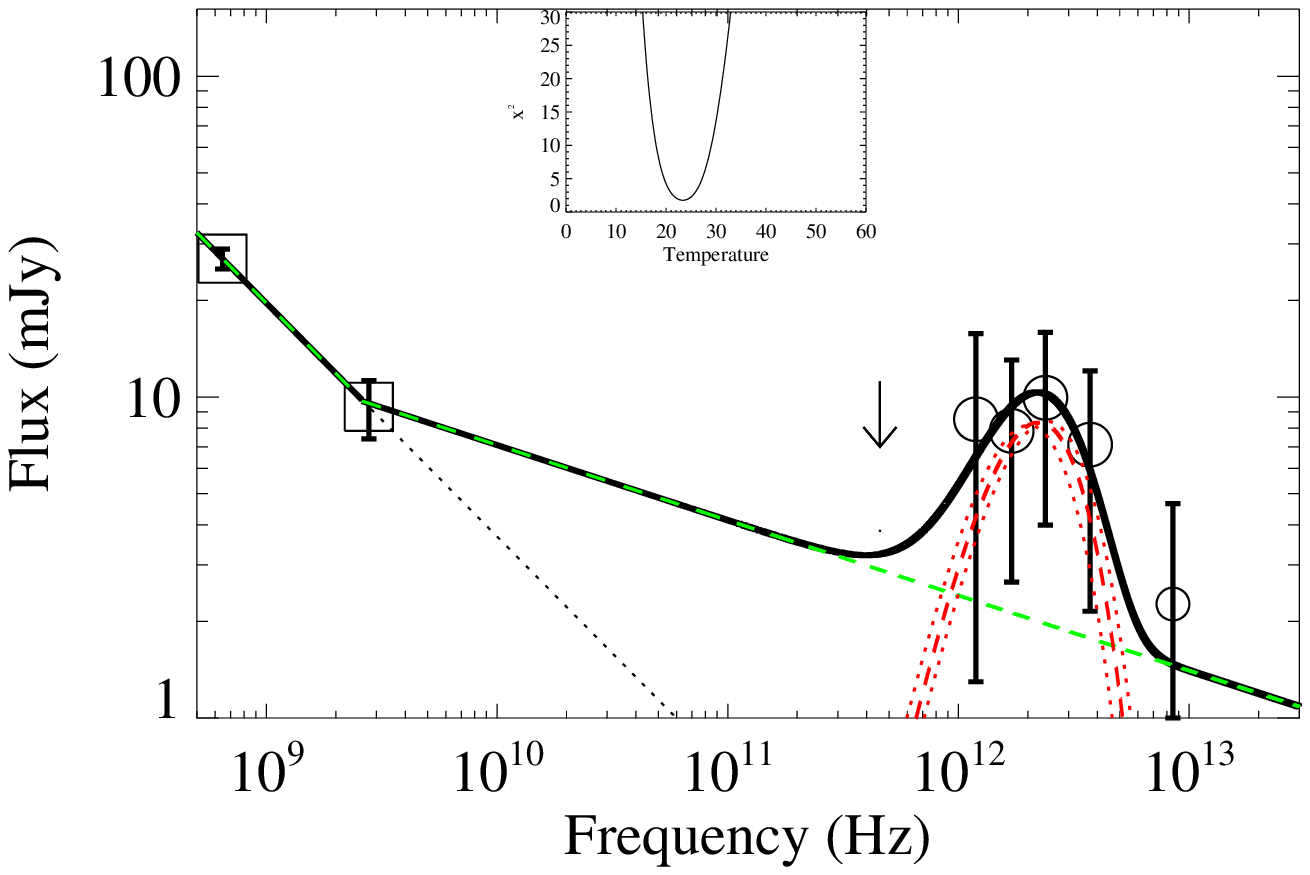} & 
\includegraphics[trim=46 50 27 25, clip,scale=0.3]{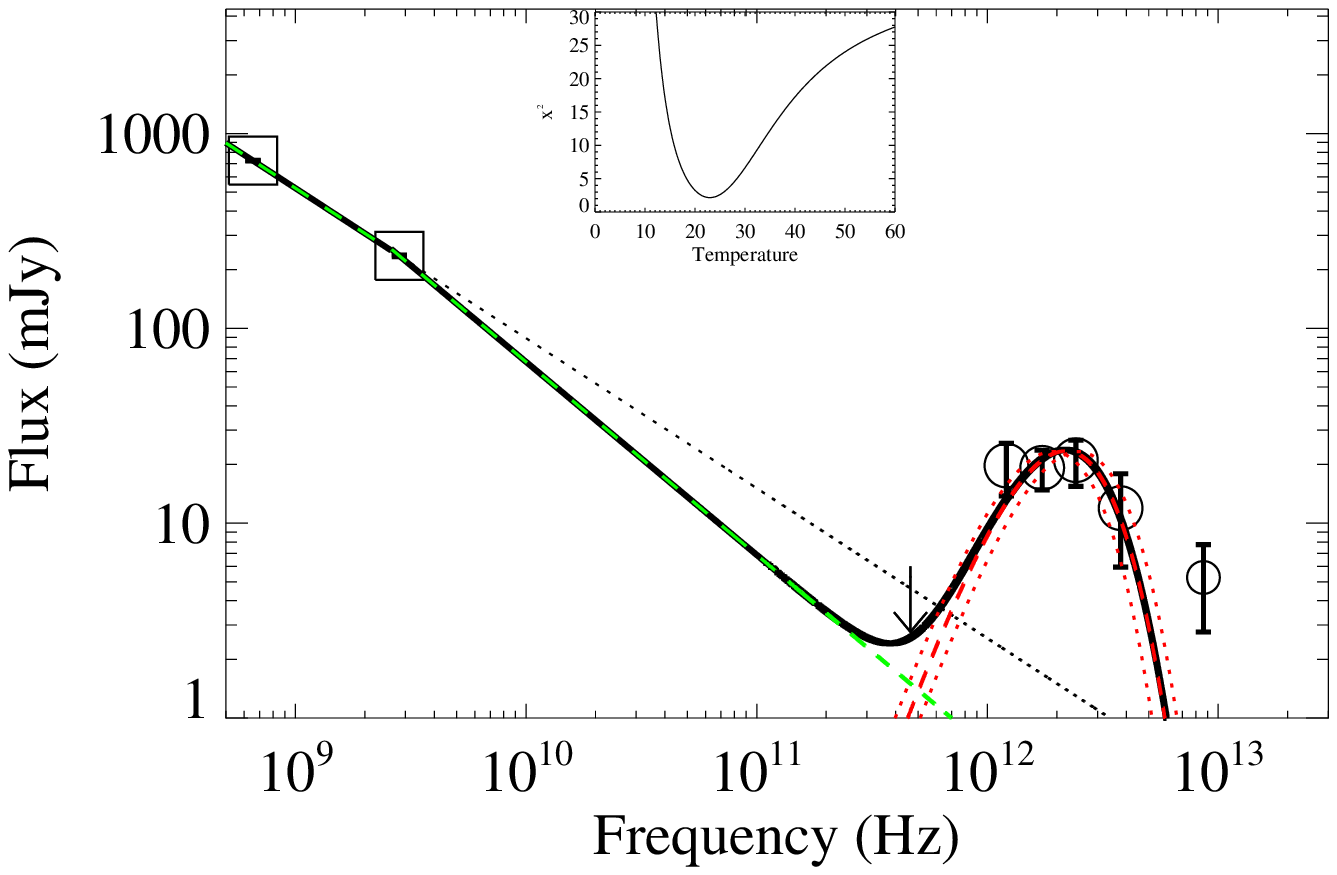} &
\includegraphics[trim=47 50 27 25, clip, scale=0.3]{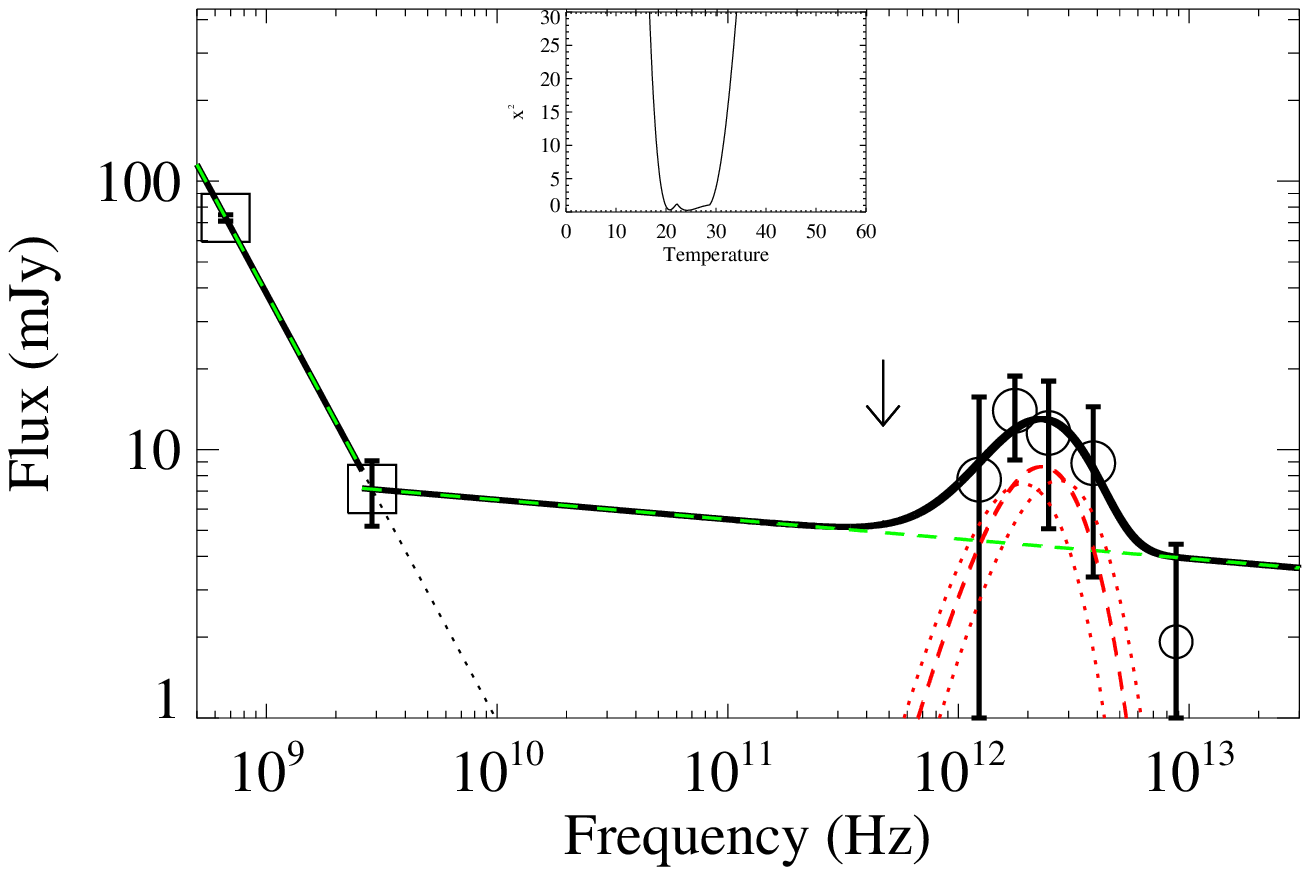} \\

	&
\includegraphics[trim=25 1 27 25, clip, scale=0.3]{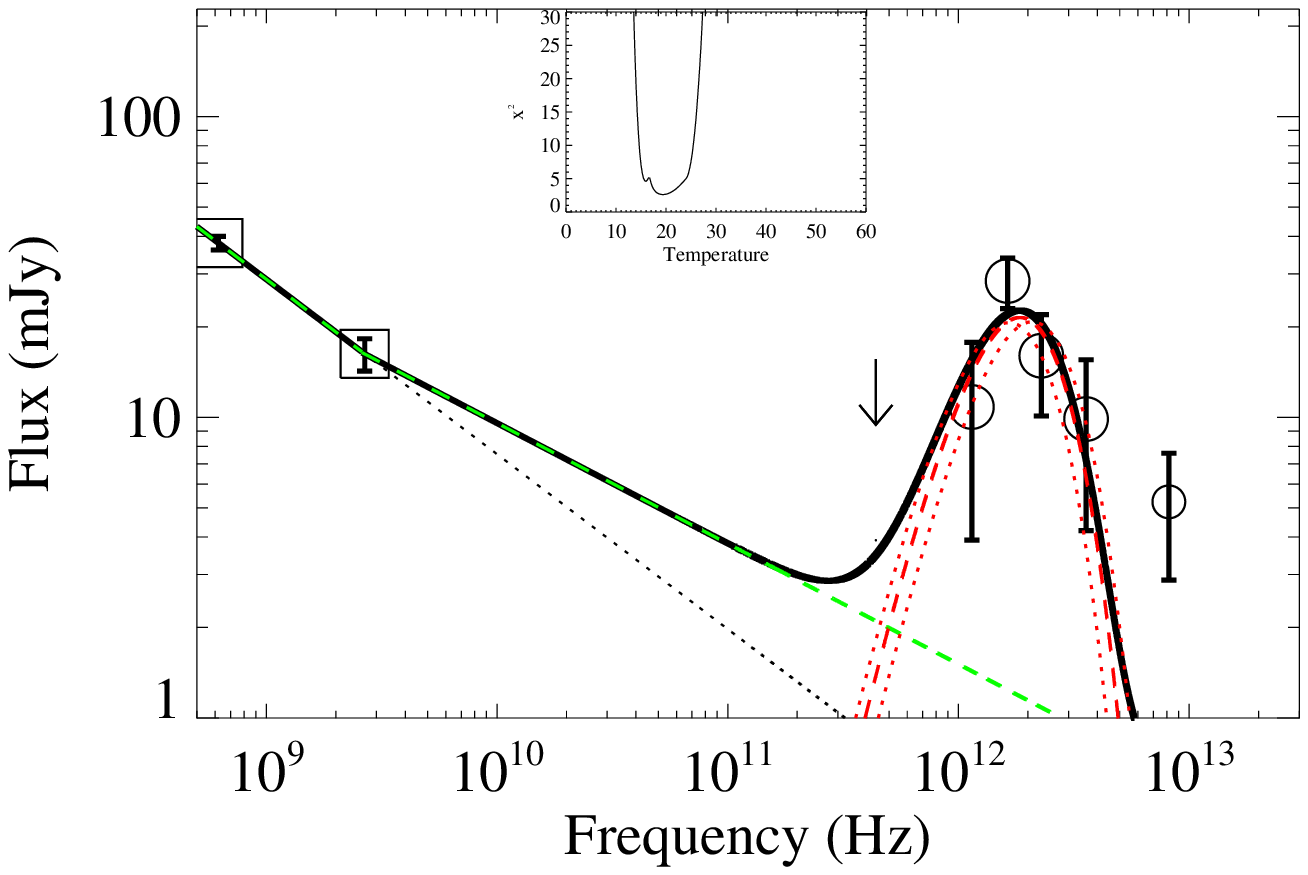} & 
\includegraphics[trim=46 1 27 25, clip,scale=0.3]{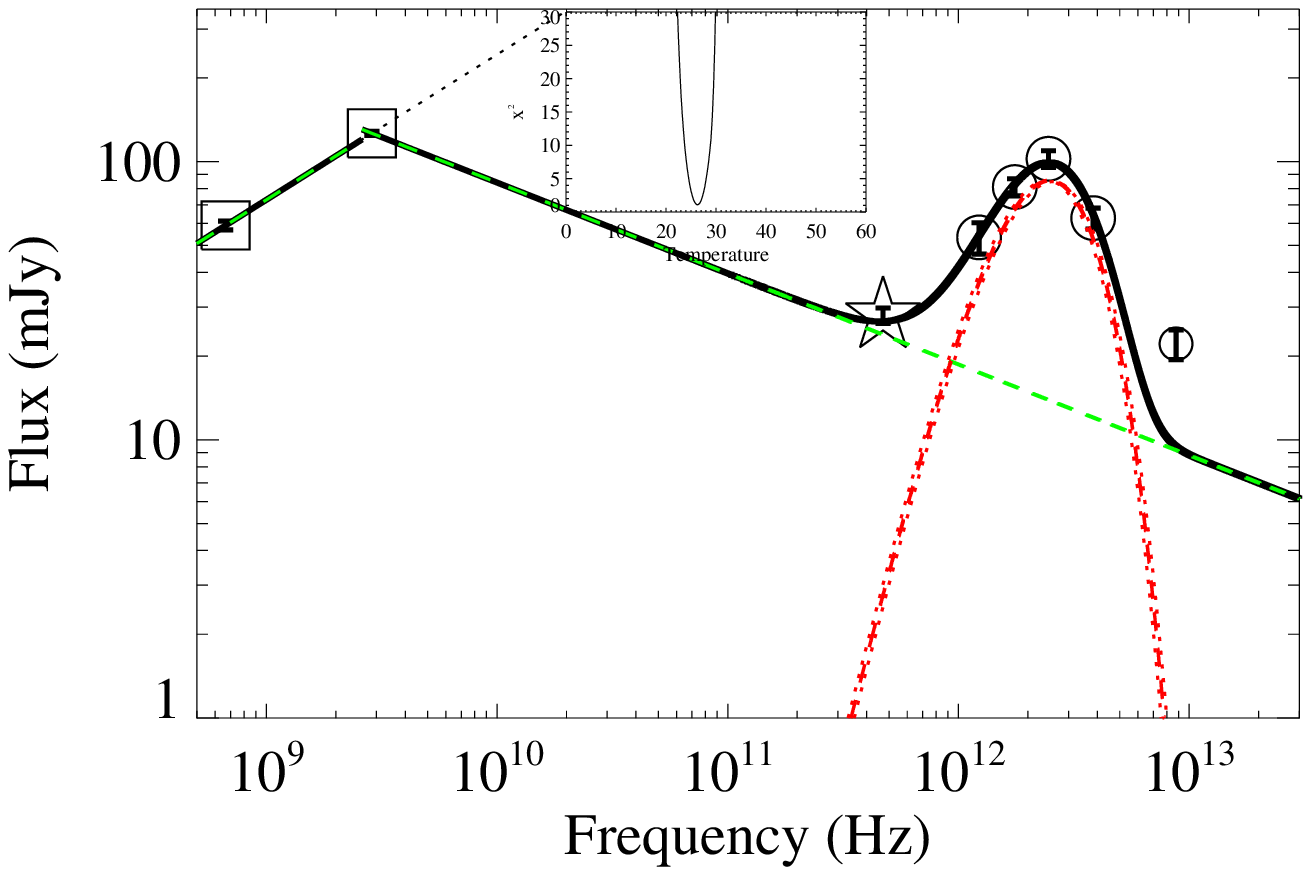} & 
\includegraphics[trim=46 1 27 25, clip,scale=0.3]{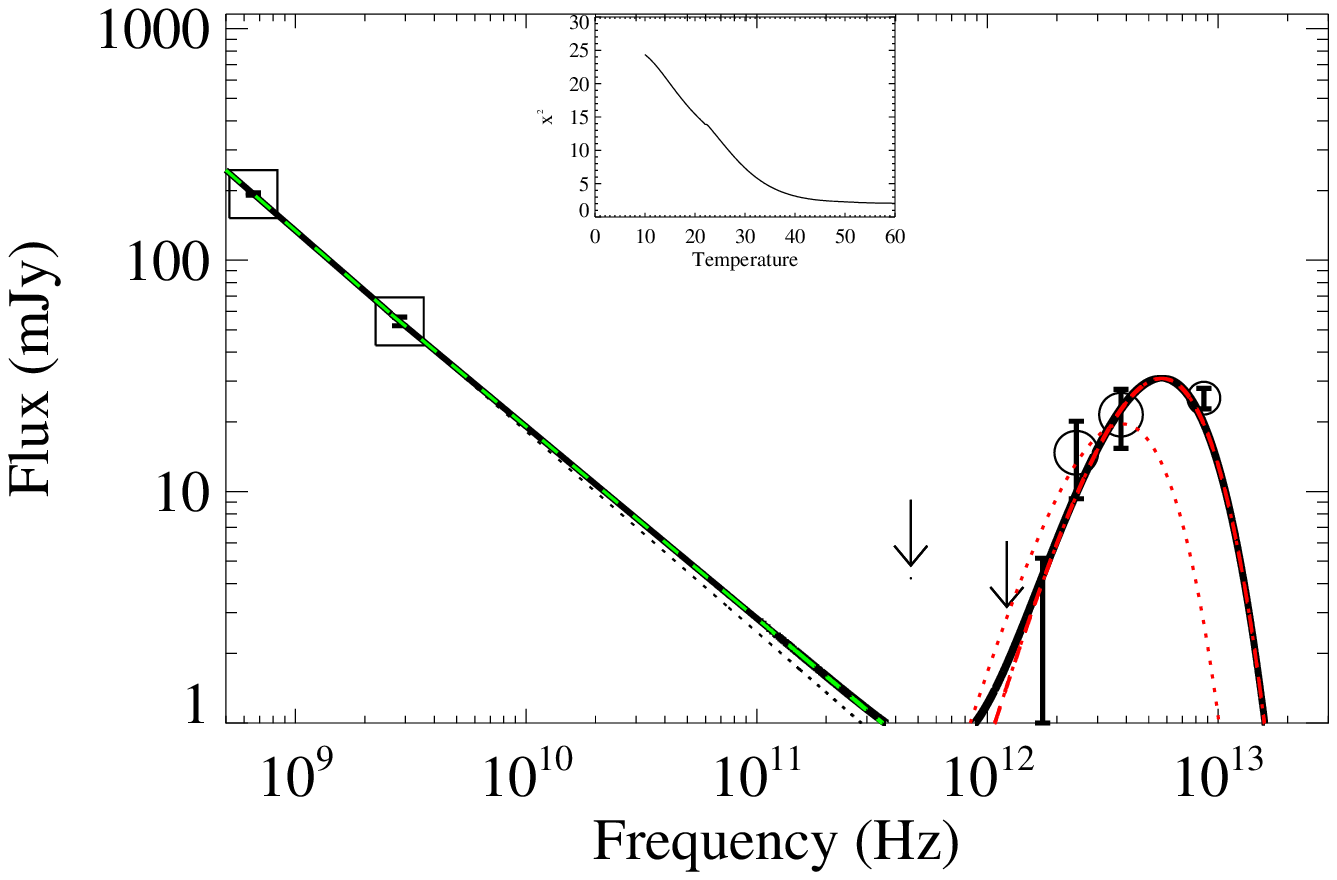} &
\includegraphics[trim=47 1 27 25, clip, scale=0.3]{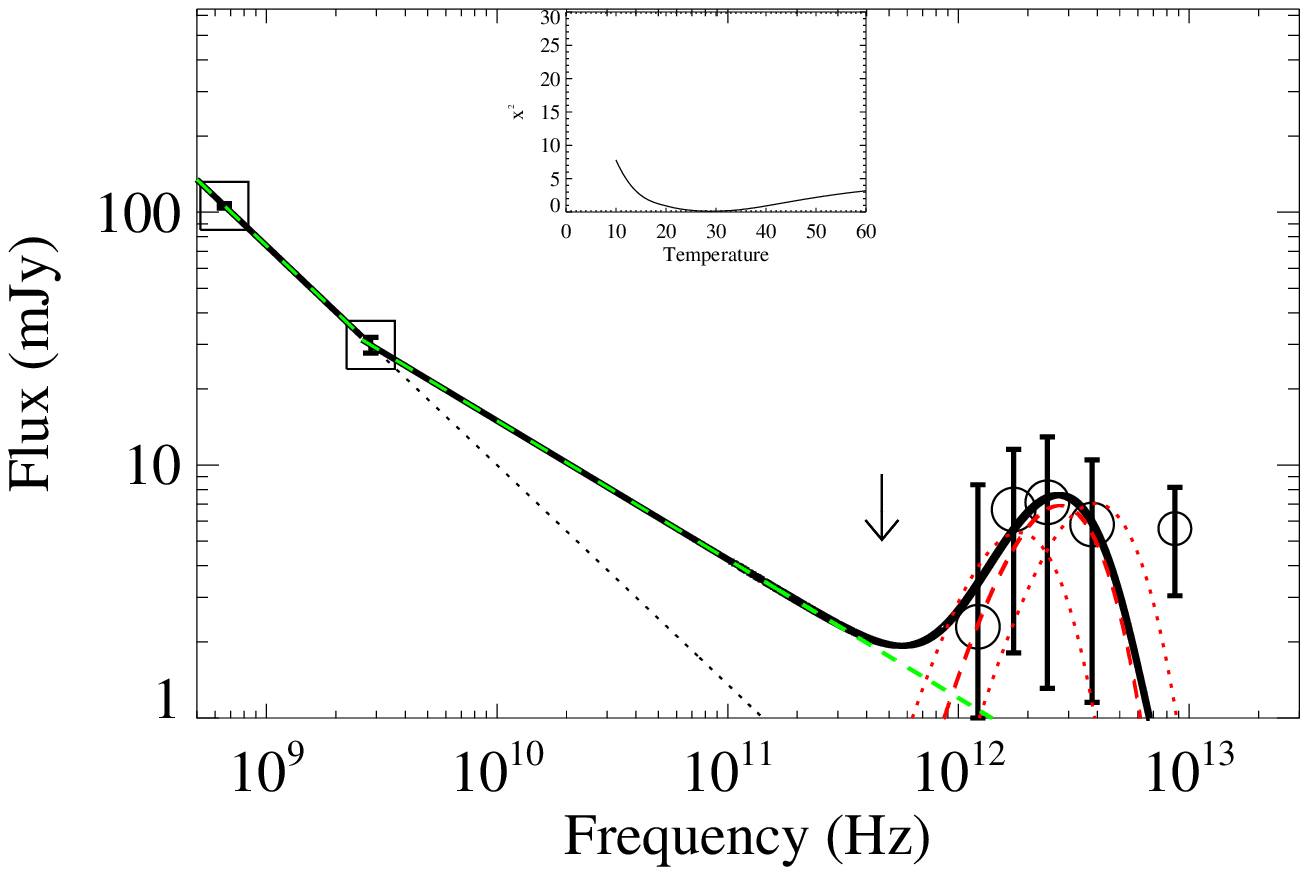} \\

\end{tabular}
\caption{Radio to FIR spectral energy distribution (SEDs, solid black) for representative RLQs observed with SMA. Open squares show the 325 (WENSS) and 1400 MHz (FIRST) radio photometry, open stars the 1300~$\mu$m SMA photometry (arrows indicate the $4\sigma$ upper limit) and open circles the SPIRE and 160~$\mu$m PACS photometry. The 70~$\mu$m PACS photometry is shown with a smaller circle and is not used for the SED fitting. Error bars correspond to $1\sigma$ photometric uncertainties. The radio photometry has been fitted with a broken power-law from 325 to 1400 MHz and from 1400 MHz to 230 GHz (or 1300~$\mu$m; green dashed line) and the FIR photometry with an optically thin modified blackbody component (red dashed line). The dotted black line shows the radio spectrum based on WENSS and FIRST radio observations only. The dotted red lines show the $1\sigma$ blackbody fitting uncertainty. The inner plot show the blackbody fitting $\chi^{2}$ value as a function of dust temperature. The SED plot are arrange in terms of non-thermal contamination at 1300~$\mu$m, from top to bottom. The top panel shows RLQs that have been rejected from our sample, the middle panel RLQs without significant synchrotron contamination at the FIR bands and the bottom panel RLQs in which the 1300~$\mu$m emission is dominated by the thermal component.} 
\label{fig:SED_appendix}
\end{figure*}

\begin{figure*}
\begin{tabular}{c c l l l}
	&
\includegraphics[trim=25 50 27 1, clip, scale=0.3]{SED123.836_27.6043.ps} & 
\includegraphics[trim=46 50 27 1, clip,scale=0.3]{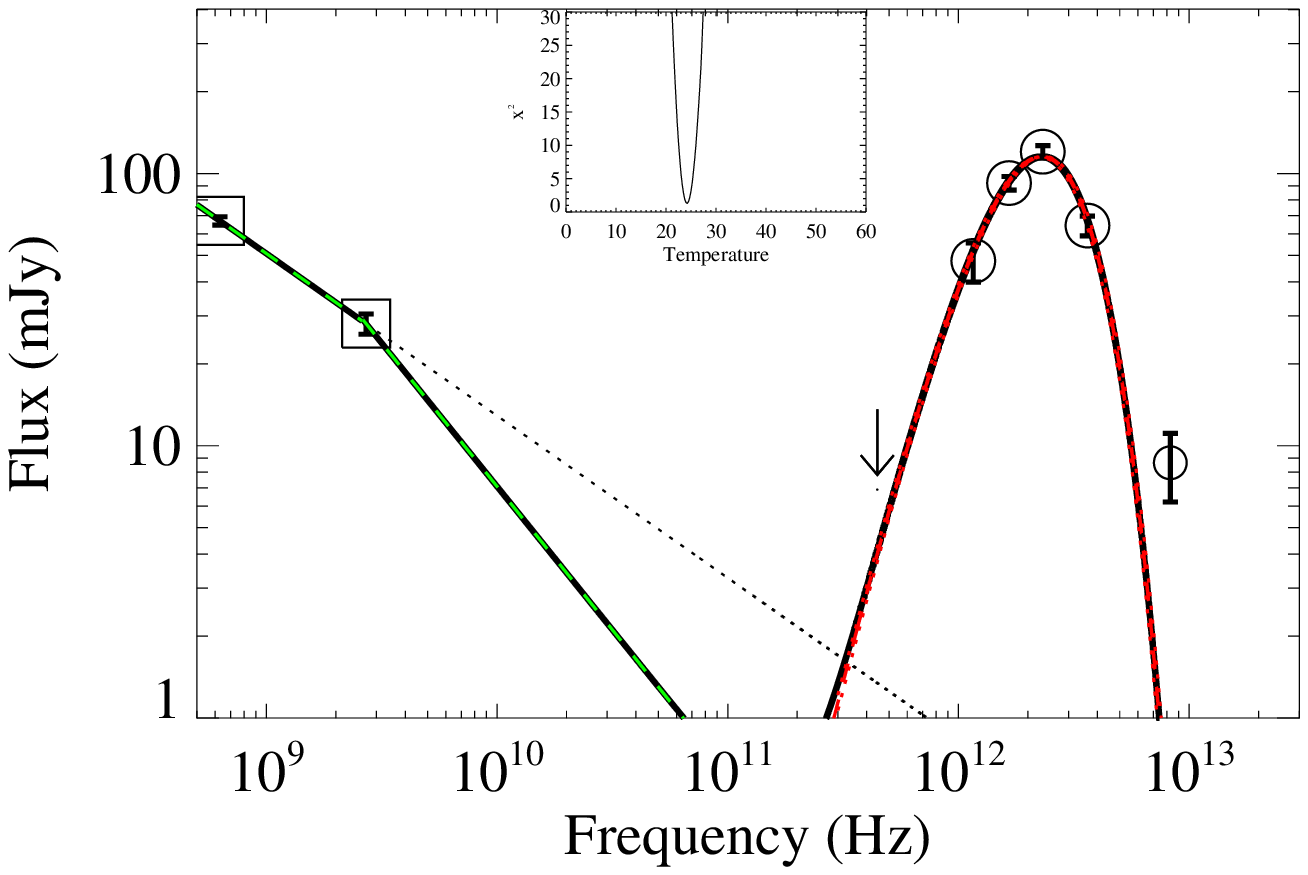} & 
\includegraphics[trim=46 50 27 1, clip,scale=0.3]{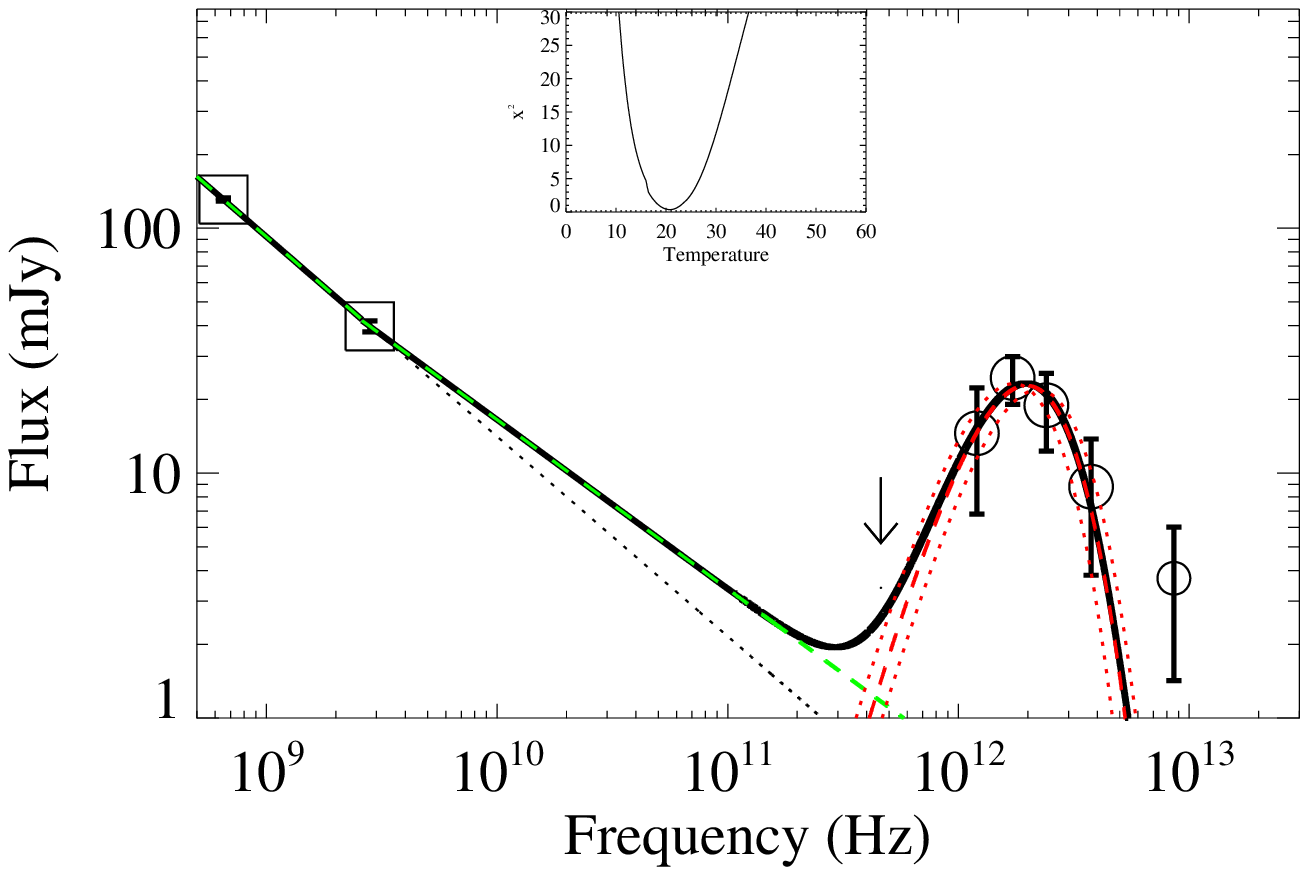} &
\includegraphics[trim=47 50 26 1, clip, scale=0.3]{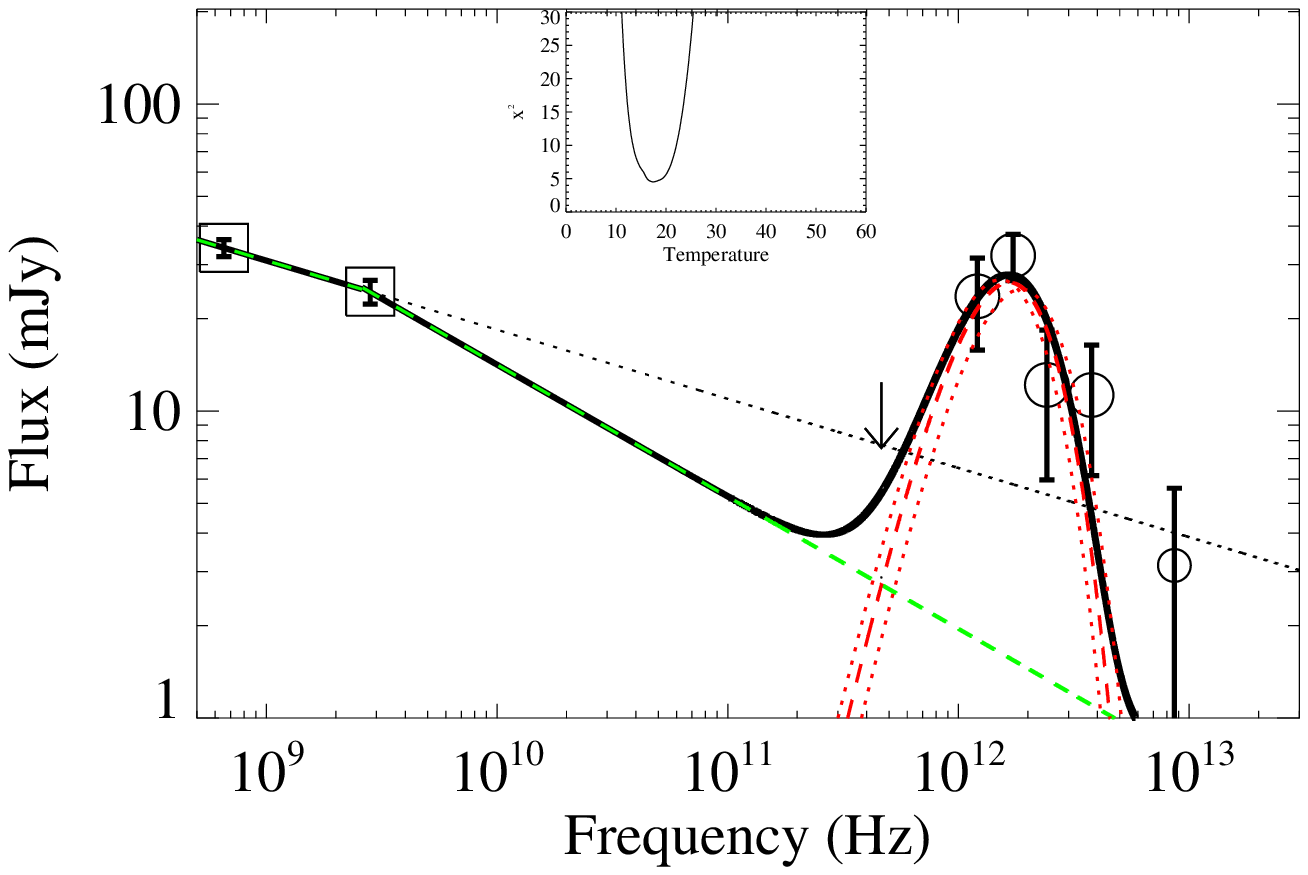} \\

\rotatebox{90}{\bf{thermal dominated}}	&
\includegraphics[trim=25 50 27 25, clip,scale=0.3]{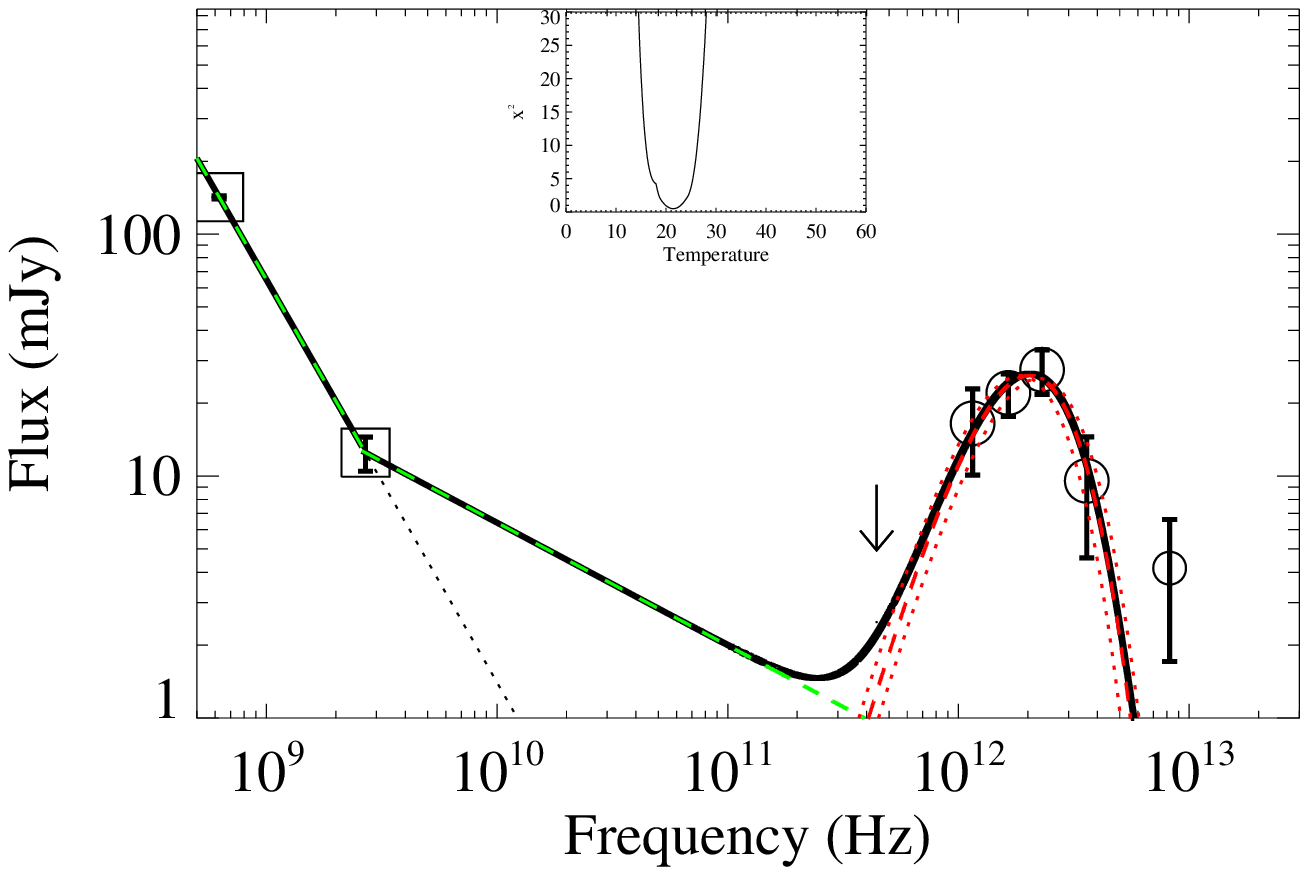} & 
\includegraphics[trim=46 50 27 25, clip,scale=0.3]{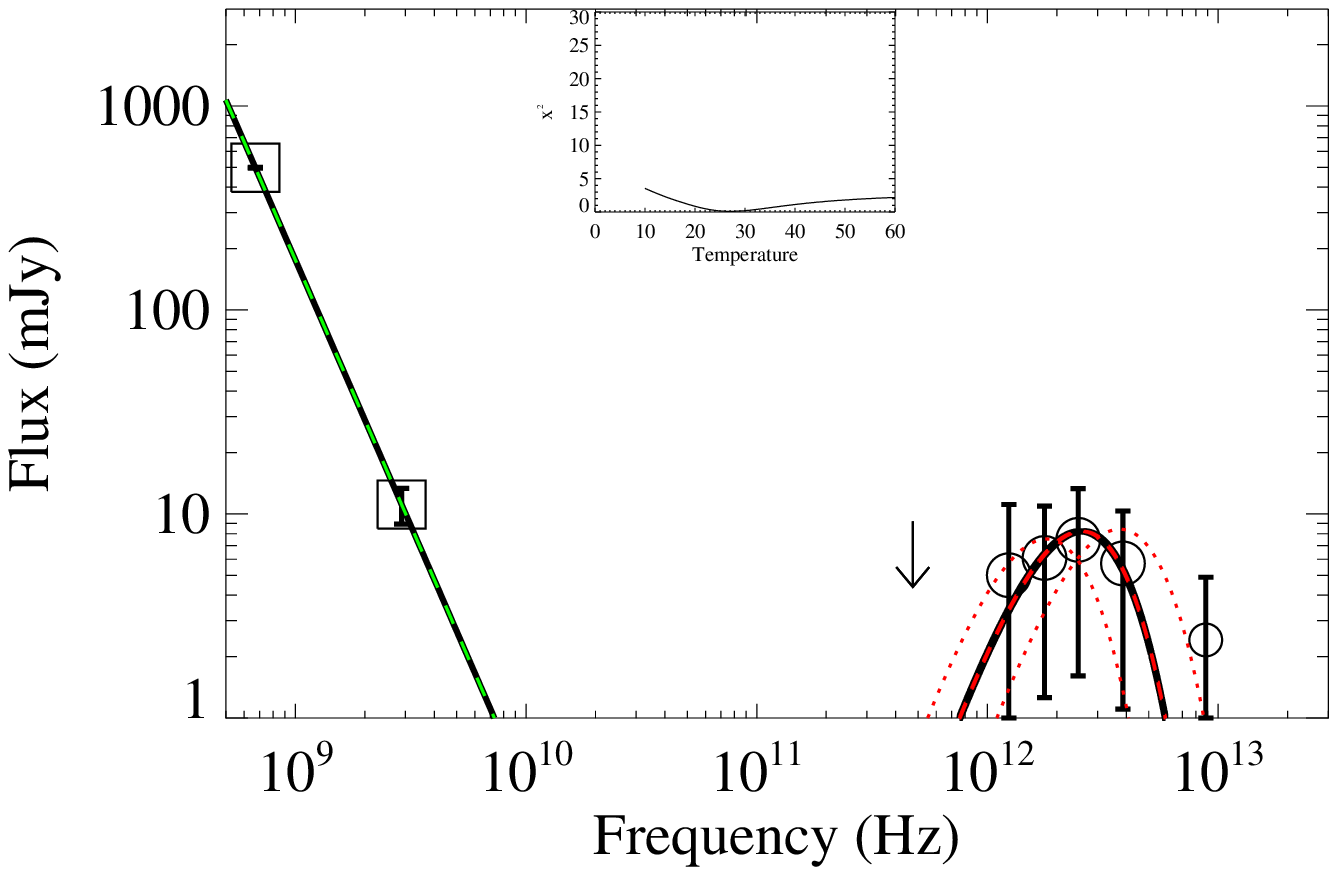} &
\includegraphics[trim=46 50 27 25, clip, scale=0.3]{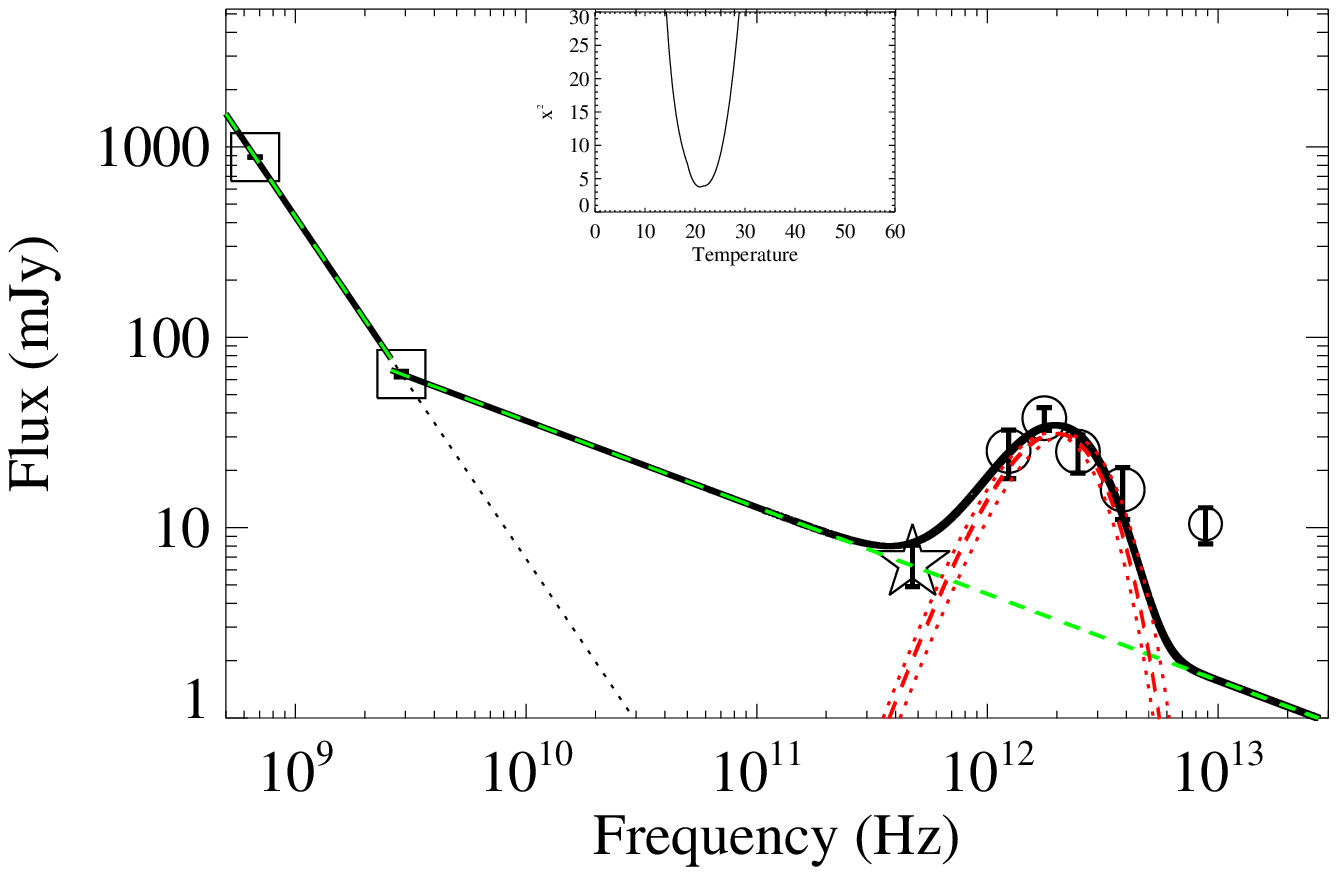} & 
\includegraphics[trim=47 50 26 25, clip, scale=0.3]{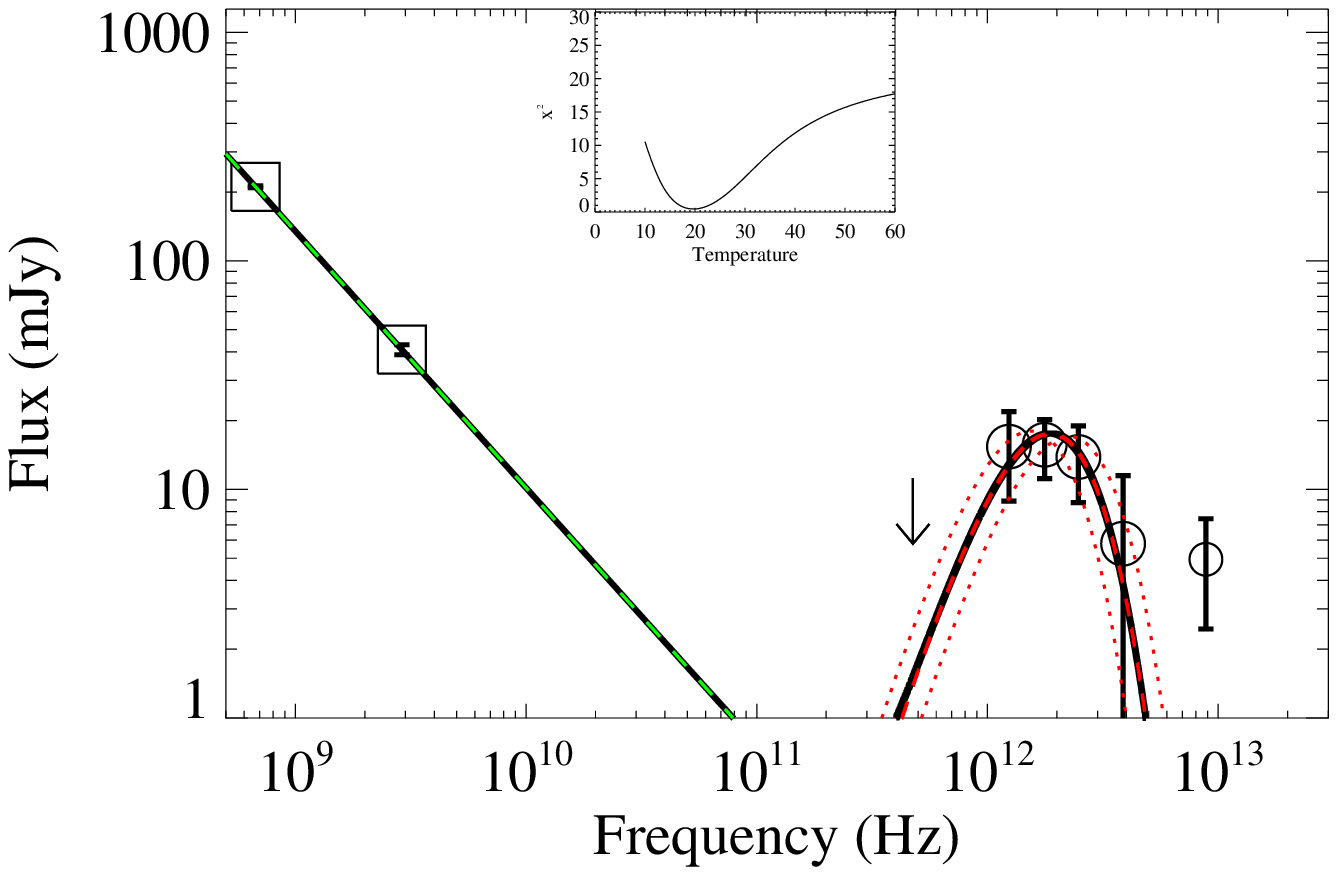}  \\

	\rotatebox{90}{\bf{Without contam.}}	&
\includegraphics[trim=25 50 27 25, clip,scale=0.3]{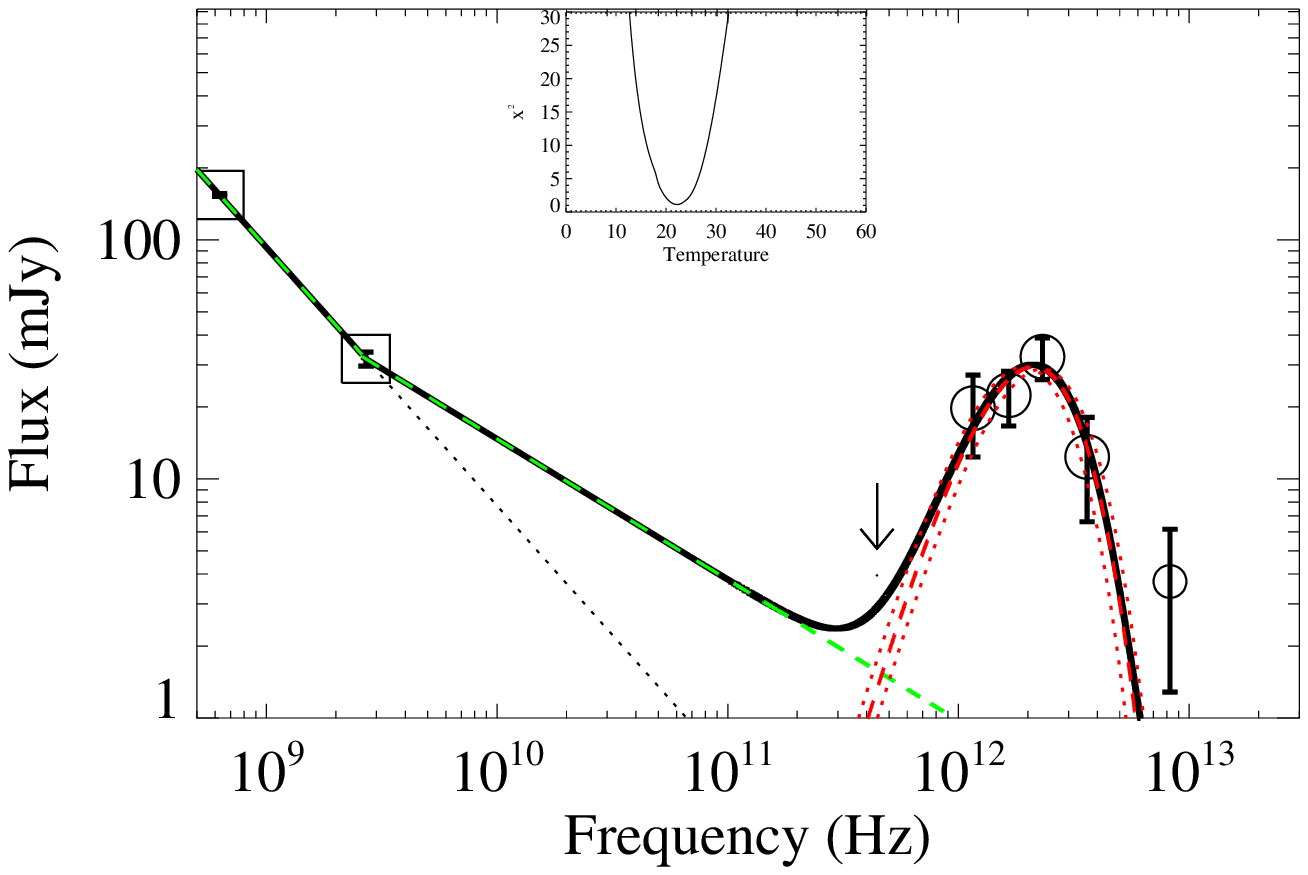} & 
\includegraphics[trim=46 50 27 25, clip,scale=0.3]{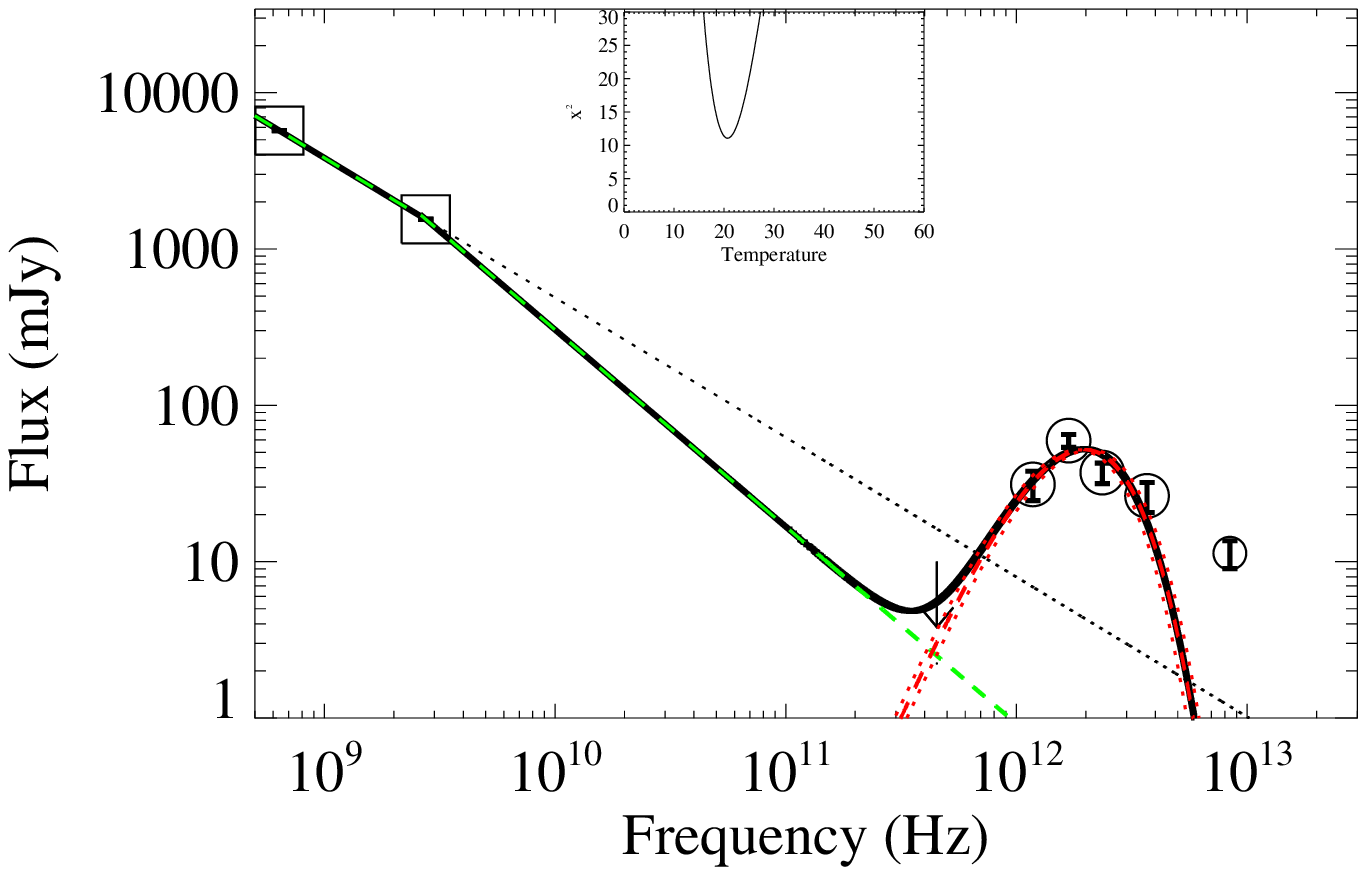} &
\includegraphics[trim=46 50 27 25, clip, scale=0.3]{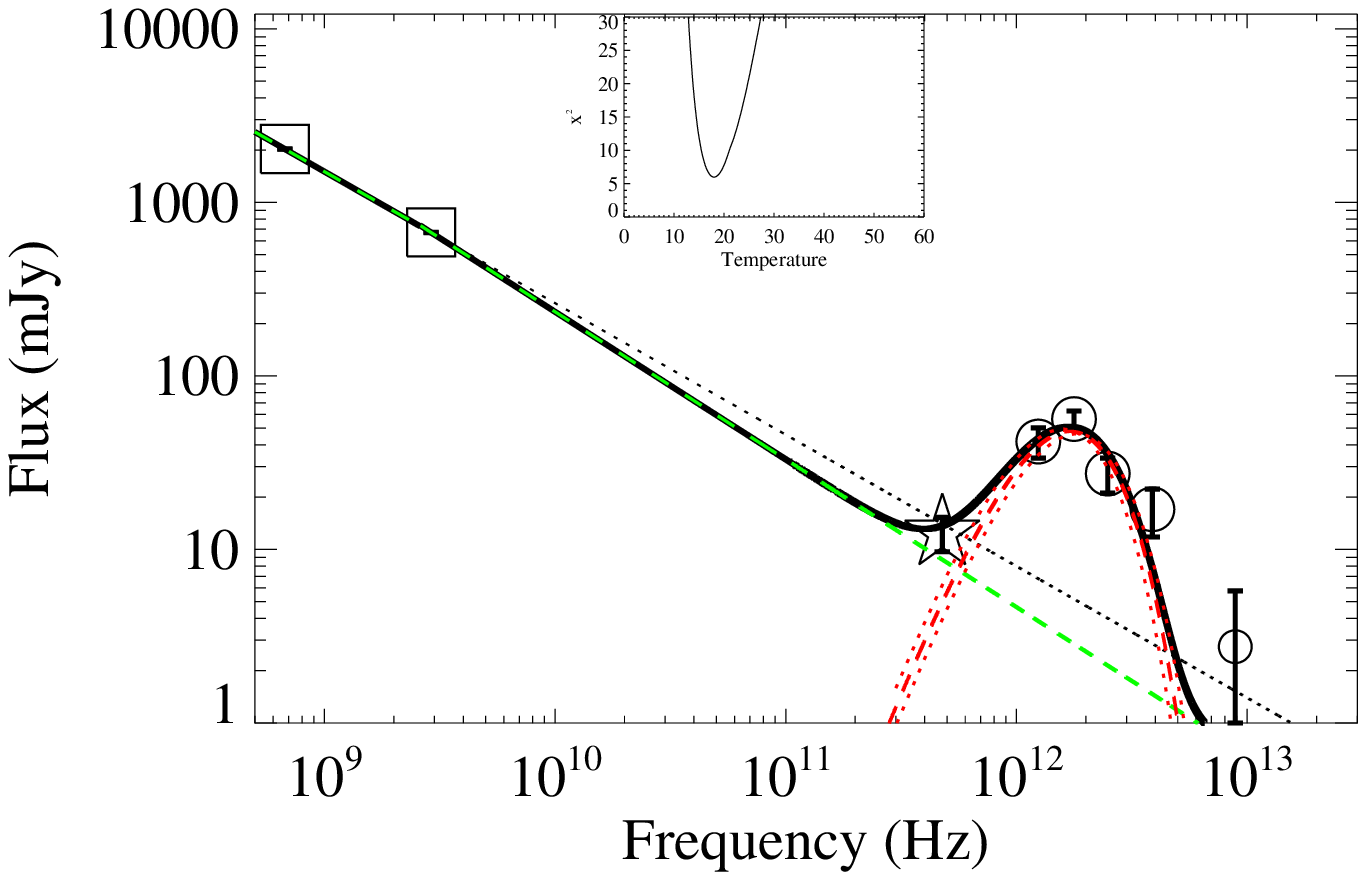} & 
\includegraphics[trim=47 50 26 25, clip,scale=0.3]{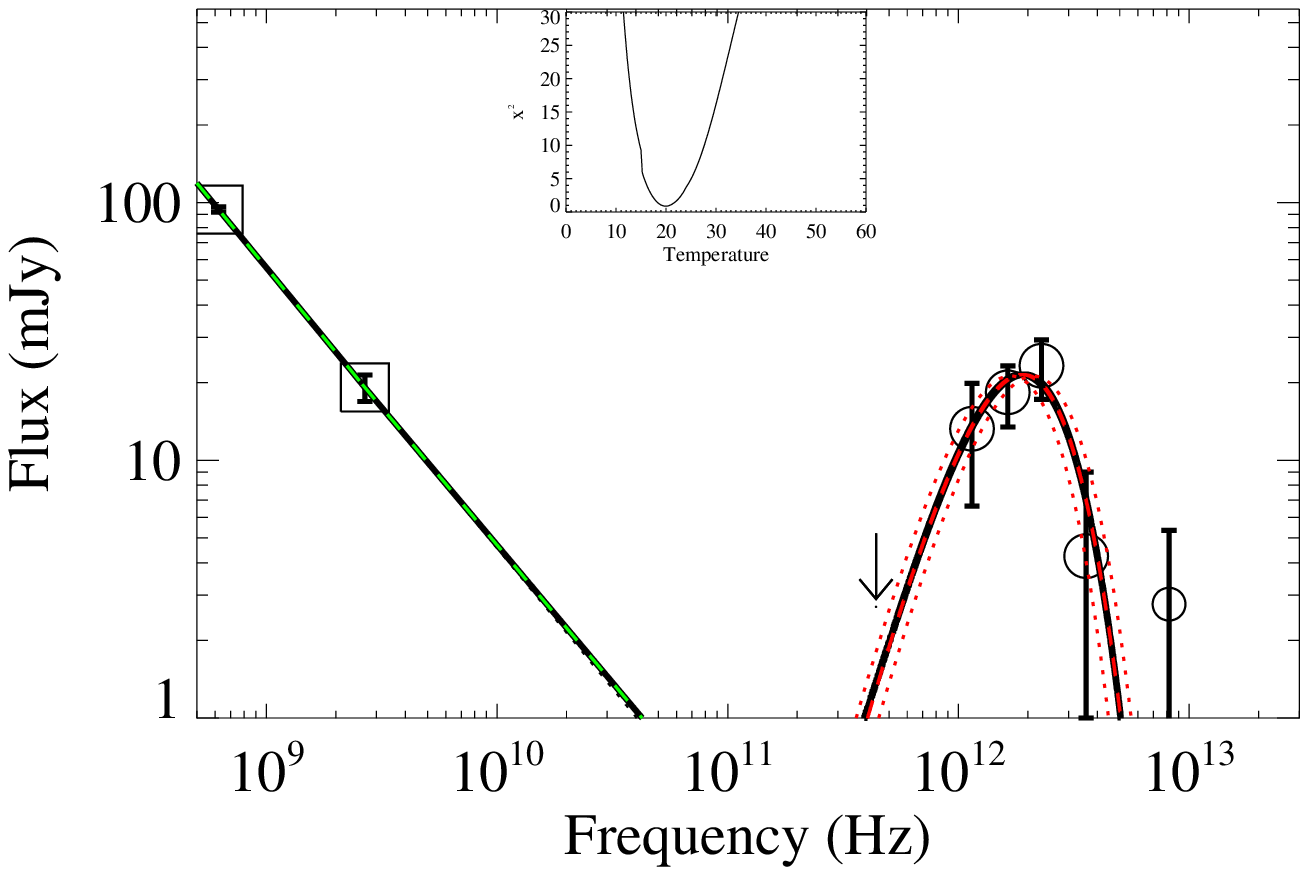} \\

	&
\includegraphics[trim=25 1 25 25, clip,scale=0.3]{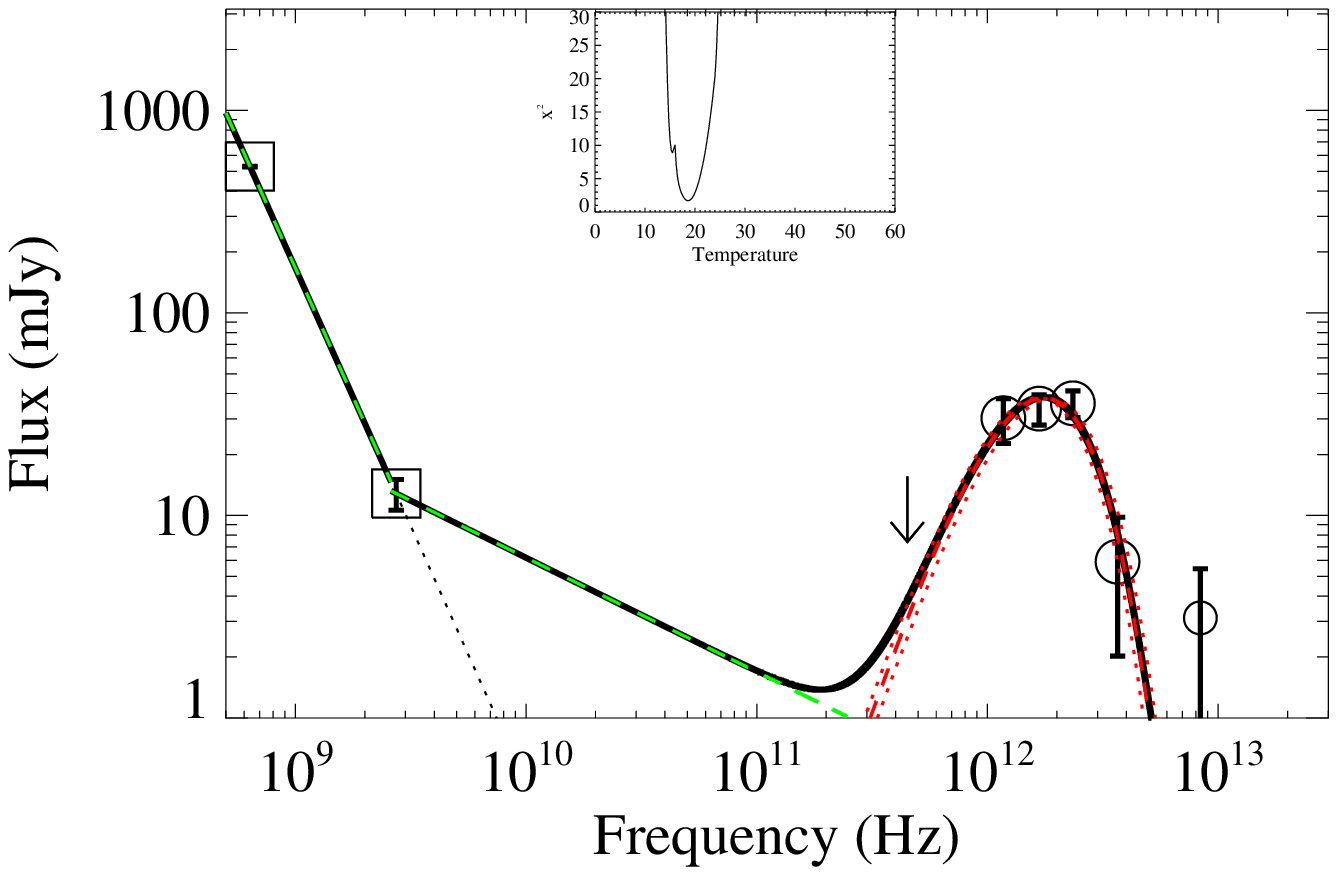}& 
\includegraphics[trim=46 1 27 25, clip,scale=0.3]{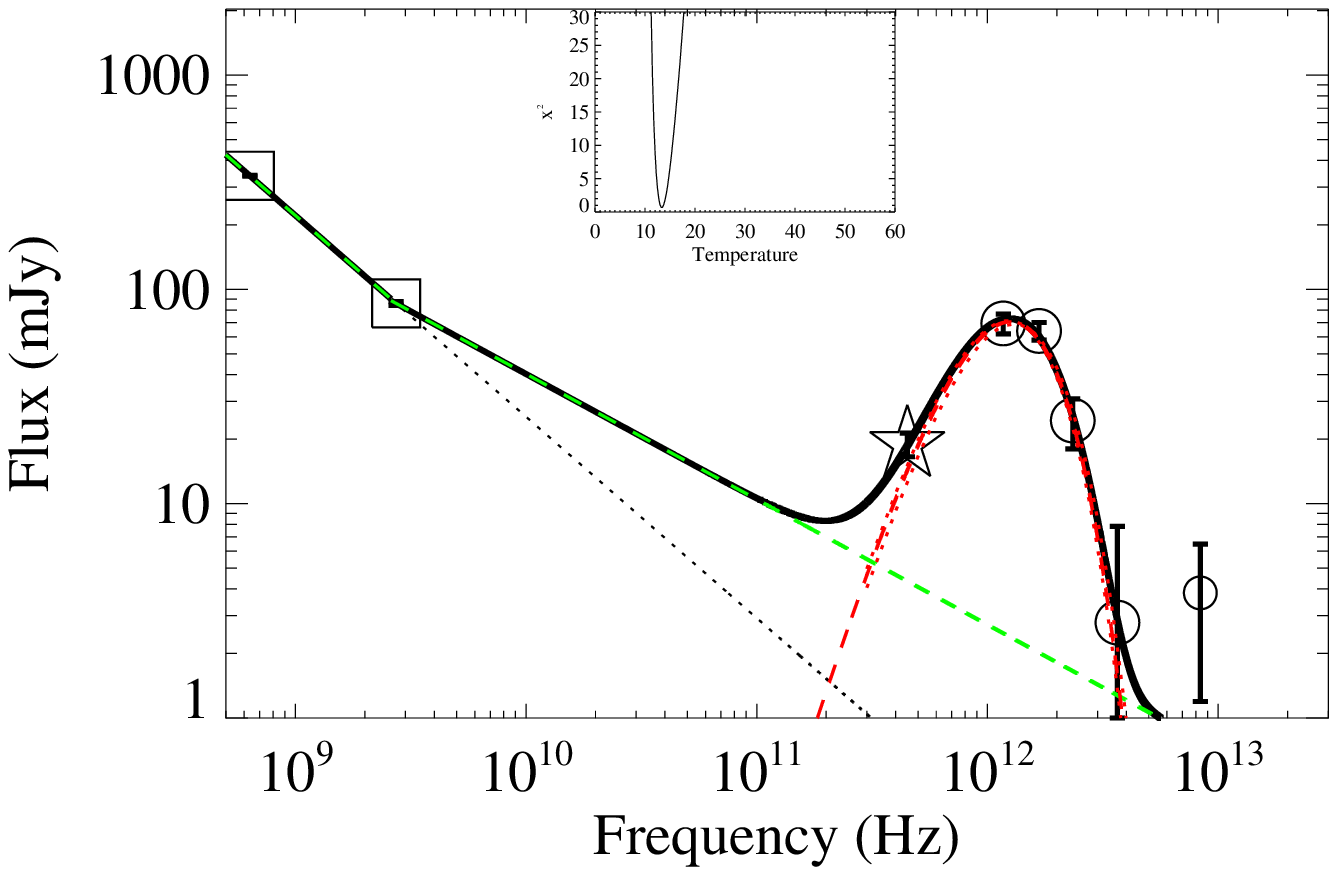} \\ %&

\end{tabular}
\caption*{A1. continued}
\label{fig:SED_appendix}
\end{figure*}

\bsp

\label{lastpage}

\end{document}